\documentclass[10pt,journal,compsoc]{IEEEtran}
\usepackage{amsthm}
\usepackage{cite}
\usepackage{graphicx}
\usepackage{balance}  
\usepackage{color}
\usepackage{array}
\usepackage[table]{xcolor}
\usepackage{algorithmicx}
\usepackage{algpseudocode}
\usepackage{gensymb}
\usepackage{amsmath,amssymb,amsfonts,bm}
\usepackage{makecell}
\usepackage{subfigure}
\usepackage{multirow}
\usepackage[ruled,vlined]{algorithm2e}
\usepackage{booktabs}
\usepackage{hyperref}
\usepackage{threeparttable}
\usepackage{flushend}
\usepackage{upgreek}
\usepackage{array}
\usepackage{colortbl}
\usepackage{url}

\newtheorem{definition}{Definition}

\newcommand{\up}{\vspace*{-0.1in}}

\begin{document}
\title{\huge{Spatio-Temporal Trajectory Similarity Measures: A Comprehensive Survey and Quantitative Study}}
%

\author{Danlei Hu,
        Lu Chen,
        Hanxi Fang,
        Ziquan Fang,
        Tianyi Li,
        Yunjun Gao,~\IEEEmembership{Member,~IEEE}

\IEEEcompsocitemizethanks{
        \IEEEcompsocthanksitem D. Hu, L. Chen (Corresponding Author), Z. Fang and Y. Gao are with the College of Computer Science, Zhejiang University, Hangzhou 310027, China, E-mail:\{dlhu, luchen, zqfang, gaoyj\}@zju.edu.cn.
        \IEEEcompsocthanksitem H. Fang is with the School of Earth Science, Zhejiang University, Hangzhou 310027, China, E-mail:hanxif@zju.edu.cn. \IEEEcompsocthanksitem T. Li is with the Department of Computer Science, Aalborg University, Denmark, E-mail:tianyi@cs.aau.dk.
}

}

\IEEEtitleabstractindextext{
\begin{abstract}
Spatio-temporal trajectory analytics is at the core of smart mobility solutions, which offers unprecedented information for diversified applications such as urban planning, infrastructure development, and vehicular networks. Trajectory similarity measure, which aims to evaluate the distance between two trajectories, is a fundamental functionality of trajectory analytics. In this paper, we propose a comprehensive survey that investigates all the most common and representative spatio-temporal trajectory measures. First, we provide an overview of spatio-temporal trajectory measures in terms of three hierarchical perspectives: Non-learning vs. Learning, Free Space vs. Road Network, and Standalone vs. Distributed. Next, we present an evaluation benchmark by designing five real-world transformation scenarios. Based on this benchmark, extensive experiments are conducted to study the effectiveness, robustness, efficiency, and scalability of each measure, which offers guidelines for trajectory measure selection among multiple techniques and applications such as trajectory data mining, deep learning, and distributed processing.
\end{abstract}

}

\maketitle

\IEEEdisplaynontitleabstractindextext

\IEEEpeerreviewmaketitle

\IEEEraisesectionheading{\section{Introduction}\label{sec:intro}}

\IEEEPARstart{W}{ith} the proliferation of GPS-equipped devices and mobile computing services, massive spatio-temporal trajectory data of moving objects such as people, vessels, and vehicles are being captured~\cite{15overview,16reviewapplication}. For example, people share visited places (e.g., POIs) on social networks by their smartphones to generate ``check-in'' trajectories; according to the AIS project~\cite{AISProject}, millions of vessels connected with the AIS services continuously report locations to ensure the sailing safety; 
and the world's largest ridesharing company Uber collects up to 17 million vehicle trips daily~\cite{Uber}.  

\begin{figure}[tbp]
 	\centering
 	\includegraphics[width=0.48\textwidth]{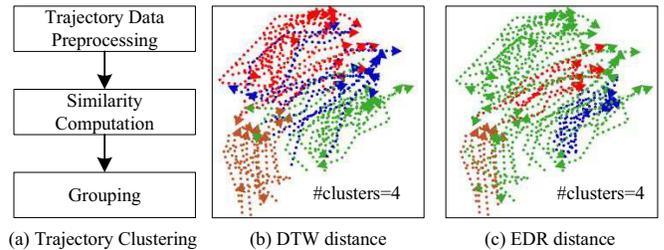}
 	\vspace{-2mm}
 	\caption{Example of Trajectory Clustering}
 	\vspace{-6mm}
 	\label{fig:clustering}
\end{figure}

Trajectory data with its analytics benefit a broad range of real-life applications across different fields such as urban computing~\cite{13trafficplanning}, transportation~\cite{08zhengyuTransportation}, behavior study~\cite{11mininganimal}, and public security~\cite{21Compression}, to name but a few. A fundamental functionality of most trajectory analysis is to evaluate the relationship/distance between two trajectories, i.e., trajectory similarity measurement. With an accurate and efficient trajectory similarity measure, downstream trajectory analytics involving retrieval~\cite{Retrieval1,Retrieval2}, clustering~\cite{07traclus,Cluster1}, classification~\cite{19LSTM,Classification2}, and mobility pattern mining~\cite{pattern1,pattern2,deepmining} tasks can be well-supported to serve upper applications. For instance, Xie et al.~\cite{DFT} propose two distance measures to support similarity queries and joins in a large-scale trajectory dataset. Wang et al.~\cite{LORS} design a road network oriented distance measure for vehicle trajectory similarity computation, based on which, $k$-means clustering is studied~\cite{WangVLDB} to detect hot/popular traveling paths in a city. In both cases, the trajectory similarity measure plays a fundamental role, and a different measure selection may result in totally different query results and clustering quality. Take the trajectory clustering as an example, clustering aims to group similar trajectories into clusters, where similarity computation is a fundamental task of clustering as shown in Figure~\ref{fig:clustering}(a); while different similarity measures may affect the clustering quality. As shown in Figure~\ref{fig:clustering}, although the numbers of clusters are the same (i.e., $\#$clusters=4), the clustering results are quite different when applying different measures (i.e., DTW and EDR). Hence, the selection of similarity measure is important. In addition, more trajectory similarity based analysis tasks can refer to~\cite{WangSurvey,16reviewapplication}.


\begin{table*}[]\tiny
\hspace{-8mm}
\vspace{-6mm}
\caption{Spatio-Temporal Trajectory Similarity Measures. (Here, $m$ and $n$ represent the lengths of two trajectories, $d$ denotes the dimension of embedded vector, and the dark backdrop indicates the studied scope in the state-of-the-art survey \cite{VLDBJSurvey})}
\vspace{-4mm}
\small
\hspace{-1mm}
\begin{tabular}{|ccc|c|c|c|c|c|c|c|}
\hline
\multicolumn{3}{|c|}{\textbf{Category}}                                                                                                                                                                                                                                  & \makebox[0.05\textwidth][c]{\textbf{Measure}}                         &\makebox[0.04\textwidth][c]{\textbf{Complexity}}                            &\makebox[0.035\textwidth][c]{\textbf{Metric}}                           & \makebox[0.05\textwidth][c]{\begin{tabular}[c]{@{}c@{}}\textbf{Unequal}\\ \textbf{Length}\end{tabular}}                                            & \makebox[0.07\textwidth][c]{\begin{tabular}[c]{@{}c@{}}\textbf{Parameter}\\ \textbf{Free}\end{tabular}} & \makebox[0.055\textwidth][c]{\begin{tabular}[c]{@{}c@{}} \textbf{Noise}\\ \textbf{Sensitive}\end{tabular}} & \textbf{\makebox[0.04\textwidth][c]{\begin{tabular}[c]{@{}c@{}}\textbf{
Repre-}\\ \textbf{sentation}\end{tabular}}}   \\ \hline
\multicolumn{1}{|c|}{}                                                                           &\multicolumn{1}{c|}{}                                                                         &                                                                               & \cellcolor{black!10} ED                & \cellcolor{black!10}$\textit{O(n)}$         & \cellcolor{black!10}\checkmark & \cellcolor{black!10}$\times$                               & \cellcolor{black!10}$\times$                               & \cellcolor{black!10}$\times$                         &\cellcolor{black!10} Point                                \\ \cline{4-10}
\multicolumn{1}{|c|}{}                                                                           & \multicolumn{1}{c|}{}                                                                         &                                                                               & \cellcolor{black!10}DTW~\cite{DTW}         & \cellcolor{black!10}$\textit{O(mn)}$        & \cellcolor{black!10}$\times$                          & \cellcolor{black!10}\checkmark        & \cellcolor{black!10}\checkmark        & \cellcolor{black!10}$\times$                          & \cellcolor{black!10}Point                                   \\ \cline{4-10}
\multicolumn{1}{|c|}{}                                                                           & \multicolumn{1}{c|}{}                                                                         &                                                                               & \cellcolor{black!10} LCSS~\cite{LCSS}     & \cellcolor{black!10}$\textit{O(mn)}$        & \cellcolor{black!10}$\times$                          & \cellcolor{black!10}\checkmark        & \cellcolor{black!10}$\times$                                 & \cellcolor{black!10}\checkmark & \cellcolor{black!10}Point                                   \\ \cline{4-10}
\multicolumn{1}{|c|}{}                                                                           & \multicolumn{1}{c|}{}                                                                         &                                                                               &\cellcolor{black!10} EDR~\cite{EDR}  & \cellcolor{black!10}$\textit{O(mn)}$        & \cellcolor{black!10}$\times$                          & \cellcolor{black!10}\checkmark        & \cellcolor{black!10}$\times$                                 & \cellcolor{black!10}\checkmark & \cellcolor{black!10}Point                                   \\ \cline{4-10}
\multicolumn{1}{|c|}{}                                                                           & \multicolumn{1}{c|}{}                                                                         &                                                                               &\cellcolor{black!10} EDwP~\cite{EDwP}         & \cellcolor{black!10}$\textit{O(mn)}$        & \cellcolor{black!10}$\times$                          & \cellcolor{black!10}\checkmark        & \cellcolor{black!10}\checkmark        & \cellcolor{black!10}\checkmark & \cellcolor{black!10}Segment                                 \\ \cline{4-10}
\multicolumn{1}{|c|}{}                                                                           & \multicolumn{1}{c|}{}                                                                         &                                                                               &\cellcolor{black!10} ERP\cite{ERP}       & \cellcolor{black!10}$\textit{O(mn)}$        & \cellcolor{black!10}\checkmark & \cellcolor{black!10}\checkmark        & \cellcolor{black!10}$\times$                                 & \cellcolor{black!10}$\times$                          & \cellcolor{black!10}Point                                   \\ \cline{4-10}
\multicolumn{1}{|c|}{}                                                                           & \multicolumn{1}{c|}{}                                                                         &                                                                               &\cellcolor{black!10}Hausdorff~\cite{Hausdorff}    & \cellcolor{black!10}$\textit{O(mn)}$        &\cellcolor{black!10} \checkmark &\cellcolor{black!10} \checkmark        &\cellcolor{black!10} \checkmark        &\cellcolor{black!10} $\times$                          &\cellcolor{black!10} Point                                   \\ \cline{4-10}
\multicolumn{1}{|c|}{}                                                                           & \multicolumn{1}{c|}{}                                                                         &                                                                               & \cellcolor{black!10}Frechet~\cite{Frechet}       & \cellcolor{black!10}$\textit{O(mn)}$        &\cellcolor{black!10} $\times$                          &\cellcolor{black!10} \checkmark        &\cellcolor{black!10} \checkmark        &\cellcolor{black!10} $\times$                          &\cellcolor{black!10} Point                                   \\ \cline{4-10}
\multicolumn{1}{|c|}{}                                                                           & \multicolumn{1}{c|}{}                                                                         &                                                                               &\cellcolor{black!10} LIP~\cite{LIP}       & \cellcolor{black!10}\begin{footnotesize}$\textit{O((m+n)log(m+n))}$\end{footnotesize} &\cellcolor{black!10} $\times$                          &\cellcolor{black!10} \checkmark        &\cellcolor{black!10} \checkmark        &\cellcolor{black!10} $\times$                          &\cellcolor{black!10} Segment                                   \\ \cline{4-10}
\multicolumn{1}{|c|}{}                                                                           & \multicolumn{1}{c|}{}                                                                         &                                                                               & \cellcolor{black!10} OWD~\cite{OWD}          & \cellcolor{black!10}$\textit{O(mn)}$        & \cellcolor{black!10}$\times$                          &\cellcolor{black!10} \checkmark        &\cellcolor{black!10} $\times$                                 & \cellcolor{black!10}$\times$                          &\cellcolor{black!10}Segment                                   \\ \cline{4-10}
\multicolumn{1}{|c|}{}                                                                           & \multicolumn{1}{c|}{}                                                                         & \multirow{-10}{*}{Standalone} & Seg-Frechet~\cite{DFT}                        & $\textit{O(mn)}$                                &$\times$                                                   & \checkmark                                & \checkmark                                &$\times$                                                   & Segment                                                         \\ \cline{3-10}
\multicolumn{1}{|c|}{}                                                                           & \multicolumn{1}{c|}{}                                                                         &                                                                               & DFT~\cite{DFT}                                                        & $\textit{O(mn)}$                                &$\times$                                                   & \checkmark                                & \checkmark                                &$\times$                                                   & Segment                                                         \\ \cline{4-10}
\multicolumn{1}{|c|}{}                                                                           & \multicolumn{1}{c|}{}                                                                         &                                                                               & DITA~\cite{18DITA}                                         & $\textit{O(mn)}$                                &$\times$                                                   & \checkmark                                & \checkmark                                &$\times$                                                   & Point                                                           \\ \cline{4-10}
\multicolumn{1}{|c|}{}                                                                           & \multicolumn{1}{c|}{\multirow{-14}{*}{\begin{tabular}[c]{@{}c@{}}Free\\ Space\end{tabular}}}  & \multirow{-3}{*}{Distributed}                                                 & REPOSE~\cite{REPOSE}                            & $\textit{O(mn)}$                  &$\times$                    & \checkmark                                & \checkmark                                &$\times$                                                   & Point                                                           \\ \cline{2-10}
\multicolumn{1}{|c|}{}                                                                           & \multicolumn{1}{c|}{}                                                                         &                                                                               & NetDTW~\cite{NetDTW}                                     & $\textit{O(mn)}$                                &$\times$                                                   & \checkmark                                & \checkmark                                &$\times$                                                   & Point                                                           \\ \cline{4-10}
\multicolumn{1}{|c|}{}                                                                           & \multicolumn{1}{c|}{}                                                                         &                                                                               & NetLCSS~\cite{ST2Vec}                                          & $\textit{O(mn)}$                                &$\times$                                                   & \checkmark                                &$\times$                                                          & \checkmark                         & Point                                                           \\ \cline{4-10}
\multicolumn{1}{|c|}{}                                                                           & \multicolumn{1}{c|}{}                                                                         &                                                                               & LORS~\cite{LORS}                                           & $\textit{O(mn)}$                                &$\times$                                                   & \checkmark                                & \checkmark                                &$\times$                                                   & Segment                                                         \\ \cline{4-10}
\multicolumn{1}{|c|}{}                                                                           & \multicolumn{1}{c|}{}                                                                         &                                                                               & TP~\cite{TP}                                              & $\textit{O(mn)}$                                &$\times$                                                   & \checkmark                                &$\times$                                                          &$\times$                                                   & Point                                                           \\ \cline{4-10}
\multicolumn{1}{|c|}{}                                                                           & \multicolumn{1}{c|}{}                                                                         &                                                                               & NetERP~\cite{NetERP}                                          & $\textit{O(mn)}$                                & \checkmark                         & \checkmark                                &$\times$                                                          &$\times$                                                   & Point                                                           \\ \cline{4-10}
\multicolumn{1}{|c|}{}                                                                           & \multicolumn{1}{c|}{}                                                                         &                                                                               & NetEDR~\cite{NetERP}                                          & $\textit{O(mn)}$                                &$\times$                                                   & \checkmark                                &$\times$                                                          & \checkmark                         & Point                                                           \\ \cline{4-10}
\multicolumn{1}{|c|}{}                                                                           & \multicolumn{1}{c|}{}                                                                         & \multirow{-7}{*}{Standalone}  & LCRS~\cite{LCRS}                                 & $\textit{O(mn)}$                                &$\times$                                                   & \checkmark                                & \checkmark                                &$\times$                                                   & Segment                                                         \\ \cline{3-10}
\multicolumn{1}{|c|}{\multirow{-22}{*}{\begin{tabular}[c]{@{}c@{}}Non-\\ learning\end{tabular}}} & \multicolumn{1}{c|}{\multirow{-8}{*}{\begin{tabular}[c]{@{}c@{}}Road\\ Network\end{tabular}}} & Distributed                                                                   & DISON~\cite{LCRS}                                               & $\textit{O(mn)}$                                &$\times$                                                   & \checkmark                                &$\times$                                                          &$\times$                                                   & Segment                                                         \\ \hline
\multicolumn{1}{|c|}{}                                                                           & \multicolumn{1}{c|}{}                                                                         &                                                                               & NEUTRAJ~\cite{19NEUTRAJ}                                    & $\textit{O(m+n)}$                               & /                                                 & \checkmark                                &$\times$                                                          & \checkmark                         & Vector                                                          \\ \cline{4-10}
\multicolumn{1}{|c|}{}                                                                           & \multicolumn{1}{c|}{\multirow{-2}{*}{\begin{tabular}[c]{@{}c@{}}Free\\ Space\end{tabular}}}   & \multirow{-2}{*}{Standalone}  & Traj2SimVec~\cite{20Traj2SimVec}                          & $\textit{O(m+n)}$                               & /                                                 & \checkmark                                &$\times$                                                          & \checkmark                         & Vector                                                          \\ \cline{2-10}
\multicolumn{1}{|c|}{}                                                                           & \multicolumn{1}{c|}{}                                                                         &                                                                               & GTS~\cite{21GTS}                         & $\textit{O(d)}$                                 & /                                                 & \checkmark                                &$\times$                                                          &$\times$                                                   & Vector                                                          \\ \cline{4-10}
\multicolumn{1}{|c|}{\multirow{-4}{*}{Learning}}                                                 & \multicolumn{1}{c|}{\multirow{-2}{*}{\begin{tabular}[c]{@{}c@{}}Road\\ Network\end{tabular}}} & \multirow{-2}{*}{Standalone}                                                 & ST2Vec~\cite{ST2Vec}                              & $\textit{O(d)}$                                 & /                                                 & \checkmark                                &$\times$                                                          &$\times$                                                   & Vector                                                          \\ \hline
\end{tabular}
\label{tab:measures}
\vspace{-4mm}
\end{table*}

Unlike isolated spatial points or one-dimensional time series where the distance definition is straightforward, it is non-trivial to define the distance between continuous two-dimensional trajectories. It also needs to consider following four trajectory characteristics: (i) Different data sources, i.e., free space vs. road network space. In the latter case, a proper trajectory measure should take the road topology into account, as people and vehicles cannot travel like vessels without spatial constraints~\cite{TP}. (ii) Various sampling rates and lengths. Unlike time series that generally feature constant and high sampling rates~\cite{18t2vec}, spatio-temporal trajectory data is usually generated via varying samplings, resulting in variable lengths. (iii) The effect of noise. The noise points commonly exist, especially due to strength attenuation and interference in urban cities~\cite{20TKDE}. (iv) Complex shapes. Compared to private-car trajectories that are usually inaccessible caused by privacy principles, taxi trajectories are widely studied in the community~\cite{VehicleSurvey,Surveycrowd,SurveyMining}. However, taxi trajectories exhibit much more diverse, complex, and flexible geometric, because of various pick-pop demands. To deal with the above spatio-temporal characteristics, tremendous amounts of research efforts are devoted to designing dozens of spatio-temporal trajectory similarity measures.

Being faced with a huge amount of trajectory measures, researchers are often too exhausted to select a proper one. On the one hand, there are too many trajectory measures, which were proposed under different scenarios, e.g., learning based or non-learning based, free-space oriented or road network oriented, as well as standalone or distributed processing. In each scenario, various measures exist. Consequently, users need to spend tons of time and efforts to explore the specific details of each measure and the relations/differences among them. On the other hand, the evaluation on various trajectory measures is still not well organized. For instance, some measures only focus on efficiency, while others may put more attention on effectiveness and robustness. To address the problems mentioned above, a comprehensive survey, benchmark, and evaluation will be a great help for researchers involved in this important topic. Specifically, considering three-dimensional aspects, we classify the existing representative spatio-temporal trajectory measures proposed from 1995 to 2022 year in a hierarchical way, i.e., \textbf{Non-learning} vs. \textbf{Learning} (first hierarchy), \textbf{Free Space} vs. \textbf{Road Network} (second hierarchy), and \textbf{Standalone} vs. \textbf{Distributed} (third hierarchy). Table~\ref{tab:measures} summarizes trajectory similarity measures.

Although previous systematic surveys have made some efforts, they mostly focus on non-learning, free-space, or standalone based measures and thus significantly narrow the studied scope of trajectory similarity community. For example, Gudmundsson and Toohey et al.~\cite{Gudmundsson} only review four basic trajectory distance measures including Euclidean distance (ED), DTW~\cite{DTW}, LCSS~\cite{LCSS}, and Fréchet~\cite{Frechet}. Based on it, Sousa et al.~\cite{VehicleSurvey} append partial vehicle trajectory similarity measures, e.g., LORS~\cite{LORS}, LCRS~\cite{LCRS}, and NetEDR~\cite{NetERP}. Note that, experimental evaluation of these measures are not studied in the surveys. Su et al.~\cite{VLDBJSurvey} conduct the state-of-the-art survey and experimental evaluation for 10 trajectory similarity measures under the non-learning, free-space, and standalone contexts. In contrast, we conduct a much more systematic review and evaluation of all the most common and representative spatio-temporal trajectory measures in a three-dimensional hierarchical way as depicted in Table~\ref{tab:measures}. We conduct a three-dimensional survey due to three following motivations:

(i) As deep learning has made great success in AI community, many studies~\cite{18t2vec, 19NEUTRAJ, 20Traj2SimVec, 21GTS, ST2Vec} start leveraging the powerful approximation capabilities of neural networks to attempt to replace the traditional handcrafted trajectory measures with learning-based models. Besides, increasing efforts have been devoted to embracing learning-based techniques as an integral part of trajectory data management and analytics, such as deep clustering~\cite{DeepCluster}, deep mobility pattern mining~\cite{Co-Movement}, and deep path recommendation~\cite{TimeTravel}, to name just a few. As such, we believe that we have reached an imperative point to systematically study the contribution and explore the relations, differences, and pros/cons among the emerging learning-based and classic non-learning based trajectory measures.


(ii) In the early stage, most of the trajectory measures~\cite{VLDBJSurvey} are proposed for objects that move freely in the Euclidean space, e.g., bird or vessel trajectories. Recently, the proliferation of vehicle navigation systems and location based services (LBSs) enable the massive collection of vehicle and people trajectories in road networks. In that case, the free-space oriented trajectory measures cannot reflect the true distance between moving objects in a moving-constrained road network. In view of this, many network-aware trajectory measures~\cite{LCRS,LORS,TP,NetERP} are designed. To the best of our knowledge, there is no previous work that has given a systematic review and experimental evaluation of those network-based trajectory distance measures.

(iii) Since the storage capacity and processing ability of a single machine can no longer support a large scale of trajectory data, another popular line of trajectory similarity study is designing efficient and scalable frameworks upon distributed processing platforms (e.g., Spark) for large-scale trajectory similarity analytics. Towards this, several system-level frameworks are also developed. Hence, we are inspired to present a sufficient review and performance evaluation of distributed-based similarity computation studies.

Motivated by the observations above, we conduct a most unprecedented survey on all the most representative spatio-temporal trajectory similarity measures proposed in the literature. Specifically, we review 25 similarity measures from the following hierarchical perspectives, namely: (i) Model architecture, i.e., non-learning based vs. learning-based; (ii) Space context, i.e., free space oriented vs. road network oriented; and (iii) Computational mechanism, i.e., standalone machine vs. distributed platform. To the best of our knowledge, none of the previous studies have reached such a wide scope of trajectory measure evaluation. It is also significant to conduct objective and sufficient evaluation to study each measure by using the same tasks, datasets, and experimental settings. In light of this, we provide a standard benchmark\footnote{publicly available at https://github.com/ZJU-DAILY/TSM} in a real setting by introducing five different scenarios, i.e., length shift, shape shift, noise shift, sampling shift, and cardinality shift, on four real datasets. Then, we extensively explore and compare the effectiveness, robustness, efficiency, and scalability for all measures via quantitative and qualitative analysis. With our benchmark, (i) users can quickly grasp the technical details about trajectory similarity measures; and (ii) users can easily choose or design suitable measures and utilize them as the baselines. Overall, this paper makes the contributions below.

\begin{itemize}\setlength{\itemsep}{-\itemsep}
    \item We conduct a concise but concrete review to evaluate and compare spatio-temporal trajectory similarity measures qualitatively and quantitatively in three dimensions, i.e., learning-based or not, road network oriented or not, and distributed-based or not, providing a reference for measure selection among traditional similarity measures, deep learning models, and distributed processing technologies.
    \item We provide a standard evaluation benchmark with five types of trajectory transformations under five typical trajectory analytics scenarios. Based on this, we conduct a comprehensive evaluation of effectiveness, robustness, efficiency, and scalability performance of 25 representative measures.
    \item We have several key observations according to experimental results, based on which, we offer insights about trajectory measure selection in specific application scenarios. According to several issues that remained to be solved, we also present detailed potential future directions.
\end{itemize}

Section~\ref{sec:problem} gives the preliminaries. Sections~\ref{sec:traditional} and~\ref{sec:learning} present the non-learning-based and learning-based trajectory similarity measures respectively, where free-space vs. network oriented and standalone vs. distributed measures are further detailed. Section~\ref{sec:benchmark} provides the benchmark. The evaluation results with insights are reported in Section~\ref{sec:exe}. Section~\ref{sec:futurework} puts forward interesting future directions. Finally, Section~\ref{sec:conclusion} concludes the paper.

\section{Preliminaries}
\label{sec:problem}
In this section, we present the definitions related to trajectory similarity measure.

\begin{definition}\label{defn:trajectory}
    {\bf (Trajectory)} \textit{A trajectory $T$ is an ordered sequence that consists of GPS sampling points, i.e., $T=\langle p_1, p_2, ..., p_n\rangle$, where $p_i \left( 1 \leq i \leq n\right)$ is an observed GPS location in d-dimensional tuple to describe the mobility of $T$.}
\end{definition}


Here, $n$ denotes the length of a trajectory. We use $T_i$ to denote a trajectory whose id is $i$, and use ${T_i^j}$ to denote the $j$-th point in $T_i$. The points observed by GPS equipments could contain various information.
For simplicity, we assume each sampling point is two-dimension or three-dimension (i.e., $(\textit{latitude, longitude})$ or $(\textit{latitude, longitude, timestamp})$). Although trajectory data is similar to time series data, time series data typically lacks of spatial information. In particular, for vehicle trajectories, they are physically constrained to the road network. Thus, directly applying similarity measures for time series in free space is not always suitable. The trajectories with road network constraints are taken into consideration. The road network is defined as follows.

\begin{definition}\label{defn:roadnetwork}
    {\bf (Road Network)} \textit{A road network is represented as a directed graph $G=(V, E)$, where $V$ is a set of road intersections (i.e., road vertices) in the road network, and $E$ is a set of edges with direction of road segments.}
\end{definition}

Here, $v_i = (lat_i, lon_i) \in V$ is a road intersection, where $i$ is the id of $v_i$, while $(lat_i, lon_i)$ denotes the latitude and longitude of $v_i$. Meanwhile, an edge $e_{i,j} \in E$ represents a directed road segment from $v_i$ to $v_j$. Existing map matching methods~\cite{mapmatching, mapmatching2} are proposed to transform a trajectory of GPS points into a road network constrained trajectory. A trajectory in a road network can be represented as $T = \left \langle v_1, v_2, ..., v_c \right \rangle (c \leq n)$ or $T = \left \langle e_1, e_2, ..., e_{c-1} \right \rangle (c \leq n)$, where $c$ is the number of road vertices in $T$.


\begin{definition}\label{defn:distance}
    {\bf (Similarity Measure)} \textit{A similarity measure is a function $f(T_i,T_j)$ to measure the distance between $T_i$ and $T_j$.}
\end{definition}

\begin{definition}\label{defn:metric}
    {\bf (Metric Measure)} \textit{Given a similarity measure f and any three trajectories $T_i$, $T_j$, and $T_a$, we call f as a metric measure if f satisfies the following conditions: (i) Uniqueness: $f(T_i, T_j)=0$ $\leftrightarrow T_i=T_j$;
   (ii) Nonnegativity: $f(T_i, T_j) \geq 0$; (iii) Triangle Inequality: $f(T_i, T_a)\leq f(T_i, T_j)+f(T_j, T_a)$;
        and (iv) Symmetry: $f(T_i, T_j) = f(T_j, T_i)$.}
\end{definition}

As shown in Table~\ref{tab:measures}, a similarity measure can be classified according to i) whether it is metric or not as defined in Definition~\ref{defn:metric} (``$\checkmark$'' denotes it is a metric, ``$\times$'' denotes it is not a metric, ``$/$'' denotes learning methods do not use a manual function to compute similarity), ii) whether it can support trajectories with different lengths,  iii) whether it is parameter-free, and iv) whether it is noise sensitive. 
Since trajectory similarity query is a prerequisite task in upper-level applications~\cite{DFT,Query1}, we perform Top-$k$ trajectory similarity query to evaluate the capabilities of all similarity measures in Section~\ref{sec:exe}, which is defined below.

\begin{definition}\label{defn:search}
    {\bf (Top-$k$ Similarity Query)} \textit{Given a query trajectory $QT$, a set of trajectories $\mathcal{T}=\{T_1, T_2, ..., T_{|\mathcal{T}|}\}$, and a similarity measure $f(\cdot)$, the Top-$k$ similarity query returns $k$ trajectories in $\mathcal{T}$ that are most similar to $QT$ $(\mathcal{S} = \{ \mathcal{S} \subseteq T \wedge |\mathcal{S}| = k \wedge \forall T_s \in \mathcal{S}, \forall T_o \in \mathcal{T}-\mathcal{S} ( f(QT,T_s) \text{ } |= f(QT, T_o))\})$. Here, ``$|=$'' means ``$\geq$'' for LCSS while  denotes ``$\leq$'' for others.}
\end{definition}

\section{Non-learning based Measures}
\label{sec:traditional}


In this section, we introduce 22 non-learning based measures (as shown in the upper part of Table~\ref{tab:measures}), and further detail each measure in terms of space context (i.e., free space and road network).  

\begin{figure}[tbp]
 	\centering
 	\includegraphics[width=0.48\textwidth]{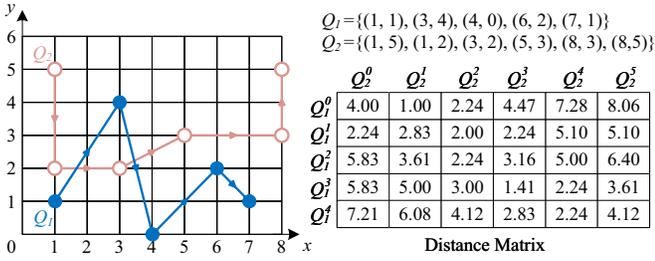}
 	\vspace{-2mm}
 	\caption{Running Example of Trajectory Similarity Computation}
 	\vspace{-6mm}
 	\label{fig:example}
\end{figure}

\subsection{Trajectory Similarity in Free Space}
We review 14 free space oriented measures. Each measure is point-based or segment-based. Here, we provide an running example in Figure~\ref{fig:example}, in order to illustrate how to compute the similarity between two trajectories $Q_1$ and $Q_2$ using different measures, where the distance matrix is used to show the distances between points of two trajectories.


\subsubsection{Point-based Measures} Point-based measures are widely utilized to compute the similarity among trajectories consisting of sampling points. 

{\bf Euclidean Distance (ED).} Euclidean distance is a well-known measure, which is used to calculate the distance between time series with the same length~\cite{PDTW}.
When applied to trajectory data, sampling points in two series are aligned in order~\cite{EDTraj}. Formally, ED between two trajectories $T_1$ and $T_2$ is defined as:

 \begin{small}
 \begin{align}
     dis(p_1, p_2)=&\sqrt{(p_1.lat-p_2.lat)^2+(p_1.lon-p_2.lon)^2} \\
     &\operatorname{ED}(T_1, T_2)=\frac{\sum^N_{i=1} dis(T_1^i, T_2^i)}{n}
 \end{align}
 \end{small}Here, $dis(p_1, p_2)$ denotes the distance between two points $p_1$ and $p_2$, $lat$ and $lon$ denote the latitude and longitude of the point, while $n$ denotes the length of $T_1$ and $T_2$.
ED is a metric and parameter-free measure with the time complexity of $\textit{O(n)}$. However, it assumes that trajectories are with the same length, which is not realistic in real life.
Thus, we do not focus on evaluating ED. Note that, in Figure~\ref{fig:example}, $Q_1$ and $Q_2$ are with different lengths, and thus, ED cannot be applied to compute the similarity in this running example.

\begin{figure*}[tbp]
 	\centering
 	\includegraphics[width=0.98\textwidth]{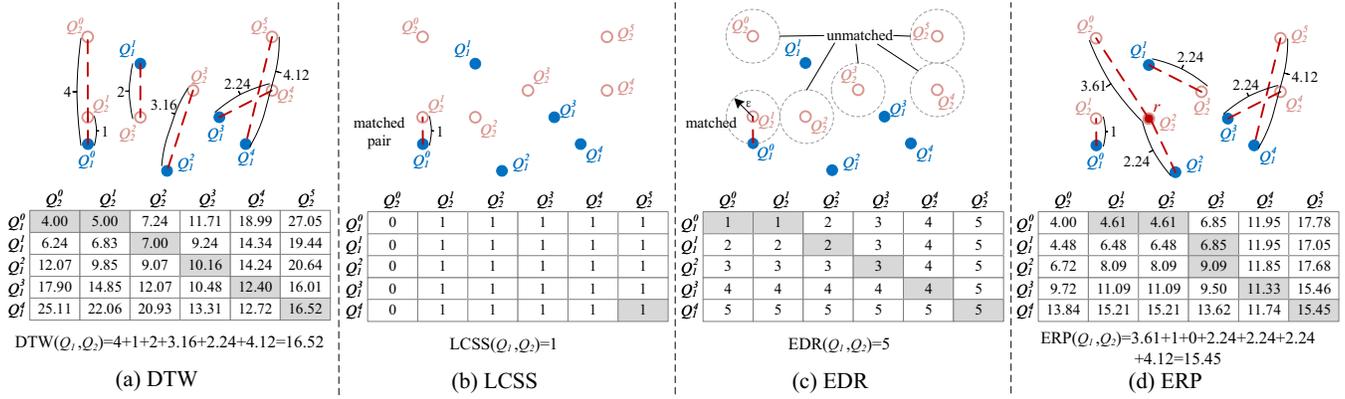}
 	\vspace{-2mm}
 	\caption{Running Example in Free Space vs. DTW, LCSS, EDR, and ERP}
 	\vspace{-6mm}
 	\label{fig:Non-learning}
\end{figure*}

{\bf Dynamic Time Warping Distance (DTW).} Similar to ED, DTW is also a measure designed for time series and can be applied to trajectory data~\cite{DTWSeries1,DTWSeris2}. However, different from ED, DTW can align a point of one trajectory to one or more consecutive points of another trajectory. 
The time complexity of DTW is $\textit{O(mn)}$ where $m$ and $n$ denote the length of two comparing trajectories respectively. Although DTW is non-metric, it is most commonly used for trajectory similarity computation. Various techniques are proposed to improve the efficiency of DTW~\cite{PDTW, FastDTW}. DTW distance between two trajectories $T_1$ and $T_2$ is defined as:

\begin{footnotesize}
\vspace{-4mm}
\hspace{-2mm}
\begin{equation}
\operatorname{DTW}(T_1,\! T_2)\!=\!
\begin{cases} 0 & \text {if} \hspace{1mm} m=n=0 \\
\infty & \text {if} \hspace{1mm} {m=0}\\& \text{or $n = 0$} \\
{dis}\left(T_1^{m}, T_2^{n}\right)+\\\min\begin{cases}\! \operatorname{DTW}\left(Head(T_1), T_2\right),\\
\operatorname{DTW}\left(T_1, Head\left(T_2\right)\right),\\
\operatorname{DTW}\left(Head(T_1), Head(T_2)\right)\end{cases} & \text{otherwise}\end{cases}
\end{equation}
\end{footnotesize}where $Head(\cdot)$ is to get the head sampling points except the last one point (i.e., $Head(T)=\langle p_1,p_2,...,p_{n-1}\rangle$). Take $Q_1$ and $Q_2$ in Figure~\ref{fig:example} as an example, we are able to compute the DTW distance matrix in Figure~\ref{fig:Non-learning}(a) and obtain DTW($Q_1$, $Q_2$) = 16.52.

{\bf Longest Common Subsequence (LCSS).} LCSS distance~\cite{LCSS,LCSS1} between two trajectories $T_1$ and $T_2$ is defined as the size of the longest common subsequence of $T_1$ and $T_2$. Given a threshold parameter $\epsilon$ to determine whether a pair of points $(p_1, p_2)$ matches, if $dis(p_1, p_2)\leq \epsilon$, $(p_1, p_2)$ is a match pair. As shown in Figure~\ref{fig:Non-learning}(b),  when $\epsilon=1.0$, the pair $(Q_1^0, Q_2^1)$ is a match pair, as $dis(Q_1^0, Q_2^1)\leq \epsilon$.
Actually, LCSS distance finds all match pairs in two trajectories, and it is non-metric but noise sensitive compared to ED. Moreover, its time complexity is $\textit{O(mn)}$. Formally, LCSS distance is defined as:

\begin{footnotesize}
\vspace{-4mm}
\begin{equation}
\operatorname{LCSS}_\epsilon(T_1, T_2)= \begin{cases}
0 &\hspace{-2mm} \text { if } m=0\quad \\ & or \quad n = 0 \\
1+LCSS_\epsilon(Head(T_m), Head(T_n)) & \hspace{-2mm}\text { if } dis(T_1^m, T_2^n)\leq \epsilon \\
\max\begin{cases}
\operatorname{LCSS}_\epsilon\left(Head(T_1), T_2\right),\\ \operatorname{LCSS}_\epsilon\left(T_1, Head(T_2)\right)\end{cases} &\hspace{-2mm} \text { otherwise }\end{cases}
\end{equation}
\end{footnotesize}As shown in Figure~\ref{fig:Non-learning}(b), when $\epsilon$ = 1.0, there is only one matched pair (i.e., ($Q_1^0, Q_2^1$)). Thus, LCSS($Q_1$, $Q_2$) = 1.




{\bf Edit Distance on Real Sequence (EDR).} Similar to LCSS, EDR distance~\cite{EDR,EDR1,EDR2,07traclus} is also one of the edit distance-based measures with the time complexity of $\textit{O(mn)}$, where edit distance equals to the number of necessary operations (i.e., insertion, deletion and replacement) when transforming one trajectory into another one. However, different from LCSS, EDR sets a parameter  $cost$ to measure the cost of an unmatched pair of points. Formally, EDR is defined as:

\begin{footnotesize}
\vspace{-4mm}
\begin{equation}
\operatorname{EDR}(T_1,\! T_2)= \begin{cases}
n &\hspace{-12mm} \text { if } m=0 \\
m &\hspace{-12mm} \text { if } n=0 \\
\min\begin{cases}
\operatorname{EDR}\left(Head(T_1), T_2\right)+1,\\
\operatorname{EDR}\left(T_1, Head(T_2)\right)+1,\\
\operatorname{EDR}\left(Head(T_1), Head(T_2)\right)+cost\end{cases} & \hspace{-12mm} \text{otherwise}
\end{cases}
\end{equation}
\end{footnotesize}Here, if $dis(T_1^m, T_2^n)\leq\epsilon$, then $cost=0$; otherwise $cost=1$. As depicted in Figure~\ref{fig:Non-learning}(c), if $\epsilon$ = 1.0, there are 1 matched point and 5 unmatched point in $Q_2$, denoting that 5 necessary operations to transform $Q_1$ into $Q_2$ are needed. Thus, EDR($Q_1$, $Q_2$) = 5.

{\bf Edit Distance with Real Penalty (ERP).} The above edit distance-based measures are basically non-metric. However, ERP~\cite{ERP,ERP1,ERP2} is a metric measure that can be used for indexing and pruning. Different from other edit distance-based measures such as LCSS and EDR, ERP does not require a threshold parameter, but sets a reference point (i.e., gap point) for measuring. ERP is commonly used in trajectory similarity computation with the time complexity of $\textit{O(mn)}$ because of its metricity. Formally, ERP distance between two trajectories $T_1$ and $T_2$ is defined as:

\begin{footnotesize}
\vspace{-2mm}
\begin{equation}
\operatorname{ERP}(T_1, T_2)= \begin{cases}
\sum_{i=1}^n dis\left(T_2^i, r\right) &\hspace{-15mm} \text { if } {m=0} \\
\sum_{i=1}^m dis\left(T_1^i, r\right) &\hspace{-15mm} \text { if } {n=0} \\
\min\begin{cases}
\operatorname{ERP}\left(Head(T_1), Head(T_2)\right)+dis\left(T_1^m, T_2^n\right),\\
\operatorname{ERP}\left(T_1, Head(T_2)\right)+dis\left(T_2^n, r\right),\\
\operatorname{ERP}\left(Head(T_1), T_2\right)+dis\left(T_1^m, r\right)
\end{cases} &\hspace{-15mm} \text { otherwise }\end{cases}
\end{equation}
\end{footnotesize}where $r$ denotes the reference point. As shown in Figure~\ref{fig:Non-learning}(d), we set the reference point $r$ as (3, 2), and get ERP($Q_1$, $Q_2$) = 15.45.


\begin{figure}[tbp]
 	\centering
 	\includegraphics[width=0.49\textwidth]{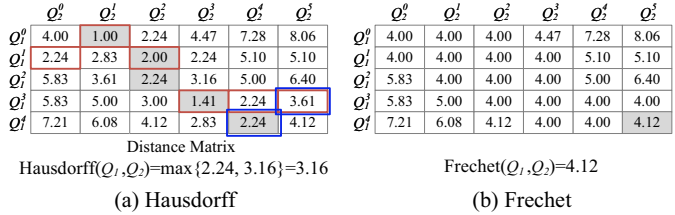}
 	\vspace{-6mm}
 	\caption{Running Example in Free Space vs. Hausdorff and Frechet}
 	\vspace{-6mm}
 	\label{fig:exampleFH}
\end{figure}

{\bf Hausdorf Distance.} Hausdorff distance~\cite{Hausdorff} is a metric measure,
which measures the maximum distance of all distance values from a point of one trajectory to the nearest point in another trajectory. The time complexity of Hausdorff distance is  $\textit{O(mn)}$, which is a parameter-free measure. Formally, Hausdorf distance is defined as:

\begin{footnotesize}
\vspace{-2mm}
\begin{equation}
\operatorname{Hausdorff }\left(T_{1}, T_{2}\right)=\max\begin{cases}
\max _{1\leq i\leq m} \min _{1\leq j\leq n} dis\left(T_{1}^{i}, T_{2}^{j}\right), \\
\max _{1\leq j\leq n} \min _{1\leq i\leq m} dis\left(T_{1}^{i}, T_{2}^{j}\right)
\end{cases}
\end{equation}
\end{footnotesize}In Figure~\ref{fig:exampleFH}(a), we find i) for each point of $Q_1$, its nearest distance to trajectory $Q_2$ (denoted as grey shaded rectangles), and ii) for each point of $Q_2$, its nearest distance to trajectory $Q_1$ (denoted as red rectangles). Based on this, we get two maximum distances (denoted as blue rectangles), and have Hausdorff($Q_1$, $Q_2$) = 3.61.

\begin{figure}[tbp]
	\centering
	\includegraphics[width=0.48\textwidth]{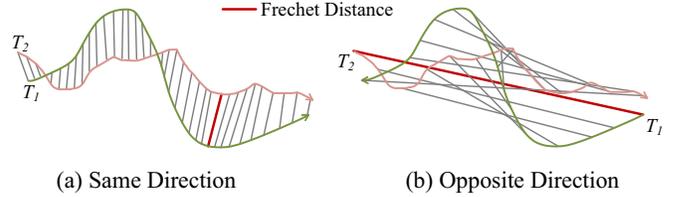}
	\vspace{-4mm}
	\caption{Frechet Distance}
	\vspace{-6mm}
	\label{fig:Frechet}
\end{figure}

{\bf Frechet Distance.} Frechet distance~\cite{Frechet} is similar to the Hausdorff distance. The main difference is that Hausdorff distance does not consider for the direction of two trajectories, while Frechet distance does as depicted in Figure~\ref{fig:Frechet}.
Fréchet~\cite{Frechet} is proposed by taking an example of walking a dog. Suppose that a man walks his dog with a leash. Though the man and his dog may have different trajectories, they move in the same direction. Frechet distance between two trajectories of man and dog is the shortest length of the leash required.
Frechet distance is sensitive to noises with the time complexity of $\textit{O(mn)}$. However, it is commonly used for trajectories with different lengths or sampling rates. Since Frechet distance is non-metric, many variants are proposed to make it be metric~\cite{VehicleSurvey}. The most common one is discrete Frechet distance~\cite{discretFrechet} with the time complexity of $\textit{O(mn)}$. Thus, we use discrete Frechet distance in our experiments for its metricty. For simplicity, we use ``Frechet" distance to denote discrete Frechet distance in the rest of this paper. Formally, discrete Frechet distacne is defined as:

\begin{footnotesize}
\hspace{-4mm}
\vspace{-2mm}
\begin{equation}
\operatorname{Frechet}(T_1, T_2)=\begin{cases}
\max_{1\leq i\leq m} dis\left(T_1^i, T_2^1\right) &\hspace{-15mm} \text { if } {n=1} \\
\max_{1\leq i\leq n} dis\left(T_1^1, T_2^i\right) &\hspace{-15mm} \text { if } {m=1} \\
\max\begin{cases}
dis(T_{1}^{m}, T_{2}^{n}),\\
\min\begin{cases}
\operatorname{Frechet}\left(T_{1}, Head(T_{2})\right),\\
\operatorname{Frechet}\left(Head(T_{1}), T_{2}\right),\\
\operatorname{Frechet}\left(Head(T_{1}), Head(T_{2})\right) \end{cases}
\end{cases} &\hspace{-15mm} \text { otherwise }\end{cases}
\end{equation}
\end{footnotesize}In Figure~\ref{fig:exampleFH}(b), we have Frechet($Q_1$,$Q_2$)=4.12.

\noindent {\bf \textit{Distributed settings.}} To accelerate the trajectory similarity computation, distributed techniques~\cite{18DITA,REPOSE,DFT,LCRS} are  developed, where DITA~\cite{18DITA} and REPOSE~\cite{REPOSE} are two representative methods designed for point-based measures.

{\bf DITA.} DITA~\cite{18DITA} is a distributed in-memory trajectory analytics system, using several classical point-based measures such as DTW and Frechet. DITA chooses STR Partition algorithm~\cite{STR} to partition trajectory points. It uses R-tree~\cite{RTree} as the local index, and designs a trie-like index as the global index. Accordingly, a trie-like partition algorithm is also used in the local index, and several pruning optimization methods are developed to improve the efficiency of similarity search and join in the distributed environment.

{\bf REPOSE.} REPOSE~\cite{REPOSE} is a distributed in-memory system destined for trajectory similarity search, which supports multiple distance measures, including Hausdorff, Frechet, DTW, LCSS, EDR and ERP. Zheng et al~\cite{REPOSE} discretize trajectories into reference trajectories and construct a trie-like index (i.e., RP-Tree) on reference points of reference trajectories. It is worth mentioning that, REPOSE tends to divide similar trajectories into different partitions as much as possible to achieve load balancing. 
\vspace{-2mm}
\subsubsection{Segment-based Measures} Segment-based measures compute the similarity among trajectories, which are consisted of line segments. A line segment is a pair of points $(p_1, p_2)$, and its length is the Euclidean distance between $p_1$ and $p_2$. Based on this, we introduce 5 segment-based measures (i.e., EDwP, LIP, OWD, Seg-Frechet and Seg-Hausdorff) and a distributed method (i.e., DFT) designed for segment-based measures.

\begin{figure}[tbp]
 	\centering
 	\includegraphics[width=0.48\textwidth]{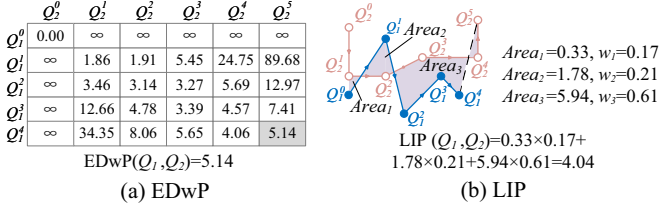}
 	\vspace{-3mm}
 	\caption{Running Example in Free Space vs. EDwP and LIP}
 	\vspace{-6mm}
 	\label{fig:exampleEL}
\end{figure}

{\bf Edit Distance with Projections (EDwP).} EDwP distance~\cite{EDwP,EDwP1,EDwP2} employs a parameter-free approach and adapts to non-uniform sampling rates through dynamic interpolation and projections. It transforms trajectory points into line segments, and uses insertion and replacement operations to compute the edit distance. The time complexity of EDwP is  $\textit{O(mn)}$, and EDwP distance between two trajectories $T_1$ and $T_2$ is defined as:

\begin{footnotesize}
\vspace{-5mm}
\begin{equation}
\label{eq:EDwP}
\operatorname{EDwP}(T_1, T_2)=\begin{cases}
0 &\hspace{-14mm} \text { if } m=n=0 \\
\infty &\hspace{-14mm} \text { if } {m=0}\\&\hspace{-14mm}  \text { or $n = 0$ } \\
\min\begin{cases}
\operatorname{EDwP}\left(T_1, ins(T_2, T_1)\right),\\ \operatorname{EDwP}\left(ins(T_1, T_2), T_2\right),\\
\operatorname{EDwP}\left(Head(T_1), Head(T_2)\right)+\\(rep(T_1^1, T_2^1)\times Coverage(T_1^1, T_2^1)) \end{cases} &\hspace{-14mm} \text { otherwise }\end{cases}
\end{equation}
\end{footnotesize}where $rep(\cdot,\cdot)$ and $ins(\cdot,\cdot)$ denote the operations of replacement and insertion respectively. Specifically, $rep(l_1, l_2)=dis(l_1.s, l_2.s)+dis(l_1.e, l_2.e)$, where $l_i.s$ and $l_i.e$ represent the first and last point of line segment $l_i$, respectively. $Coverage(l_1, l_2) = dis(l_1.s, l_1.e)+dis(l_2.s, l_2.e)$ is used to measure the cost of replacing. 
In addition, $ins(l_1, l_2)$ effectively divides the line segment $l_1$ into two segments by projecting $l_2$ onto it. In Figure~\ref{fig:exampleEL}(a), we calculate the EDwP distance matrix and have EDwP($Q_1$,$Q_2$)=5.14.


{\bf Locality In-between Polylines (LIP).}
LIP~\cite{LIP,LIP1} is a shape-based measure, which considers the areas of polygons formulated by intersection points of two trajectories. As shown in Figure~\ref{fig:exampleEL}(b), polygons have different areas (i.e., $Area_{\textit{1,2,3}}$). LIP sets a weight $w$ (i.e., the average perimeters of the polygons) for each area to adjust the effect of polygons. Nikos et al.~\cite{LIP} extend LIP to a spatio-temporal (i.e., time-aware) distance called STLIP, and also propose a speed-pattern spatio-temporal distance called SPSTLIP. The time complexity of LIP, STLIP and SPSTLIP are all $\textit{O((m+n)log(m+n))}$. Formally, the LIP is defined as:

\begin{footnotesize}
\vspace{-2mm}
\begin{align}
w_{i}=&\frac{{Perimeter}_{T_1}\left(I_{i}, I_{i+1}\right)+{Perimeter}_{T_2}\left(I_{i}, I_{i+1}\right)}{{Perimeter}_{T_1}+{Perimeter}_{T_2}}\\
\operatorname{LIP}&(T_1, T_2)=\sum _{\forall Polygon_i}Area_i \cdot w_i
\end{align}
\end{footnotesize}where $I_i$ denotes the intersection point. In Figure~\ref{fig:exampleEL}(b), the areas and weights of all polygons are calculated. Then, we have LIP($Q_1$,$Q_2$)=4.04.

\begin{figure}[tbp]
 	\centering
 	\includegraphics[width=0.47\textwidth]{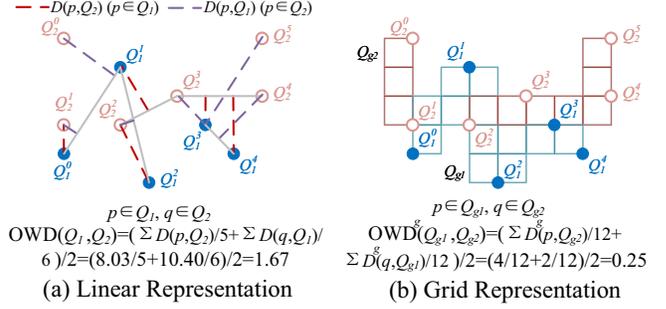}
 	\vspace{-3mm}
 	\caption{Running Example in Free Space vs. OWD}
 	\vspace{-6mm}

 	\label{fig:exampleO}
\end{figure}

{\bf Ont-way Distance (OWD).} OWD~\cite{OWD,OWD1,OWD2} supports two representations of trajectories (i.e., linear representation and grid representation) as shown in Figure~\ref{fig:exampleO}. Lin et al.~\cite{OWD} propose OWD based on linear representation, which transforms point-based trajectory data into segment-based. It measures the average minimal distance from each point in one trajectory to the other trajectory. OWD is an asymmetric measure according to Definition~\ref{defn:metric}. Formally,  the distance $D(p, T)$ between a point $p$ and a trajectory $T$, and the OWD distance of linear representation are defined as below.

\begin{footnotesize}
\vspace{-2mm}
\begin{align}
D(p,T)&=\min _{q \in T} ED(p, q)\\
\operatorname{OWD}\left(T_{1}\rightarrow T_{2}\right)&=\frac{1}{\left|T_{1}\right|}\left(\int_{p \in T_{1}} D\left(p, T_{2}\right) d p\right)\\
\operatorname{OWD}\left(T_{1}, T_{2}\right)=\frac{1}{2}&(\operatorname{OWD}\left(T_{1}\rightarrow T_{2}\right)+\operatorname{OWD}\left(T_{2}\rightarrow T_{1}\right))
\end{align}
\end{footnotesize}Considering the high computation cost $\textit{O(mn)}$ of linear representation, OWD is extended to grid representation called $\operatorname{OWD^g}$, where trajectory points are mapped into grid cells according to their spatial information. Thus, OWD only computes the distances between grid cells (instead of sample points) and grid-based trajectories. The time complexity is reduced to $\textit{O(}m^{\prime} n^{\prime}\textit{)}$, where $m^{\prime}$ and $n^{\prime} $ denote the numbers of grid cells occupied by two trajectories, respectively. Compared to $\operatorname{OWD}$, $\operatorname{OWD^g}$ is more popular due to its low time complexity~\cite{VLDBJSurvey}. Formally, $\operatorname{OWD^g}$ between two grid-based trajectory $T_{\operatorname{g1}}$ and $T_{\operatorname{g2}}$ is defined as:

\begin{footnotesize}
\begin{align}
D^{\operatorname{g}}\left(\operatorname{g}, T_{\operatorname{g}}\right)=&\min _{\operatorname{g}^{\prime} \in T_{\operatorname{g}}} d^{\operatorname{g}} \left(\operatorname{g}, \operatorname{g}^{\prime}\right)\\
\operatorname{OWD^g}\left(T_{\operatorname{g1}}\rightarrow T_{\operatorname{g2}}\right)&=\frac{1}{\left|T_{\operatorname{g1}}\right|} \sum_{p \in T_{\operatorname{g1}}} D^{\operatorname{g}}\left(p, T_{\operatorname{g2}}\right)\\
\operatorname{OWD^g}\left(T_{\operatorname{g1}}, T_{\operatorname{g2}}\right)=&\frac{\operatorname{OWD^g}\left(T_{\operatorname{g1}}\rightarrow T_{\operatorname{g2}}\right) + \operatorname{OWD^g}\left(T_{\operatorname{g2}}\rightarrow T_{\operatorname{g1}}\right)}{2}
\end{align}
\end{footnotesize}where $d^{\operatorname{g}}$(·) computes the distance of two grid cells. As depicted in Figure~\ref{fig:exampleO}(a), we have $\operatorname{OWD}$($Q_1$,$Q_2$)=1.67. In addition, $Q_1$ and $Q_2$ are extended to grid representations $Q_{\operatorname{g1}}$ and $Q_{\operatorname{g2}}$ (consisted of 12 grid cells respectively) as shown in Figure~\ref{fig:exampleO}(b), where $\operatorname{OWD^g}$($Q_{\operatorname{g1}}$,$Q_{\operatorname{g2}}$)=0.25. Note that, when dealing with sparse data with a large spatial span, the space overhead of $\operatorname{OWD^g}$ is quite large.

{\bf Seg-Frechet and Seg-Hausdorff.} Inspired by Frechet and Hausdorff distances, Seg-Frechet and Seg-Hausdorff distances~\cite{DFT} are proposed for segment-based trajectories, by changing points to line segments when computing Frechet and Hausdorff.

\noindent {\bf \textit{Distributed setting.}} Xie et al.~\cite{DFT} develop a distributed framework (i.e., DFT) based on Spark that supports Seg-Frechet and Seg-Hausdorff measures. DFT uses STR partition algorithm~\cite{STR} to partition segments, while constructs a dual R-Tree as the global index and a general R-Tree as the local index. DFT is the first distributed method to support fast trajectory similarity computation.

\subsection{Trajectory Similarity in Road Network}
In this subsection, we review 7 distance measures designed for road network constrained trajectories in terms of point-based and segment-based measures. 
Different from free space, trajectories in road network are usually denoted by road vertices or road segments. Here, Figure~\ref{fig:Non-learning-RN} gives the running example in road network, where the grid (denoted as grey lines) is assumed as the underlying road network. The spatial and temporal distance matrices are used to store the road network distances and temporal distances between vertices of two trajectories $Q_1$ and $Q_2$, respectively.


\subsubsection{Point-based Measures} Point-based measures in road network are utilized to compute the similarity among trajectories that are consisted of road vertices. We mainly introduce 5 point-based measures (i.e., NetERP, NetEDR, NetDTW, NetLCSS and TP) below.

\begin{figure}[tbp]
 	\centering
 	\includegraphics[width=0.48\textwidth]{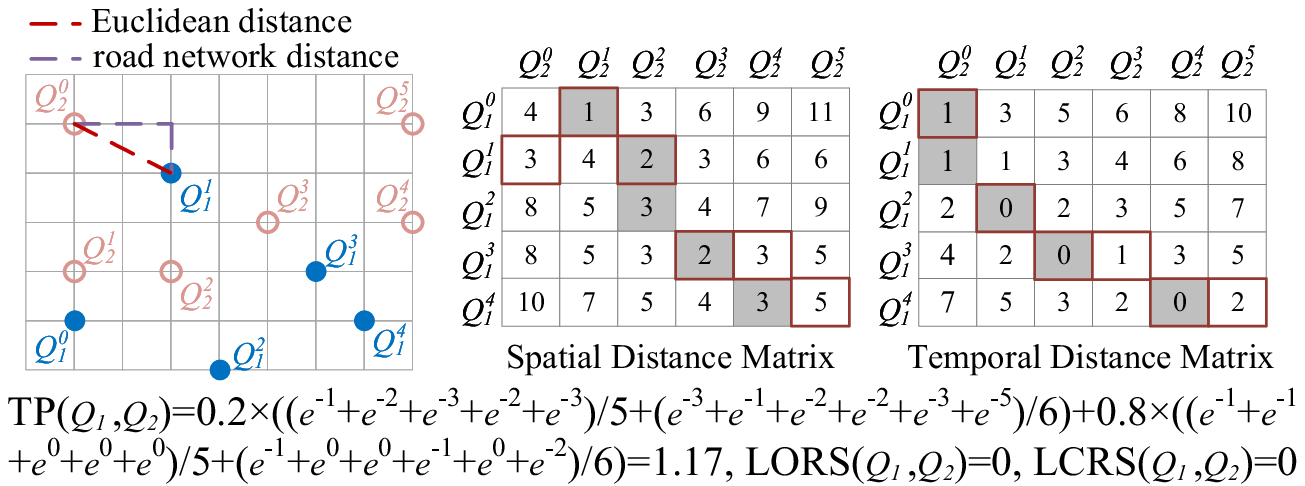}
 	\vspace{-3mm}
 	\caption{Running Example in Road Network}
 	\vspace{-6mm}
 	\label{fig:Non-learning-RN}
\end{figure}

{\bf NetERP, NetEDR, NetDTW and NetLCSS.} NetERP~\cite{NetERP}, NetEDR~\cite{NetERP}, NetDTW~\cite{NetDTW} and NetLCSS~\cite{ST2Vec} are expanded from classic measures in free space. They first map original trajectories into road network paths that consist of vertices or segments. Then, they define similarity measures based on classic distance measures such as ERP, EDR, DTW and LCSS, generally by aggregating the distances between road vertices or segments of two trajectories. Note that, these measures employ the shortest path distance (instead of Euclidean distance) between two road vertices in the graph. Thus, NetERP, NetEDR, NetDTW and NetLCSS have the same features as corresponding classic measures, and can be used in road network. Consider the running example in Figure~\ref{fig:Non-learning-RN}, we have NetERP($Q_1$,$Q_2$)=20.00, NetEDR($Q_1$,$Q_2$)=4, NetDTW($Q_1$,$Q_2$)=19.00, and NetLCSS($Q_1$,$Q_2$)=1.

{\bf TP.} TP~\cite{TP} is a measure that considers both the spatial and temporal similarities.
Shang et al.~\cite{TP} assume that the point is 3-dimension, i.e., $v=(p, t)=((latitude, longitude), timestamp)$. The spatial and temporal similarities (i.e., $Sim_S$ and $Sim_T$) are:

\begin{footnotesize}
\vspace{-2mm}
\begin{align}
&D'(v.p,T)=\min _{q \in T} d'(v.p, q.p)\\
&D'(v.t,T)=\min _{q \in T} |v.t-q.t|\\
{Sim}_{S}\left(T_{1}, T_{2}\right) &=\frac{\sum_{T_1^i \in T_{1}} e^{-D'\left(T_1^i \cdot p, T_{2}\right)}}{\left|T_{1}\right|}+\frac{\sum_{T_2^j \in T_{2}} e^{-D'\left(T_2^j \cdot p, T_{1}\right)}}{\left|T_{2}\right|} \\
{Sim}_{T}\left(T_{1}, T_{2}\right) &=\frac{\sum_{T_1^i \in T_{1}} e^{-D'\left(T_1^i \cdot t, T_{2}\right)}}{\left|T_{1}\right|}+\frac{\sum_{T_2^j \in T_{2}} e^{-D'\left(T_2^j \cdot t, T_{1}\right)}}{\left|T_{2}\right|}
\end{align}
\end{footnotesize}Here, $d'(\cdot, \cdot)$ denotes the shortest path distance between two road vertices in the graph. The spatial and temporal similarities are combined linearly to obtain the spatio-temporal similarity, i.e., $ Sim_{ST}(T_1, T_2)$ $=\lambda\cdot Sim_S(T_1, T_2) + (1-\lambda)\cdot Sim_T(T_1, T_2)$, where $\lambda$ $\left(0\leq\lambda\leq 1\right)$ is a parameter to adjust the relative importance of spatial and temporal similarities. As shown in Figure~\ref{fig:Non-learning-RN}, since TP needs to utilize the temporal information of trajectories, we randomly add timestamps to points of $Q_1$ and $Q_2$, i.e., $Q_1 = {(1,1,1),(3,4,3),(4,0,4),(6,2,6),(7,1,9)}$, $Q_2 = {(1,5,2),(1,2,4),(3,2,6),(5,3,7),(8,3,9),(8,5,11)}$. Based on this, when $\lambda$=0.2, we have TP($Q_1$, $Q_2$)=1.17.



\subsubsection{Segment-based Measures.} Segment-based measures in road network measure the similarity among trajectories consisted of road segments. We mainly introduce 2 segment-based measures (i.e., LORS and LCRS) and a distributed framework (i.e., DISON).

{\bf Longest Overlapping Road Segments (LORS).} LORS~\cite{LORS} is a distance measure based on the length of overlapped edges (i.e., road segments) with the time complexity of $\textit{O(mn)}$. 
Given two trajectories $T_1$ and $T_2$, where the $i$-th ($1\leq i\leq m$) segment of $T_1$ is denoted as $T_1^i$, LORS measure is defined formally as:

\begin{footnotesize}
\vspace{-2mm}
\begin{equation}
\label{eq:LORS}
\operatorname{LORS}(T_1, T_2)=\begin{cases}
0 &\hspace{-3mm} \text { if } {m=0}  \text { or $n = 0$ } \\
length(T_1^m)\\+\operatorname{LORS}(Head(T_1), Head(T_2)) &\hspace{-3mm} \text { if } T_1^m=T_2^n \\
\max\begin{cases} \operatorname{LORS}\left(T_1, Head(T_2)\right),\\ \operatorname{LORS}\left(Head(T_1), T_2\right),\end{cases} &\hspace{-3mm} \text { otherwise }\end{cases}
\end{equation}
\end{footnotesize}where $length$(·) denotes the length of the edge. According to Eq.~\ref{eq:LORS}, LORS does not need to compute the shortest path distance in the graph, which is relatively efficient. In Figure~\ref{fig:Non-learning-RN}, we have LORS($Q_1$, $Q_2$)=0, as $Q_1$ and $Q_2$ have no overlapping segments.

{\bf Longest Common Road Segments (LCRS).} Similar to LCSS, LCRS~\cite{LCRS} finds the longest common road segment between two trajectories, and can be used in several applications such as carpooling. The time complexity of LCRS is also $\textit{O(mn)}$, and LCRS satisfies symmetry. It is formally defined as:

\begin{footnotesize}
\vspace{-2mm}
\begin{align}
&\operatorname{LCRS}(T_1, T_2)=\frac{\operatorname{LORS}(T_1, T_2)}{m+n-\operatorname{LORS}(T_1, T_2)}
\end{align}
\end{footnotesize}As shown in Figure~\ref{fig:Non-learning-RN}, LCRS($Q_1$, $Q_2$)=0.00, as $Q_1$ and $Q_2$ have no common segments.


\begin{figure*}[tbp]
 	\centering
 	\includegraphics[width=0.96\textwidth]{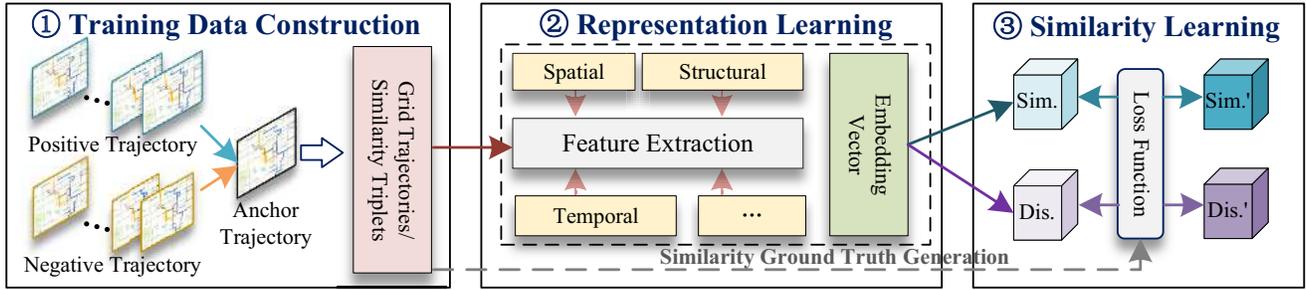}
 	 \vspace{-2mm}
 	\caption{The Processing Pipeline of Learning-based Models}
 	\vspace{-6mm}
 	\label{fig:learning}
\end{figure*}

\noindent {\bf \textit{Distributed setting.}} Yuan et al.~\cite{LCRS} extend standalone LCRS to a distributed framework called DISON, which supports road-network constraint similarity search and join. DISON first uses STR partition algorithm to partition the road segments, and then it builds a two-layer index with hashmap as the global index and an inverted index as the local index. Note that, DISON is unique distributed framework for similarity measure in road network.


\section{Learning Based Measures}
\label{sec:learning}

As mentioned in Section~\ref{sec:traditional}, non-learning measures generally have a time complexity of $\textit{O(mn)}$, which is high. To address this, learning-based methods are developed to reconstruct the input data of high-dimension into a new representation of low-dimension~\cite{word2vec,DeepLRepresentationCluster,18t2vec}.

t2vec~\cite{18t2vec} designs a trajectory deep learning framework to achieve robustness to low sample rates and noises. It maps a trajectory into a $d$-dimensional embedding vector. Note that, t2vec is not a similarity measure, but a representation method to transform trajectories into vectors. 

Inspired by t2vec, many studies employ different deep learning frameworks in free space and road network, to learn approximate distance functions for non-learning measures. Generally, as shown in Figure~\ref{fig:learning}, they first select similar and dissimilar trajectories (i.e., positive and negative trajectories) of anchor trajectories (origin trajectories) to obtain grid-based trajectories~\cite{19NEUTRAJ} or similarity triplets~\cite{20Traj2SimVec, 21GTS, ST2Vec}. Then, features (in terms of spatial, temporal, structural and so on) are extracted, while trajectory embedding vectors are generated using deep representation learning. Finally, the learning (dashed rectangle in Figure~\ref{fig:learning}) can stop  until the trajectory similarities and dissimilarities evaluated on the embedding vectors are close to the ground-truth. Here, results calculated using non-learning measures are taken as ground-truth. Note that, the time complexity of similarity learning among vectors is linear. In view of this, we review 4 representative learning-based measures.

{\bf NEUTRAJ.} The first learning-based trajectory similarity measure is NEUTRAJ~\cite{19NEUTRAJ}, which is based on neutral metric learning. Specifically, NEUTRAJ first maps trajectories into grid-trajectories. Then, it samples trajectories as seeds, and uses their pair-wise similarities and dissimilarities as guidance. Finally, NEUTRAJ uses Long Short-Term Memory (LSTM) model to generate the embedding vectors, and approximates various non-learning similarity computations (as depicted in Figure~\ref{fig:learning}) with the complexity of $\textit{O(m+n)}$. Note that, embedding vectors generated by NEUTRAJ can preserve the spatial information of trajectories. In addition, in order to improve the performance of deep representation learning methods on long trajectories, TrajGAT~\cite{TrajGAT} is proposed, which is based on graph attention networks (GATs), Transformer and a quad-tree index for effectively embedding trajectories.

{\bf Traj2SimVec.} Although NEUTRAJ has greatly reduced the time complexity, it needs a pre-training process to compute the similarity among all seed trajectories, which incurs a quadric time complexity during training. Thus, Traj2SimVec~\cite{20Traj2SimVec} is proposed to improve the training efficiency by simplifying training trajectories into triplet training samples. And the training time complexity is reduced to $\textit{O(logn)}$, where $n$ denotes the average length of training trajectories. Similar to Traj2SimVec, T3S~\cite{21T3S} and TMN~\cite{TMN} are proposed based on attention networks, where the former considers dissimilar trajectories while the latter focuses on the inter-information between trajectories.

{\bf GTS.} Different from the above two methods that are designed in free space, a Graph Neural Network (GNN) based approach named GTS~\cite{21GTS} is designed to measure the trajectory similarity in road network. In order to reflect the information on the road network, GTS learns the representation of point-of-interest (POI) in spatial network and the relationship between trajectories, between POIs, or between trajectories and POIs.

{\bf ST2Vec.}  ST2Vec~\cite{ST2Vec} is also proposed for spatio-temporal trajectory similarity computation in road network. Considering the complex temporal information in trajectory data~\cite{VLDBJSurvey}, ST2Vec with a time complexity of $\textit{O(d)}$ captures both spatial and temporal features, and fusions these features to obtain spatio-temporal embedding vectors based on GNN and LSTM. Also, ST2Vec applys training triplets samples to reduce the training cost, and sets spatial and temporal weights that can be adjusted for flexible analysis. 

Note that, due to the similarity among different learning-based methods, we mainly evaluate NEUTRAJ~\cite{19NEUTRAJ}, Traj2SimVec~\cite{20Traj2SimVec}, GTS~\cite{21GTS}, ST2Vec~\cite{ST2Vec} in Section~\ref{sec:exe} for simplicity.

\section{Evaluation Benchmark}
\label{sec:benchmark}

\textbf{Motivation.} Since the evaluation on such a big family of trajectory similarity measures has never been organized, we are inspired to propose the first standard evaluation framework to examine the function of each measure. On the one hand, the primary goal of a distance measure is being effective, robust, efficient, and scalable. These four desiderata are often in tension with each other. In view of this, the evaluation framework must enable to study the capability of each measure in terms of multi-aspects. On the other hand, trajectory data usually have various characteristics such as variable lengths, complex shapes (especially for vehicle trajectories), varying samplings, outlier noises, etc. Specifically, the variable lengths and complex shapes are intrinsic properties~\cite{15overview}, while the varying samplings and outlier noises are usually caused by other factors~\cite{VLDBJSurvey} such as battery power and urban building effect. Overall, we build an evaluation framework to benchmark the trajectory measures under as many real-life settings as possible, objectively studying the capability of each measure in terms of four performance aspects.
Table~\ref{tab:transformation} summarizes all the evaluation dimensions and corresponding scenarios. According to Table~\ref{tab:transformation}, we evaluate each trajectory similarity measure from four dimensions as follows.

(i) \textit{Effectiveness vs. length and shape.} Length and shape are considered as intrinsic properties of trajectory data~\cite{07outlierdetection}, which requires an effective measure. However, as the similarity of two trajectories does not have a ground truth~\cite{07traclus}, we cannot determine whether the value computed by trajectory measures is correct or not. It is also meaningless to cross-compare the distance values returned by different measures. To this end, we adopt a qualitative way to visualize the Top-$k$ similarity query results returned by different trajectory measures, for the same query trajectory. Thus, we give an intuitive analysis when the query trajectory is selected with different lengths or shapes to show the effectiveness of each measure.

Specifically, we use parameter $L$ to change the trajectory length. For example, $\textit{L=}20$ means to select the first $20\%$ points from the original trajectory as a deformed trajectory. Such length-oriented evaluation enables us to see how each distance measure performs when the length of the trajectory changes from short to long, which is useful in an online setting that involves trajectory evolution. In terms of varying shapes, we select four typical geometrics, i.e., straight line, polyline without overlaps, polyline with overlaps, and round line, to explore how each similarity measure performs when the query trajectories are of different shapes.

(ii) \textit{Robustness vs. sampling and noise.} Quality issues commonly exist during data collection. For example, a taxi driver may alter the default device sampling rate to reduce the power consumption~\cite{powerconsumption}. In this case, even for the same trajectory, this may result in sparse and dense segments. In addition, noise may occur in GPS points, which is usually caused by low satellite visibility or urban canyons that affect signal quality. Consequently, given a query trajectory, a robust trajectory similarity measure is expected to retrieve the same query results for datasets with different sampling rates and noise ratios.

\begin{table}[]
\caption{Evaluation Benchmark}
\small
\vspace{-3.5mm}
\begin{tabular}{|c|c|c|}
\hline
\makebox[0.08\textwidth][c]{\textbf{Dimension}}                      & \makebox[0.10\textwidth][c]{\textbf{Varying Cases}}      & \makebox[0.13\textwidth][c]{\textbf{Adjustable Parameters}} \\ \hline
\multirow{2}{*}{\textbf{Effectiveness}} & vs. \textit{length}     & $L$ (\%): 20, 60, 100                              \\ \cline{2-3}
                                        & vs. \textit{shape}  & Four typical geometrics                              \\ \hline
\multirow{2}{*}{\textbf{Robustness}}    & vs. \textit{sampling}   & $S$ (\%): 10, 20, 40                             \\ \cline{2-3}
                                        & vs. \textit{noise}              & $N$ (\%): 13, 16, 19/10, 20, 30                              \\ \hline
\textbf{Efficiency}                     & Top-$k$ query       & /                              \\ \hline
\multirow{2}{*}{\textbf{Scalability}}   & vs. \textit{length}     & $L$ (\%): 20, 60, 100          \\ \cline{2-3}
                                        & vs. \textit{cardinality} & $O_r$ (\%): 20, 60, 100      \\ \hline
\end{tabular}
\label{tab:transformation}
\vspace{-6mm}
\end{table}

We use parameter $S$ to denote the percentage of sampling points in each trajectory. For example, $\textit{S=}20$ means that we sample $20\%$ points from a trajectory as a deformed trajectory. Note that, sampling transformation is meaningful and useful for free space trajectories, while meaningless and useless for road network constrained trajectories after map matching. We use parameter $N$ to represent the ratio of noises in a whole trajectory. For instance, $\textit{N=}10$ denotes to sample $10\%$ points from a trajectory, and then add an outlier for each point. Here, we utilize Gaussian noise~\cite{GaussNoise} to add an outlier $p^{\prime}$ for each filtered point $p$, i.e., $p^{\prime}=p+\Delta \cdot p \sim Gaussian(0,1)$, where $\Delta$ is the radius range of outliers.

(iii) \textit{Efficiency vs. Top-$k$ query.} An efficient trajectory similarity measure is expected to efficiently compute the distance value between any pair of trajectories, which especially is important for online trajectory applications. Therefore, we directly use the Top-$k$ similarity query as the evaluation task.

(iv) \textit{Scalability vs. length and cardinality.} Despite how the average trajectory length or the data cardinality (i.e., the number of moving objects) of the trajectory dataset changes, a scalable measure should show stable inter-trajectory distance computation performance. Specifically, we proceed to use $L (\%)$ to control the average length of trajectories in the dataset. In addition, we use another parameter $O_r$ to control the percentage of moving objects w.r.t. all the objects, i.e., the trajectory dataset to be queried/processed.

\section{Experimental Evaluation}
\label{sec:exe}

We conduct extensive experiments aimed at offering insights to the performance of trajectory similarity measures summarized in Table~\ref{tab:measures}. First, we present experimental settings. Next, based on the evaluation benchmark proposed in Section 5, we study trajectory measures in terms of four aspects: effectiveness, robustness, efficiency, and scalability. 
Finally, we conduct a study of metric similarity measures to show the effect of indexing and pruning on them. 

\subsection{Experimental Setup}
\label{subsec:setting}

\begin{figure} [tb]
	\centering
	\hspace{-2mm}
	\subfigure[AIS (Free Space)]{
		\includegraphics[width=0.23\textwidth]{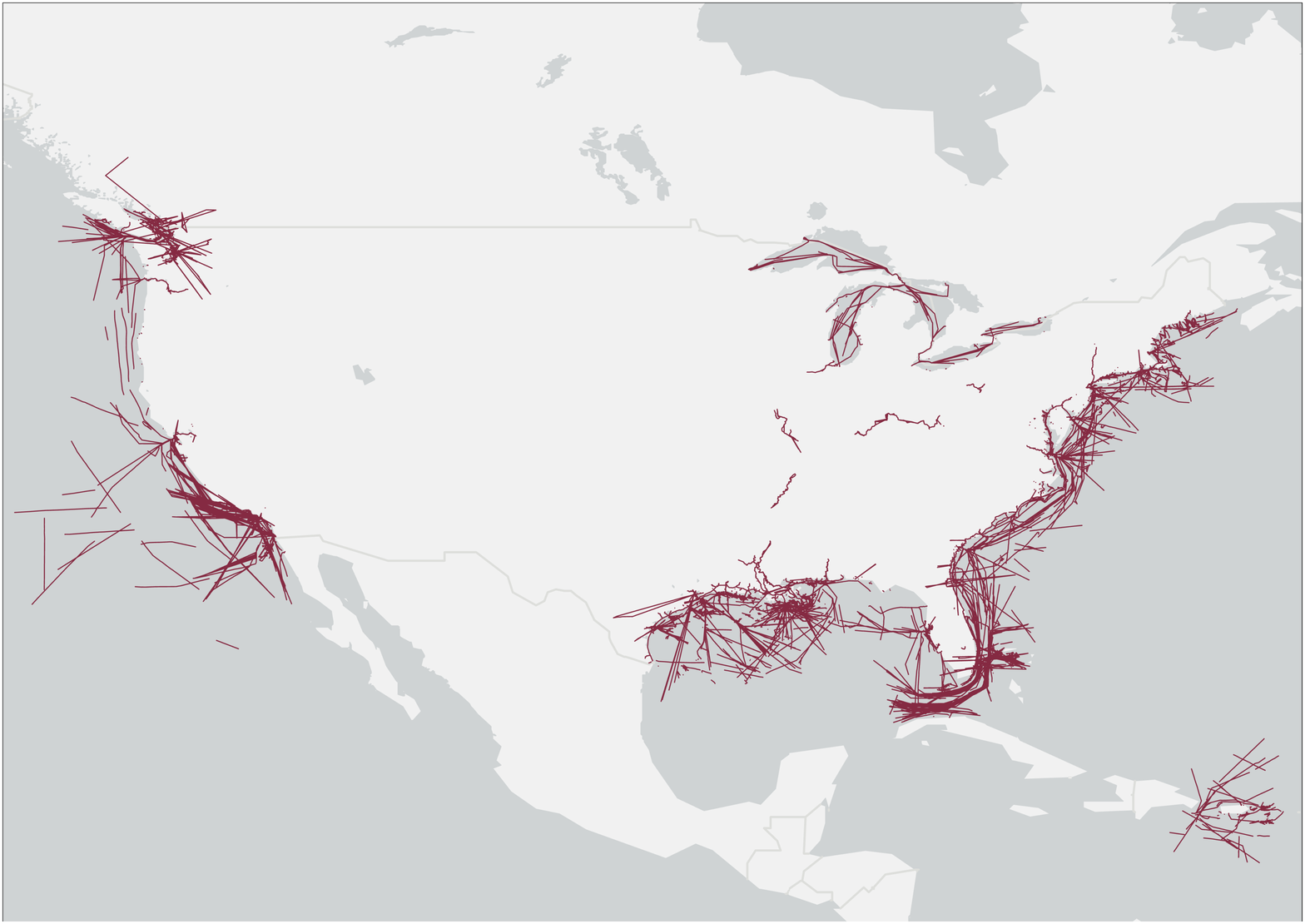}}
	\subfigure[Geolife (Free Space)]{
	    \includegraphics[width=0.23\textwidth]{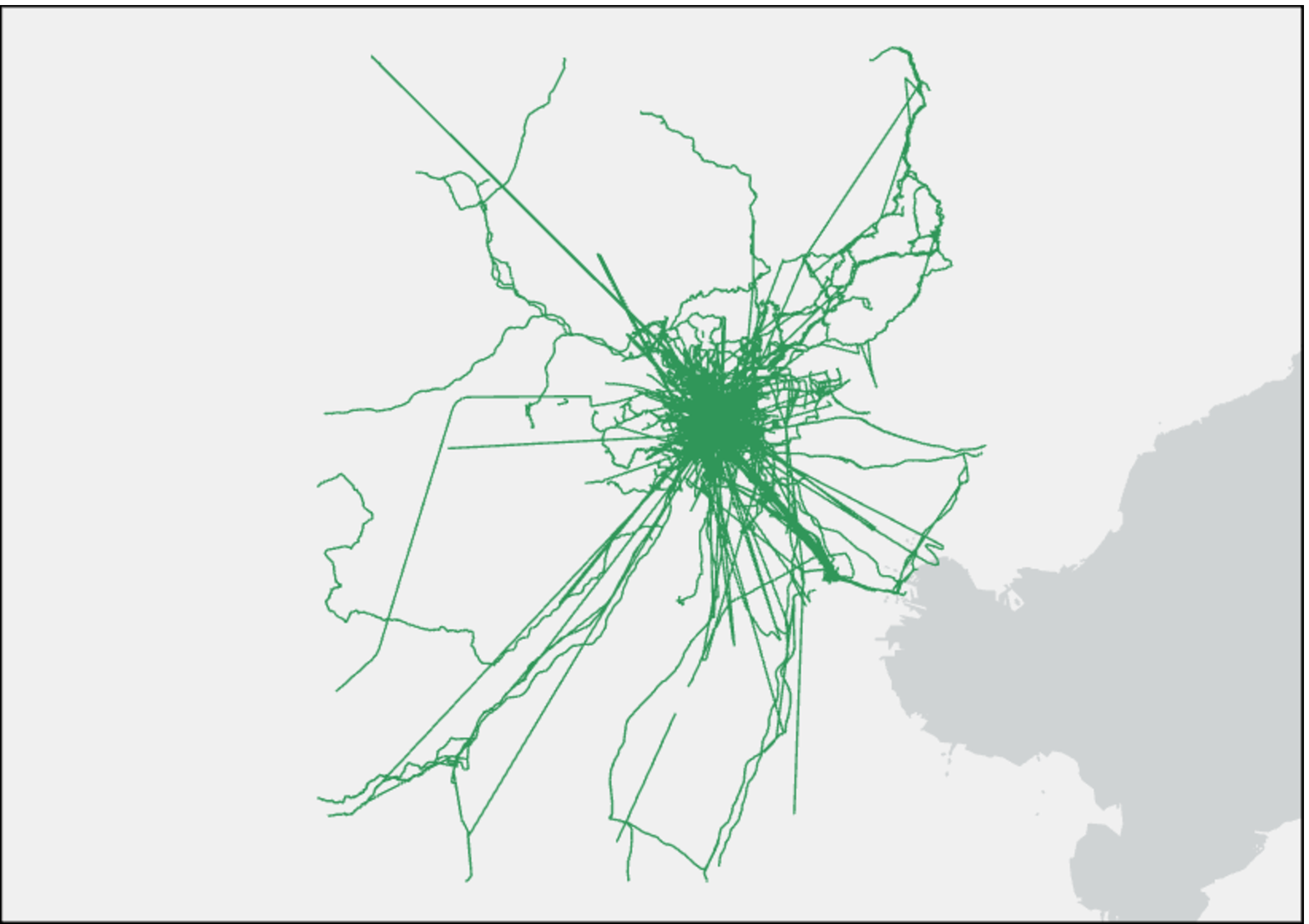}}\\
	\vspace{-1mm}
	\subfigure[Tdrive (Road Network Space)]{
		\includegraphics[width=0.23\textwidth]{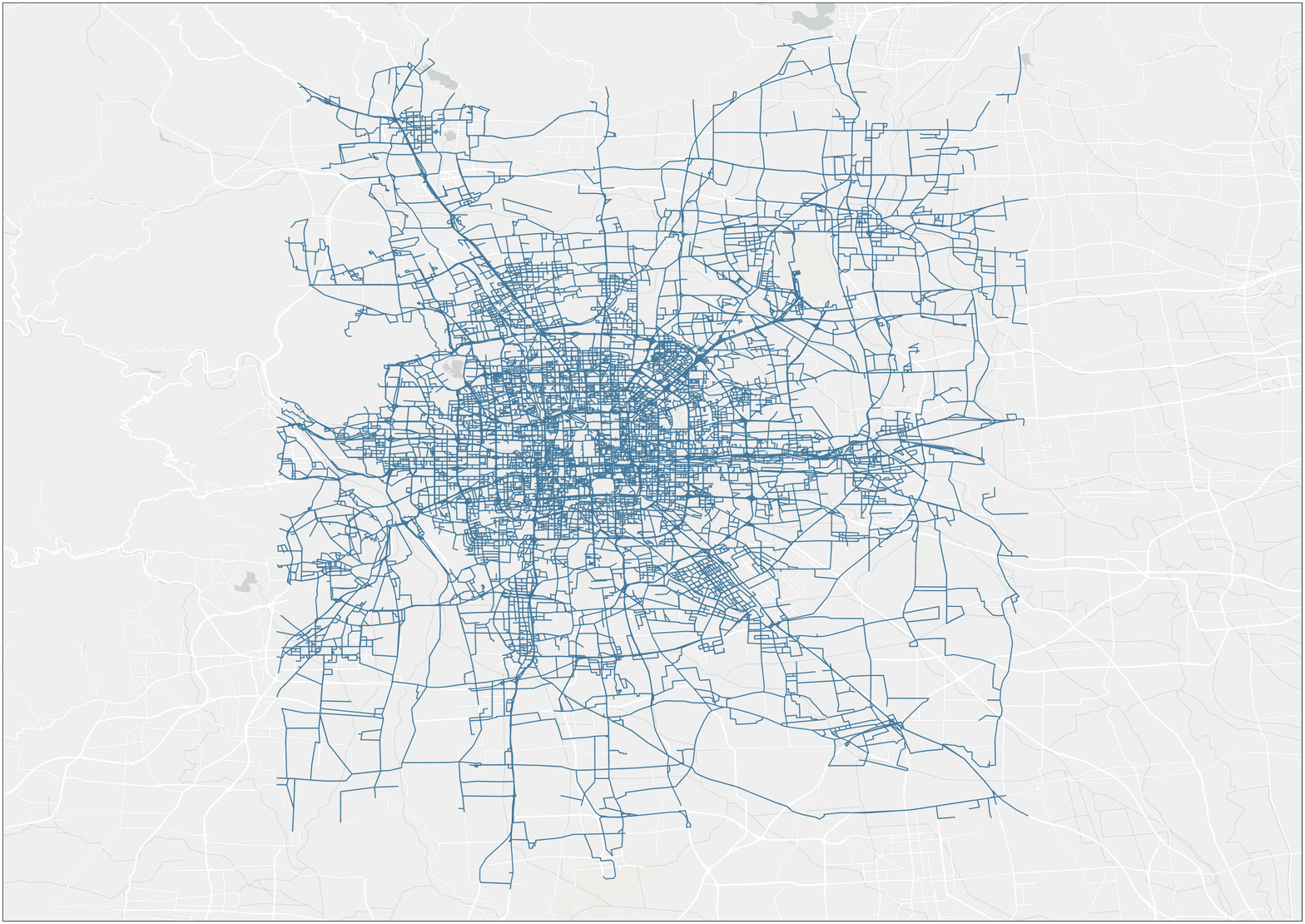}}
	\subfigure[Porto (Road Network Space)]{
	    \includegraphics[width=0.23\textwidth]{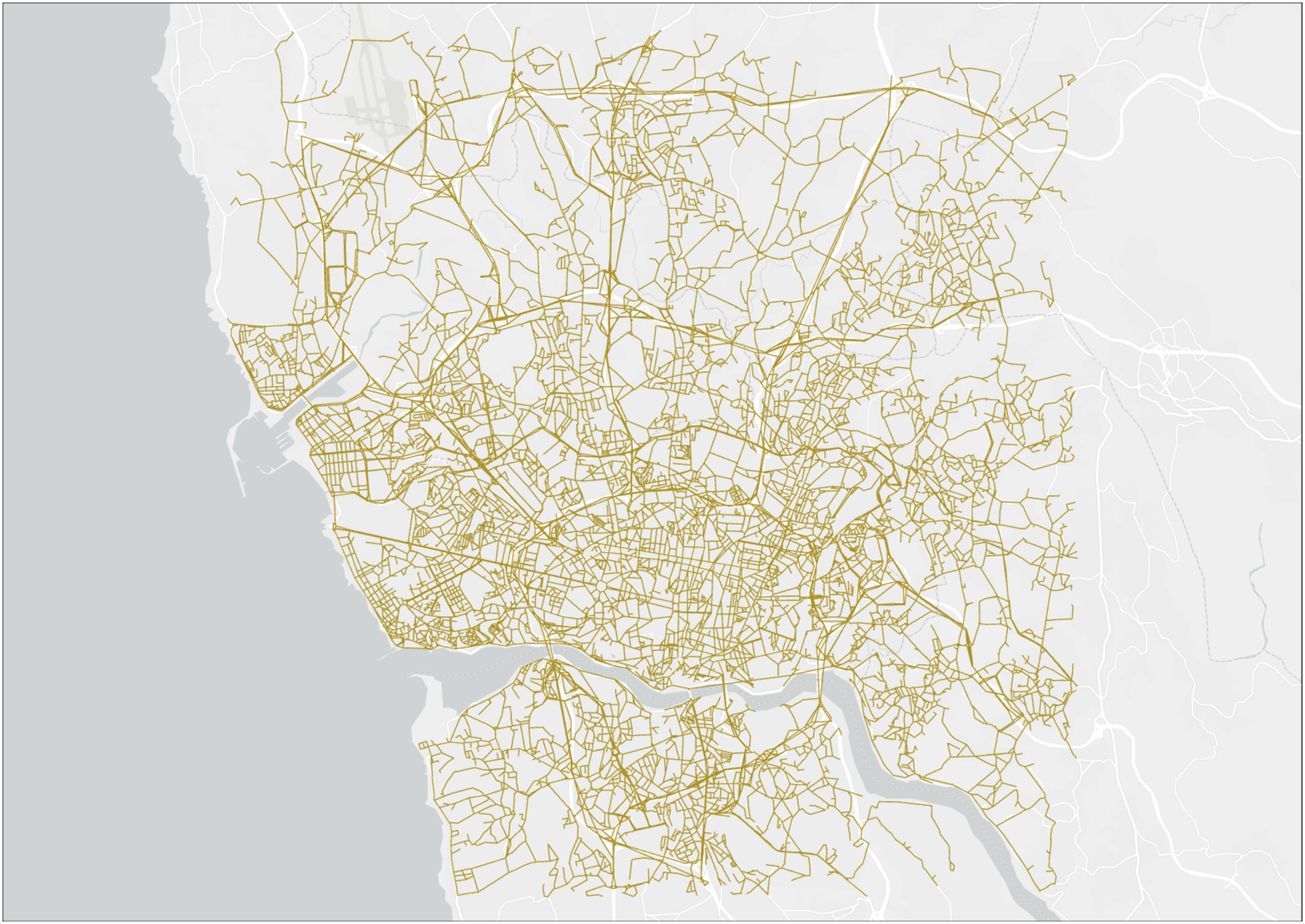}}\\
    \up
	\caption{Visualization of The Datasets Used}
	\label{fig:dataset}
	\vspace{-6mm}
\end{figure}

\begin{figure*} [tb]
	\centering
	\hspace{-4mm}
	\includegraphics[width=0.6\textwidth]{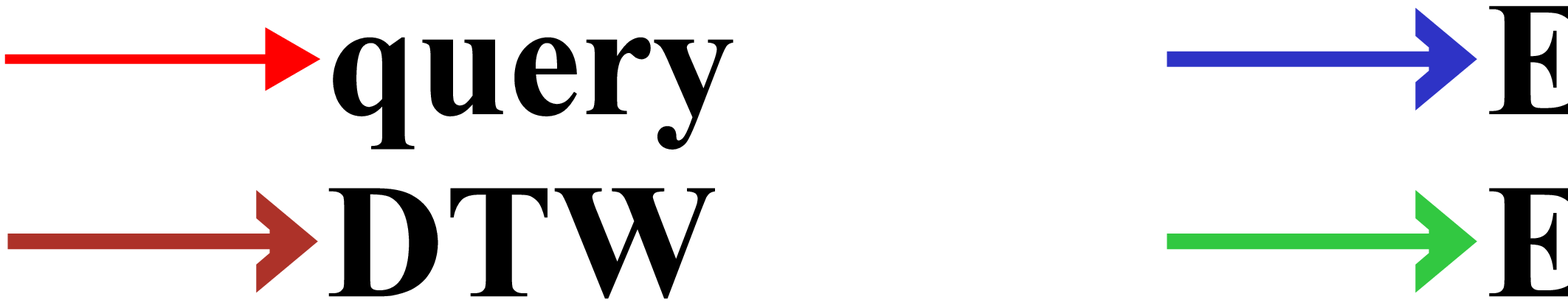}\\
	\hspace{-2mm}
	\subfigure[$L$=20\% (AIS)]{
		\includegraphics[width=0.19\textwidth]{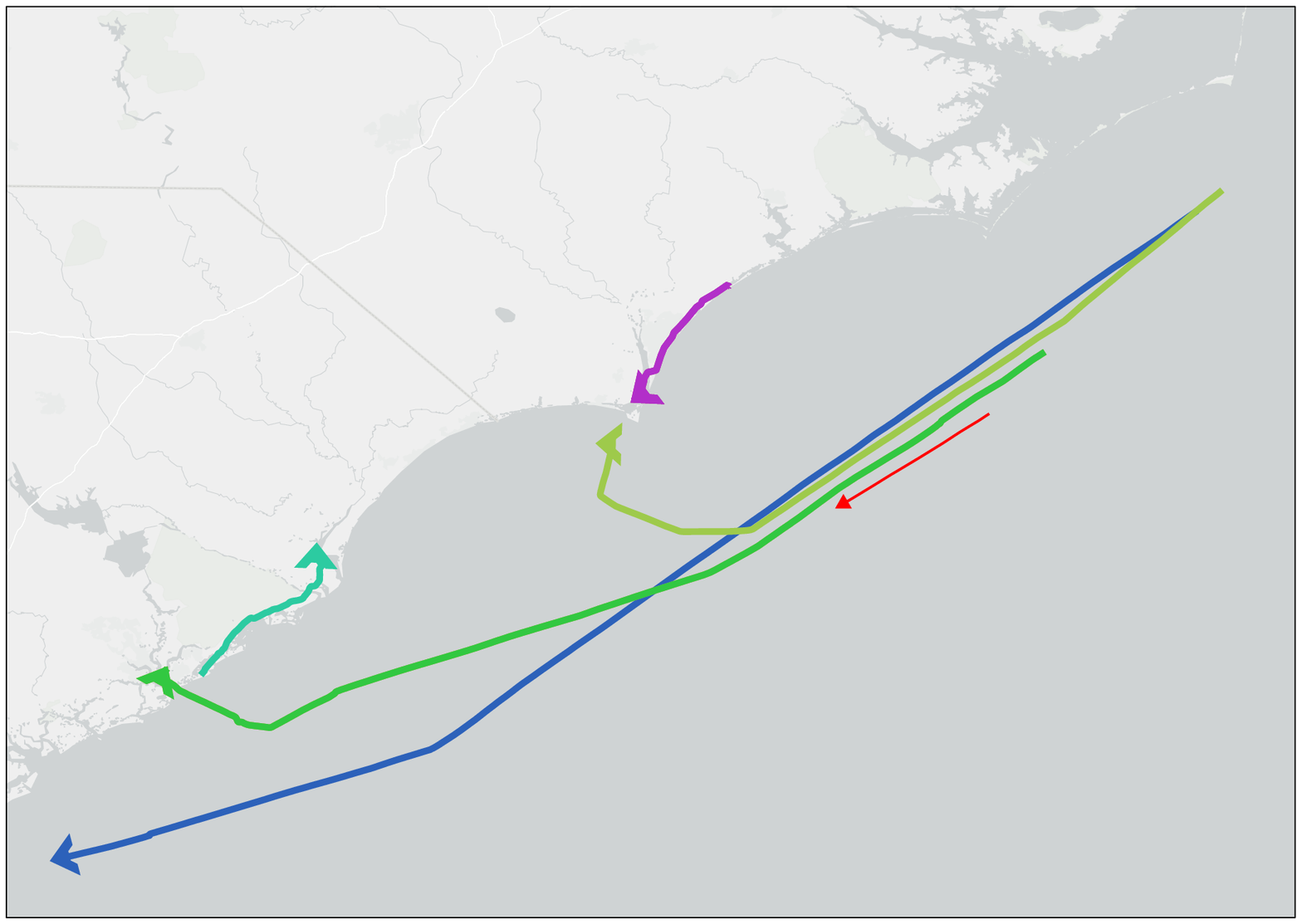}}
	\subfigure[$L$=40\% (AIS)]{
		\includegraphics[width=0.19\textwidth]{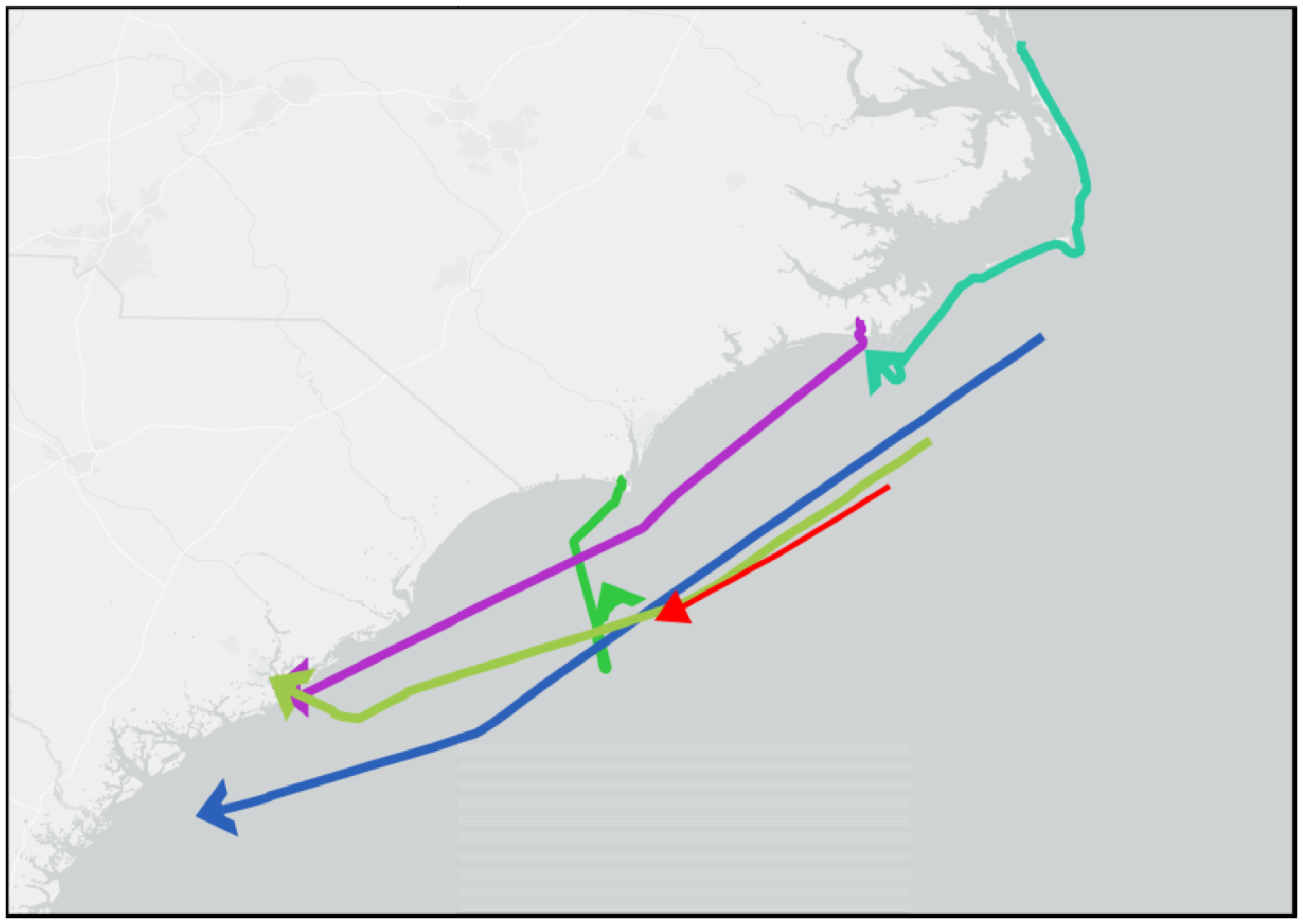}}
	\subfigure[$L$=60\% (AIS)]{
		\includegraphics[width=0.19\textwidth]{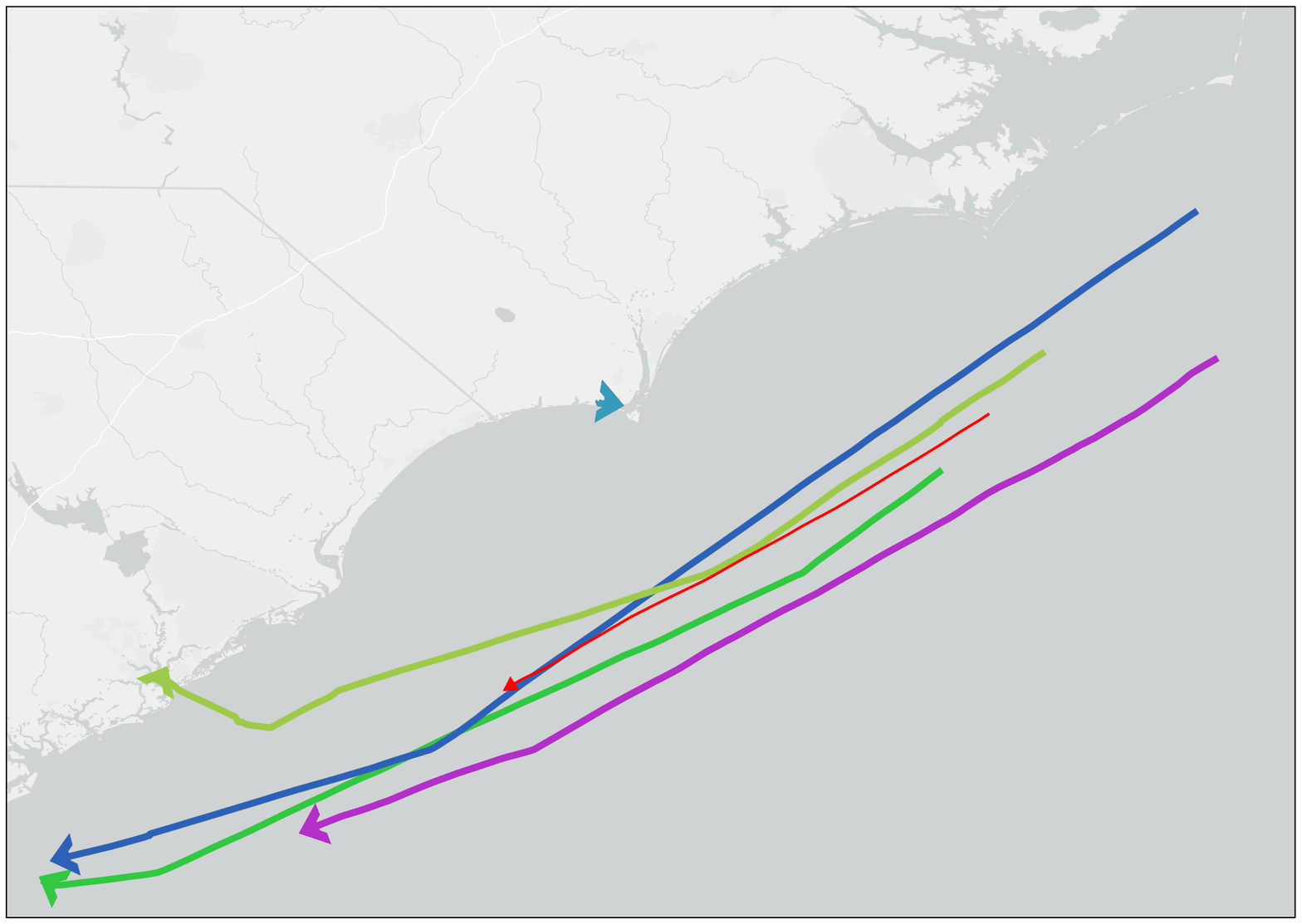}}
	\subfigure[$L$=80\% (AIS)]{
		\includegraphics[width=0.19\textwidth]{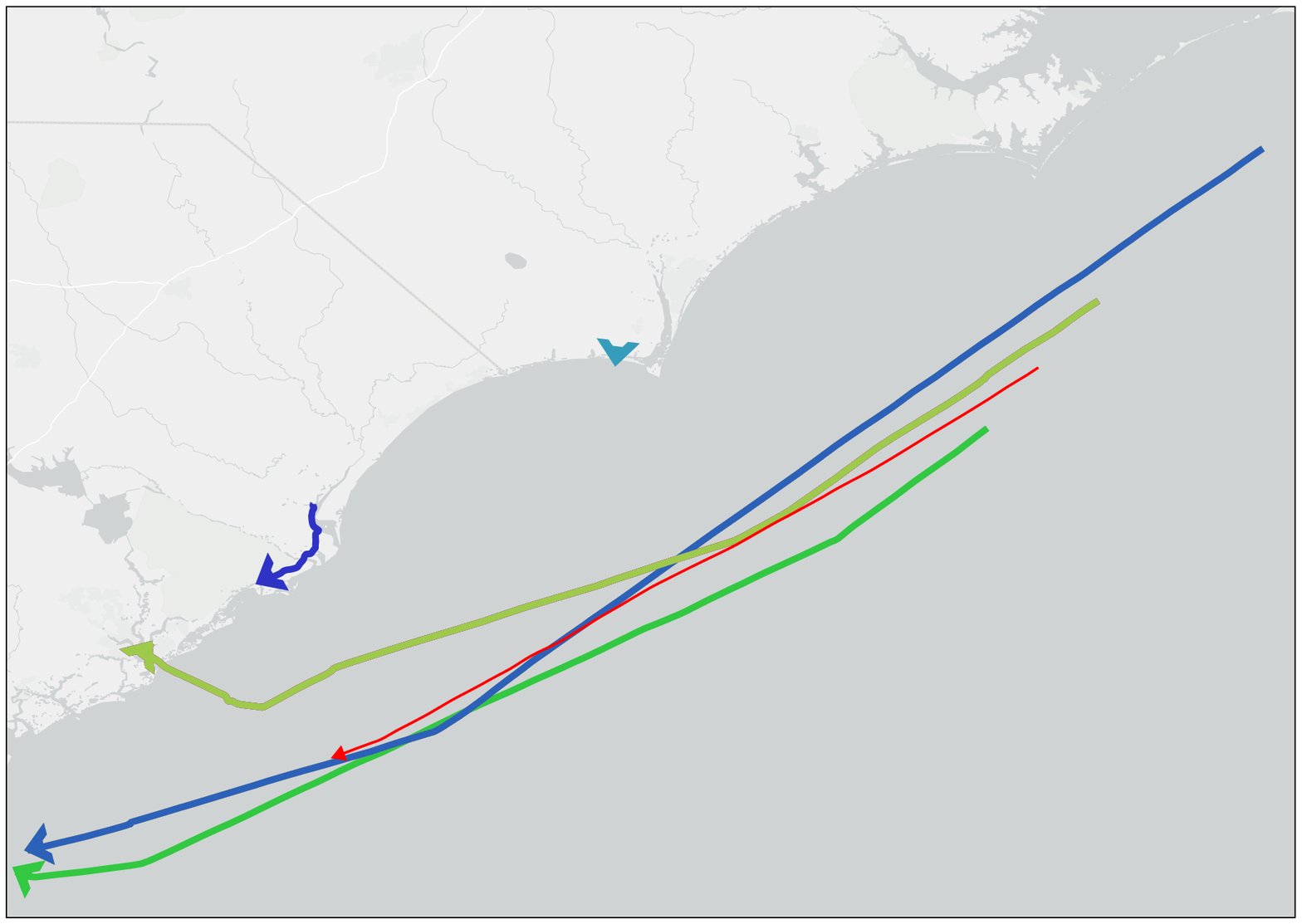}}
	\subfigure[$L$=100\% (AIS)]{
		\includegraphics[width=0.19\textwidth]{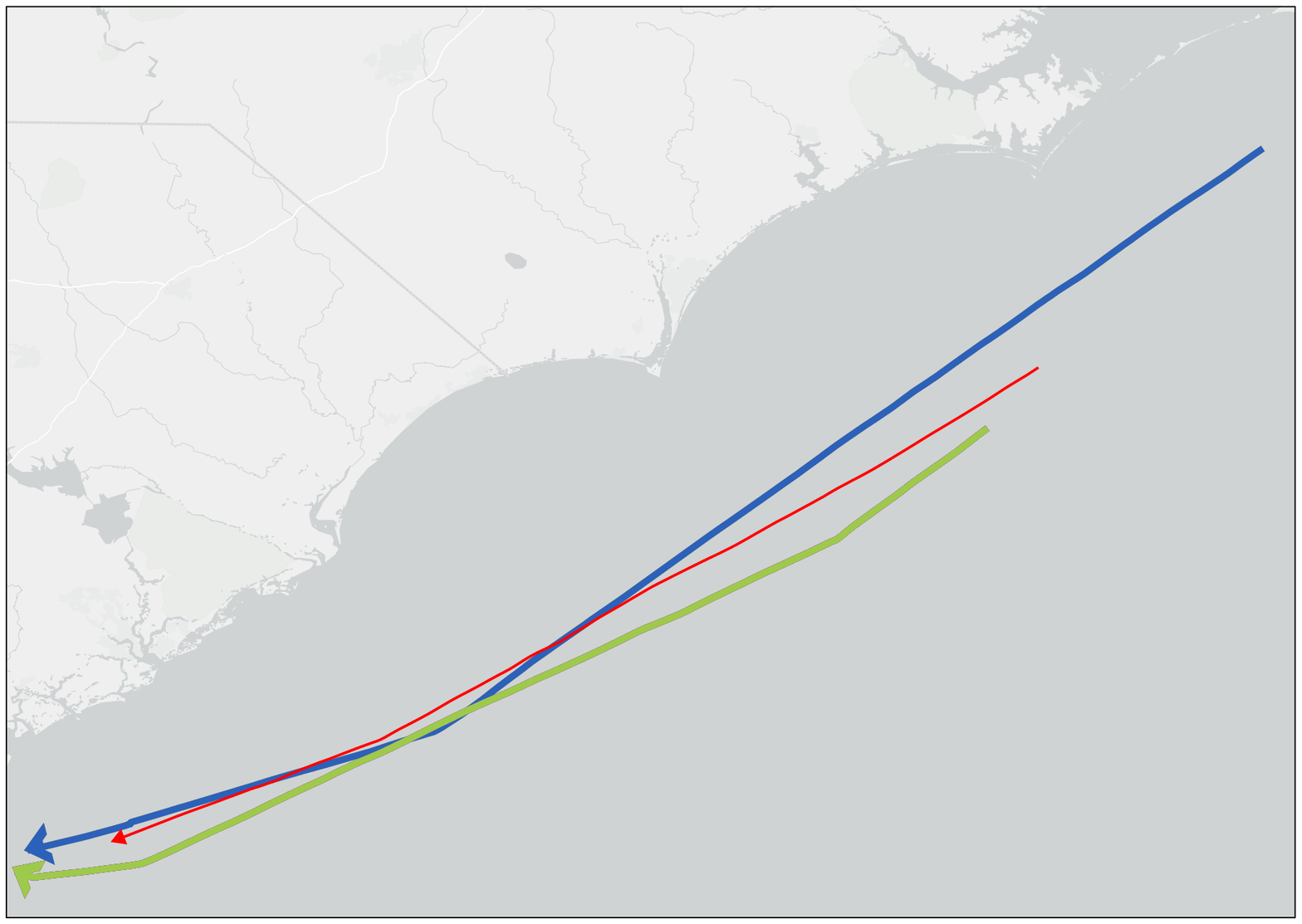}}\\
		
	\vspace{-2mm}
	\hspace{-2mm}
	\subfigure[$L$=20\% (Geolife)]{
		\includegraphics[width=0.19\textwidth]{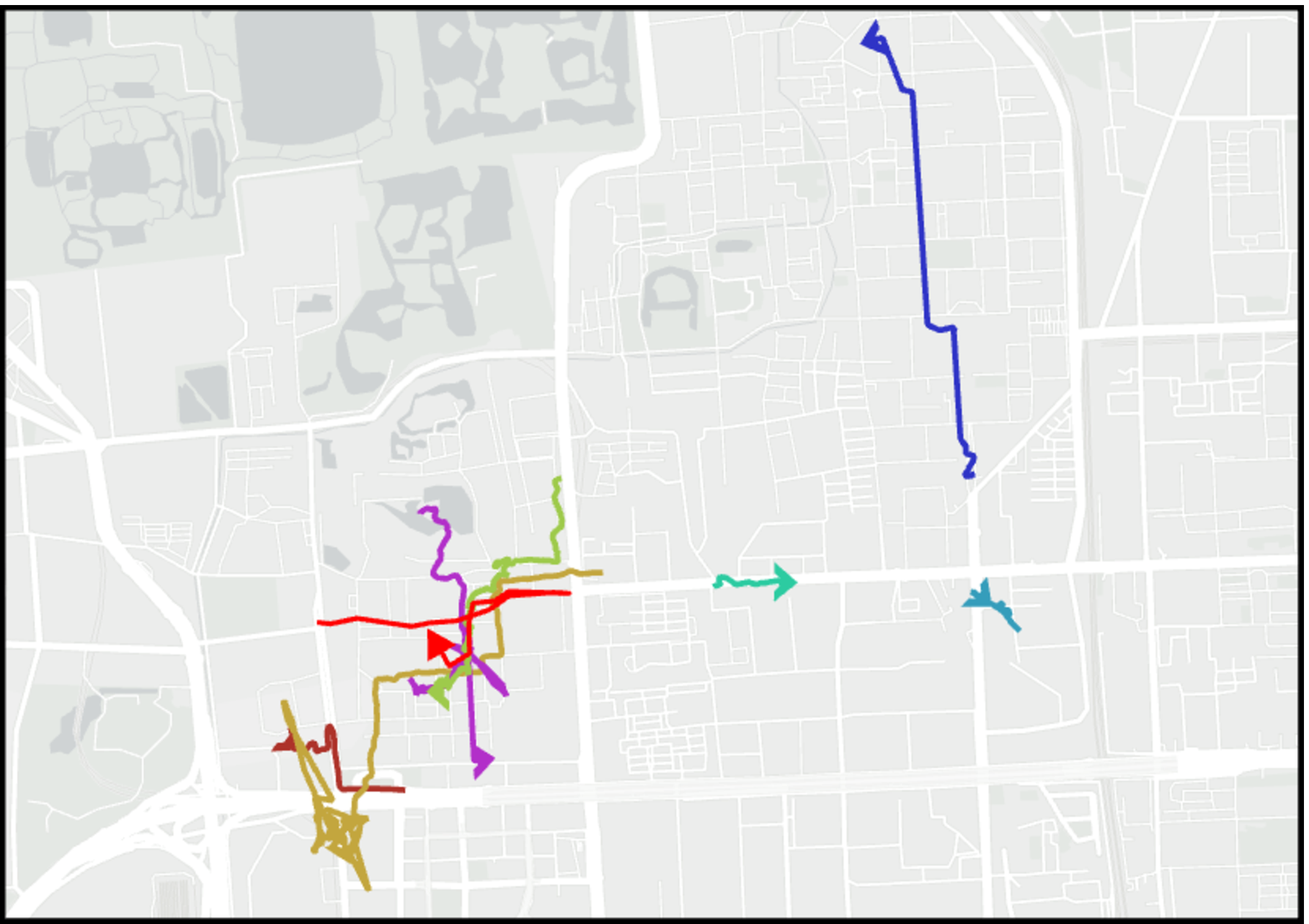}}
	\subfigure[$L$=40\% (Geolife)]{
		\includegraphics[width=0.19\textwidth]{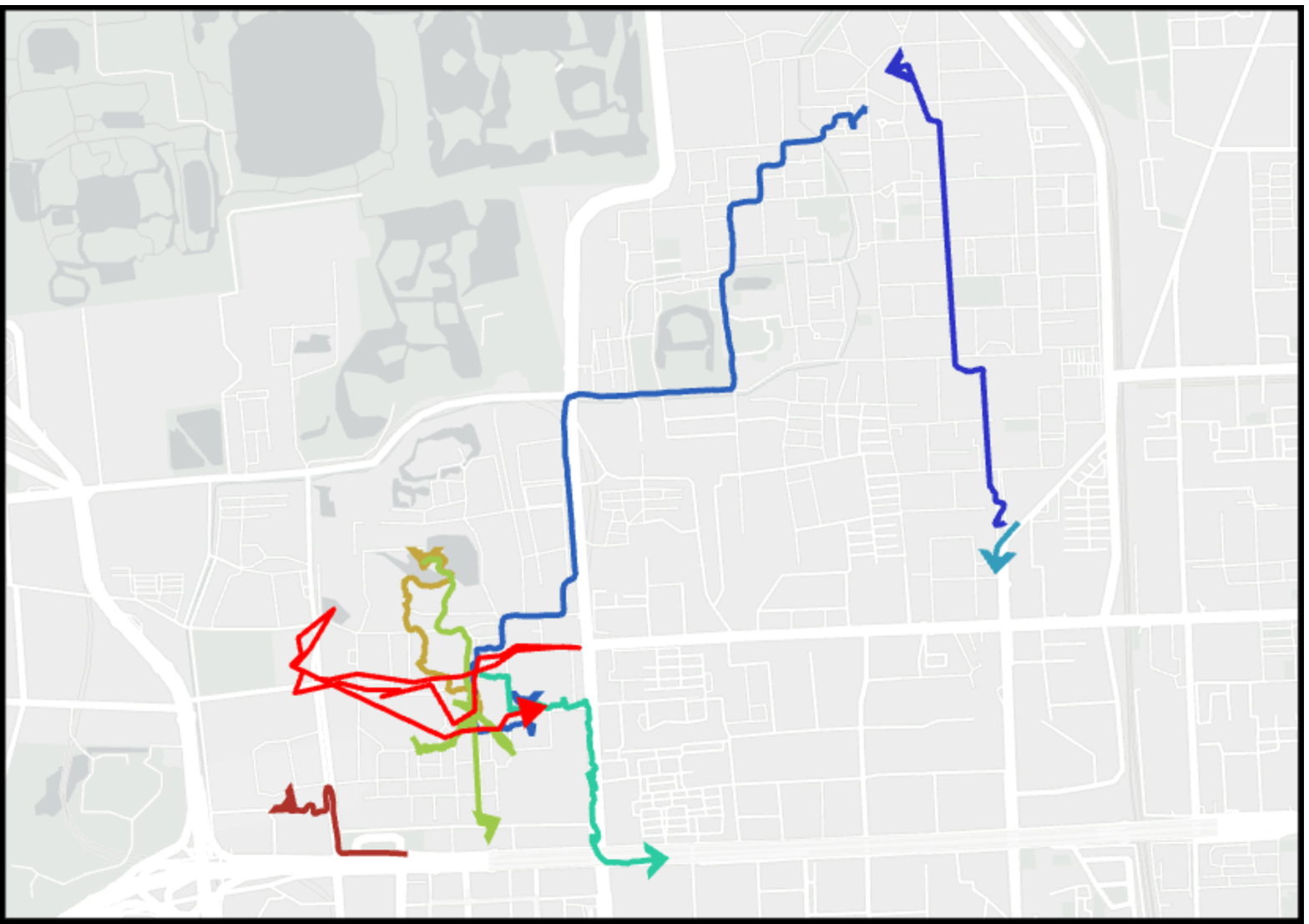}}
	\subfigure[$L$=60\% (Geolife)]{
		\includegraphics[width=0.19\textwidth]{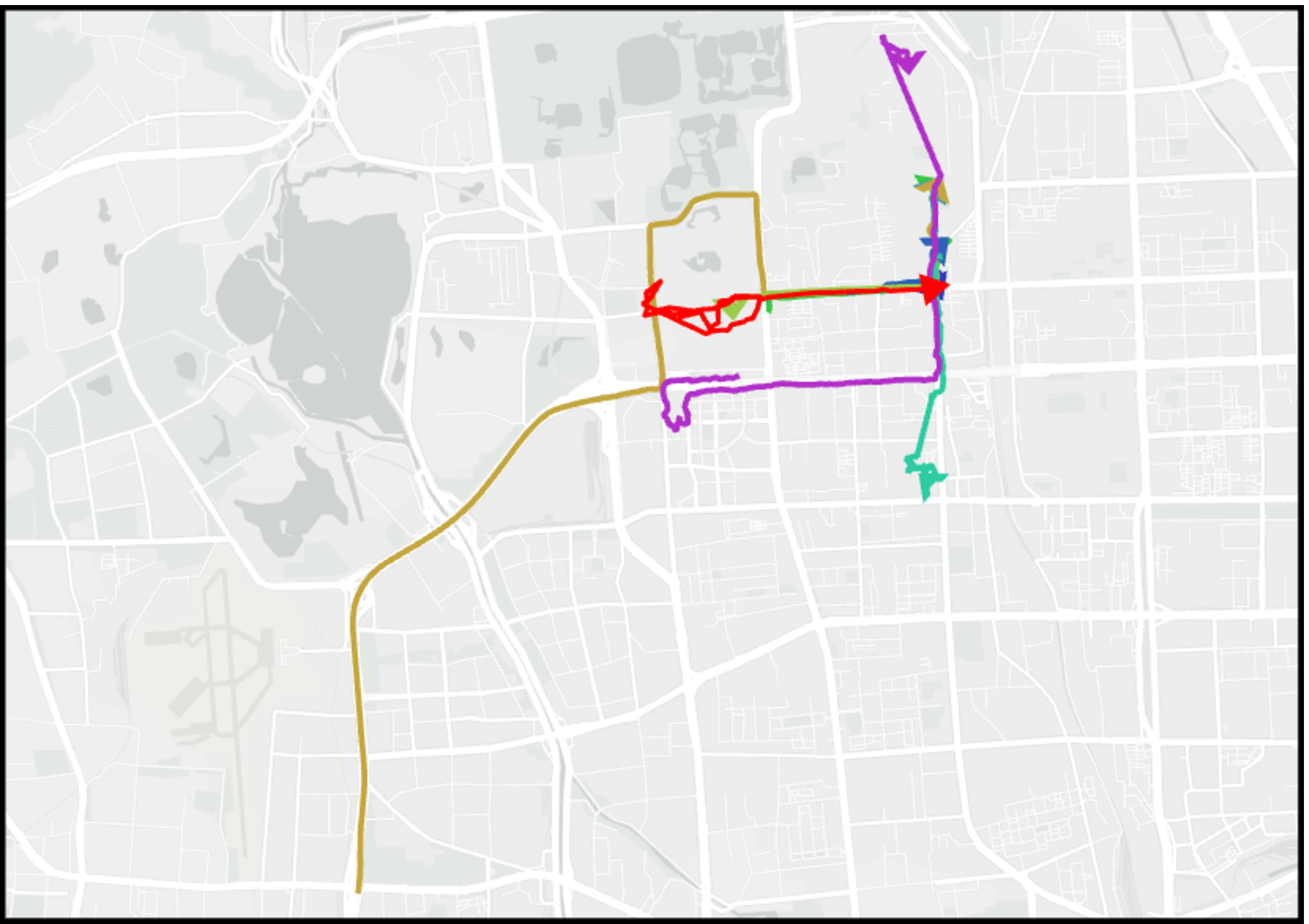}}
	\subfigure[$L$=80\% (Geolife)]{
		\includegraphics[width=0.19\textwidth]{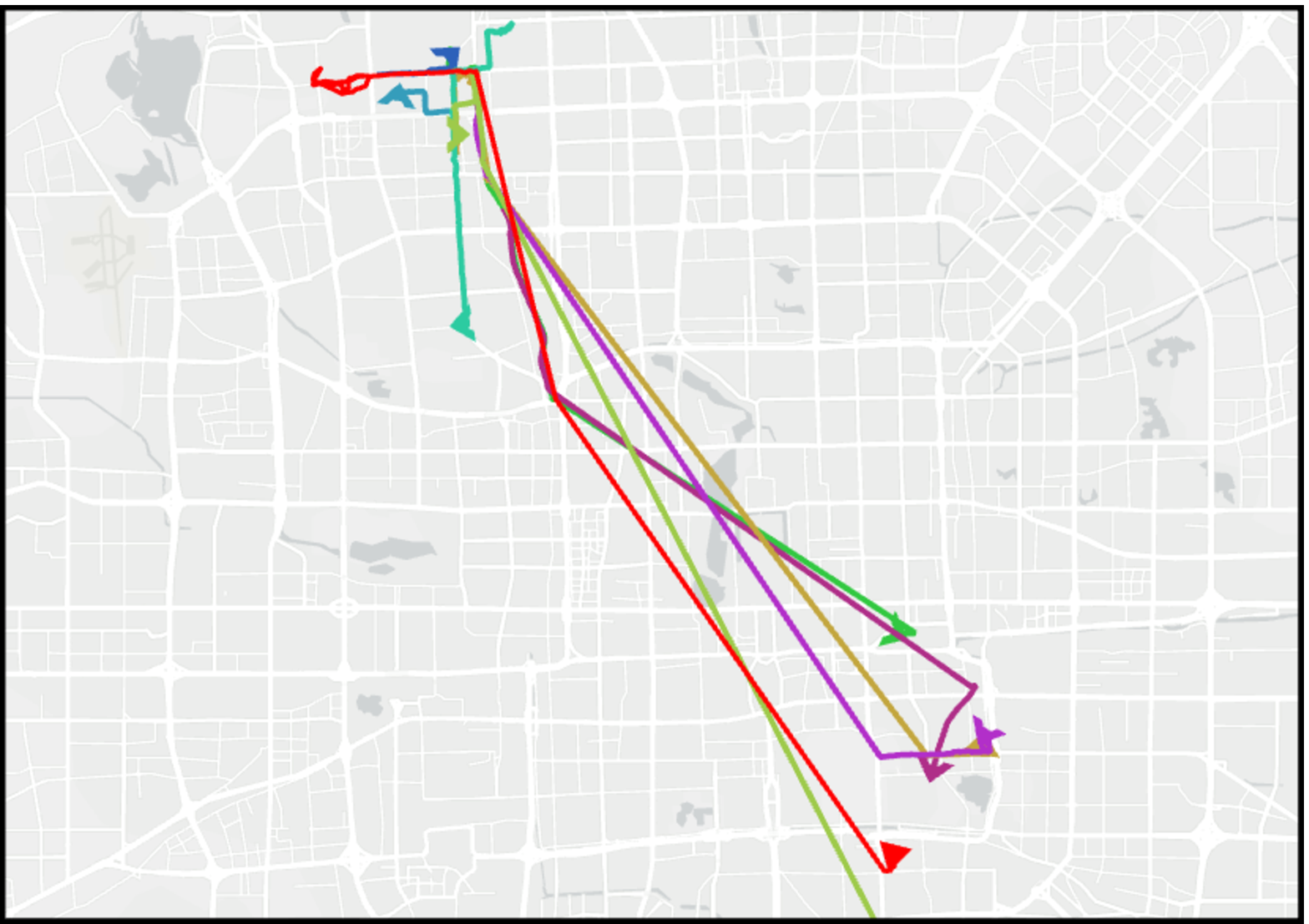}}
	\subfigure[$L$=100\% (Geolife)]{
		\includegraphics[width=0.19\textwidth]{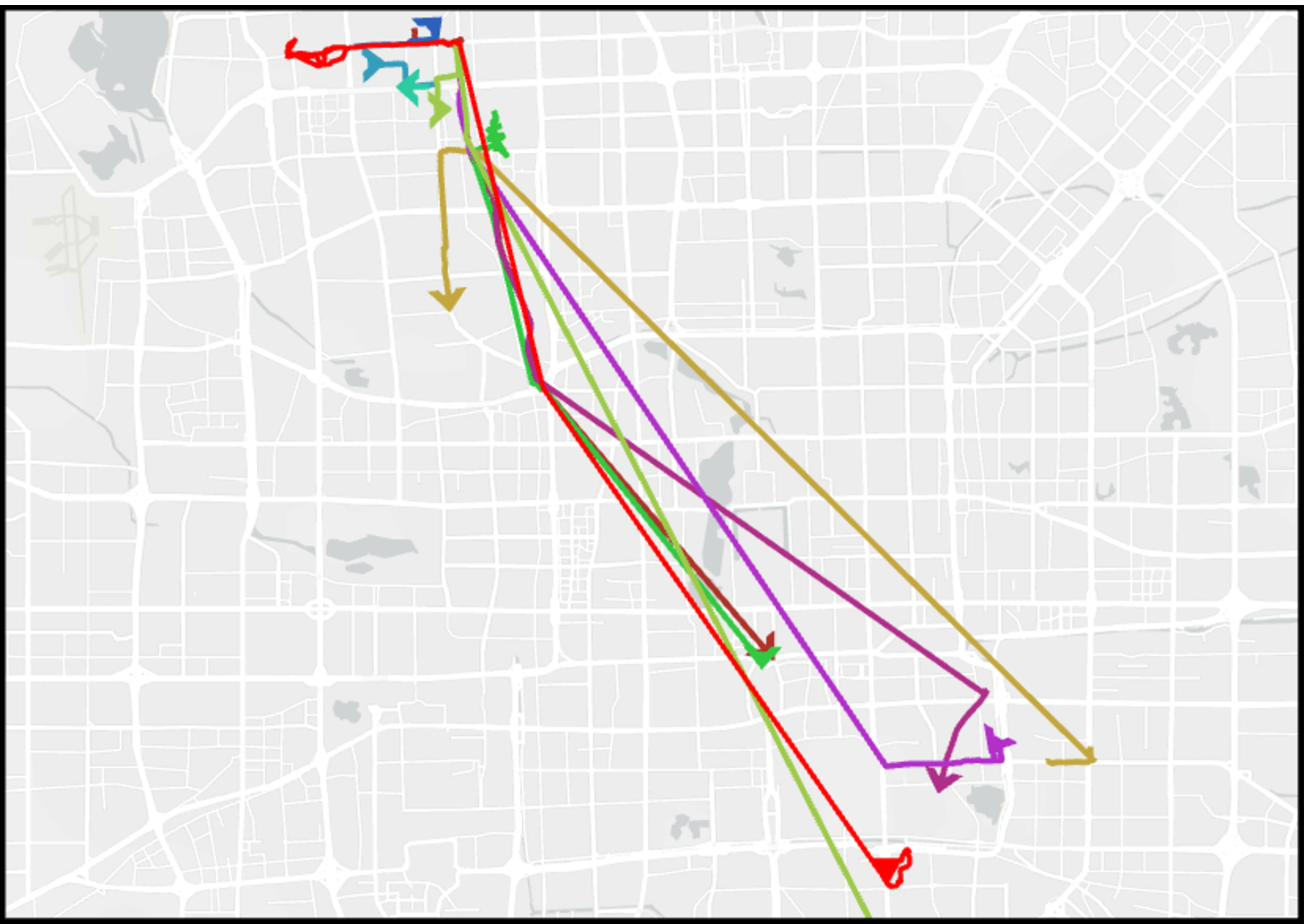}}\\
    \up
	\caption{Effectiveness Study of Non-learning based Measures vs. Trajectory Length} 
	\label{fig:length}
	\vspace{-4mm}
\end{figure*}

\begin{figure*} [tb]
	\centering
	\hspace{-4mm}
	\includegraphics[width=0.6\textwidth]{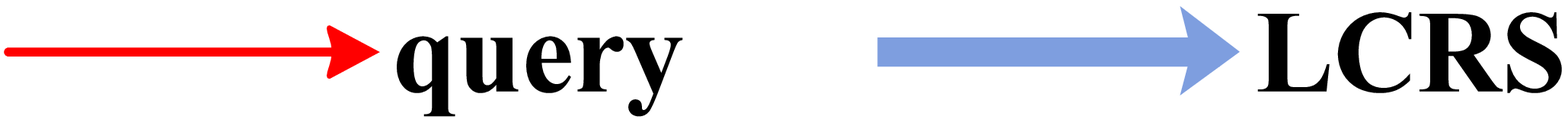}\\
	\hspace{-2mm}
	\subfigure[$QT_{s}$ (T-Drive)]{
		\includegraphics[width=0.23\textwidth]{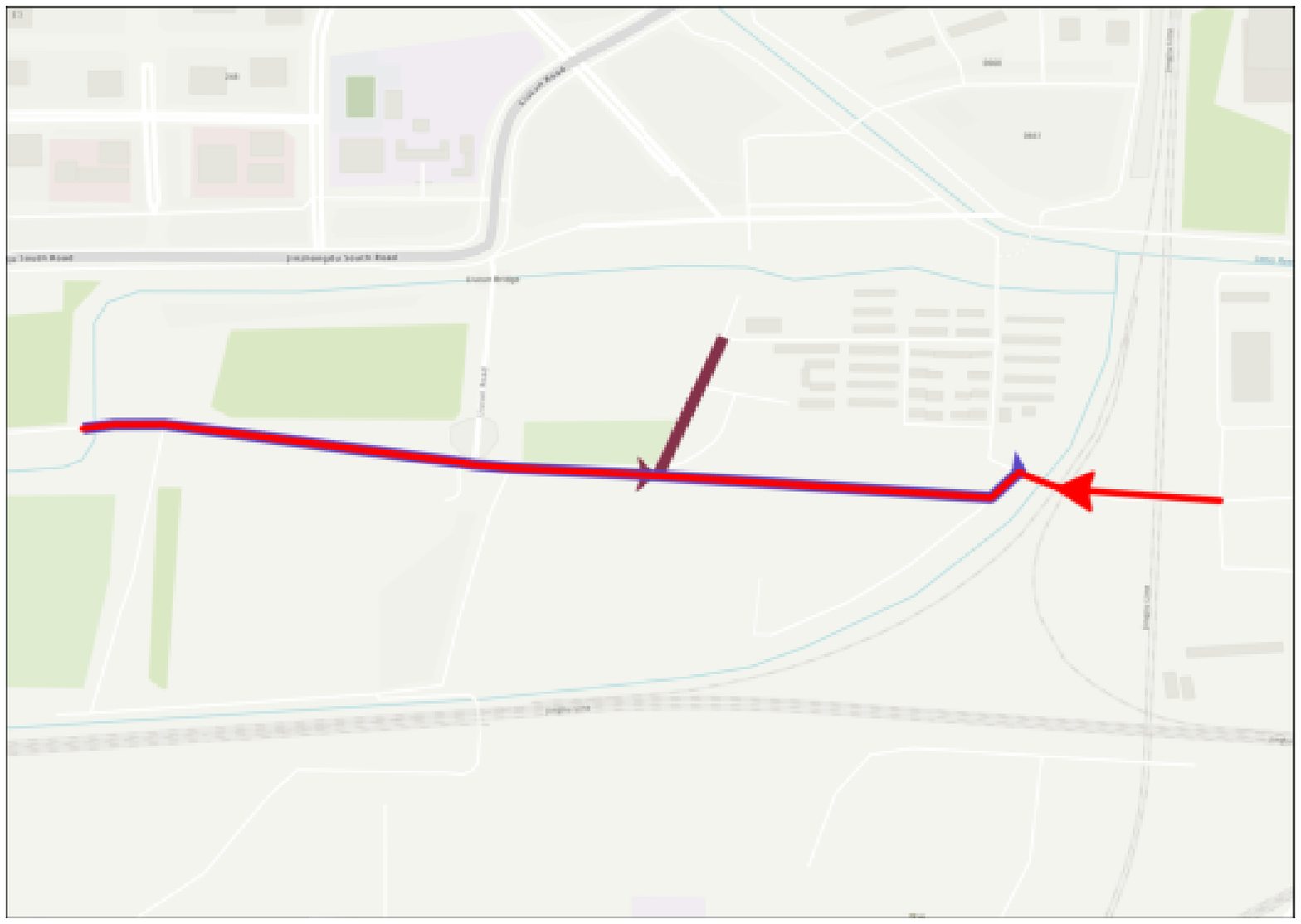}}
	\subfigure[$QT_{o1}$ (T-Drive)]{
	    \includegraphics[width=0.23\textwidth]{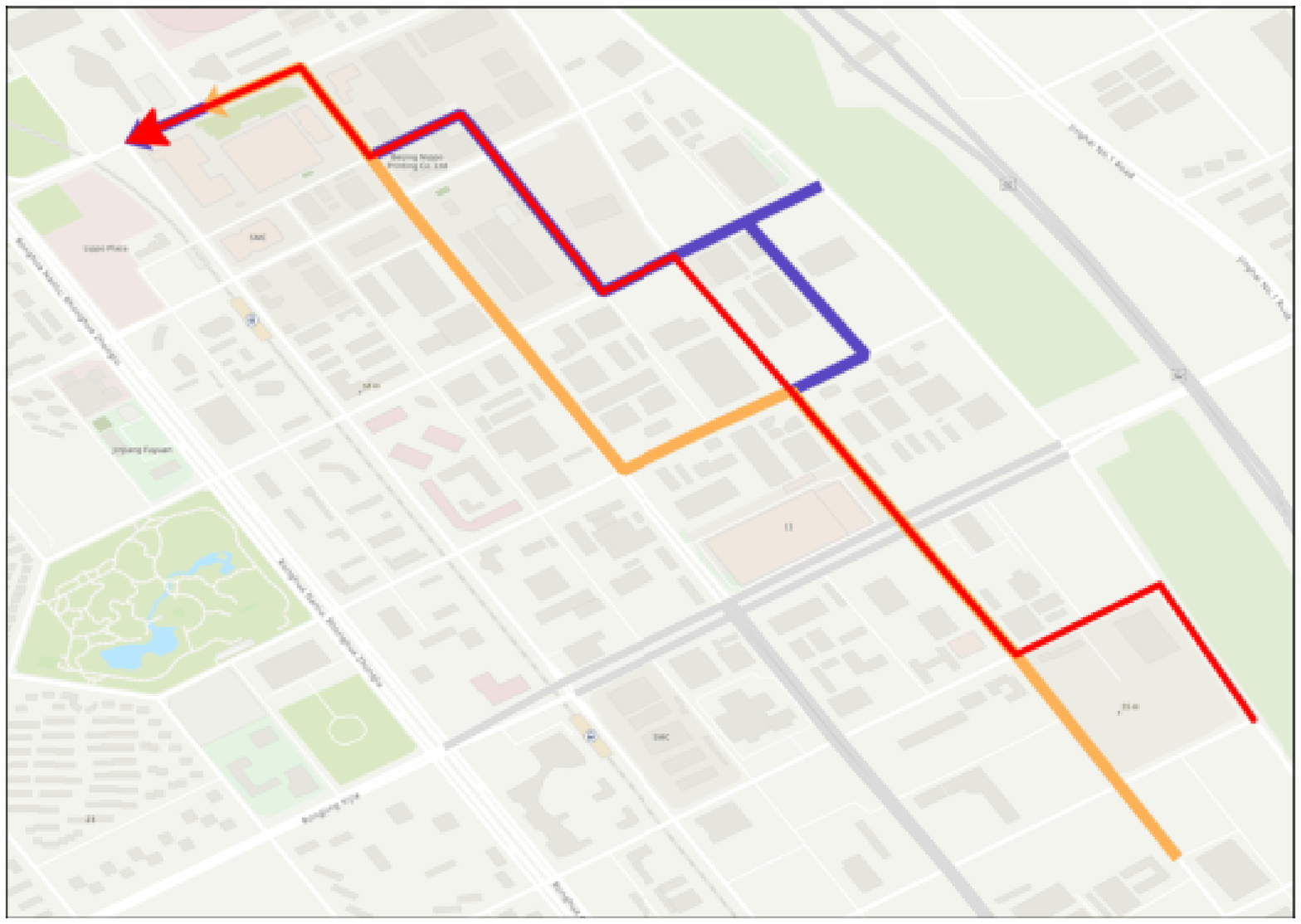}}
	\subfigure[$QT_{o2}$ (T-Drive)]{
		\includegraphics[width=0.23\textwidth]{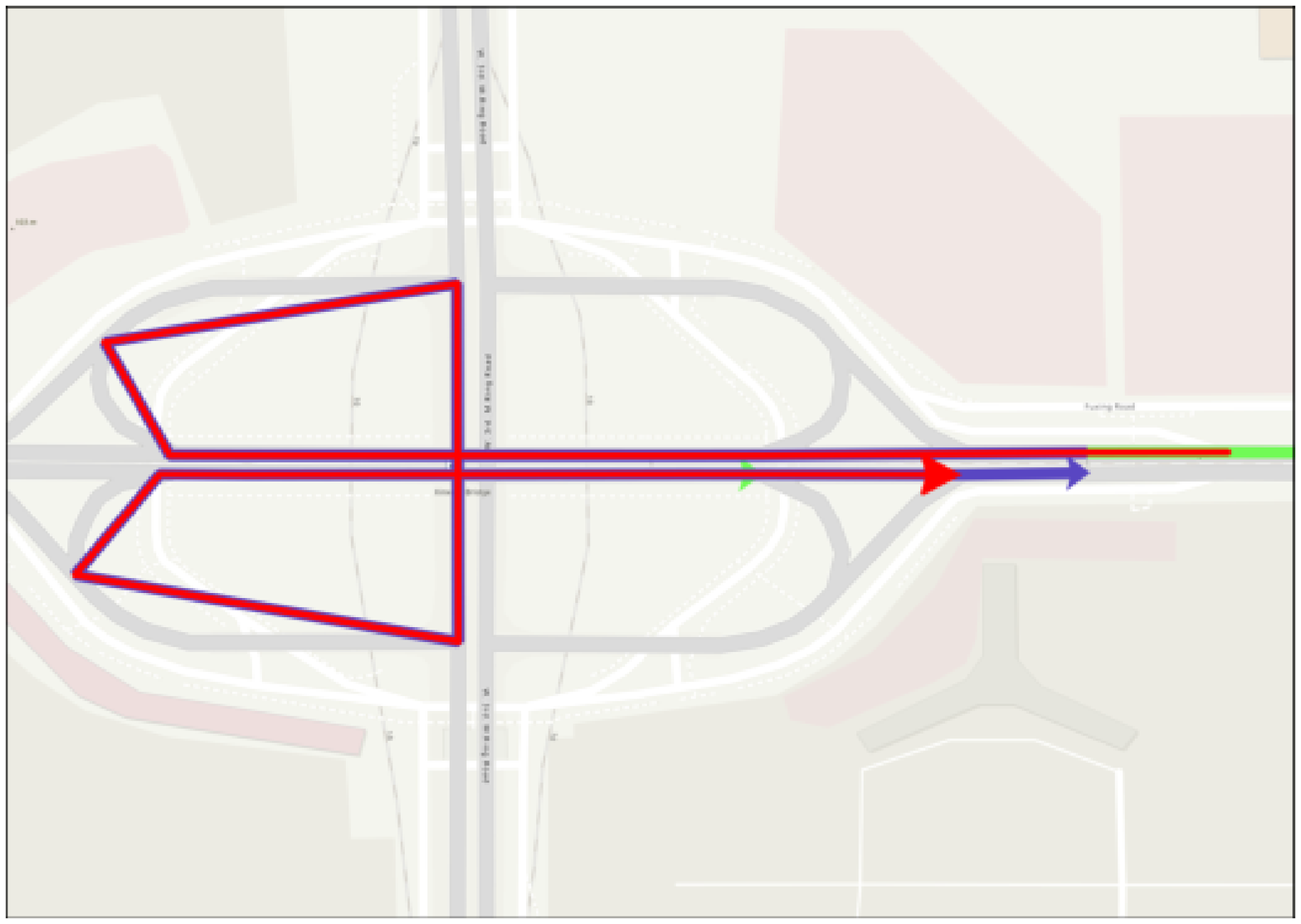}}
	\subfigure[$QT_{r}$ (T-Drive)]{
	    \includegraphics[width=0.23\textwidth]{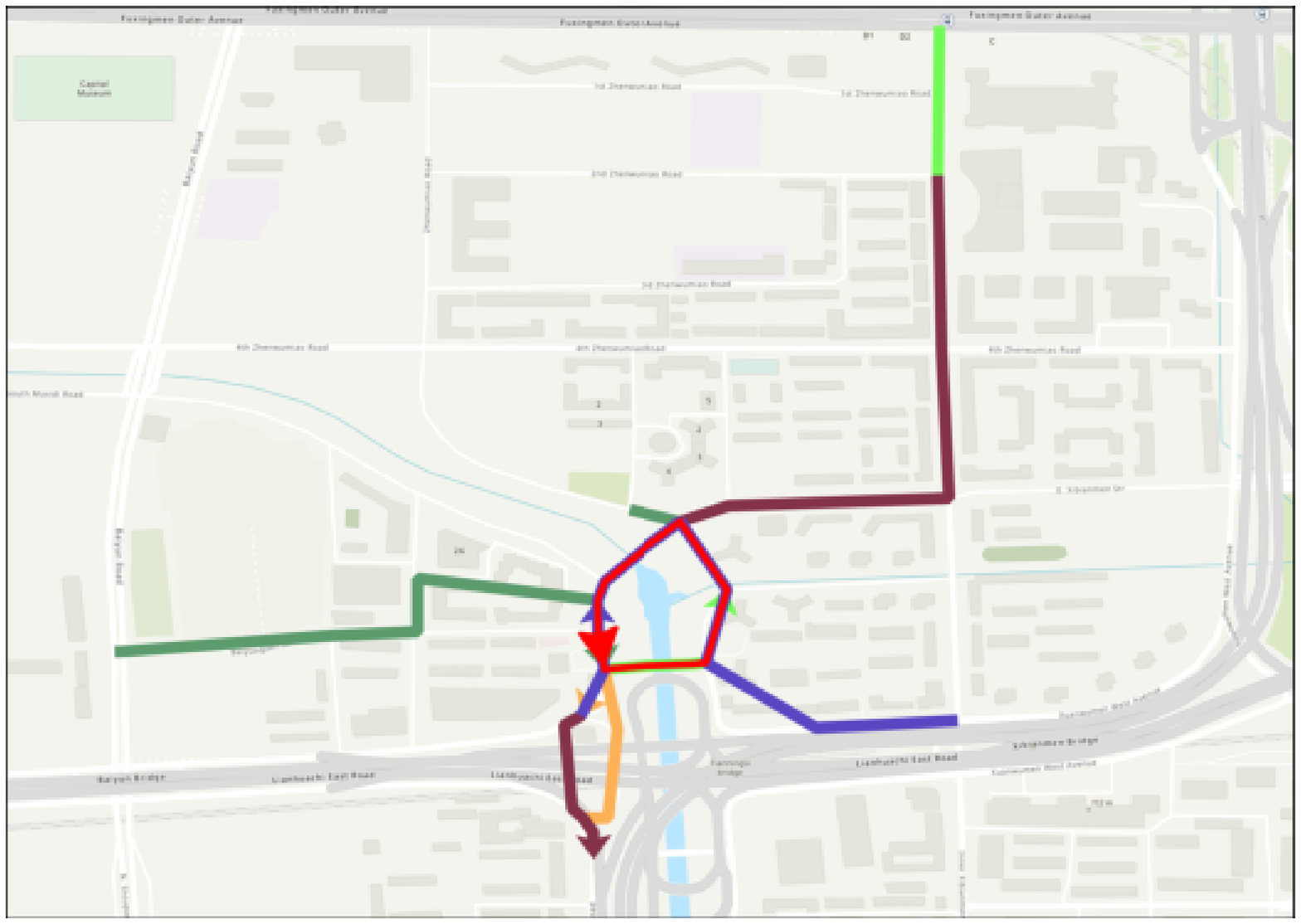}}\\
	
	\hspace{-2mm}
	\subfigure[$QT_{s}$ (Porto)]{
		\includegraphics[width=0.23\textwidth]{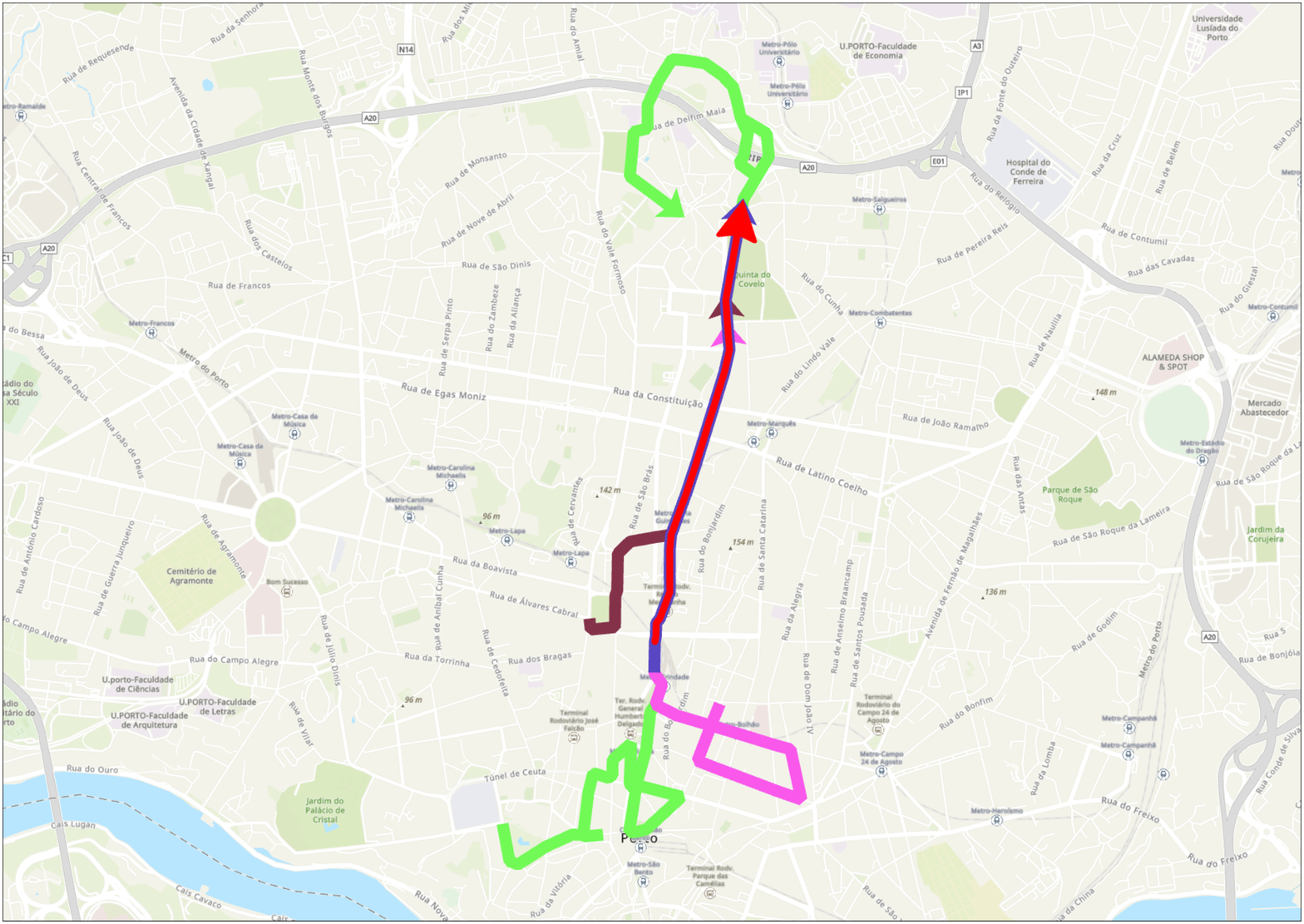}}
	\subfigure[$QT_{o1}$ (Porto)]{
	    \includegraphics[width=0.23\textwidth]{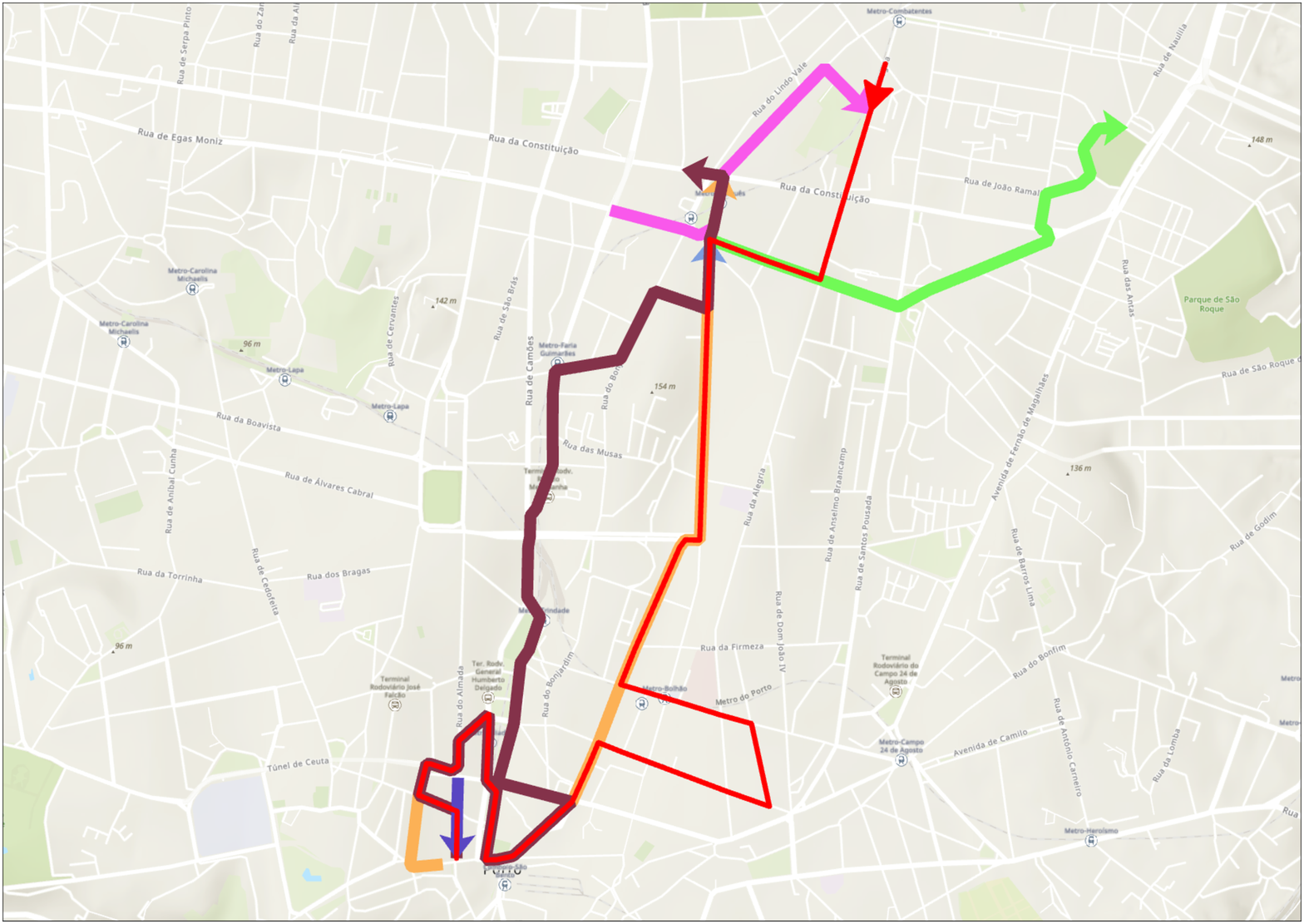}}
	\subfigure[$QT_{o2}$ (Porto)]{
		\includegraphics[width=0.23\textwidth]{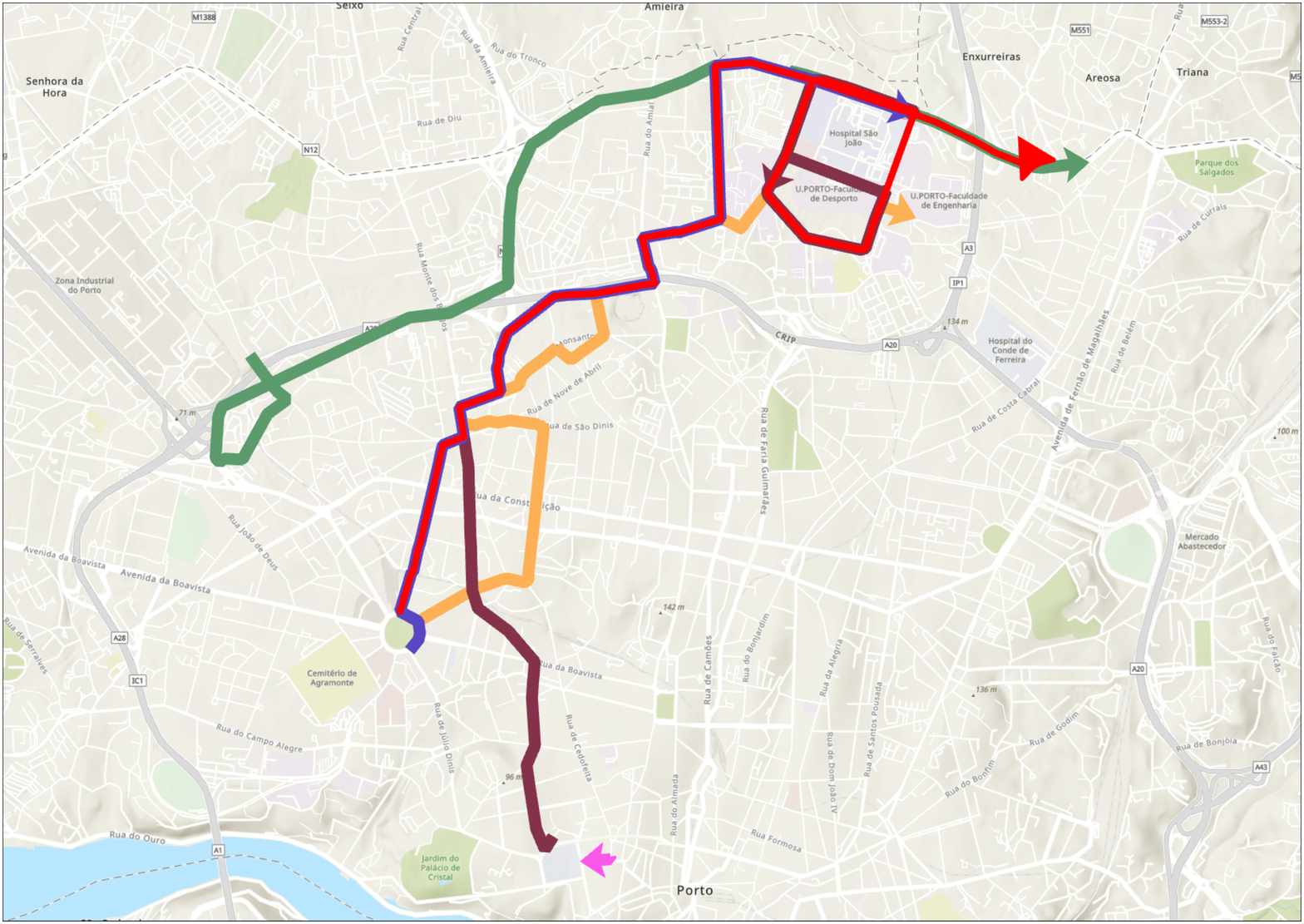}}
	\subfigure[$QT_{r}$ (Porto)]{
	    \includegraphics[width=0.23\textwidth]{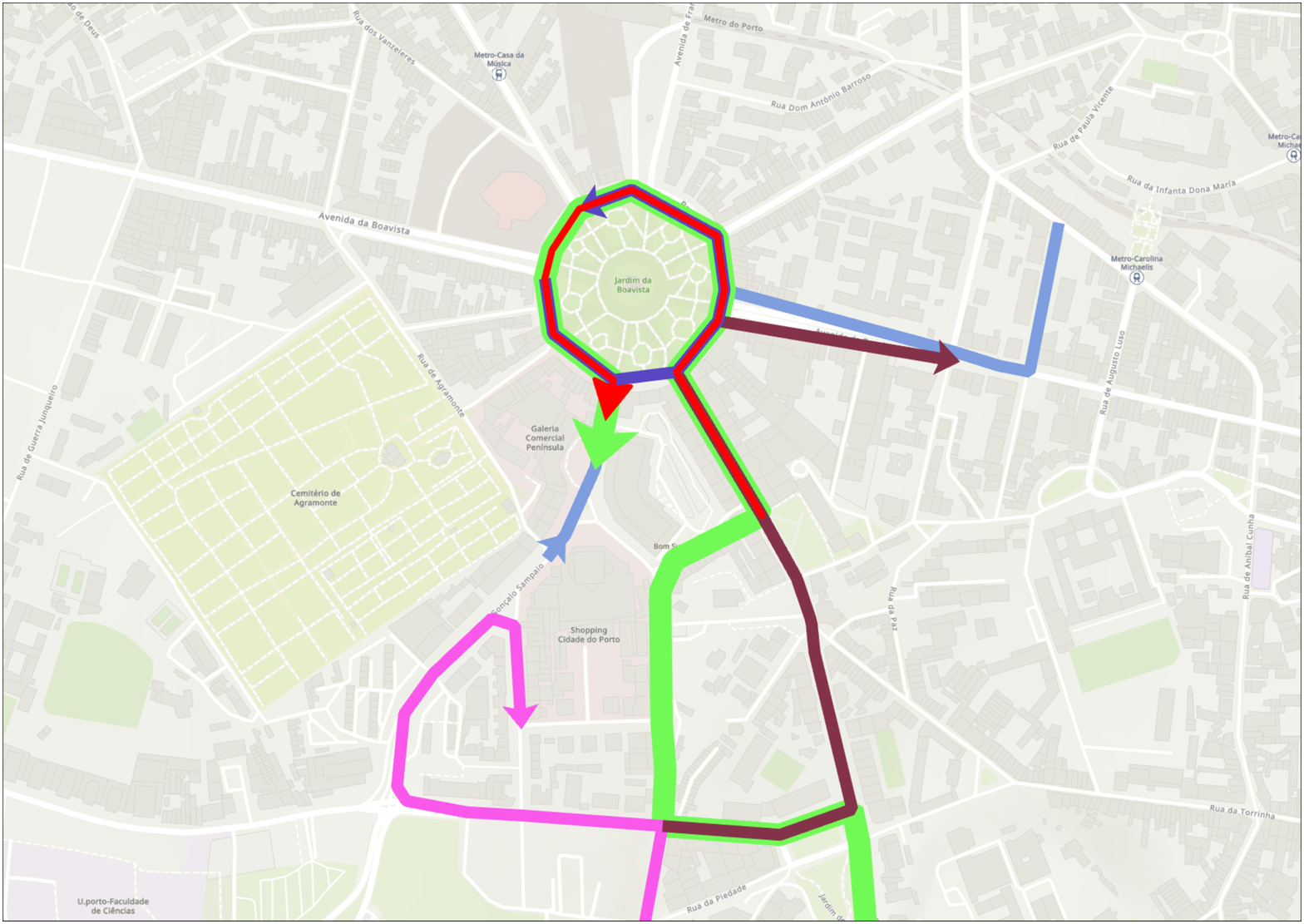}}\\
    \up
	\caption{Effectiveness Study of Non-learning based Measures vs. Shape}
	\label{fig:shape}
	\vspace{-4mm}
\end{figure*}

\begin{figure*} [tb]
	\centering
	\hspace{-2mm}
	\subfigure[AIS (Free Space)]{
		\includegraphics[width=0.23\textwidth]{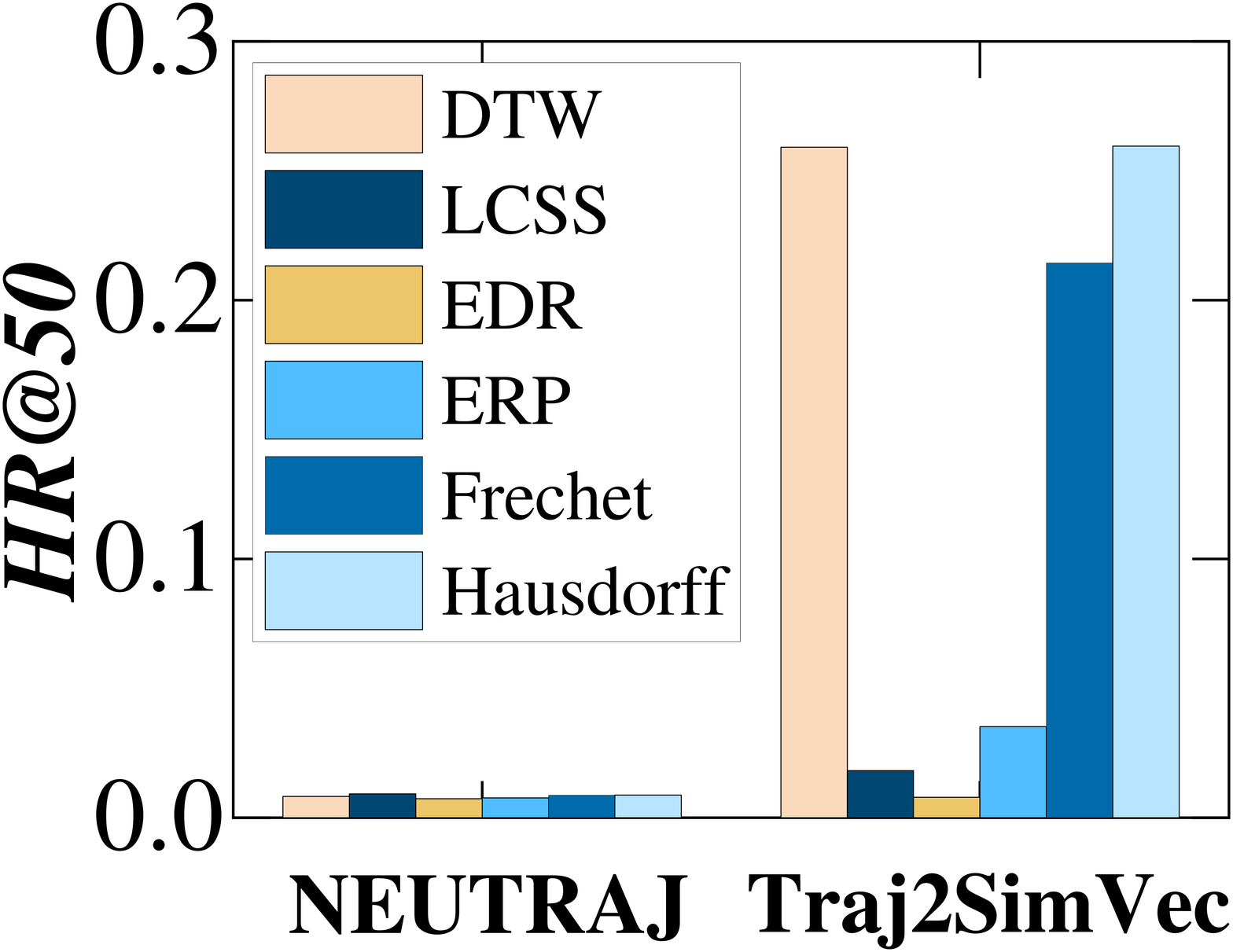}}
	\subfigure[Geolife (Free Space)]{
		\includegraphics[width=0.23\textwidth]{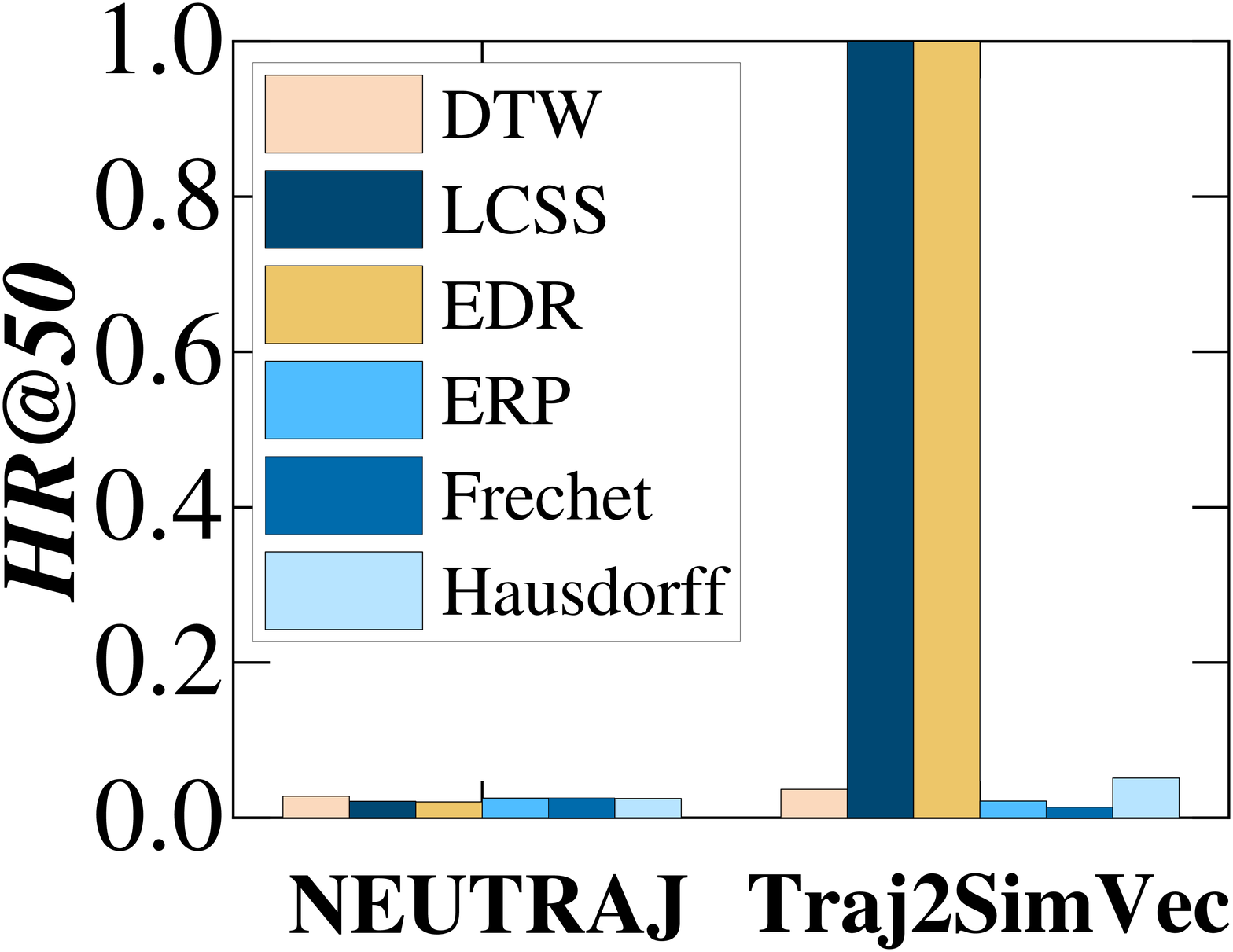}}
	\subfigure[T-Drive (Road Network)]{
		\includegraphics[width=0.23\textwidth]{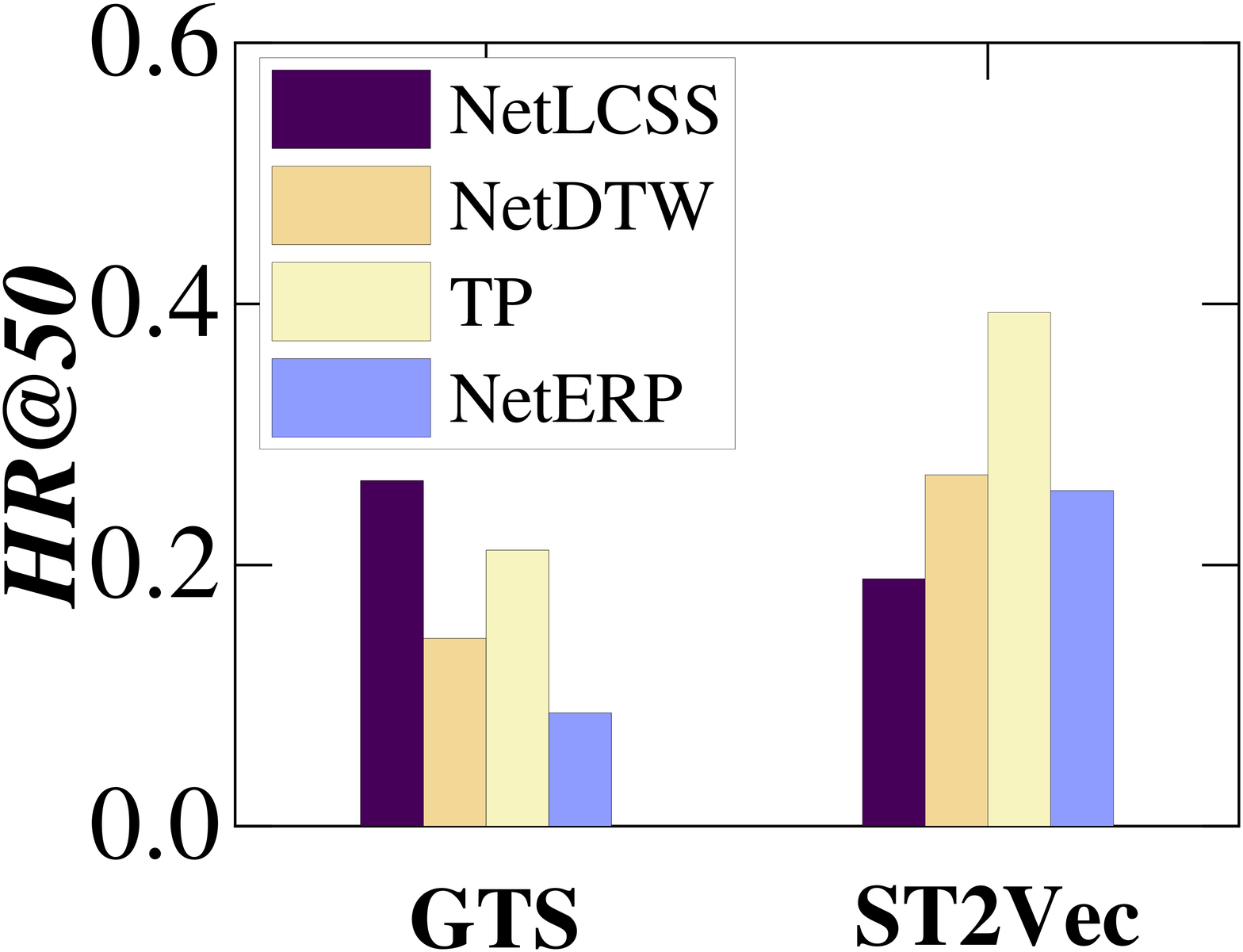}}
	\subfigure[Porto (Road Network)]{
		\includegraphics[width=0.23\textwidth]{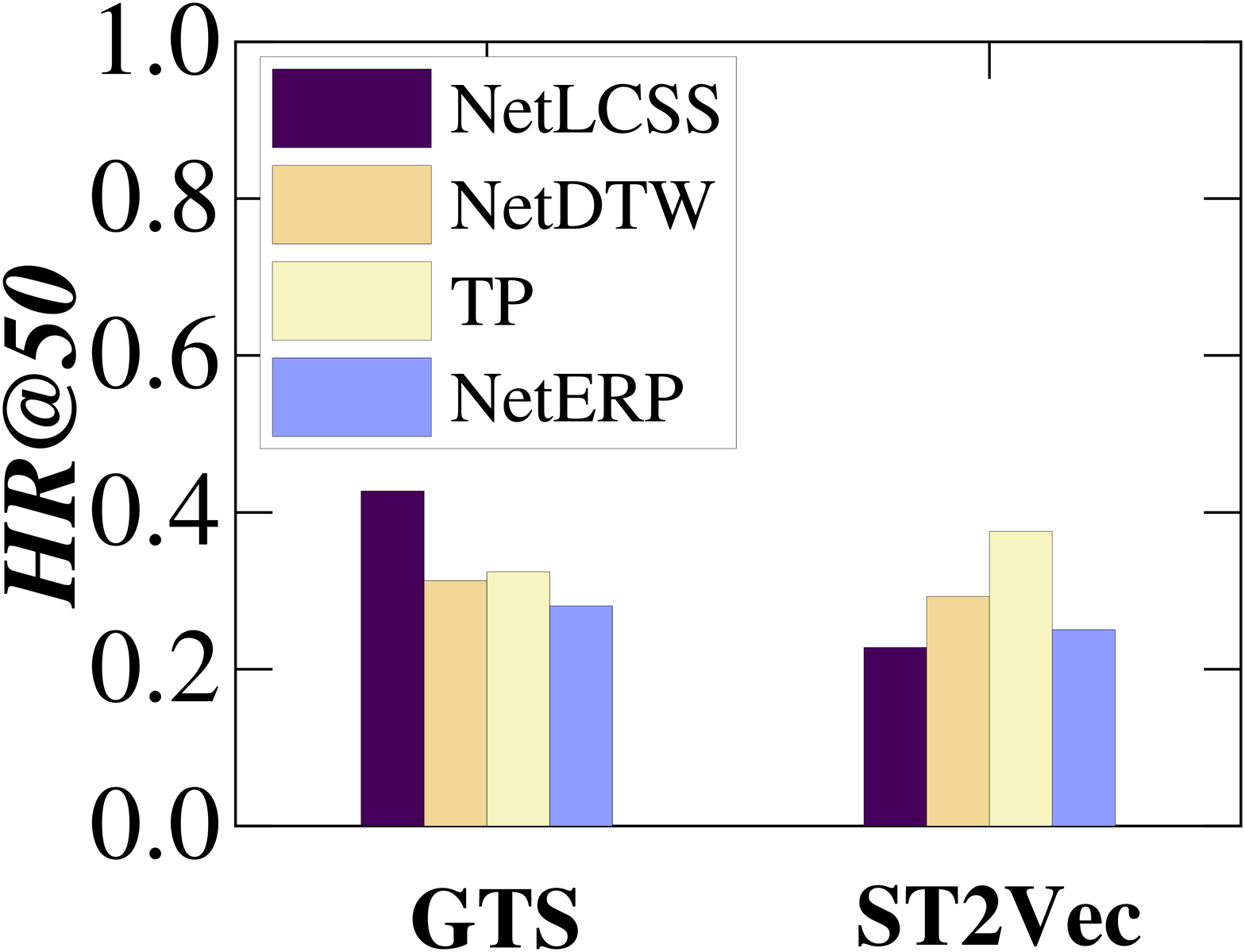}}\\
    \up
	\caption{Effectiveness of Learning-based Measures}
	\label{fig:origin}
	\vspace{-4mm}
\end{figure*}

\textbf{Datasets.} We employ four real-world trajectory datasets: AIS~\cite{AISProject}, Geolife~\cite{Geolife}, T-Drive~\cite{Tdrive}, and Porto~\cite{Porto}, all of which are widely used in previous trajectory similarity studies~\cite{NetERP, LORS, Porto19TKDE, 18t2vec, 19NEUTRAJ, 20Traj2SimVec, 21T3S, ST2Vec, VLDBJSurvey, 21GTS}.
\begin{itemize}\setlength{\itemsep}{-\itemsep}
    \item \textbf{AIS}  records the locations of vessels in U.S. and
international waters. We use the data from Jan. 1 to Dec. 1, 2019. The sample interval varies between 1 and 90 seconds.
  \item \textbf{Geolife} contains about 25 million  GPS points collected from 182 pedestrian in Beijing, from Apr. 2007 to Aug. 2012. The sample interval is 2 seconds. Geolife records different transportation modes of users, and thus, the trajectories' speeds vary significantly.
    \item \textbf{T-Drive} includes 1.5 million GPS points generated by 10,357 taxis in Beijing, from Feb. 2 to Feb. 8, 2008. The sample interval varies between 1 and 177 seconds.
   \item \textbf{Porto} contains 1.7 million GPS points generated by 442 taxis in Porto, Portugal, from Aug. 2011 to Apr. 2012. The sample interval is 15 seconds.
\end{itemize}

The visualization of four datasets is shown in Figure~\ref{fig:dataset}. It suggests that the GPS points collected by AIS have a wider spatial coverage, and are distributed more unevenly than those collected by Geolife.
Since vessels and pedestrian move in free space and taxis move along road networks, we use AIS and Geolife for evaluating free space oriented trajectory measures, and use T-drive and Porto for evaluating road network constrained trajectory measures.
We remove trajectories that contain less than 5 GPS points, and map match~\cite{ST-matching} all the trajectories on T-Drive and Porto to the corresponding road networks extracted from OpenStreetMap~\cite{OSM}. We randomly select 50,000 trajectories for evaluation in the distributed settings. The average lengths of AIS, Geolife, T-Drive, and Porto are 512, 343, 150, and 58, respectively. Due to the high complexity of similarity computation and the limitations of a single machine, we randomly selected 10,000 trajectories out of a total of 50,000 trajectories for evaluation in standalone processing mode.


\vspace{1mm}
\noindent
\textbf{Experimental design.} We verify trajectory similarity measures summarized in Table~\ref{tab:measures} by the evaluation benchmark depicted in Table~\ref{tab:transformation}.

\textit{(i) Non-learning based measures.} We evaluate 17 standalone measures: DTW, LCSS, EDR, EDwP, ERP, Hausdorff, Frechet, LIP, OWD, Seg-Frechet, LORS, TP, NetERP, NetEDR, NetDTW, NetLCSS, and LCRS in terms of  effectiveness, robustness, efficiency, and scalability. However, we evaluate four distributed measures: DFT, DITA, REPOSE, and DISON only in terms of efficiency and scalability. This is because the distributed implementations only improve time performance by techniques such as parallelization and partitioning, without modifying the measures themselves.


\textit{(ii) Learning-based measures.}
Since all learning-based measures, i.e., NEUTRAJ~\cite{19NEUTRAJ}, Traj2SimVec~\cite{20Traj2SimVec}, GTS~\cite{21GTS}, and ST2Vec~\cite{ST2Vec} are standalone-based, we evaluate them in terms of effectiveness, robustness, efficiency, and scalability. Note that, the target of learning-based models is to approximate the inter-trajectory similarity computed by non-learning measures and all existing learning-based measures can only support point-based similarity learning. Therefore, for each learning-based measure, we verify its capability of approximating point-based trajectory similarity measures, including DTW, LCSS, EDR, EDwP, ERP, Hausdorff, and Frechet in free space, and  NetDTW, NetLCSS, NetERP, and TP in road networks.

\vspace{1mm}
\noindent
\textbf{Experimental settings.}
To ensure a fair and objective comparison, we do not implement any additional indexing structures when evaluating the standalone measures, and do not modify the indexing structures inherent in each of the distributed measures during evaluation.

For EDR and LCSS used in free space, we set the similarity threshold used in similarity computation to 30km. For NetEDR and NetLCSS used in road network, we set the similarity threshold to 1km. Moreover, we set the gap point of ERP and NetERP to the centroids of datasets. To vary noise ratios in benchmark settings, we set
$\Delta{\textit{lat}}=\Delta{\textit{lon}}=0.1\degree$ for AIS, $\Delta{\textit{lat}}= \Delta{\textit{lon}}=0.0005\degree$ for Geolife, and $\Delta{\textit{lat}}=0.008\degree$ and $\Delta{\textit{lon}}=0.007\degree$ for T-Drive and Porto.

For standalone measures, we conduct experiments on an 8-core Intel(R) Xeon(R) CPU E5-2640 v4 @ 2.40GHz processor. For distributed measures, we conduct experiments on a cluster consisting of 9 nodes with nine 10-core Intel(R) Xeon(R) CPU E5-2640 v4 @ 2.40GHz processors. Each node ran CentOS7 system with Hadoop 2.6.5 and Spark 2.2.0. For learning-based measures, we set the hyperparameters based on the performance of each model. Specifically, for NEUTRAJ~\cite{19NEUTRAJ}, the batch size is set to 60 and the initial learning rate is set to 0.01. For Traj2SimVec~\cite{20Traj2SimVec}, the batch size is set to 64 and the embedding size of GCN is set to 128. For GTS~\cite{21GTS} and ST2Vec~\cite{ST2Vec}, the batch size is set to 128 and the initial learning rate is set to 0.1. All of the LSTMs used in these measures have hidden sizes of 128. In addition, we split each dataset into a training set, validation set, and test set in a 3:1:6 ratio. All the experiments of learning-based methods were conducted on a server with GeForce RTX 3090, 2.40GHz GPU, and 24GB RAM.

\subsection{Effectiveness Evaluation}
\label{subsec:effective}
We evaluate the effectiveness of each trajectory similarity measure by performing Top-1 similarity queries.
\subsubsection{Non-learning based measures} 

\noindent\textbf{Effect of Lengths.} Figures~\ref{fig:length}(a)--(j) plot the Top-$1$ query results for a given query trajectory $QT$ (denoted by red lines) when the ratio of the length of the $QT$ to its entire length varies from 20\% to 100\%. Here, we only evaluate the free-space oriented measures on AIS and Geolife, because the statics show that the trajectories in free space are much longer than trajectories in road networks. The higher the similarity between the query trajectory and the corresponding Top-$1$ query result by visualization, the higher the effectiveness. The observations are below.

 i) For AIS, results computed by LCSS and Hausdorff are almost identical when $\textit{L=100\%}$; while those on DTW, Frechet, OWD, and Seg-Frechet are almost identical. For Geolife, DTW, EDwP, OWD, Hausdorff, Frechet, and Seg-Frechet  share the similar results with the same transportation mode as $QT$, while only Hausdorff returns the result with opposite direction of it. This is because Hausdorff is an symmetric measure (cf. Definition~\ref{defn:metric}), which does not take into account trajectories' directions. 
ii) When the length of $QT$ varies, DTW, EDwP, Hausdorff, and OWD always return the same results on both datasets. This suggests that they can be flexibly adapted to identify similar trajectories w.r.t. a query trajectory with various lengths.
iii) When the length of $QT$ is small, EDR performs worse than others, i.e., the results identified by EDR are quite different from $QT$ (cf. Figures~\ref{fig:length}(a)--(c) and \ref{fig:length}(f)--(g)). The reason is that the costs of different edit operations in EDR computation are the same, making it sensitive to trajectory length (i.e., the number of trajectory points). Specifically, EDR always finds the query result whose length is similar with query trajectory $QT$, even the result is spatially distant with $QT$.
iv) The results computed by LIP vary significantly with the length of $QT$. This is because the results highly relies on the shapes of polygons formed by intersection points of $QT$ and a to-be-examined trajectory, while the shapes are obviously affected by the length of $QT$.
v) Finally, results identified by all the measures on AIS are generally better (i.e., more similar to $QT$) than those on Geolife. This is because each trajectory in Geolife generally contains multiple transportation modes and tends to exhibit various features.

\begin{figure*} [tb]
	\centering
	\includegraphics[width=0.35\textwidth]{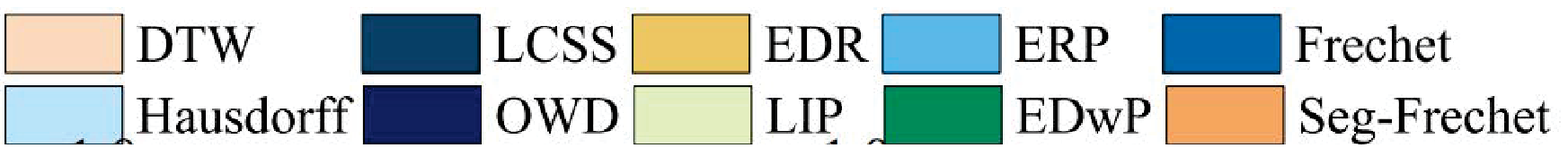}\\
	\vspace{-1mm}
	\hspace{-6mm}
	\subfigure[$S$=10\% AIS]{
		\includegraphics[width=0.16\textwidth]{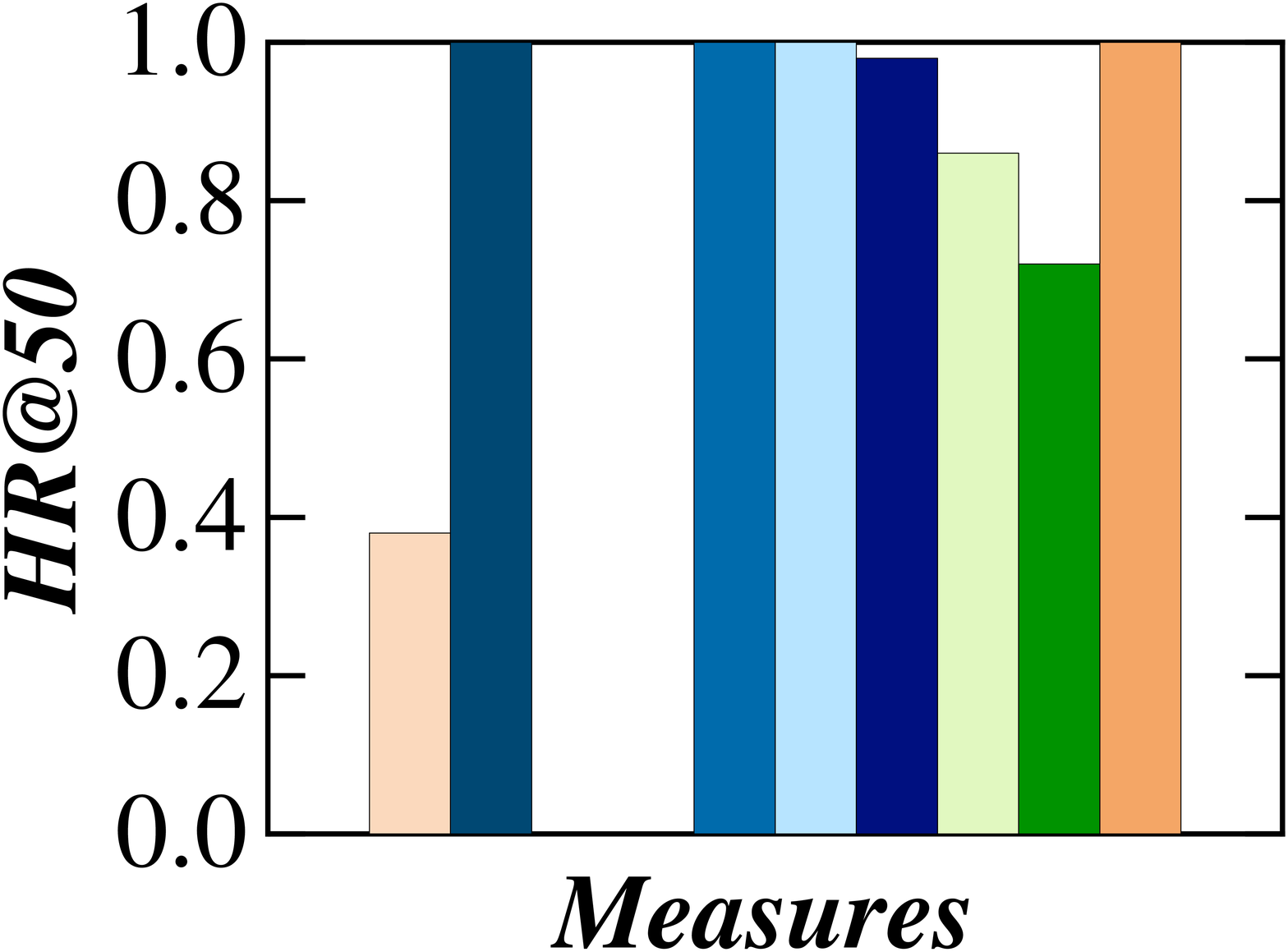}}
	\subfigure[$S$=20\% AIS]{
		\includegraphics[width=0.16\textwidth]{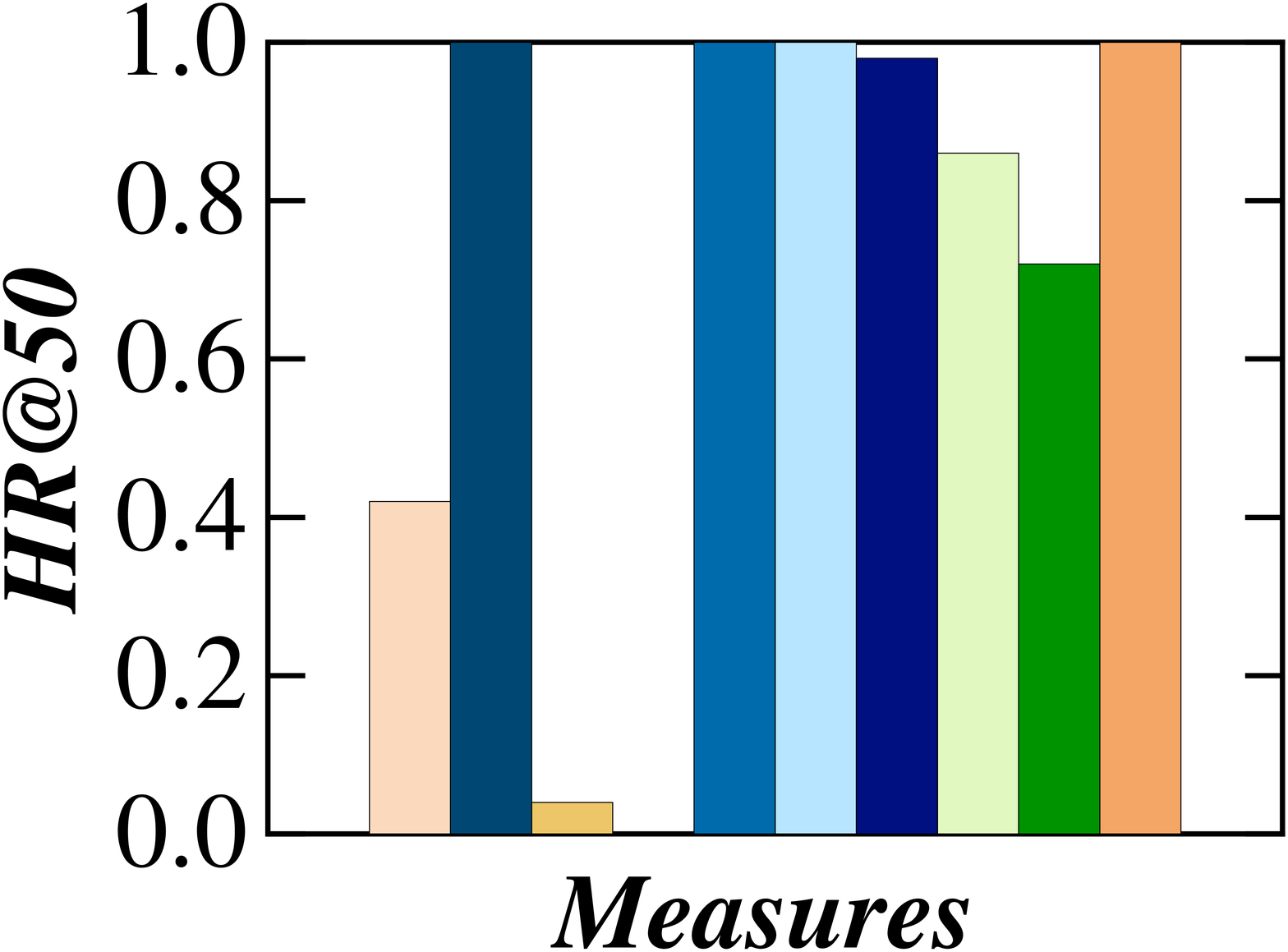}}
	\subfigure[$S$=40\% AIS]{
		\includegraphics[width=0.16\textwidth]{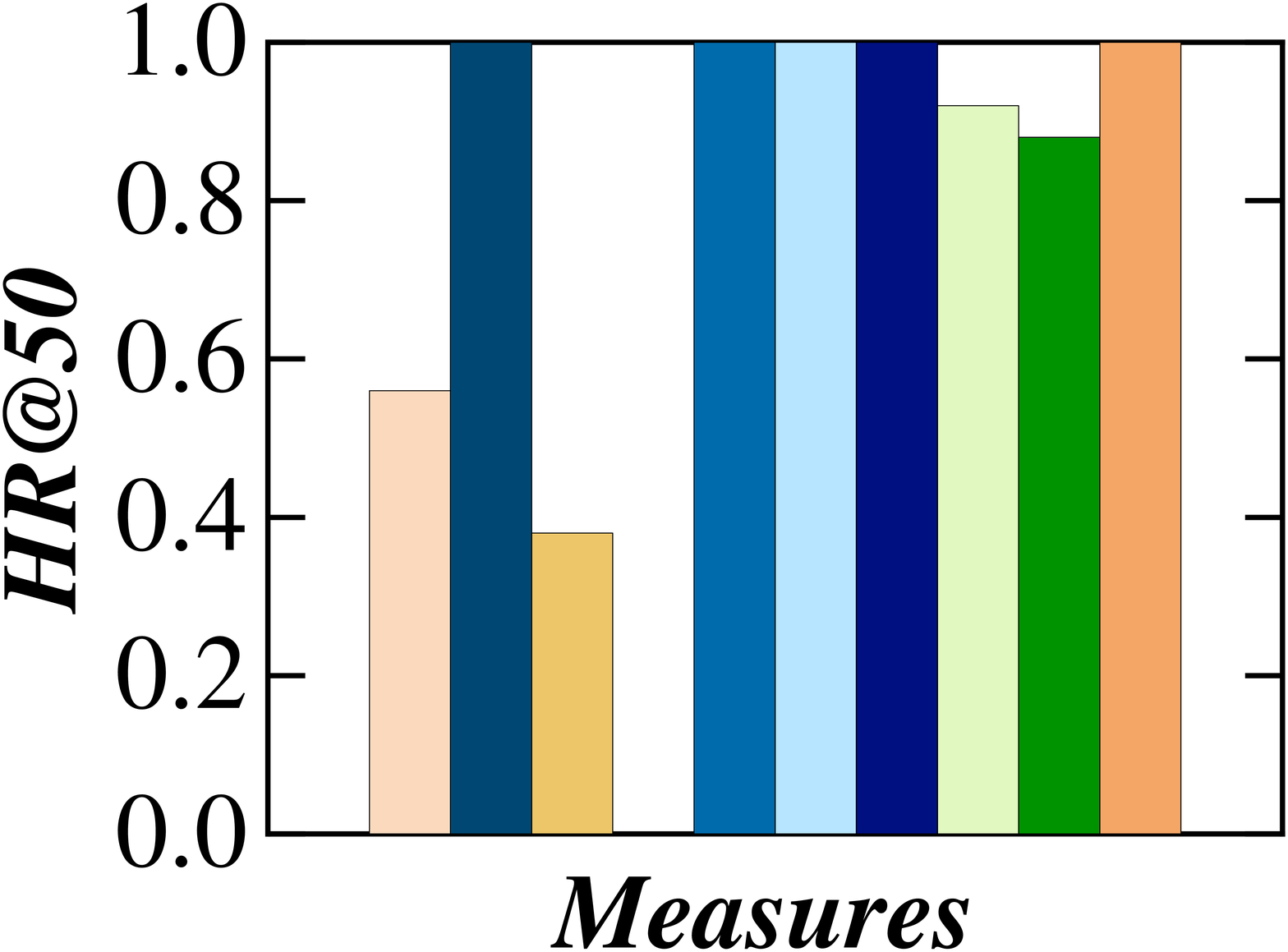}}
	\subfigure[$S$=10\% Geolife]{
		\includegraphics[width=0.16\textwidth]{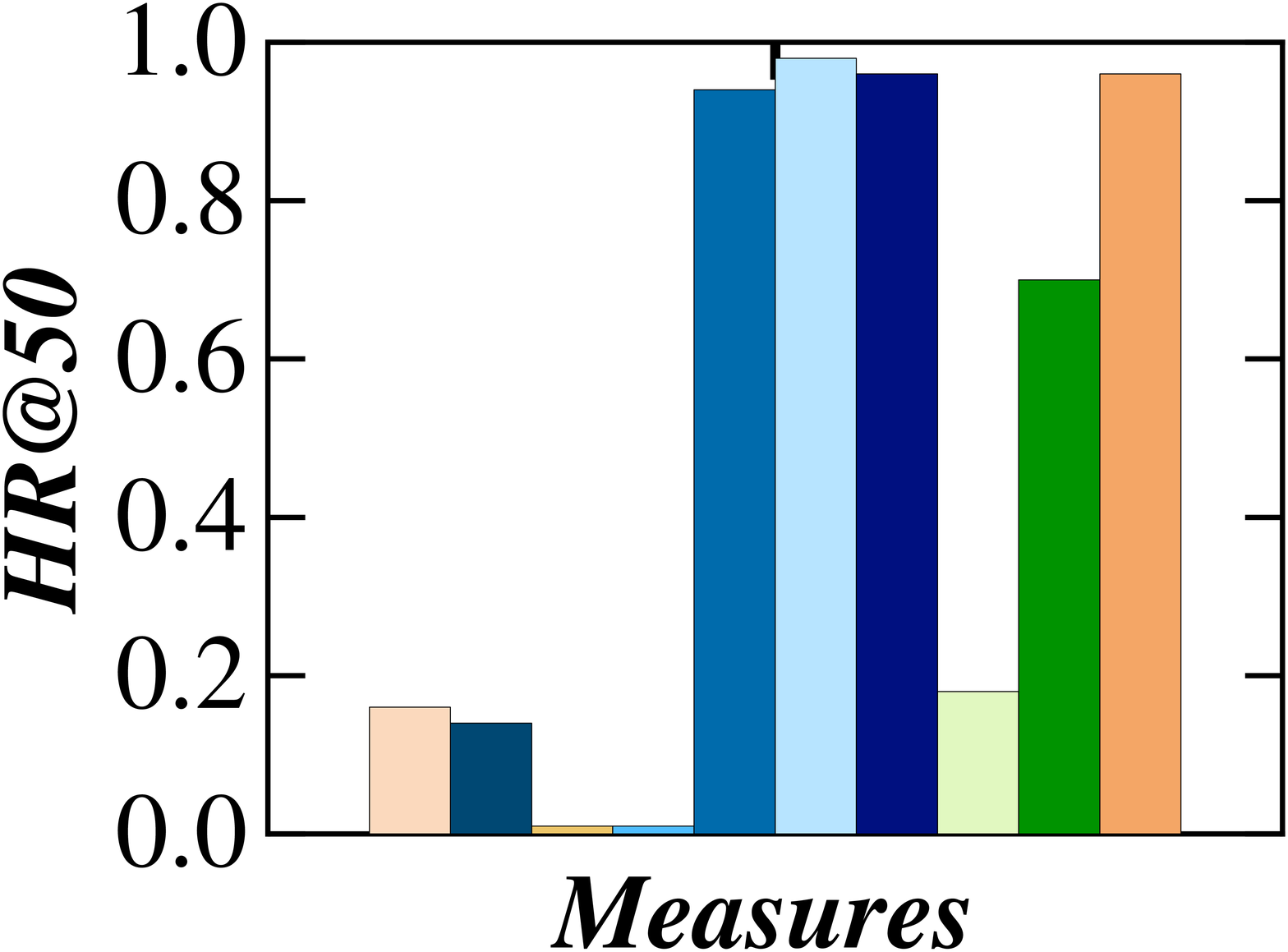}}
	\subfigure[$S$=20\% Geolife]{
		\includegraphics[width=0.16\textwidth]{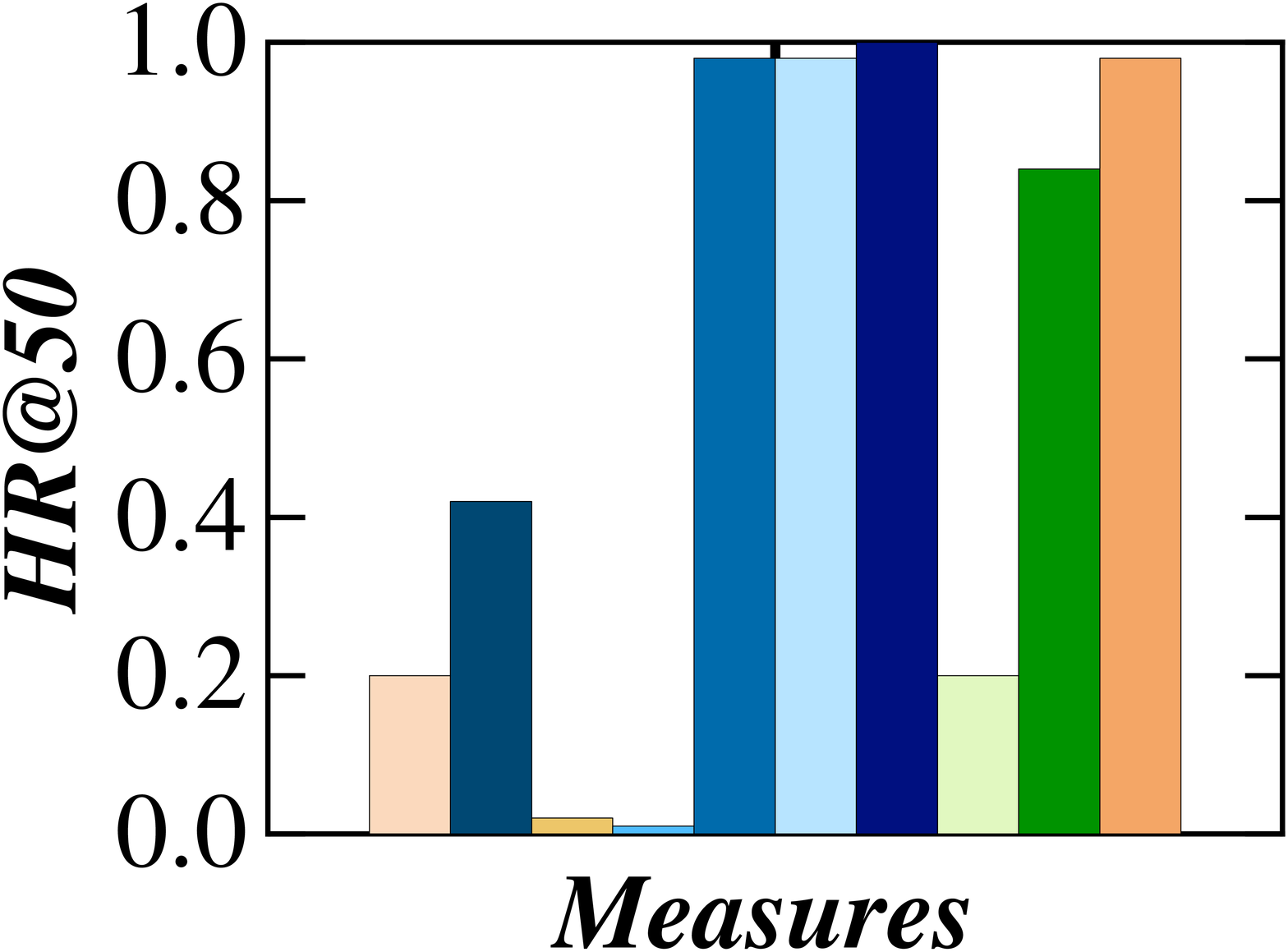}}
	\subfigure[$S$=40\% Geolife]{
		\includegraphics[width=0.16\textwidth]{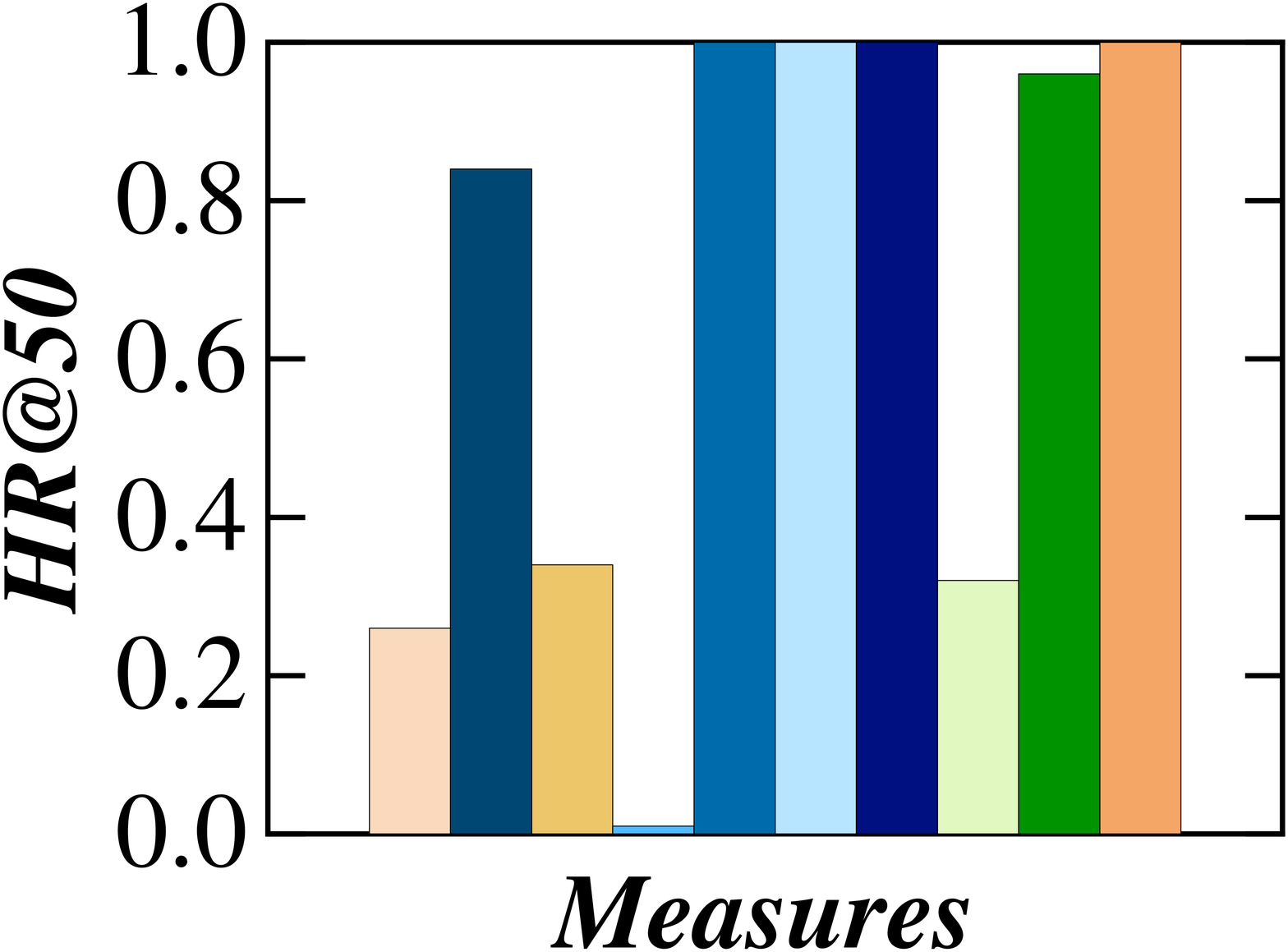}}\\
    \up
	\caption{Robustness of Non-learning Measures vs. Sampling Rate}
	\label{fig:SR}
	\vspace{-2mm}
\end{figure*}

\begin{figure*} [tb]
	\centering
	\includegraphics[width=0.6\textwidth]{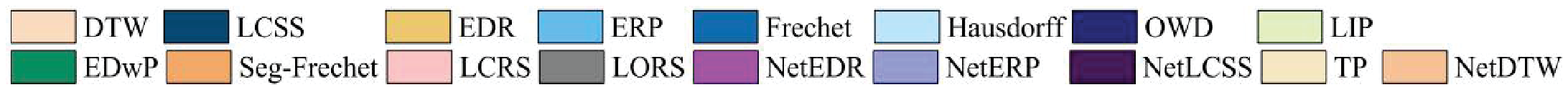}\\
	\hspace{-4mm}
	\subfigure[$N$=13\% AIS]{
		\includegraphics[width=0.16\textwidth]{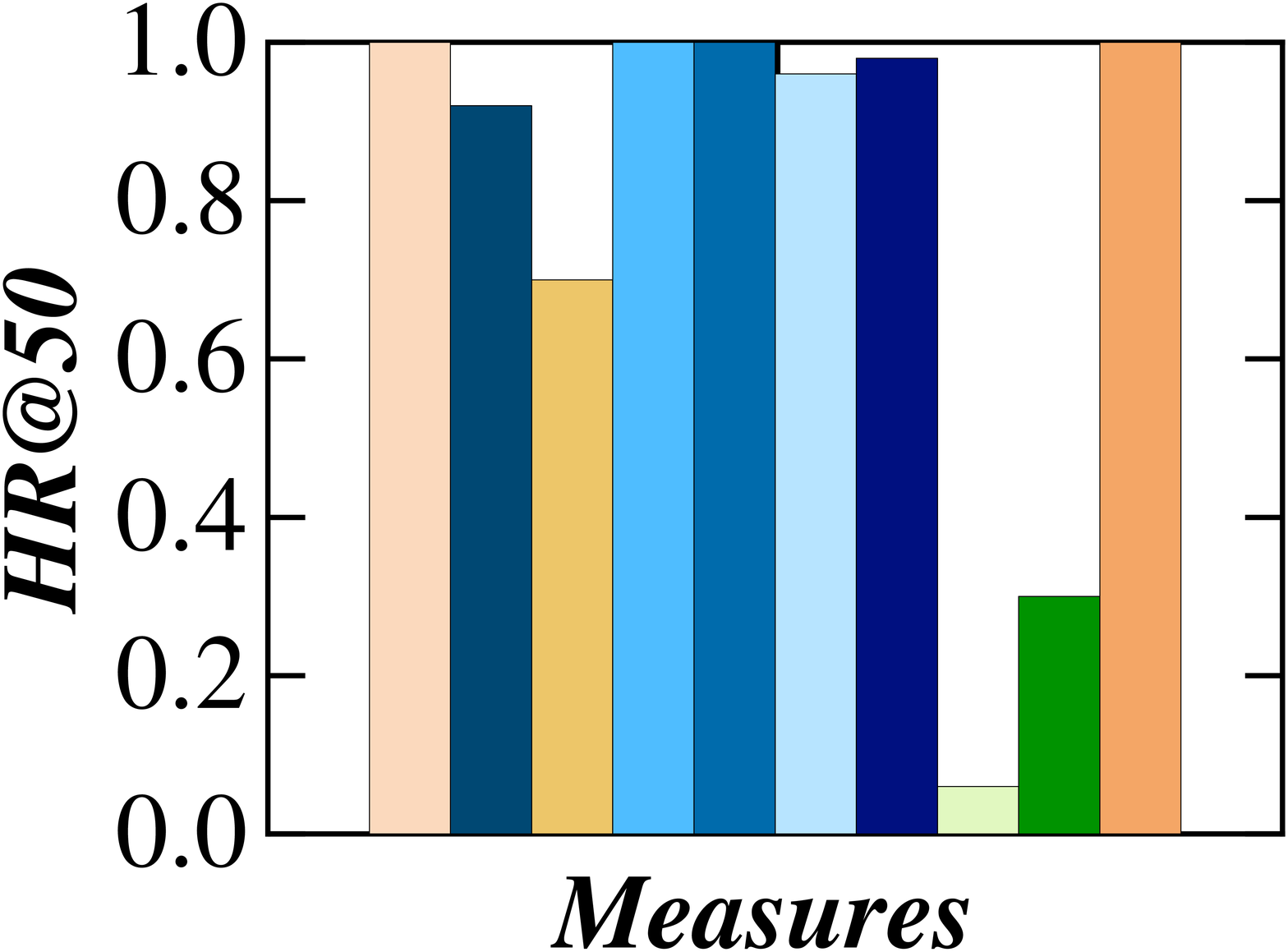}}
	\subfigure[$N$=16\% AIS]{
		\includegraphics[width=0.16\textwidth]{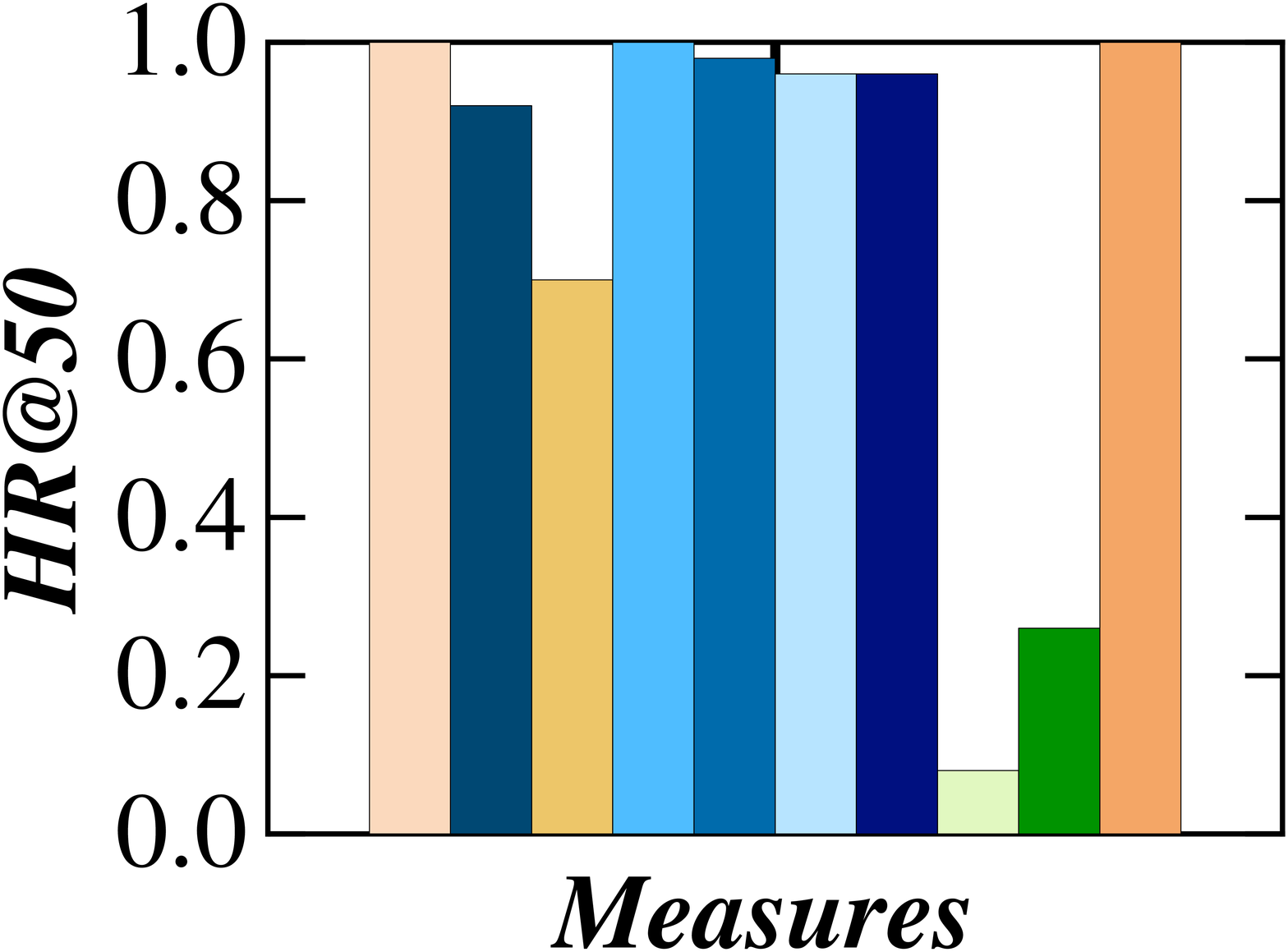}}
	\subfigure[$N$=19\% AIS]{
		\includegraphics[width=0.16\textwidth]{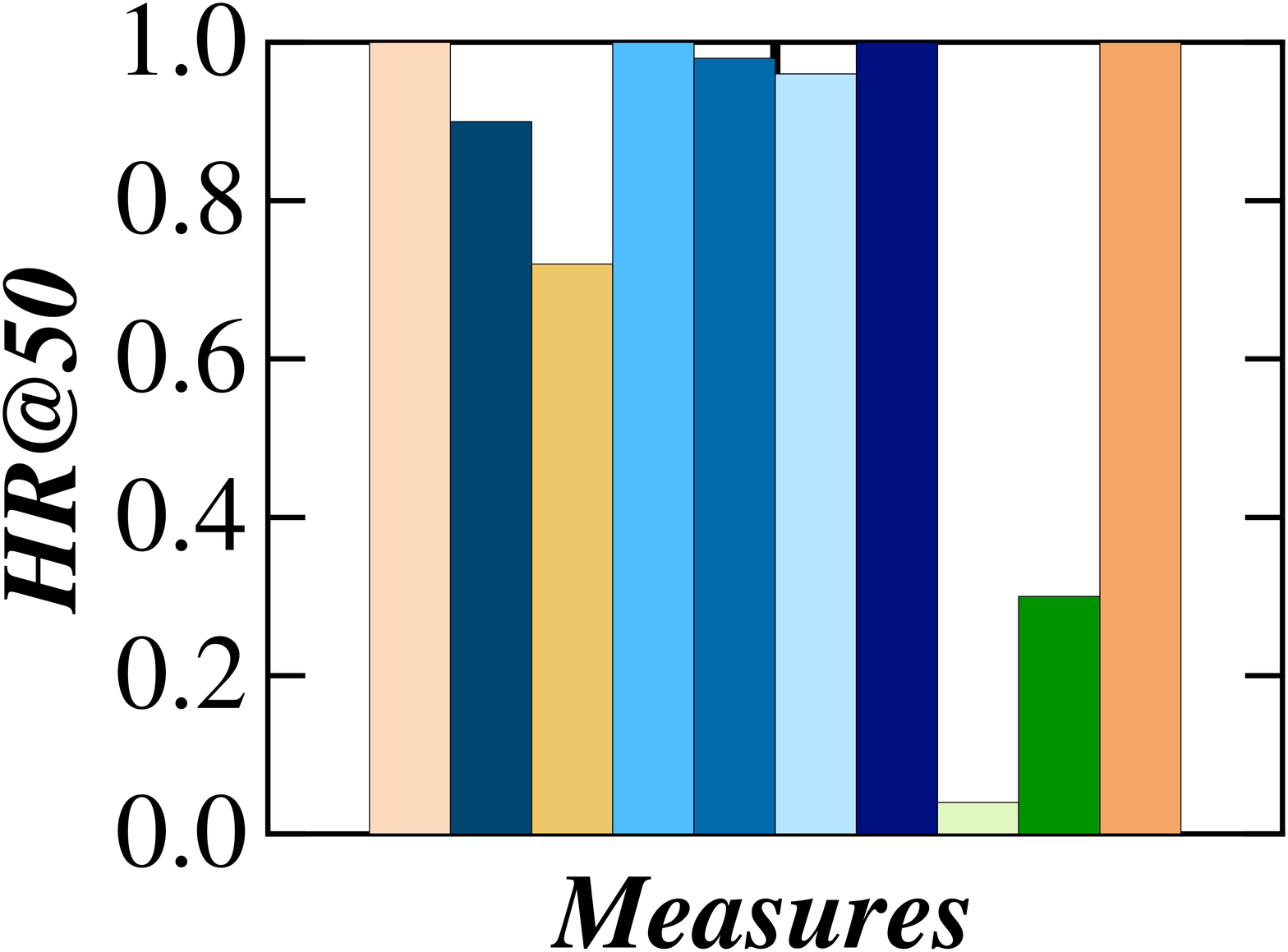}}
	\subfigure[$N$=13\% Geolife]{
		\includegraphics[width=0.16\textwidth]{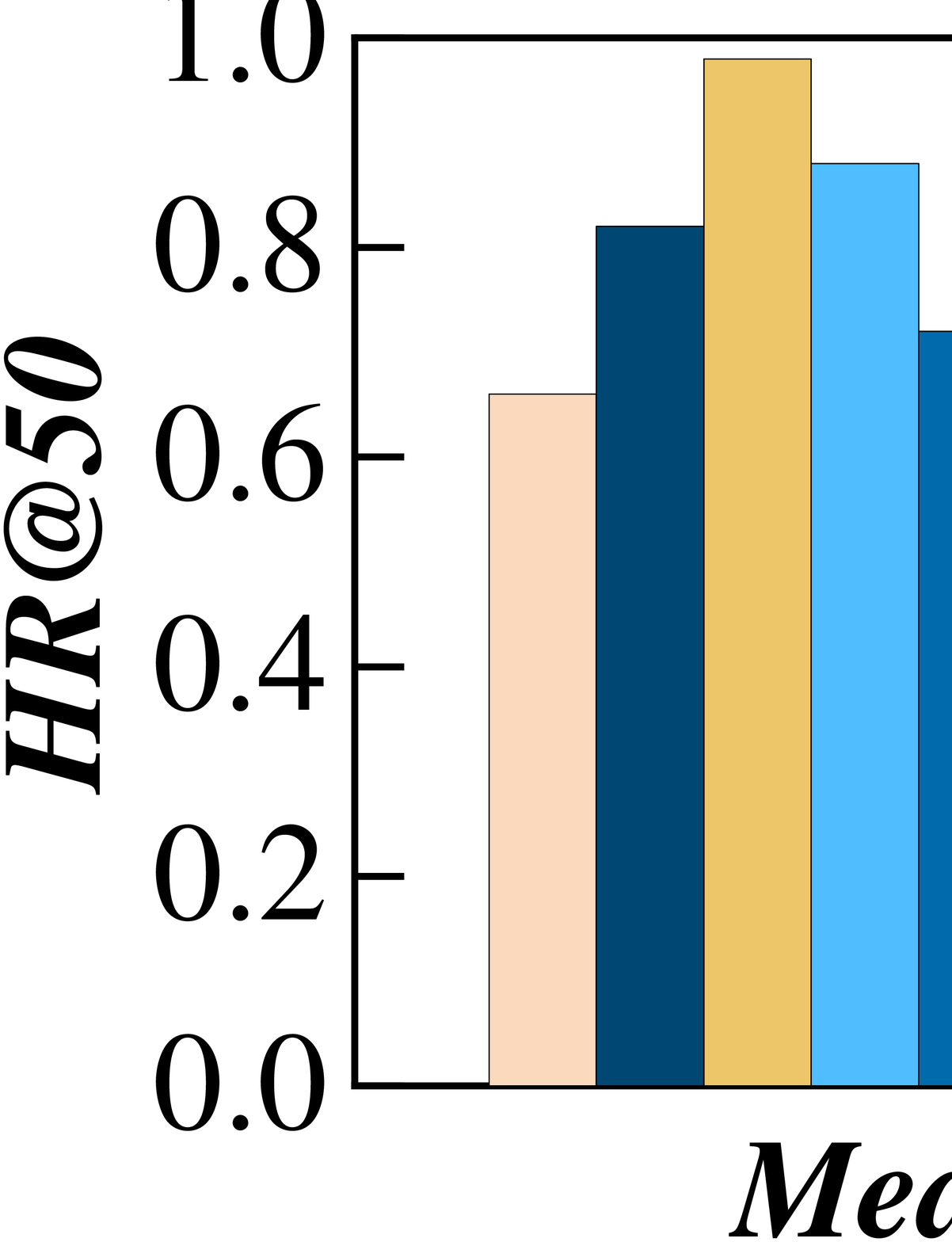}}
	\subfigure[$N$=16\% Geolife]{
		\includegraphics[width=0.16\textwidth]{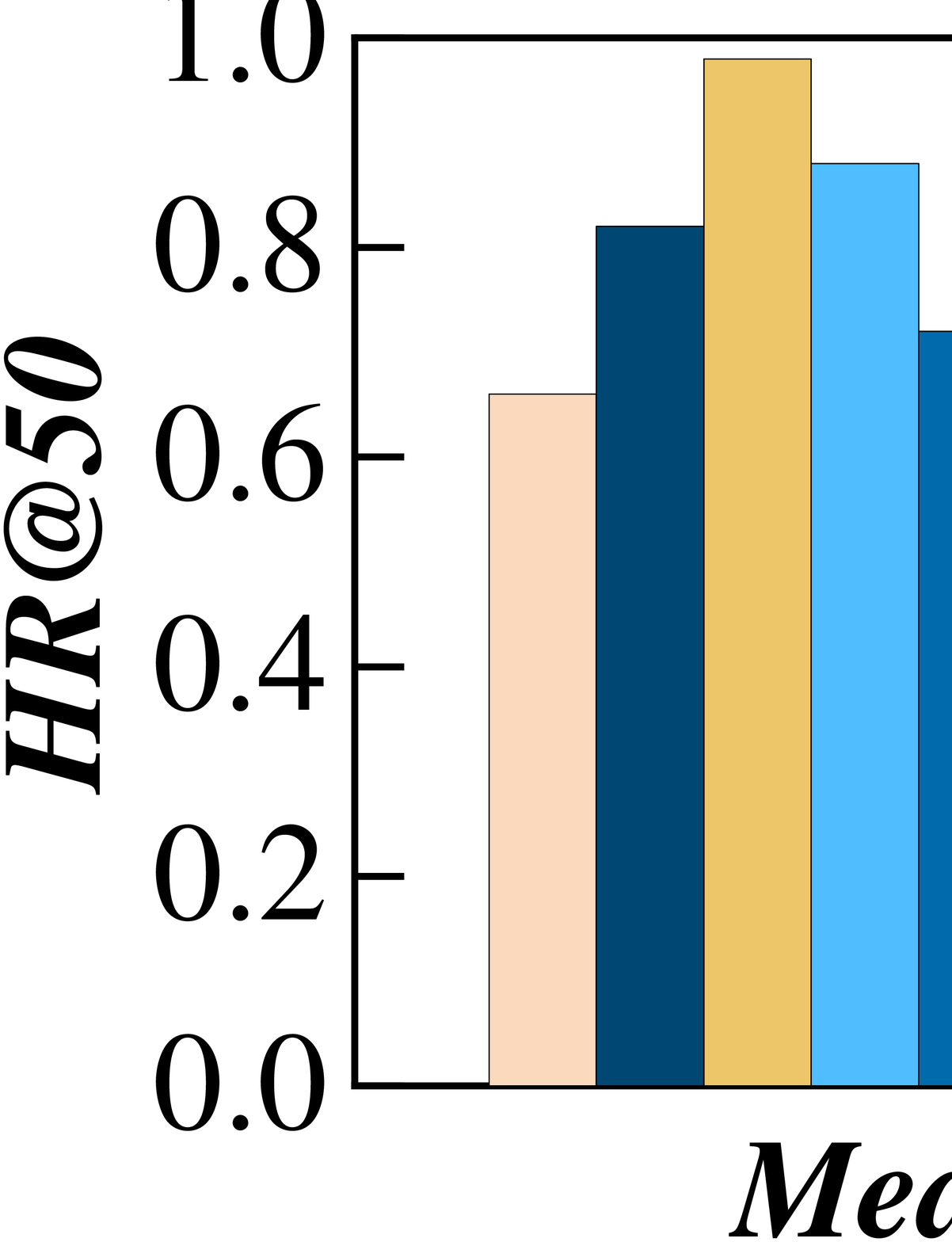}}
	\subfigure[$N$=19\% Geolife]{
		\includegraphics[width=0.16\textwidth]{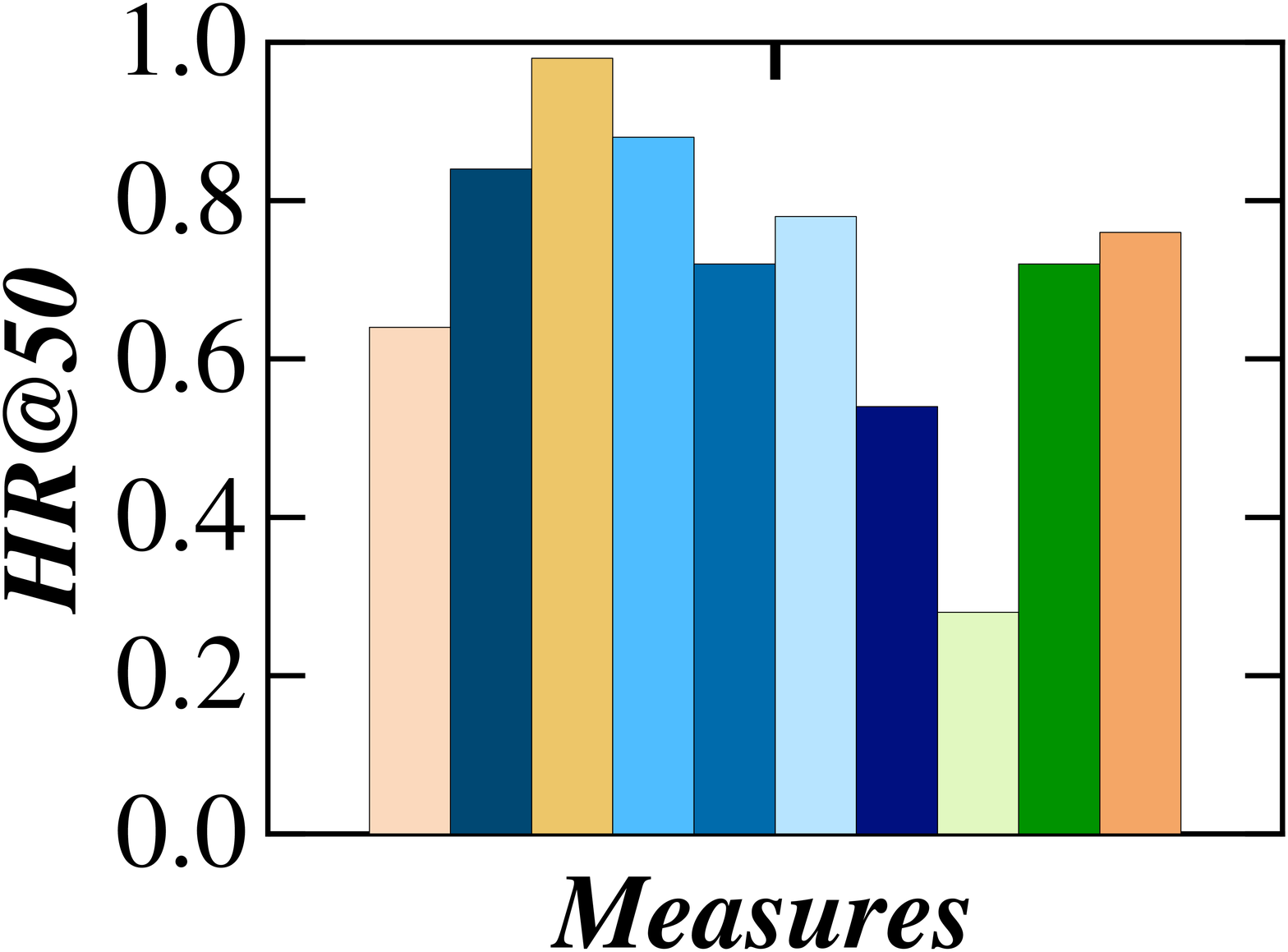}}\\	
	
	\hspace{-4mm}
	\subfigure[$N$=10\% T-Drive]{
		\includegraphics[width=0.16\textwidth]{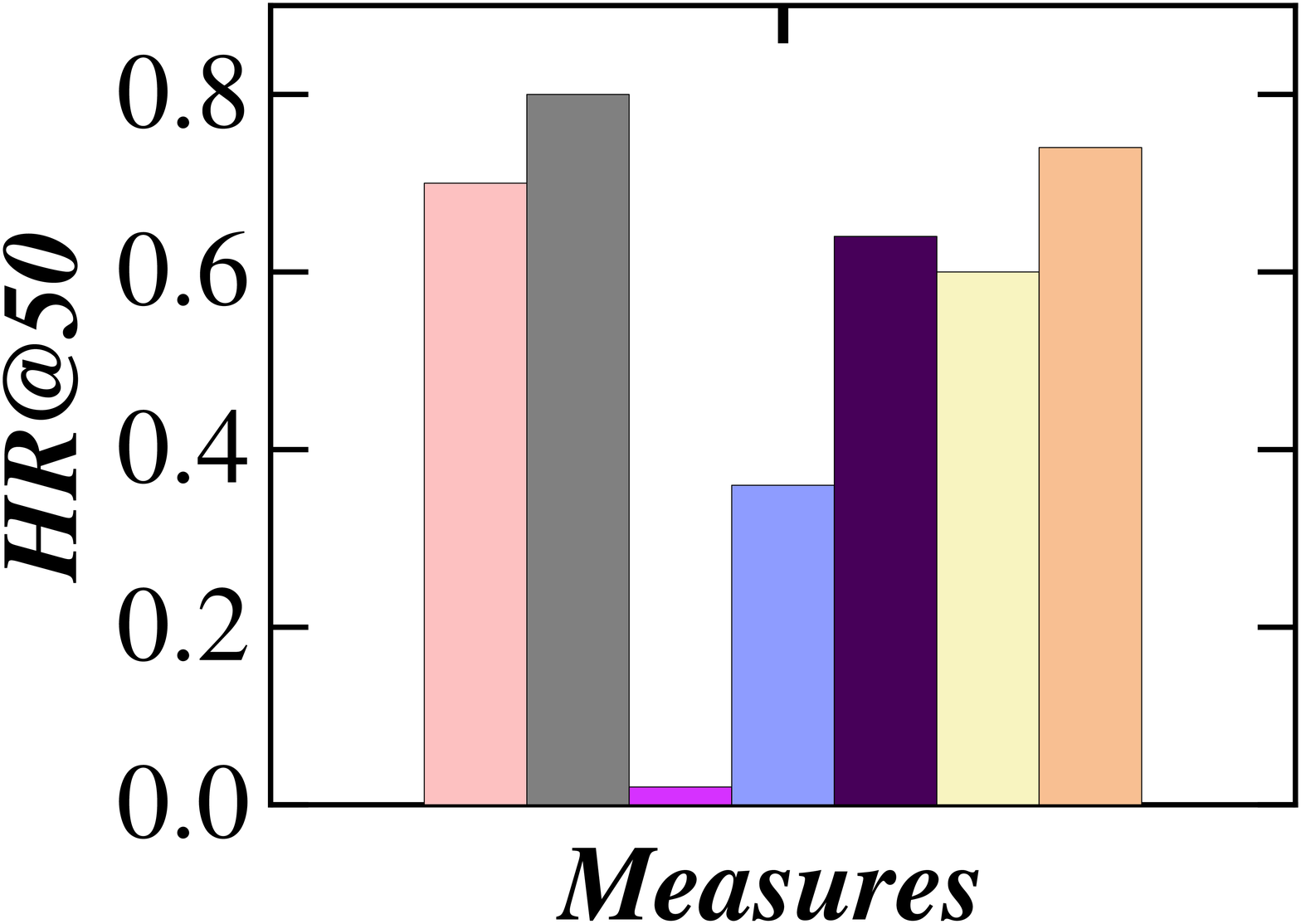}}
	\subfigure[$N$=20\% T-Drive]{
		\includegraphics[width=0.16\textwidth]{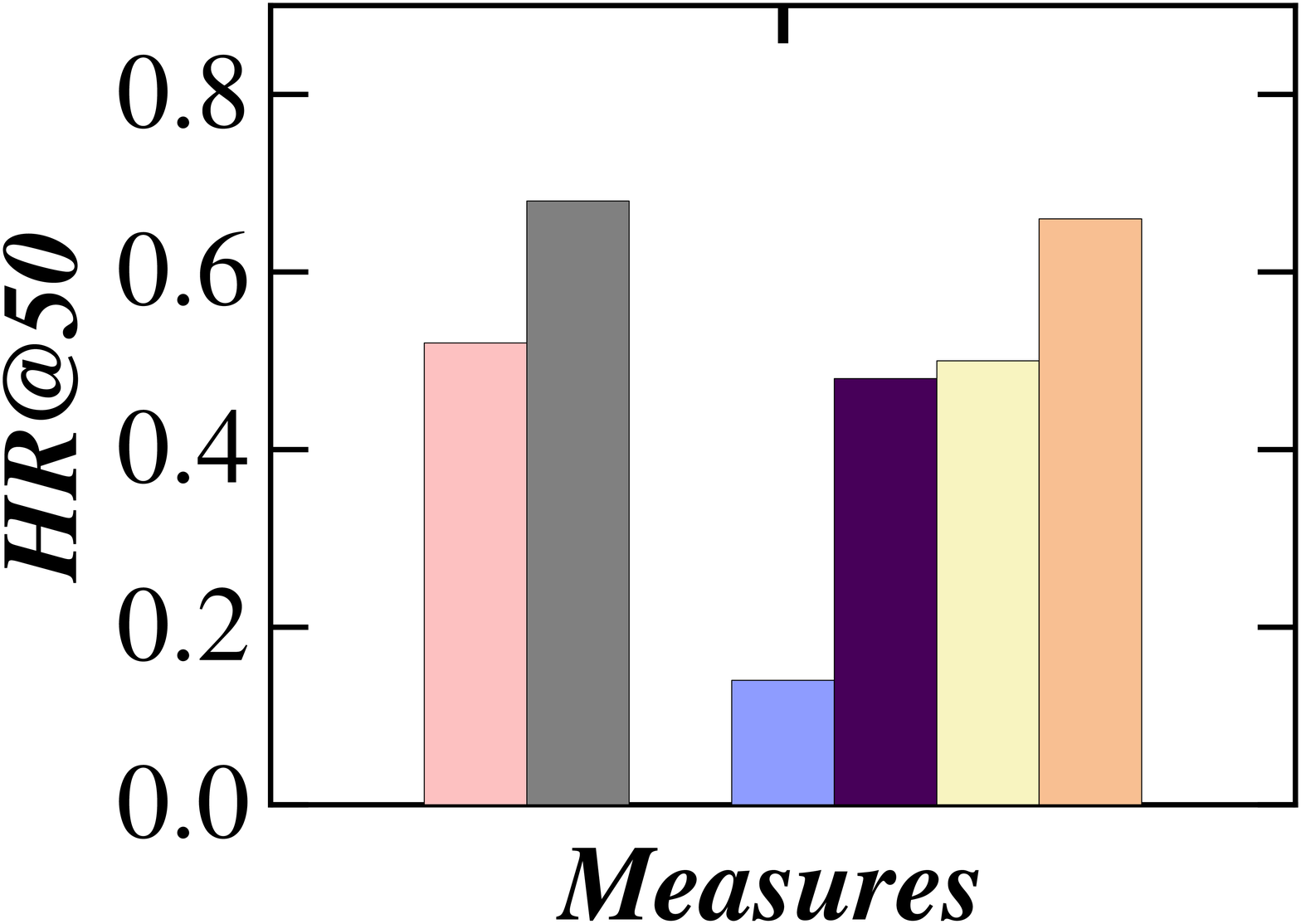}}
	\subfigure[$N$=30\% T-Drive]{
		\includegraphics[width=0.16\textwidth]{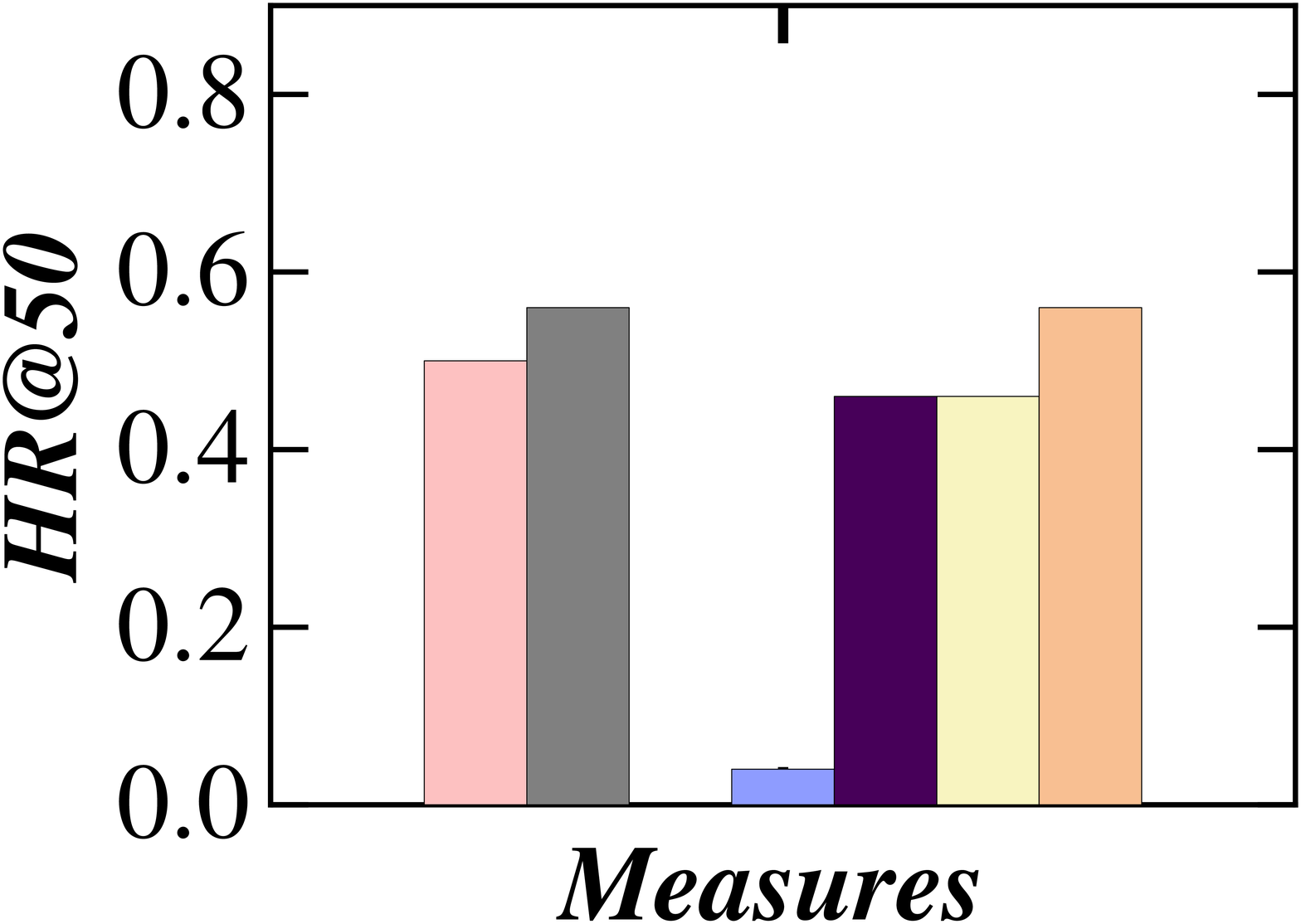}}
	\subfigure[$N$=10\% Porto]{
		\includegraphics[width=0.16\textwidth]{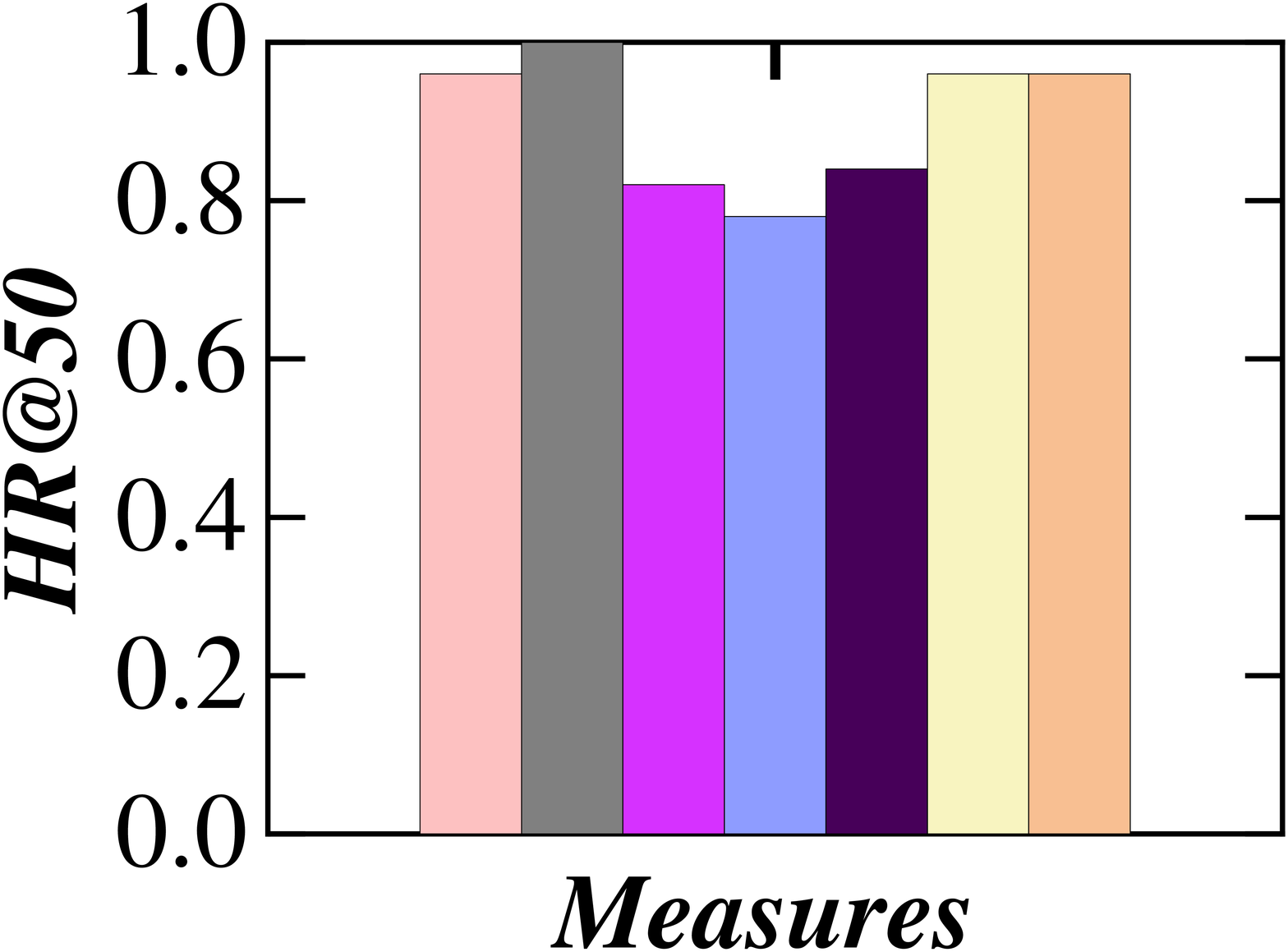}}
	\subfigure[$N$=20\% Porto]{
		\includegraphics[width=0.16\textwidth]{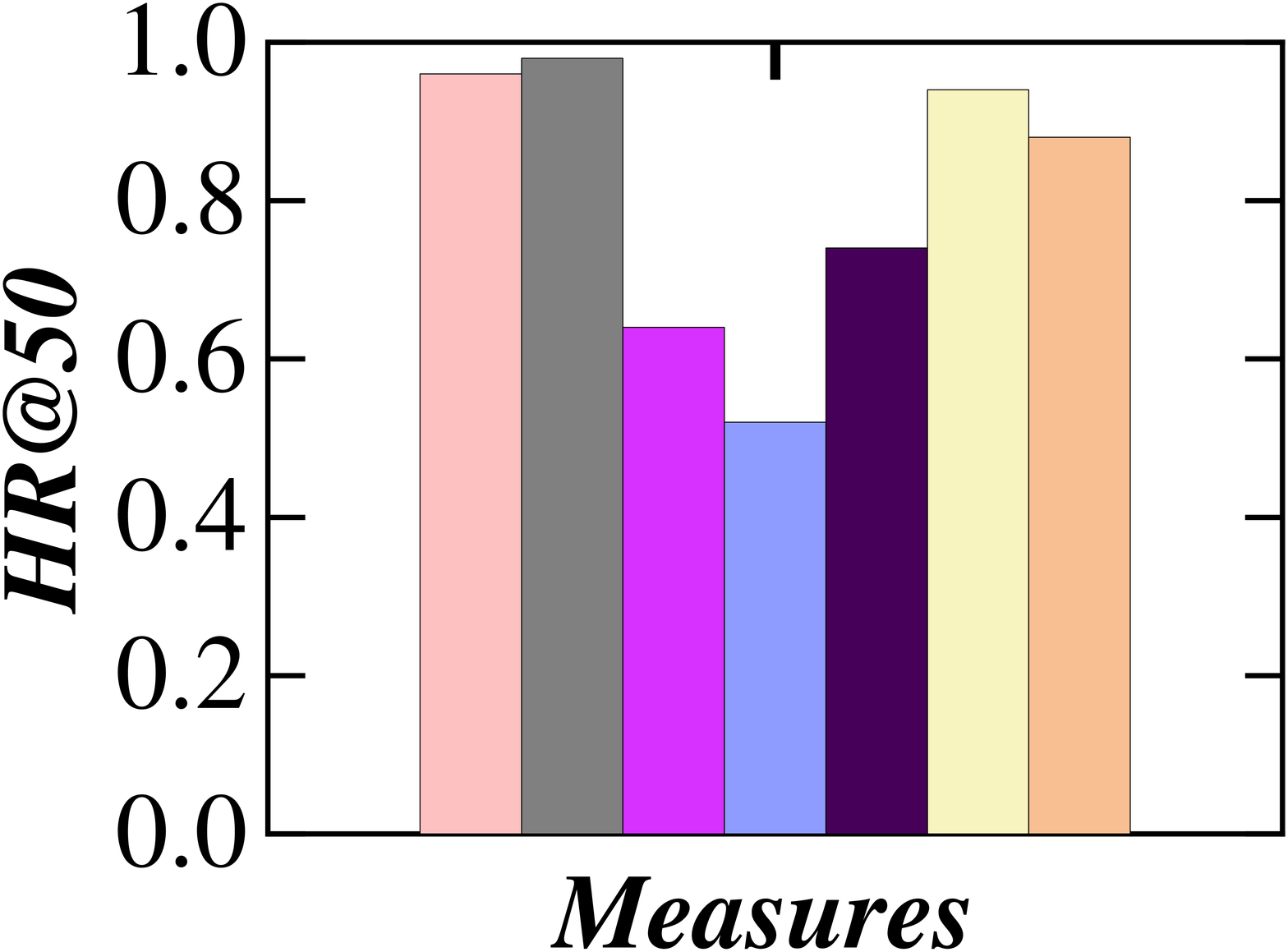}}
	\subfigure[$N$=30\% Porto]{
		\includegraphics[width=0.16\textwidth]{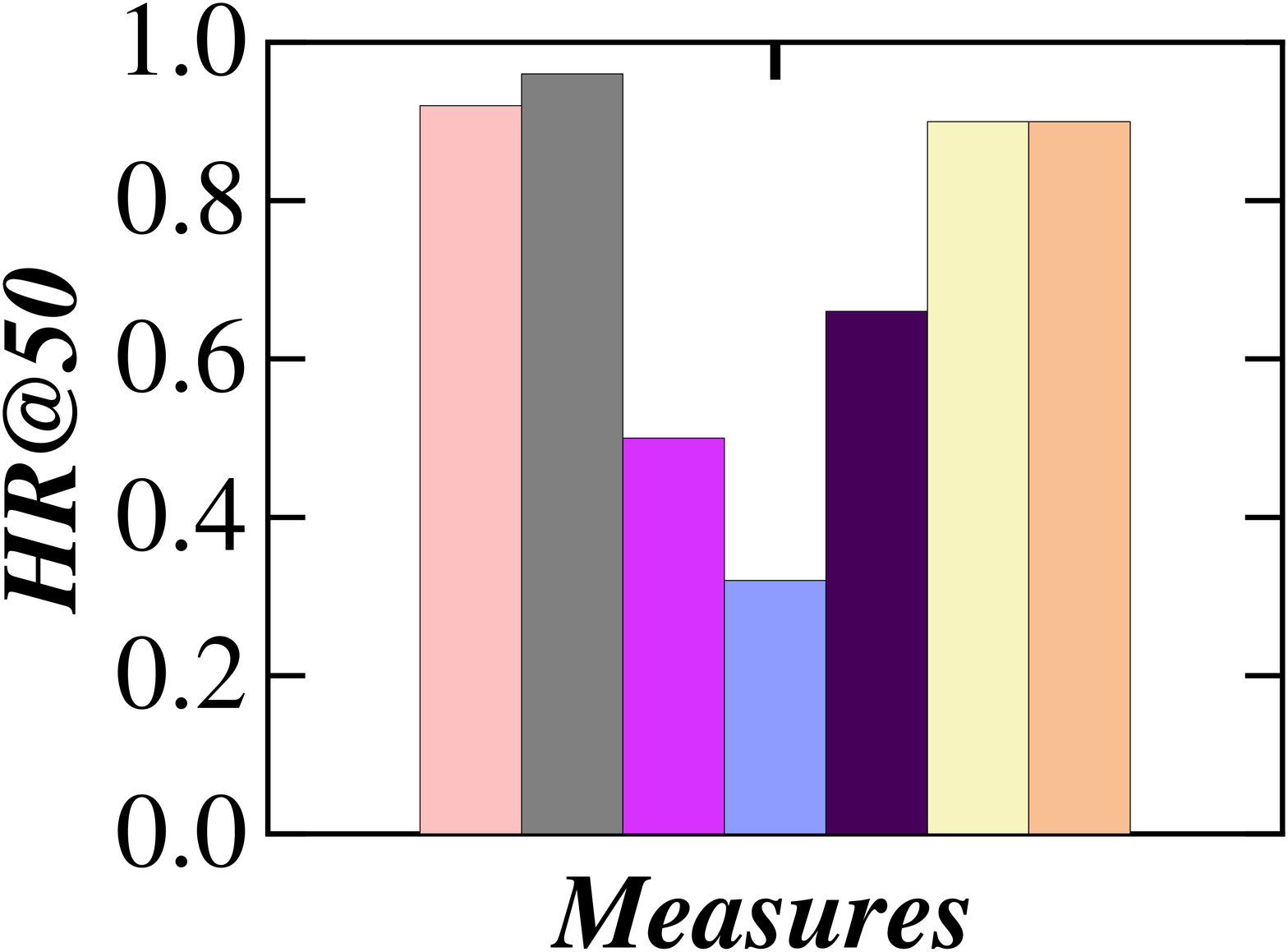}}\\
    \up
	\caption{Robustness of Non-learning based Measures vs. Noise Ratio}
	\label{fig:Noise}
	\vspace{-4mm}
\end{figure*}

\begin{table*}[]
\caption{Robustness of Learning-based Measures in Free  vs. Sampling Rate}
\footnotesize
\vspace{-3mm}
\hspace{-2mm}
\begin{tabular}{|c|c|cccccc|cccccc|}
\hline
\multirow{2}{*}{\begin{tabular}[c]{@{}c@{}}{Sampling}\\ {Rate (\%)}\end{tabular}} & \multirow{2}{*}{\makebox[0.005\textwidth][c]{Measures}} & \multicolumn{6}{c|}{\makebox[0.25\textwidth][c]{AIS}}                                                                                                                                          & \multicolumn{6}{c|}{\makebox[0.35\textwidth][c]{Geolife}}                                                                                                                                     \\ \cline{3-14}
        &                           & \multicolumn{1}{c|}{\makebox[0.0007\textwidth][c]{DTW}}     & \multicolumn{1}{c|}{\makebox[0.0007\textwidth][c]{LCSS}}   & \multicolumn{1}{c|}{\makebox[0.0007\textwidth][c]{EDR}}    & \multicolumn{1}{c|}{\makebox[0.0007\textwidth][c]{ERP}}    & \multicolumn{1}{c|}{\makebox[0.0007\textwidth][c]{Frechet}} & \makebox[0.04\textwidth][c]{Hausdorff} & \multicolumn{1}{c|}{\makebox[0.0005\textwidth][c]{DTW}}    & \multicolumn{1}{c|}{\makebox[0.0005\textwidth][c]{LCSS}}   & \multicolumn{1}{c|}{\makebox[0.0005\textwidth][c]{EDR}}    & \multicolumn{1}{c|}{\makebox[0.0005\textwidth][c]{ERP}}    & \multicolumn{1}{c|}{\makebox[0.0005\textwidth][c]{Frechet}} & \makebox[0.04\textwidth][c]{Hausdorff} \\ \hline
\multirow{2}{*}{$S$=10}               & NEUTRAJ                   & \multicolumn{1}{c|}{0.0087}  & \multicolumn{1}{c|}{0.0075} & \multicolumn{1}{c|}{0.0073} & \multicolumn{1}{c|}{0.0082} & \multicolumn{1}{c|}{0.0086}  & 0.0087    & \multicolumn{1}{c|}{0.0281} & \multicolumn{1}{c|}{0.0217} & \multicolumn{1}{c|}{0.0140} & \multicolumn{1}{c|}{0.0156} & \multicolumn{1}{c|}{0.0178}  & 0.0166    \\ \cline{2-14}
                                    & Traj2SimVec               & \multicolumn{1}{c|}{0.2515}  & \multicolumn{1}{c|}{0.0016} & \multicolumn{1}{c|}{0.0078} & \multicolumn{1}{c|}{0.0356} & \multicolumn{1}{c|}{0.2067}  & 0.2520    & \multicolumn{1}{c|}{0.0106} & \multicolumn{1}{c|}{0.3312} & \multicolumn{1}{c|}{0.3312} & \multicolumn{1}{c|}{0.0271} & \multicolumn{1}{c|}{0.0158}  & 0.0110    \\ \hline
\multirow{2}{*}{$S$=20}               & NEUTRAJ                   & \multicolumn{1}{c|}{0.0087}  & \multicolumn{1}{c|}{0.0083} & \multicolumn{1}{c|}{0.0087} & \multicolumn{1}{c|}{0.0080} & \multicolumn{1}{c|}{0.0091}  & 0.0091    & \multicolumn{1}{c|}{0.0277} & \multicolumn{1}{c|}{0.0204} & \multicolumn{1}{c|}{0.0113} & \multicolumn{1}{c|}{0.0179} & \multicolumn{1}{c|}{0.0200}  & 0.0183    \\ \cline{2-14}
                                    & Traj2SimVec               & \multicolumn{1}{c|}{0.2603}  & \multicolumn{1}{c|}{0.0019} & \multicolumn{1}{c|}{0.0077} & \multicolumn{1}{c|}{0.0352} & \multicolumn{1}{c|}{0.2151}  & 0.2610    & \multicolumn{1}{c|}{0.0094} & \multicolumn{1}{c|}{1.0000} & \multicolumn{1}{c|}{1.0000} & \multicolumn{1}{c|}{0.0245} & \multicolumn{1}{c|}{0.0158}  & 0.0110    \\ \hline
\multirow{2}{*}{$S$=40}               & NEUTRAJ                   & \multicolumn{1}{c|}{0.0090}  & \multicolumn{1}{c|}{0.0080} & \multicolumn{1}{c|}{0.0082} & \multicolumn{1}{c|}{0.0084} & \multicolumn{1}{c|}{0.0086}  & 0.0095    & \multicolumn{1}{c|}{0.0277} & \multicolumn{1}{c|}{0.0206} & \multicolumn{1}{c|}{0.0121} & \multicolumn{1}{c|}{0.0210} & \multicolumn{1}{c|}{0.0220}  & 0.0214    \\ \cline{2-14}
                                    & Traj2SimVec               & \multicolumn{1}{c|}{0.2590}  & \multicolumn{1}{c|}{0.0018} & \multicolumn{1}{c|}{0.0079} & \multicolumn{1}{c|}{0.0351} & \multicolumn{1}{c|}{0.2151}  & 0.2602    & \multicolumn{1}{c|}{0.0094} & \multicolumn{1}{c|}{1.0000} & \multicolumn{1}{c|}{1.0000} & \multicolumn{1}{c|}{0.0245} & \multicolumn{1}{c|}{0.0158}  & 0.0249    \\ \hline
\multirow{2}{*}{$S$=100}              & NEUTRAJ                   & \multicolumn{1}{c|}{0.0082}  & \multicolumn{1}{c|}{0.0094} & \multicolumn{1}{c|}{0.0074} & \multicolumn{1}{c|}{0.0076} & \multicolumn{1}{c|}{0.0087}  & 0.0088    & \multicolumn{1}{c|}{0.0277} & \multicolumn{1}{c|}{0.0212} & \multicolumn{1}{c|}{0.0204} & \multicolumn{1}{c|}{0.0252} & \multicolumn{1}{c|}{0.0258}  & 0.0249    \\ \cline{2-14}
                                    & Traj2SimVec               & \multicolumn{1}{c|}{0.25905} & \multicolumn{1}{c|}{0.0018} & \multicolumn{1}{c|}{0.0079} & \multicolumn{1}{c|}{0.0352} & \multicolumn{1}{c|}{0.2143}  & 0.2705    & \multicolumn{1}{c|}{0.0366} & \multicolumn{1}{c|}{1.0000} & \multicolumn{1}{c|}{1.0000} & \multicolumn{1}{c|}{0.0215} & \multicolumn{1}{c|}{0.0131}  & 0.0513    \\ \hline
\end{tabular}
\label{tab:SRF}
\vspace{-4mm}
\end{table*}

\noindent\textbf{Effect of Shapes.} Figure~\ref{fig:shape} shows the Top-$1$ query results for four different query trajectories (denoted by red lines), each of which is with a typical spatial shape, i.e., straight (denoted as $QT_s$), polyline without overlaps (denoted as $QT_{o1}$), polyline with overlaps (denoted as $QT_{o2}$), and round (denoted as $QT_r$). Here, we only evaluate road-network constrained similarity measures on T-drive and Porto, as the vehicle trajectories usually have much more complex geometry than free space trajectories due to the constraints of road networks.
The higher the similarity between the query trajectory and the corresponding Top-$1$ query result by visualization, the higher the effectiveness.

First, given a $QT_s$, only NetEDR and NetLCSS return back-and-forth trajectories containing a large number of matched pairs with $QT$, while all of the remaining measures return straight trajectories (cf. Figures~\ref{fig:shape}(a) and (e)). The reason is that NetEDR and NetLCSS do not consider the direction of trajectories, but rather the number of matched pairs. Specifically, the more the matched pairs, the higher the similarity (cf. Section~\ref{sec:traditional}).
Second, given a $QT_{o1}$ or a $QT_{o2}$, all of measures except for NetEDR return similar results to it (cf. Figures~\ref{fig:shape}(b), (c), (f), and (g)). This is because, NetEDR does not consider the spatial distance between two points; instead, it identifies the similar trajectories by comparing their lengths, i.e., a similar lengths results in a higher similarity.
Third, given a $QT_{r}$, only LORS, LCRS, and TP return results with round sub-trajectories, where LORS performs the best (cf. Figures~\ref{fig:shape}(d) and (h)), because LORS can identify overlapping road segments, which enables it to effectively match trajectories with round shapes.

\subsubsection{Learning based measures} We evaluate the effectiveness of learning based measures by their capabilities of approximating non-learning based measures, which are their targets. 
Specifically, given a query trajectory $QT$ and a non-learning based measure, we use its Top-50 query result calculated by the non-learning based measure as the ground truth. Next, we implement the model of a learning based measure to approximate the non-learning based measure, and apply the learned measure to compute Top-50 query result w.r.t. $QT$.
Finally, we utilize $\textit{HR}@50$ to measure the overlap between the Top-50 similarity query results returned by the learning-based measures and the ground truths. Clearly, the higher the $\textit{HR}@50$, the higher the effectiveness.

Figure~\ref{fig:origin} depicts the effectiveness of learning based measures. First, Traj2SimVec has better performance (i.e.,  higher $\textit{HR}@50$) than NEUTRAJ in free space settings (cf. Figures~\ref{fig:learning}(a) and (b)). Specifically, Traj2SimVec simplifies trajectories into triplets (anchor, similar, dissimilar) (cf. Section~\ref{sec:learning}), as such triplets enhance the similarity learning among trajectories.
Second, the performance of Traj2SimVec is unstable across different similarity measures and different datasets.
This is because (i) LCSS, EDR, and ERP are string-based, whose information cannot be preserved in embeddings; and (ii) the data distribution of AIS is uneven, and the spatio-temporal features on Geolife is complex due to various transportation modes.
Third, both $\textit{HR}@50$ of GTS and ST2Vec are stable in road network settings, and ST2Vec performs better than GTS. The reason is that the spatial information exploited by ST2Vec (i.e., road networks) is more comprehensive than that by  GTS (i.e., POIs), and ST2Vec is able to extract. In a conclusion, learning-based measures are more effective for road network settings.

\subsection{Robustness Evaluation}
\label{subsec:robust}
We study the robustness of each trajectory similarity measure by performing Top-$50$ similarity queries. Specifically, given a specific non-learning based measure and a query trajectory $QT$, we perform robustness evaluation by the following three steps. First, we compute the $QT$'s Top-$50$ similar trajectories via a similarity measure, and set the result as the ground-truth. Second, we perform down sampling or add-noise operations on the original dataset according to parameters $S$ and $N$ in
Table~\ref{tab:transformation}, resulting in a set of transformed trajectory datasets. Third, we perform Top-$50$ similarity query for the same query trajectory $QT$ in each transformed dataset to obtain the corresponding Top-$50$ result. Finally, we use $\textit{HR}@50$ to measure the overlap between the Top-50 similarity query results on transformed datasets and the ground truths. Clearly, the higher
the $\textit{HR}@50$, the higher the robustness. Note that, we compute Top-$50$ similarity results for non-learning based measures, while learn those for learning based measures.


\subsubsection{Non-learning based measures} 

\noindent\textbf{Effect of Sampling Rates.} We follow Table~\ref{tab:transformation} to perform downsampling with $\textit{S (\%)=}$10, 20, and 40, where $S$ is the sampling rate. When $\textit{S=}$100, no sampling is performed.
Here, we only verify free space oriented measures on AIS and Geolife, as the variation of sampling rates has little effect on map-matched trajectories in road networks.

The results are shown in Figure~\ref{fig:SR}. First, DTW, EDR, and ERP become less robust as the sampling rate increases. This is in contrast to the findings of a previous study~\cite{VLDBJSurvey}, where experimental datasets with small spatial coverage are used. In this case, the change of the sampling rate does not significantly impact the distances between pair points when computing DTW and ERP distance.
On the contrary, as AIS covers large spatial space, DTW and ERP that perform well in small urban areas~\cite{VLDBJSurvey} are sensitive to sampling rates.


Second,
 $\textit{HR}@50$ of EDwP varies between 0.6 and 1.0 on both AIS and Geolife, which suggests that its performance is slightly affected by sampling rates. In addition, LIP is robust on AIS, as its corresponding $\textit{HR}@50$ is larger than 0.6. However, it performs poorly on Geolife with $\textit{HR}@50$ being smaller than 0.4. Specifically, the similarity between each pair of trajectories calculated by LIP is determined by the areas and perimeters of the polygons formed by point intersections of the trajectories. Moreover, each trajectory on Geolife generally contains a variety of transportation modes, where the decreasing of sampling rates largely affects the shape of a trajectory. Thus, LIP is very sensitive to the sampling rates of trajectories on Geolife.

Finally, Frechet, Hausdorff, OWD, and Seg-Frechet are robust, as their $\textit{HR}@50$ are always larger than 0.9 even when $\textit{S=}$10$\%$. This is because these measures rely on the number of matched pairs, which remains stable when the number of sampling points changes.

\begin{table*}[]
\caption{Robustness of Learning-based Measures in Free Space vs. Noise Ratio}
\footnotesize
\vspace{-3mm}
\hspace{-2mm}
\begin{tabular}{|c|c|cccccc|cccccc|}
\hline
\multirow{2}{*}{\begin{tabular}[c]{@{}c@{}}Noise \\ Ratio (\%)\end{tabular}} & \multirow{2}{*}{\makebox[0.005\textwidth][c]{Measures}} & \multicolumn{6}{c|}{\makebox[0.25\textwidth][c]{AIS}}                                                                                                                                          & \multicolumn{6}{c|}{\makebox[0.35\textwidth][c]{Geolife}}                                                                                                                                    \\ \cline{3-14}
    &                           & \multicolumn{1}{c|}{\makebox[0.0007\textwidth][c]{DTW}}     & \multicolumn{1}{c|}{\makebox[0.0007\textwidth][c]{LCSS}}   & \multicolumn{1}{c|}{\makebox[0.0007\textwidth][c]{EDR}}    & \multicolumn{1}{c|}{\makebox[0.0007\textwidth][c]{ERP}}    & \multicolumn{1}{c|}{\makebox[0.0007\textwidth][c]{Frechet}} & \makebox[0.05\textwidth][c]{Hausdorff} & \multicolumn{1}{c|}{\makebox[0.0005\textwidth][c]{DTW}}    & \multicolumn{1}{c|}{\makebox[0.0005\textwidth][c]{LCSS}}   & \multicolumn{1}{c|}{\makebox[0.0005\textwidth][c]{EDR}}    & \multicolumn{1}{c|}{\makebox[0.0005\textwidth][c]{ERP}}    & \multicolumn{1}{c|}{\makebox[0.0005\textwidth][c]{Frechet}} & \makebox[0.05\textwidth][c]{Hausdorff} \\ \hline
\multirow{2}{*}{$N$=0}                                                       & NEUTRAJ                   & \multicolumn{1}{c|}{0.0082} & \multicolumn{1}{c|}{0.0094} & \multicolumn{1}{c|}{0.0074} & \multicolumn{1}{c|}{0.0076} & \multicolumn{1}{c|}{0.0087}  & 0.0088    & \multicolumn{1}{c|}{0.0277} & \multicolumn{1}{c|}{0.0212} & \multicolumn{1}{c|}{0.0204} & \multicolumn{1}{c|}{0.0252} & \multicolumn{1}{c|}{0.0258}  & 0.0249    \\ \cline{2-14}
                                    & Traj2SimVec               & \multicolumn{1}{c|}{0.2591} & \multicolumn{1}{c|}{0.0018} & \multicolumn{1}{c|}{0.0079} & \multicolumn{1}{c|}{0.0352} & \multicolumn{1}{c|}{0.2143}  & 0.2705    & \multicolumn{1}{c|}{0.0366} & \multicolumn{1}{c|}{1.0000} & \multicolumn{1}{c|}{1.0000} & \multicolumn{1}{c|}{0.0215} & \multicolumn{1}{c|}{0.0131}  & 0.0513    \\ \hline
\multirow{2}{*}{$N$=13}                                                      & NEUTRAJ                   & \multicolumn{1}{c|}{0.0091} & \multicolumn{1}{c|}{0.0080} & \multicolumn{1}{c|}{0.0073} & \multicolumn{1}{c|}{0.0079} & \multicolumn{1}{c|}{0.0087}  & 0.0085    & \multicolumn{1}{c|}{0.0276} & \multicolumn{1}{c|}{0.0223} & \multicolumn{1}{c|}{0.0203} & \multicolumn{1}{c|}{0.0243} & \multicolumn{1}{c|}{0.0247}  & 0.0227    \\ \cline{2-14}
                                    & Traj2SimVec               & \multicolumn{1}{c|}{0.2460} & \multicolumn{1}{c|}{0.0017} & \multicolumn{1}{c|}{0.0079} & \multicolumn{1}{c|}{0.0345} & \multicolumn{1}{c|}{0.2091}  & 0.2583    & \multicolumn{1}{c|}{0.0094} & \multicolumn{1}{c|}{1.0000} & \multicolumn{1}{c|}{1.0000} & \multicolumn{1}{c|}{0.0255} & \multicolumn{1}{c|}{0.0158}  & 0.0110    \\ \hline
\multirow{2}{*}{$N$=16}                                                      & NEUTRAJ                   & \multicolumn{1}{c|}{0.0092} & \multicolumn{1}{c|}{0.0071} & \multicolumn{1}{c|}{0.0080} & \multicolumn{1}{c|}{0.0082} & \multicolumn{1}{c|}{0.0091}  & 0.0085    & \multicolumn{1}{c|}{0.0259} & \multicolumn{1}{c|}{0.0201} & \multicolumn{1}{c|}{0.0107} & \multicolumn{1}{c|}{0.0283} & \multicolumn{1}{c|}{0.0259}  & 0.0251    \\ \cline{2-14}
                                    & Traj2SimVec               & \multicolumn{1}{c|}{0.2464} & \multicolumn{1}{c|}{0.0021} & \multicolumn{1}{c|}{0.0080} & \multicolumn{1}{c|}{0.0348} & \multicolumn{1}{c|}{0.2053}  & 0.2558    & \multicolumn{1}{c|}{0.0094} & \multicolumn{1}{c|}{1.0000} & \multicolumn{1}{c|}{1.0000} & \multicolumn{1}{c|}{0.0249} & \multicolumn{1}{c|}{0.0158}  & 0.0110    \\ \hline
\multirow{2}{*}{$N$=19}                                                      & NEUTRAJ                   & \multicolumn{1}{c|}{0.0083} & \multicolumn{1}{c|}{0.0071} & \multicolumn{1}{c|}{0.0066} & \multicolumn{1}{c|}{0.0081} & \multicolumn{1}{c|}{0.0087}  & 0.0084    & \multicolumn{1}{c|}{0.0262} & \multicolumn{1}{c|}{0.0200} & \multicolumn{1}{c|}{0.0103} & \multicolumn{1}{c|}{0.0266} & \multicolumn{1}{c|}{0.0259}  & 0.0254    \\ \cline{2-14}
                                    & Traj2SimVec               & \multicolumn{1}{c|}{0.2403} & \multicolumn{1}{c|}{0.0017} & \multicolumn{1}{c|}{0.0078} & \multicolumn{1}{c|}{0.0334} & \multicolumn{1}{c|}{0.2034}  & 0.2540    & \multicolumn{1}{c|}{0.0094} & \multicolumn{1}{c|}{1.0000} & \multicolumn{1}{c|}{1.0000} & \multicolumn{1}{c|}{0.0252} & \multicolumn{1}{c|}{0.0158}  & 0.0110    \\ \hline
\end{tabular}
\label{tab:NF}
\vspace{-2mm}
\end{table*}

\begin{table*}[]
\caption{Robustness of Learning-based Measures in Road Network vs. Noise Ratio}
\vspace{-3mm}
\hspace{-2mm}
\footnotesize
\begin{tabular}{|c|c|cccc|cccc|}
\hline
\multirow{2}{*}{Noise Ratio (\%)} & \multirow{2}{*}{\makebox[0.1\textwidth][c]{Measures}} & \multicolumn{4}{c|}{\makebox[0.3\textwidth][c]{T-Drive}}                                                                      & \multicolumn{4}{c|}{\makebox[0.3\textwidth][c]{Porto}}                                                                        \\ \cline{3-10}
                    &                           & \multicolumn{1}{c|}{\makebox[0.07\textwidth][c]{NetLCSS}} & \multicolumn{1}{c|}{\makebox[0.07\textwidth][c]{NetDTW}} & \multicolumn{1}{c|}{\makebox[0.07\textwidth][c]{TP}}     & \multicolumn{1}{c|}{\makebox[0.07\textwidth][c]{NetERP}} & \multicolumn{1}{c|}{\makebox[0.07\textwidth][c]{NetLCSS}} & \multicolumn{1}{c|}{\makebox[0.07\textwidth][c]{NetDTW}} & \multicolumn{1}{c|}{\makebox[0.07\textwidth][c]{TP}}     & \makebox[0.07\textwidth][c]{NetERP} \\ \hline
\multirow{2}{*}{$N$=0}                                                       & GTS                       & \multicolumn{1}{c|}{0.2647}  & \multicolumn{1}{c|}{0.1439} & \multicolumn{1}{c|}{0.2113} & 0.0869 & \multicolumn{1}{c|}{0.2647}  & \multicolumn{1}{c|}{0.1439} & \multicolumn{1}{c|}{0.2113} & 0.0869 \\ \cline{2-10}
                    & ST2Vec                    & \multicolumn{1}{c|}{0.1895}  & \multicolumn{1}{c|}{0.2691} & \multicolumn{1}{c|}{0.3935} & 0.2570 & \multicolumn{1}{c|}{0.2276}  & \multicolumn{1}{c|}{0.2928} & \multicolumn{1}{c|}{0.3758} & 0.2505 \\ \hline
\multirow{2}{*}{$N$=10}                                                      & GTS                       & \multicolumn{1}{c|}{0.0099}  & \multicolumn{1}{c|}{0.0097} & \multicolumn{1}{c|}{0.0097} & 0.0098 & \multicolumn{1}{c|}{0.0510}  & \multicolumn{1}{c|}{0.0501} & \multicolumn{1}{c|}{0.0502} & 00509  \\ \cline{2-10}
                    & ST2Vec                    & \multicolumn{1}{c|}{0.0100}  & \multicolumn{1}{c|}{0.0101} & \multicolumn{1}{c|}{0.0098} & 0.0102 & \multicolumn{1}{c|}{0.0950}  & \multicolumn{1}{c|}{0.0354} & \multicolumn{1}{c|}{0.0463} & 0.0340 \\ \hline
\multirow{2}{*}{$N$=20}                                                      & GTS                       & \multicolumn{1}{c|}{0.0098}  & \multicolumn{1}{c|}{0.0101} & \multicolumn{1}{c|}{0.0099} & 0.0101 & \multicolumn{1}{c|}{0.0507}  & \multicolumn{1}{c|}{0.0506} & \multicolumn{1}{c|}{0.0499} & 0.0501 \\ \cline{2-10}
                    & ST2Vec                    & \multicolumn{1}{c|}{0.0105}  & \multicolumn{1}{c|}{0.0103} & \multicolumn{1}{c|}{0.0099} & 0.0103 & \multicolumn{1}{c|}{0.0944}  & \multicolumn{1}{c|}{0.0334} & \multicolumn{1}{c|}{0.0452} & 0.0348 \\ \hline
\multirow{2}{*}{$N$=30}                                                      & GTS                       & \multicolumn{1}{c|}{0.0099}  & \multicolumn{1}{c|}{0.0102} & \multicolumn{1}{c|}{0.0105} & 0.0102 & \multicolumn{1}{c|}{0.0512}  & \multicolumn{1}{c|}{0.0495} & \multicolumn{1}{c|}{0.0495} & 0.0494 \\ \cline{2-10}
                    & ST2Vec                    & \multicolumn{1}{c|}{0.0103}  & \multicolumn{1}{c|}{0.0103} & \multicolumn{1}{c|}{0.0098} & 0.0098 & \multicolumn{1}{c|}{0.0924}  & \multicolumn{1}{c|}{0.0348} & \multicolumn{1}{c|}{0.0462} & 0.0313 \\ \hline
\end{tabular}
\label{tab:NR}
\vspace{-2mm}
\end{table*}

\begin{figure*} [tb]
	\centering
	\hspace{-4mm}
	\subfigure[AIS]{
		\includegraphics[width=0.235\textwidth]{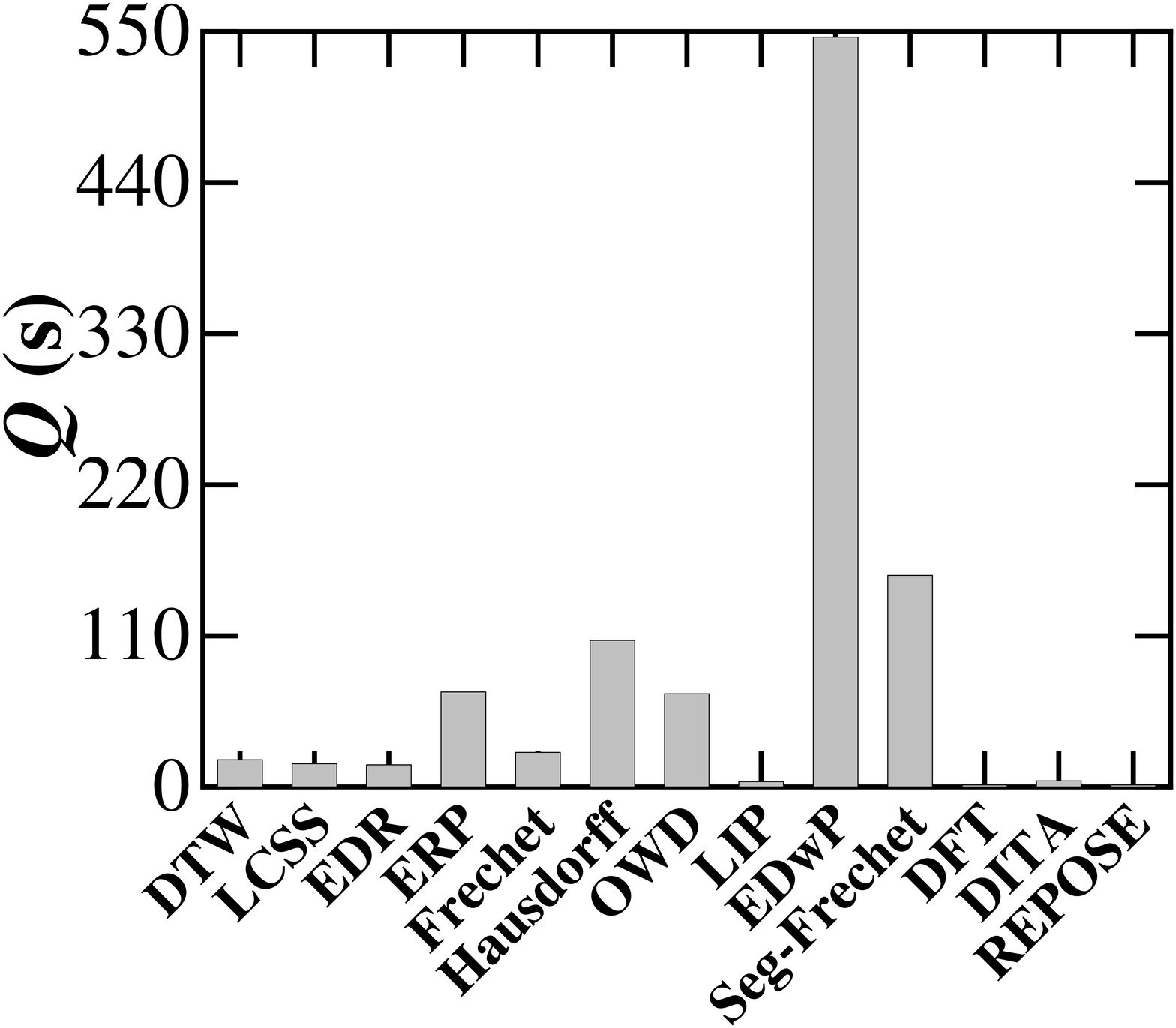}}
	\subfigure[Geolife]{
		\includegraphics[width=0.235\textwidth]{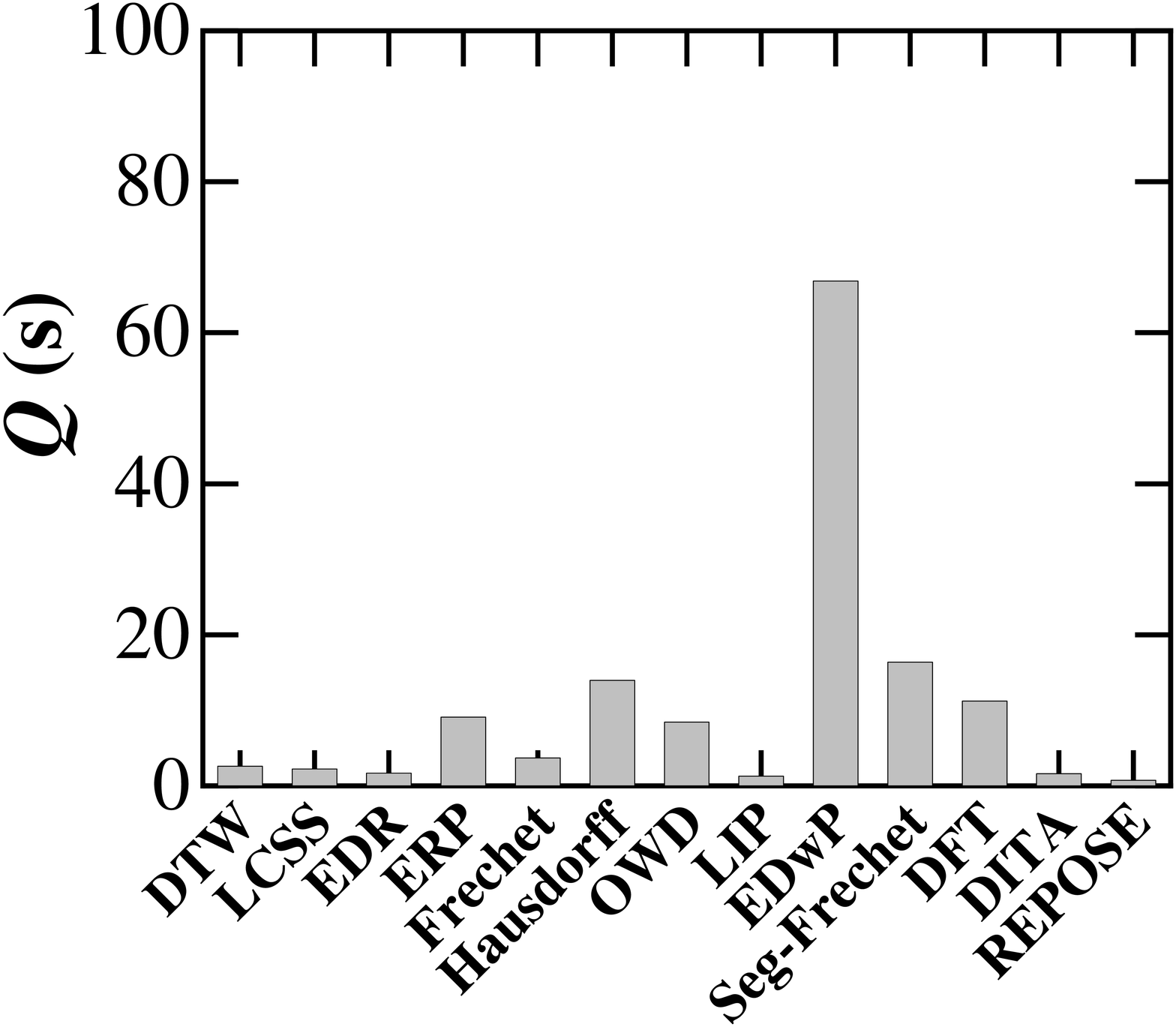}}
	\subfigure[T-Drive]{
		\includegraphics[width=0.235\textwidth]{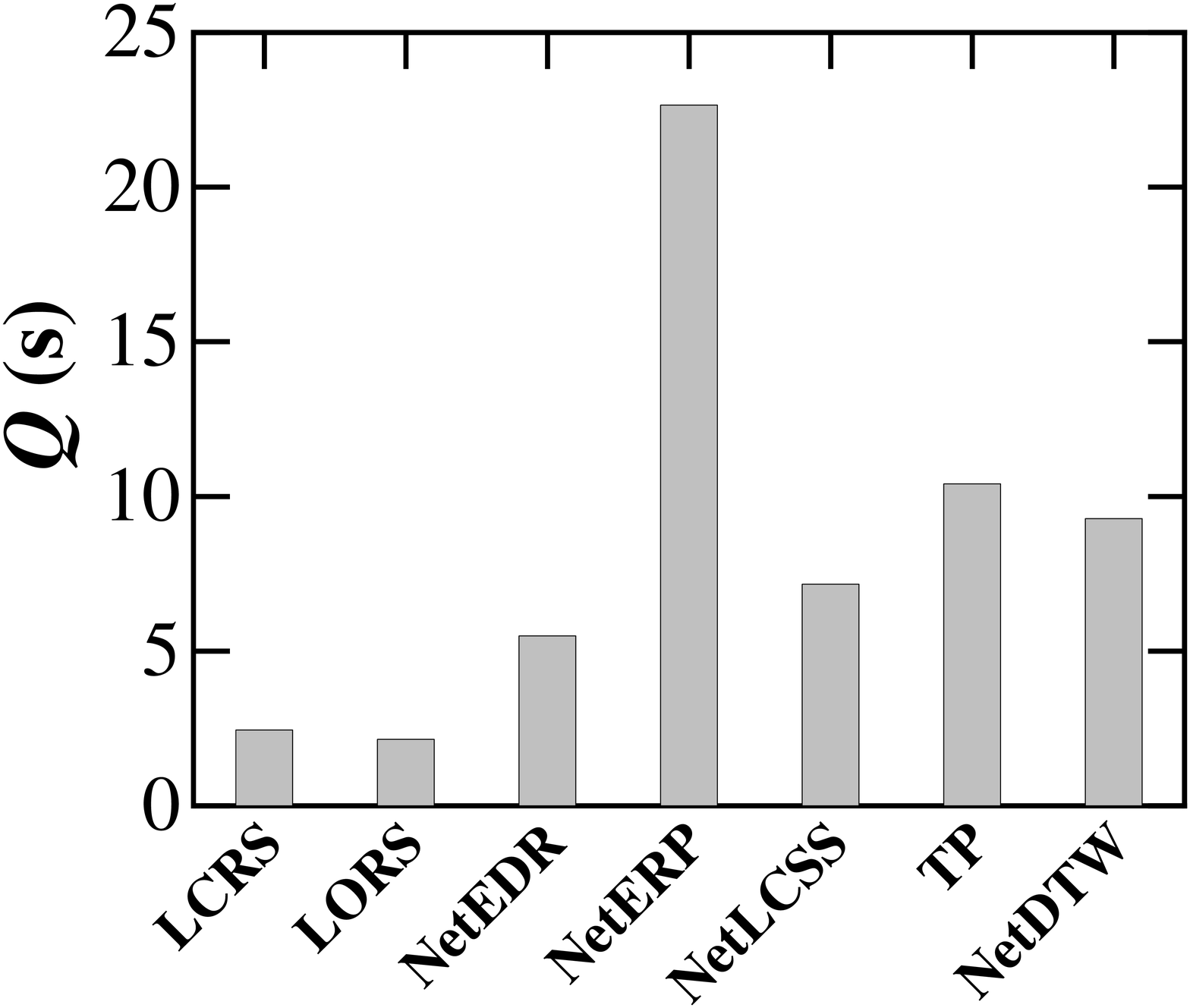}}
	\subfigure[Porto]{
		\includegraphics[width=0.235\textwidth]{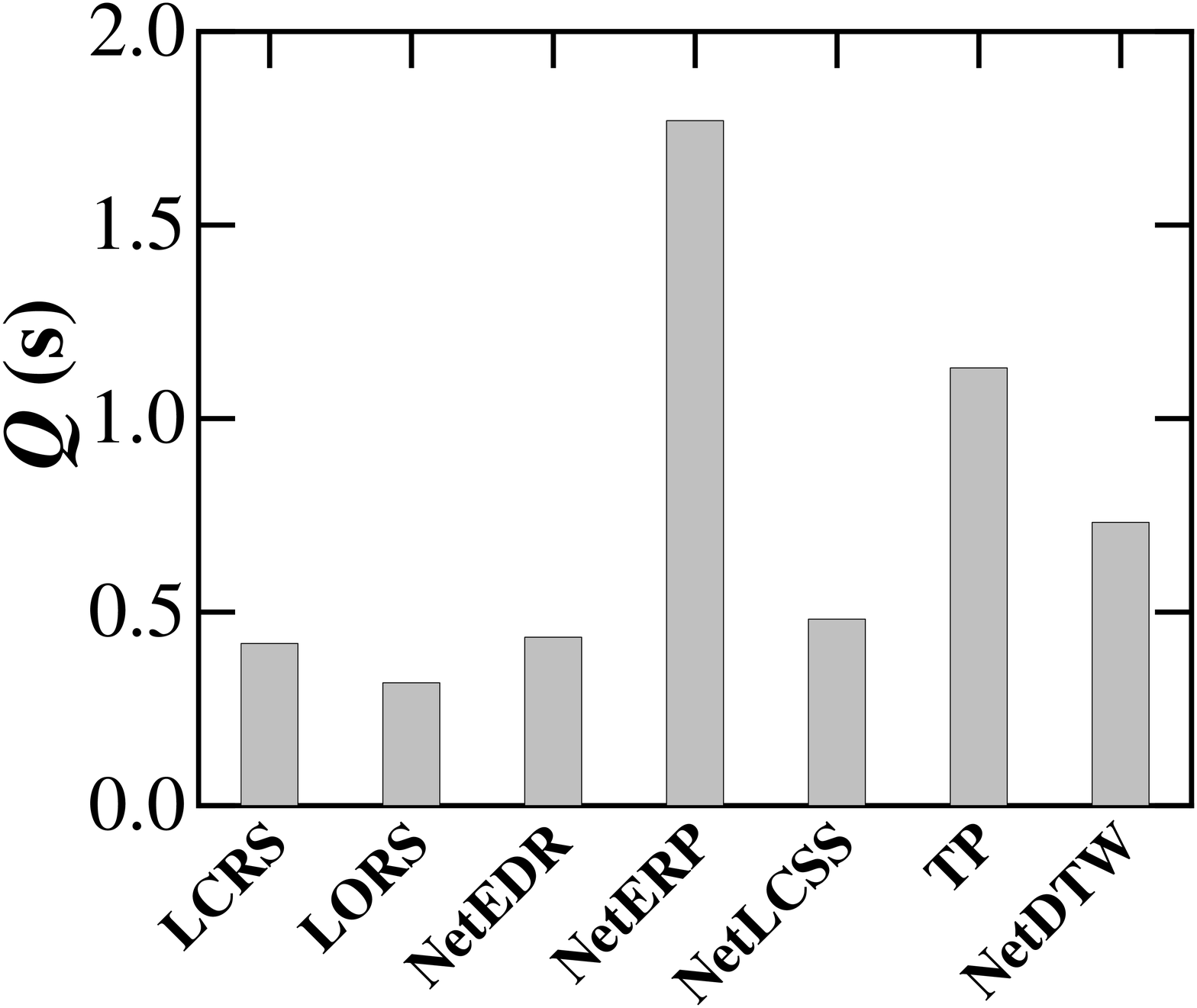}}\\
    \up
	\caption{Efficiency of Non-learning based Measures}
	\label{fig:efficiency}
	\vspace{-2mm}
\end{figure*}

\noindent\textbf{Effect of Noise Ratios.} We set the noise ratio $\textit{N (\%)}$ to 13, 16, and 19 for AIS and Geolife, and 10, 20, and 30 for T-drive and Porto, based on the data distributions of these datasets (as listed in Table~\ref{tab:transformation}). A noise ratio of $\textit{N=}$0 corresponds to the original datasets.

Figure~\ref{fig:Noise} plots the results. First, DTW, LCSS, ERP, Frechet, Hausdorff, OWD, and Seg-Frechet perform stable and good in free space settings when the noise ratio is varied (cf. Figures~\ref{fig:Noise}(a)--(f)). This is because measures such as LCSS and EDR use string distance to compute the inter-trajectory similarity, making them robust to noises.

Second, as the trajectories on AIS are gathered locally and dispersed globally, adding noises does not affect grid-trajectory mapping of OWD (cf. Figures~\ref{fig:Noise}(a)--(c)). Nevertheless, according to the results on Geolife, noises may affect the performance of OWD (cf. Figures~\ref{fig:Noise}(d)--(f)). This is because the complex features and small spatial coverage of Geolife render that the noise points severely modify the grid representation of origin trajectories.

Third, the performance of LIP is the worst among all measures on Geolife (cf. Figures~\ref{fig:Noise}(d)--(f)), which is in contrast to previous study~\cite{VLDBJSurvey}. This is because LIP computes similarity according to the polygons formed by point intersections of trajectories. However, the shapes of polygons are greatly affected by noise points, with the result that their areas and perimeters largely differ from polygons formed by original point intersections.

Next, EDwP's performance becomes worse on AIS but remains stable on Geolife, as the noise ratio increases (cf. Figures~\ref{fig:Noise}(a)--(c) and (d)--(f)). Specifically, the low sampling rate (cf. Section~\ref{sec:exe}.1) of AIS causes that the large number of noise points have a significant impact on the lengths of trajectory segments, which in turn affects the similarity computation of EDwP  (cf. Eq.~\ref{eq:EDwP}).

Finally, in road network settings, all similarity measures's performance drops drastically when adding more noises (cf. Figures~\ref{fig:Noise}(g)--(l)). Specifically, there are many paths between two locations in road networks, and thus, changing some road segments of a trajectory may result in another new trajectory with a new path. In this case, most of the measures sensitive to noise ratios. Moreover, when adapting free space oriented measures to the road networks, some noise-tolerant measures such as ERP and EDR tend to be noise-sensitive and others such as DTW and LCSS still are noise-tolerant. This indicates that some similarity measures are not suitable for being adjusted to the road networks.

\begin{table*}[]
\caption{Efficiency of Learning-based Measures}
\vspace{-3mm}
\hspace{-2mm}
\footnotesize
\begin{tabular}{|c|cccc|cccc|}
\hline
\multirow{3}{*}{\makebox[0.1\textwidth][c]{\begin{tabular}[c]{@{}c@{}}Measures in\\ Free Space\end{tabular}}}   & \multicolumn{4}{c|}{\makebox[0.42\textwidth][c]{AIS}}                                                                                         & \multicolumn{4}{c|}{\makebox[0.42\textwidth][c]{Geolife}}                                                                                     \\ \cline{2-9}
                & \multicolumn{2}{c|}{\makebox[0.1\textwidth][c]{NEUTRAJ}}                                      & \multicolumn{2}{c|}{\makebox[0.1\textwidth][c]{Traj2SimVec}}
                & \multicolumn{2}{c|}{\makebox[0.1\textwidth][c]{NEUTRAJ}}                                      & \multicolumn{2}{c|}{\makebox[0.1\textwidth][c]{Traj2SimVec}}             \\ \cline{2-9}
            & \multicolumn{1}{c|}{\makebox[0.08\textwidth][c]{$T_{tra}$(s)}}
            & \multicolumn{1}{c|}{\makebox[0.08\textwidth][c]{$Q$(ms)}}
            & \multicolumn{1}{c|}{\makebox[0.08\textwidth][c]{$T_{tra}$(s)}}
            &\multicolumn{1}{c|}{\makebox[0.05\textwidth][c]{$Q$(ms)}}
            & \multicolumn{1}{c|}{\makebox[0.08\textwidth][c]{$T_{tra}$(s)}}
            & \multicolumn{1}{c|}{\makebox[0.08\textwidth][c]{$Q$(ms)}}
            & \multicolumn{1}{c|}{\makebox[0.08\textwidth][c]{$T_{tra}$(s)}}
            & \multicolumn{1}{c|}{\makebox[0.05\textwidth][c]{$Q$(ms)}} \\ \hline
DTW                                                                                 & \multicolumn{1}{c|}{125.01}        & \multicolumn{1}{c|}{46.80}   & \multicolumn{1}{c|}{135.19}        & 6.59    & \multicolumn{1}{c|}{58.72}         & \multicolumn{1}{c|}{9.70}    & \multicolumn{1}{c|}{66.95}         & 2.96    \\ \hline
LCSS                                                                                & \multicolumn{1}{c|}{107.87}        & \multicolumn{1}{c|}{42.59}   & \multicolumn{1}{c|}{143.94}        & 5.96    & \multicolumn{1}{c|}{59.25}         & \multicolumn{1}{c|}{9.64}    & \multicolumn{1}{c|}{65.83}         & 3.36    \\ \hline
EDR                                                                                 & \multicolumn{1}{c|}{126.02}        & \multicolumn{1}{c|}{45.89}   & \multicolumn{1}{c|}{138.36}        & 6.13    & \multicolumn{1}{c|}{54.16}         & \multicolumn{1}{c|}{9.01}    & \multicolumn{1}{c|}{66.58}         & 3.35    \\ \hline
ERP                                                                                 & \multicolumn{1}{c|}{111.30}        & \multicolumn{1}{c|}{47.14}   & \multicolumn{1}{c|}{141.94}        & 5.99    & \multicolumn{1}{c|}{57.64}         & \multicolumn{1}{c|}{9.43}    & \multicolumn{1}{c|}{65.85}         & 3.94    \\ \hline
Frechet                                                                             & \multicolumn{1}{c|}{104.01}        & \multicolumn{1}{c|}{42.80}   & \multicolumn{1}{c|}{136.37}        & 7.46    & \multicolumn{1}{c|}{53.20}         & \multicolumn{1}{c|}{9.61}    & \multicolumn{1}{c|}{66.55}         & 3.65    \\ \hline
Hausdorff                                                                           & \multicolumn{1}{c|}{113.38}        & \multicolumn{1}{c|}{46.89}   & \multicolumn{1}{c|}{135.45}        & 6.61    & \multicolumn{1}{c|}{57.80}         & \multicolumn{1}{c|}{9.60}    & \multicolumn{1}{c|}{63.30}         & 3.59    \\ \hline
\multirow{3}{*}{\makebox[0.1\textwidth][c]{\begin{tabular}[c]{@{}c@{}}Measures in\\ Road Network\end{tabular}}} & \multicolumn{4}{c|}{\makebox[0.42\textwidth][c]{T-Drive}}                                                                                     & \multicolumn{4}{c|}{\makebox[0.42\textwidth][c]{Porto}}                                                                                       \\ \cline{2-9}
& \multicolumn{2}{c|}{\makebox[0.1\textwidth][c]{{GTS}}}                                          & \multicolumn{2}{c|}{\makebox[0.1\textwidth][c]{ST2Vec}}                  & \multicolumn{2}{c|}{\makebox[0.1\textwidth][c]{GTS}}                                          & \multicolumn{2}{c|}{\makebox[0.1\textwidth][c]{ST2Vec}}                  \\ \cline{2-9}
        & \multicolumn{1}{c|}{\makebox[0.08\textwidth][c]{$T_{tra}$(s)}}
        & \multicolumn{1}{c|}{\makebox[0.08\textwidth][c]{$Q$(ms)}}
        & \multicolumn{1}{c|}{\makebox[0.08\textwidth][c]{$T_{tra}$(s)}} & \multicolumn{1}{c|}{\makebox[0.05\textwidth][c]{$Q$(ms)}}
        & \multicolumn{1}{c|}{\makebox[0.08\textwidth][c]{$T_{tra}$(s)}}
        & \multicolumn{1}{c|}{\makebox[0.08\textwidth][c]{$Q$(ms)}}
        & \multicolumn{1}{c|}{\makebox[0.08\textwidth][c]{$T_{tra}$(s)}}
        & \multicolumn{1}{c|}{\makebox[0.05\textwidth][c]{$Q$(ms)}} \\ \hline
NetLCSS                                                                             & \multicolumn{1}{c|}{8.62}          & \multicolumn{1}{c|}{2.67}    & \multicolumn{1}{c|}{171.84}        & 3.61    & \multicolumn{1}{c|}{6.31}          & \multicolumn{1}{c|}{0.96}    & \multicolumn{1}{c|}{106.22}        & 1.23    \\ \hline
NetDTW                                                                              & \multicolumn{1}{c|}{8.66}          & \multicolumn{1}{c|}{2.66}    & \multicolumn{1}{c|}{166.04}        & 2.56    & \multicolumn{1}{c|}{6.41}          & \multicolumn{1}{c|}{1.33}    & \multicolumn{1}{c|}{106.80}        & 1.20    \\ \hline
TP                                                                                  & \multicolumn{1}{c|}{8.64}          & \multicolumn{1}{c|}{2.63}    & \multicolumn{1}{c|}{165.75}        & 3.14    & \multicolumn{1}{c|}{6.35}          & \multicolumn{1}{c|}{1.08}    & \multicolumn{1}{c|}{119.15}        & 1.15    \\ \hline
NetERP                                                                              & \multicolumn{1}{c|}{9.83}          & \multicolumn{1}{c|}{2.82}    & \multicolumn{1}{c|}{212.68}        & 12.32   & \multicolumn{1}{c|}{6.34}          & \multicolumn{1}{c|}{0.97}    & \multicolumn{1}{c|}{111.29}        & 1.20    \\ \hline
\end{tabular}
\label{tab:efficiency}
\vspace{-4mm}
\end{table*}

\subsubsection{Learning-based measures} 

\noindent\textbf{Effect of Sampling Rates.} We perform downsampling operations with the same $\textit{S (\%)=}$10, 20, and 40 to obtain a set of transformed datasets.
Then, we directly apply the deep models trained on the original datase (i.e., $\textit{N=}$100$\%$)  to perform the Top-$50$ similarity queries on the transformed datasets and then compute $\textit{HR}@50$. Table~\ref{tab:SRF} depicts the results. We observe that all similarity functions learned by NEUTRAJ and Traj2SimVec are stable when varying sampling rates. This confirms the high robustness of learning-based similarity, which is able to learn and accommodate various sampling rates and thus is not affected by sampling points.

\noindent\textbf{Effect of Noise Ratios.} We perform the same downsampling operations in learning settings as those in  non-learning settings to obtain a set of transformed datasets. Then, we directly apply the deep models trained on the original dataset (i.e., $\textit{N=}0\%$) to perform the Top-$50$ similarity queries on the transformed datasets and then compute $\textit{HR}@50$. Tables~\ref{tab:NF}, and \ref{tab:NR} show the results on AIS, Geolife, T-drive, and Porto, respectively. As reported in Table~\ref{tab:NF}, all similarity measures learned by NEUTRAJ and Traj2SimVec are robust to noise ratios, and the performance keeps stable when adding noises. This indicates that NEUTRAJ and Traj2SimVec are noise-tolerant, and outperform non-learning based measures significantly. However, as reported in Table~\ref{tab:NR}, the performance of all similarity measures learned by ST2Vec and GTS significantly drops when adding more noises.

\subsection{Efficiency Evaluation}
\label{subsec:efficient}
We evaluate the efficiency of each measure by performing Top-$50$ similarity queries.

\subsubsection{Non-learning based measures} Given a query trajectory $QT$, we record the query time (denoted as $Q$) of using free space oriented and road network constrained measures in Figure~\ref{fig:efficiency}. First, Figures~\ref{fig:efficiency}(a) and \ref{fig:efficiency}(b) indicate that the running time of DTW, LCSS, EDR, ERP, Frechet, OWD, and Hausdorff are similar. LIP achieves the highest efficiency (the running time is around 4 seconds), while EDwP achieves the lowest efficiency (the running time is more than 500 seconds). This is because (i) the time complexity of LIP is $\textit{O((m+n)log(m+n))}$, while that of the remaining measures is $\textit{O(mn)}$; and (ii)  EdwP maintains four dynamic processing tables for similarity computation while others only maintains one.

Next, Figures~\ref{fig:efficiency}(c) and~\ref{fig:efficiency}(d) suggest that the running time of LCRS, LORS, NetEDR, NetLCSS, and NetDTW are less than 10 seconds on T-Drive and 1 seconds on Porto, respectively, while that of TP and NetERP are much larger. Specifically, LCRS and LORS achieve the best efficiency in road network settings.
This is because TP additionally considers temporal information of trajectories, and NetERP needs to compute the sum of distances between road intersections. Note that, we need to pre-compute the road network distance matrices for evaluating most of the standalone trajectory measures, which takes more than two days. In contrast, LCRS and LORS compute the overlapping segments as inter-trajectory distance, which does not need the road network distance matrices. 

Finally, all of the distributed measures (i.e., DFT, DITA, REPOSE)
achieve much higher efficiency than standalone measures (i.e., Seg-Frechet and DTW).
Among DFT, DITA, and REPOSE, REPOSE achieves the best efficiency, and DFT performs the worst. This is because the former leverages a novel heterogeneous global partitioning strategy to achieve load balancing, while the latter traverses R trees to find a pruning boundary before querying.

\begin{figure*} [tb]
	\centering
	\includegraphics[width=0.6\textwidth]{Figs/Exp/Robustness/legendFR.eps}\\
	\hspace{-4mm}
	\subfigure[$L$=20\% (AIS)]{
		\includegraphics[width=0.16\textwidth]{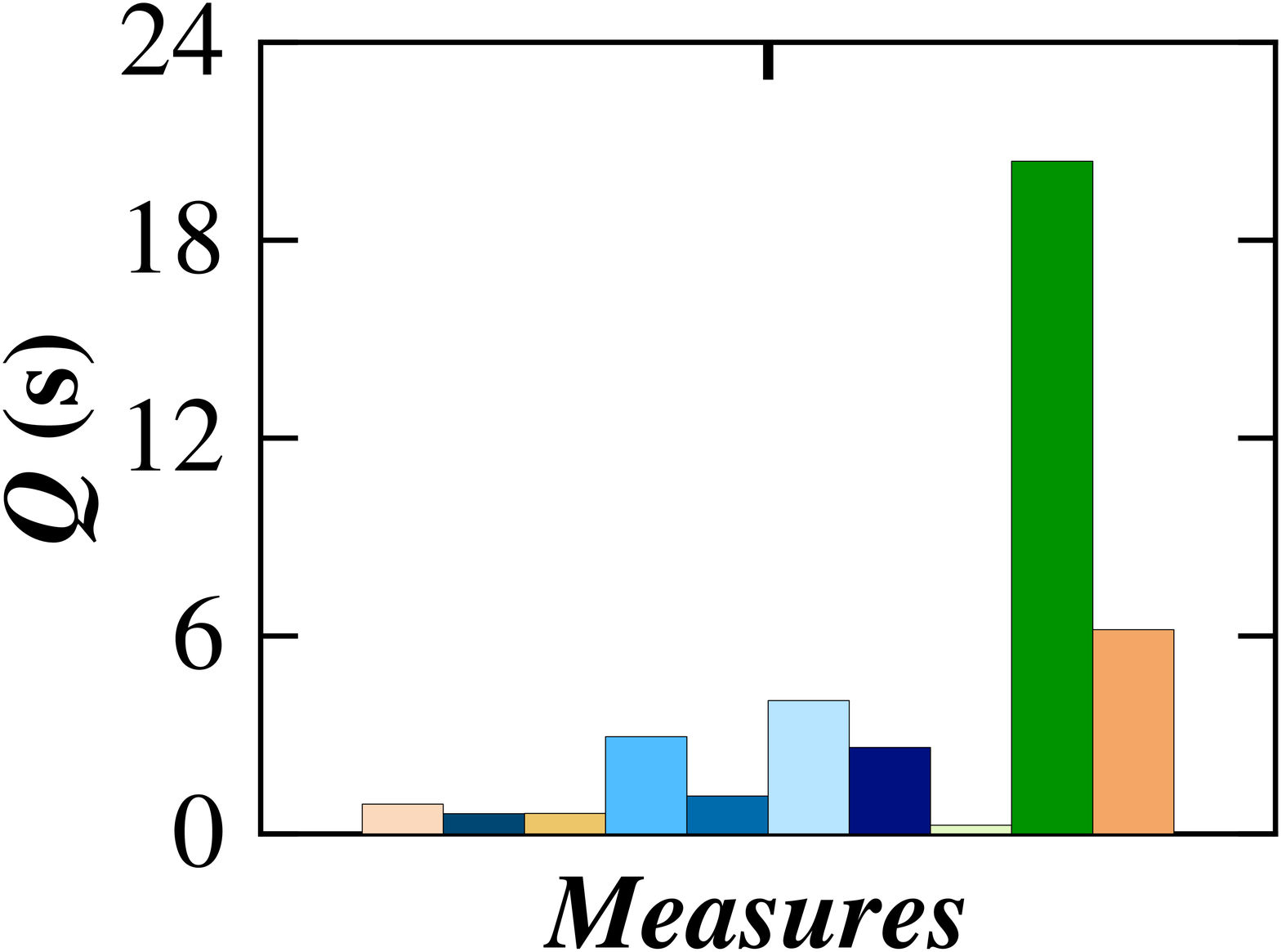}}
	\subfigure[$L$=60\% (AIS)]{
		\includegraphics[width=0.16\textwidth]{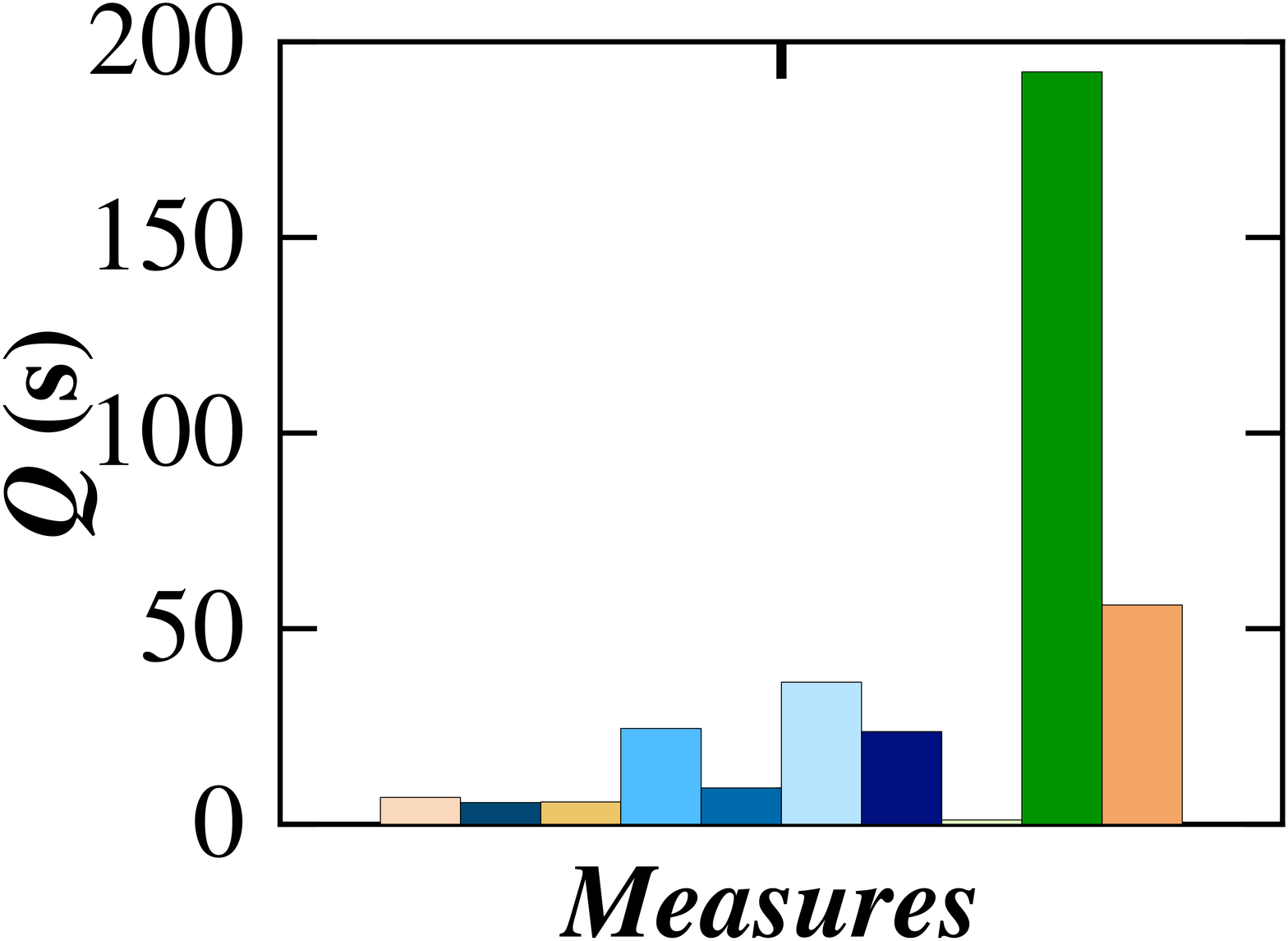}}
	\subfigure[$L$=100\% (AIS)]{
		\includegraphics[width=0.16\textwidth]{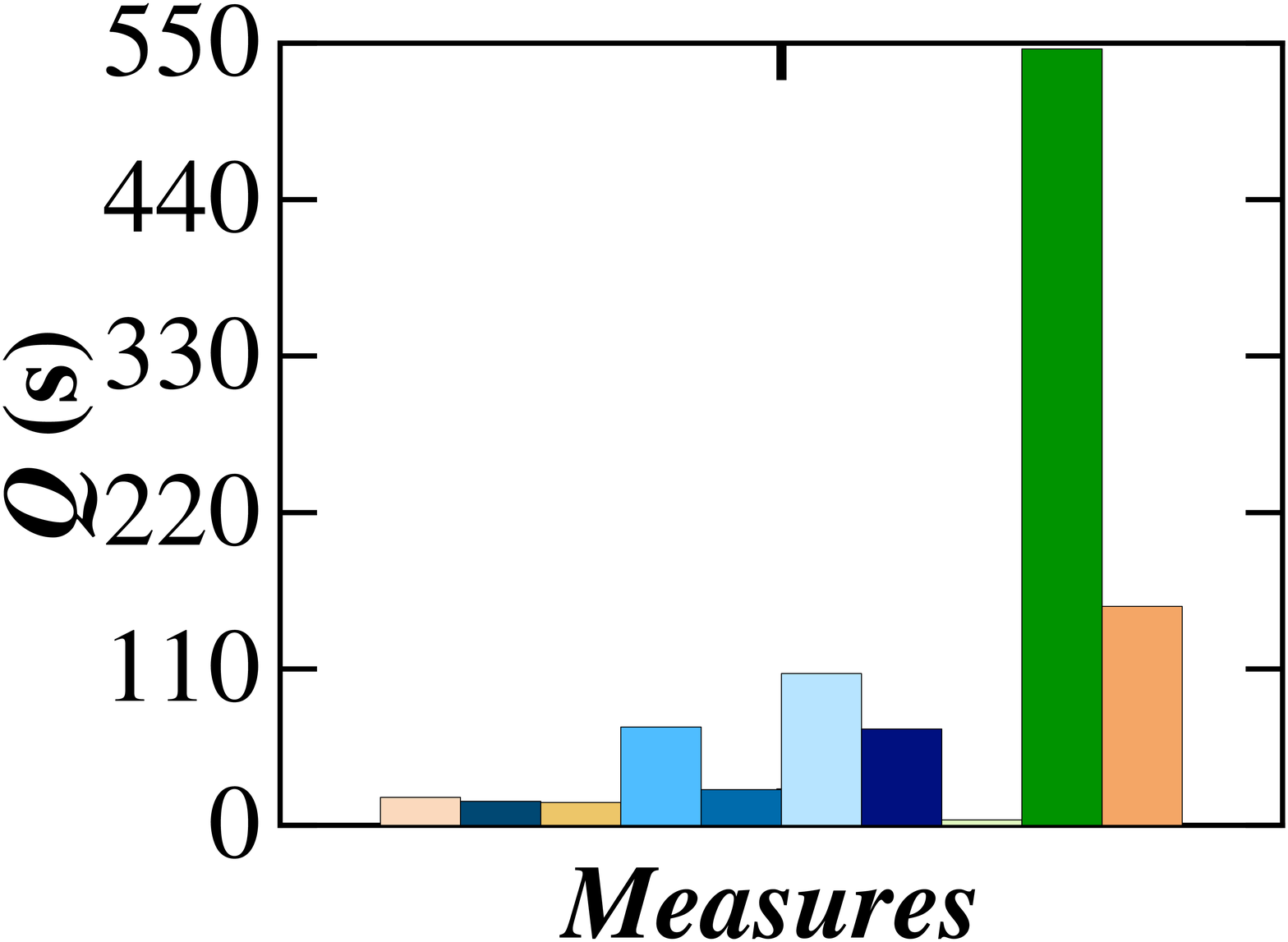}}
	\subfigure[$L$=20\% (Geolife)]{
		\includegraphics[width=0.16\textwidth]{Figs/Exp/Scalability/LF1.eps}}
	\subfigure[$L$=60\% (Geolife)]{
		\includegraphics[width=0.16\textwidth]{Figs/Exp/Scalability/LF2.eps}}
	\subfigure[$L$=100\% (Geolife)]{
		\includegraphics[width=0.16\textwidth]{Figs/Exp/Scalability/LF3.eps}}
	
	\vspace{-2mm}
	\hspace{-4mm}
	\subfigure[$L$=20\% (T-Drive)]{
		\includegraphics[width=0.16\textwidth]{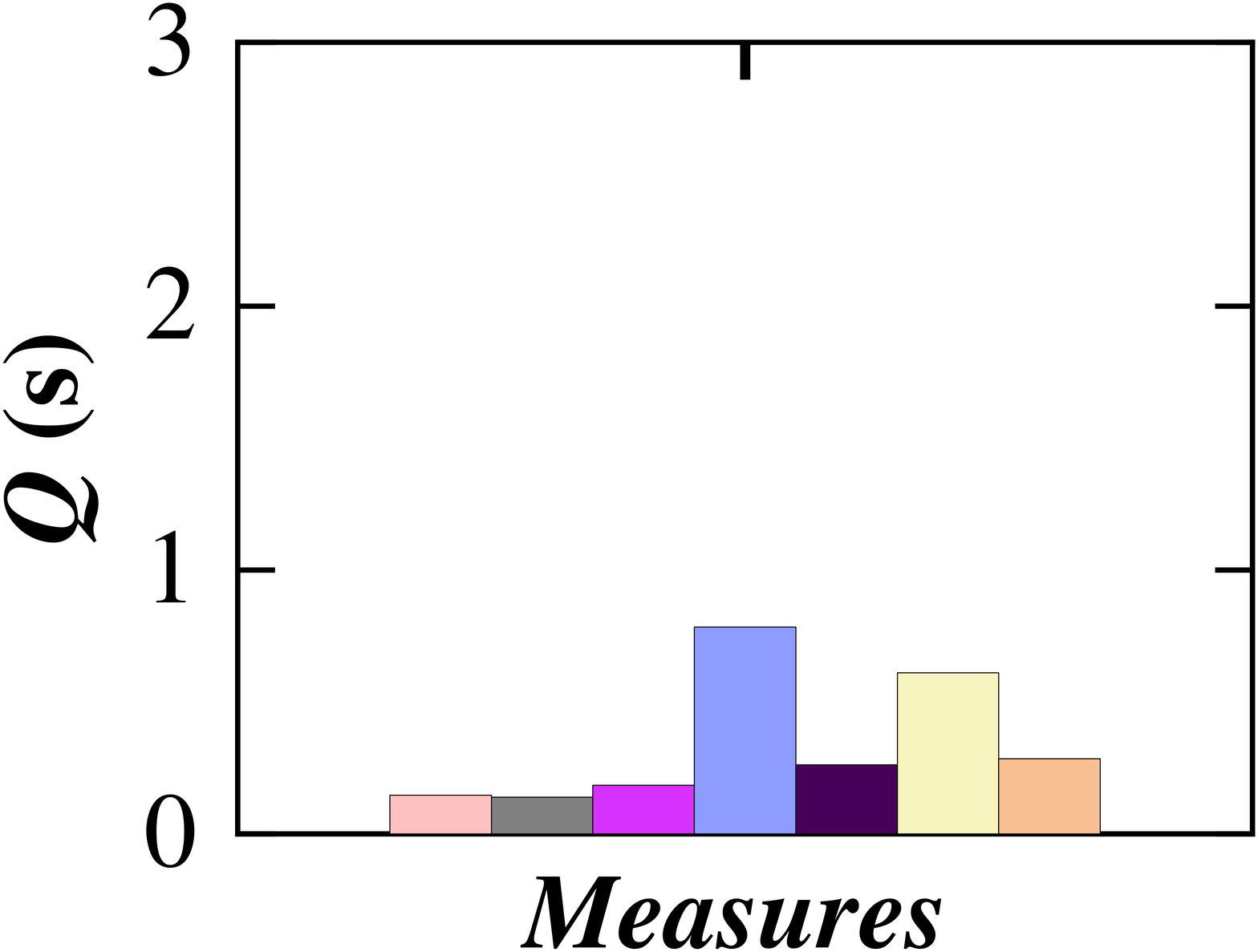}}
	\subfigure[$L$=60\% (T-Drive)]{
		\includegraphics[width=0.16\textwidth]{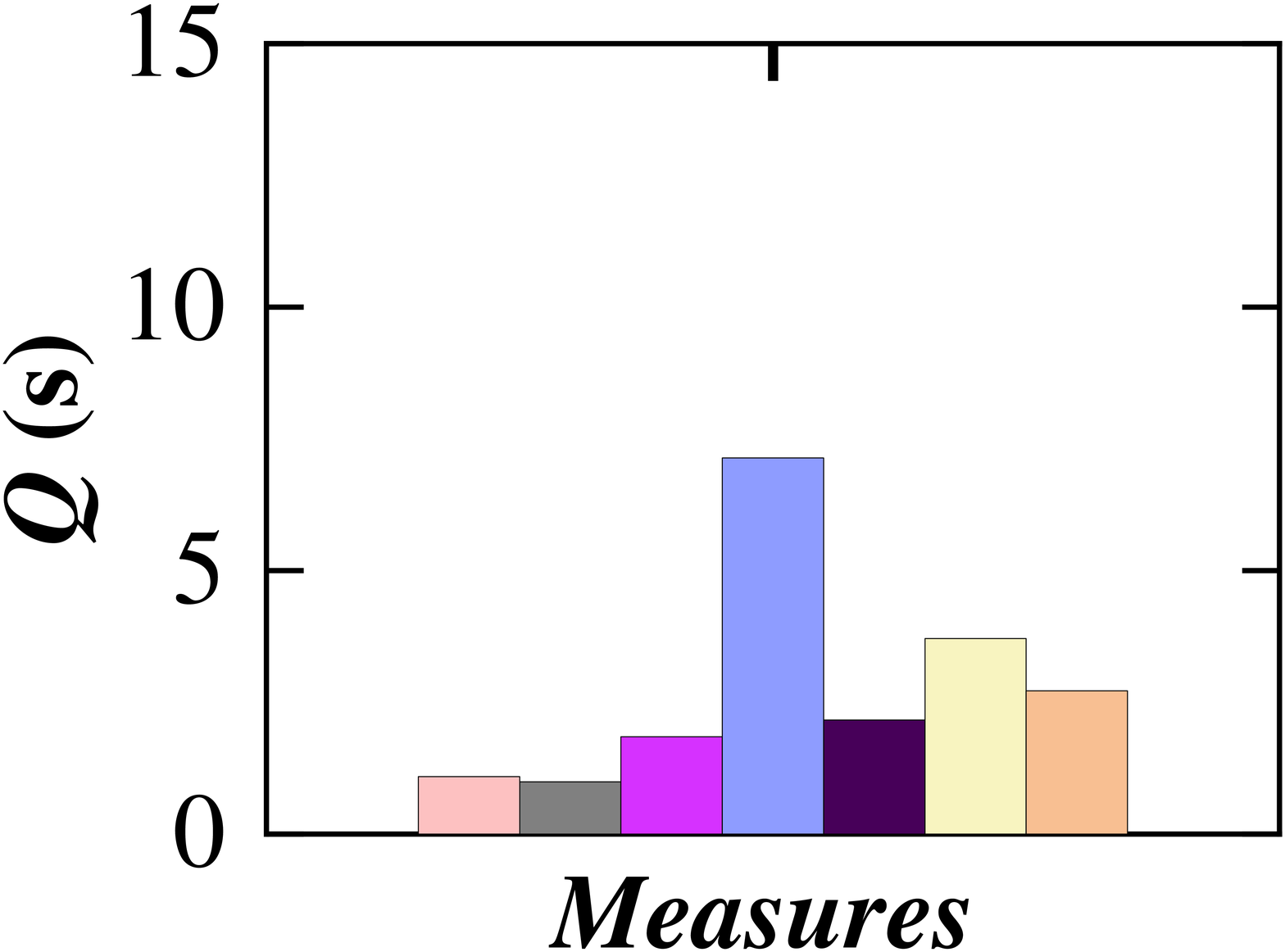}}
	\subfigure[$L$=100\% (T-Drive)]{
		\includegraphics[width=0.16\textwidth]{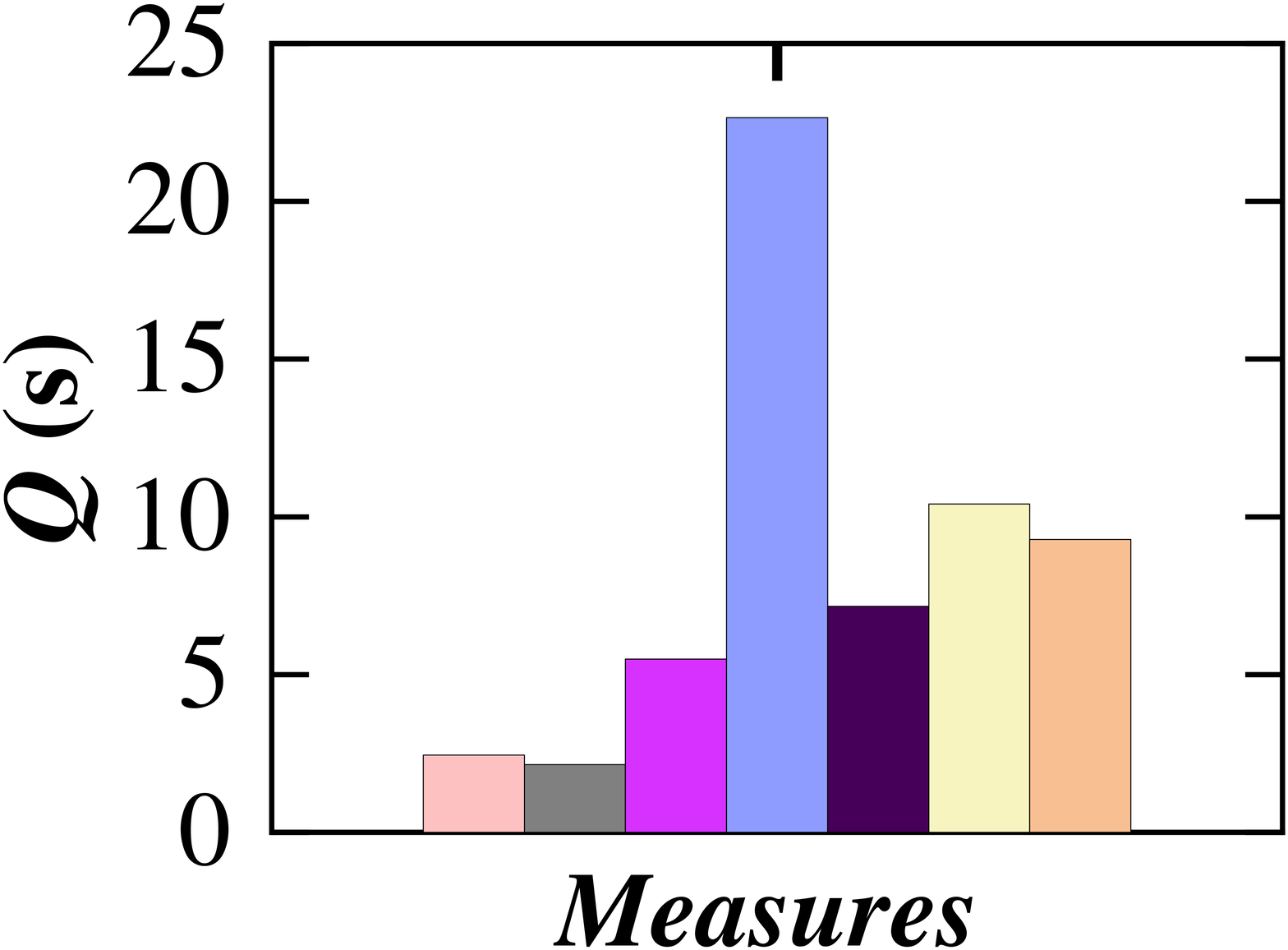}}
	\subfigure[$L$=20\% (Porto)]{
		\includegraphics[width=0.16\textwidth]{Figs/Exp/Scalability/LR1.eps}}
	\subfigure[$L$=60\% (Porto)]{
		\includegraphics[width=0.16\textwidth]{Figs/Exp/Scalability/LR2.eps}}
	\subfigure[$L$=100\% (Porto)]{
		\includegraphics[width=0.16\textwidth]{Figs/Exp/Scalability/LR3.eps}}\\
    \up
	\caption{Scalability Evaluation of Non-learning based Measures vs. Trajectory Length}
	\label{fig:scalabilityL}
	\vspace{-4mm}
\end{figure*}

\begin{figure*} [tb]
	\centering
	\vspace{-1mm}
	\hspace{-4mm}
	\subfigure[$O_r$=20\% (AIS)]{
		\includegraphics[width=0.16\textwidth]{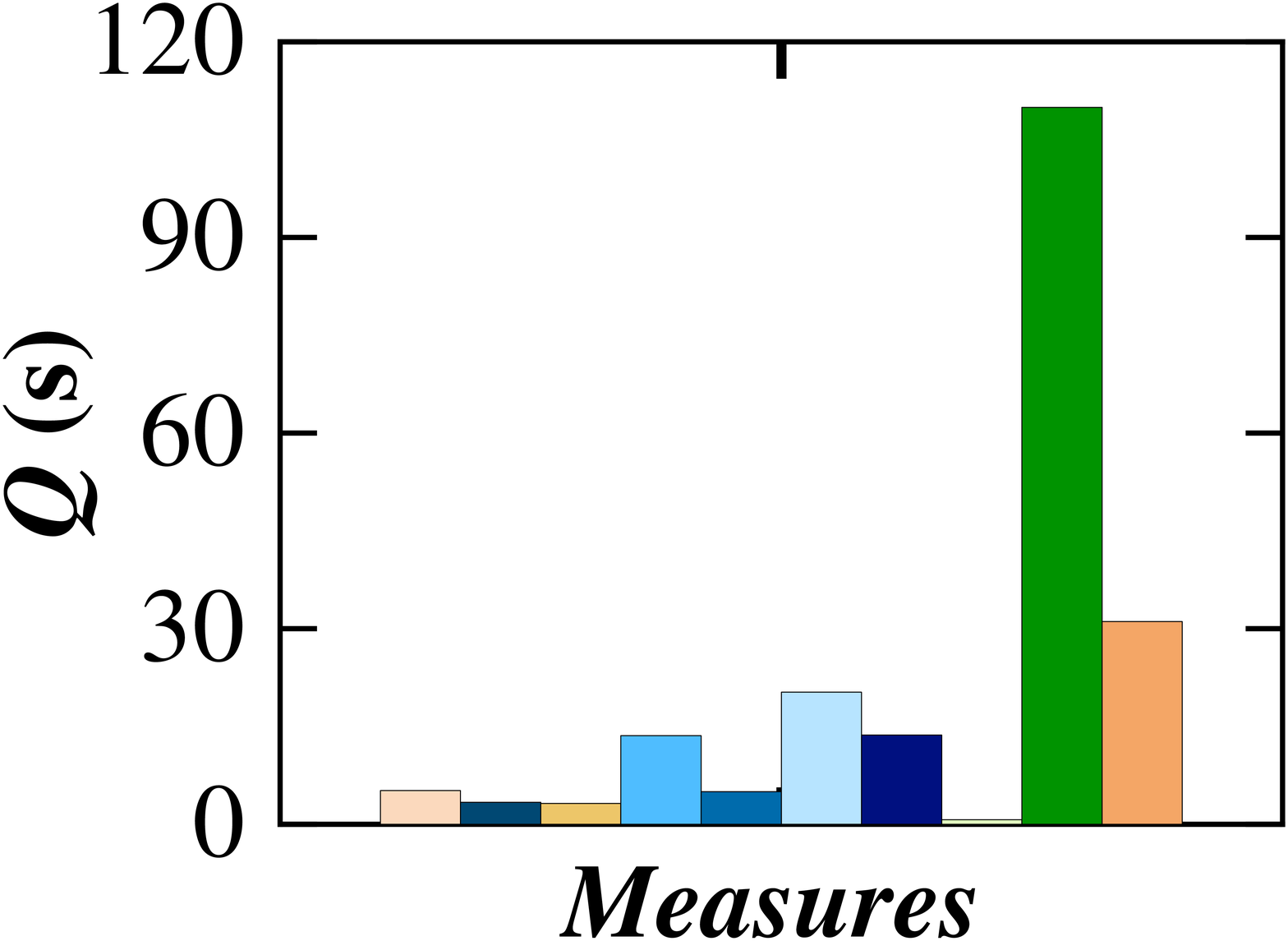}}
	\subfigure[$O_r$=60\% (AIS)]{
		\includegraphics[width=0.16\textwidth]{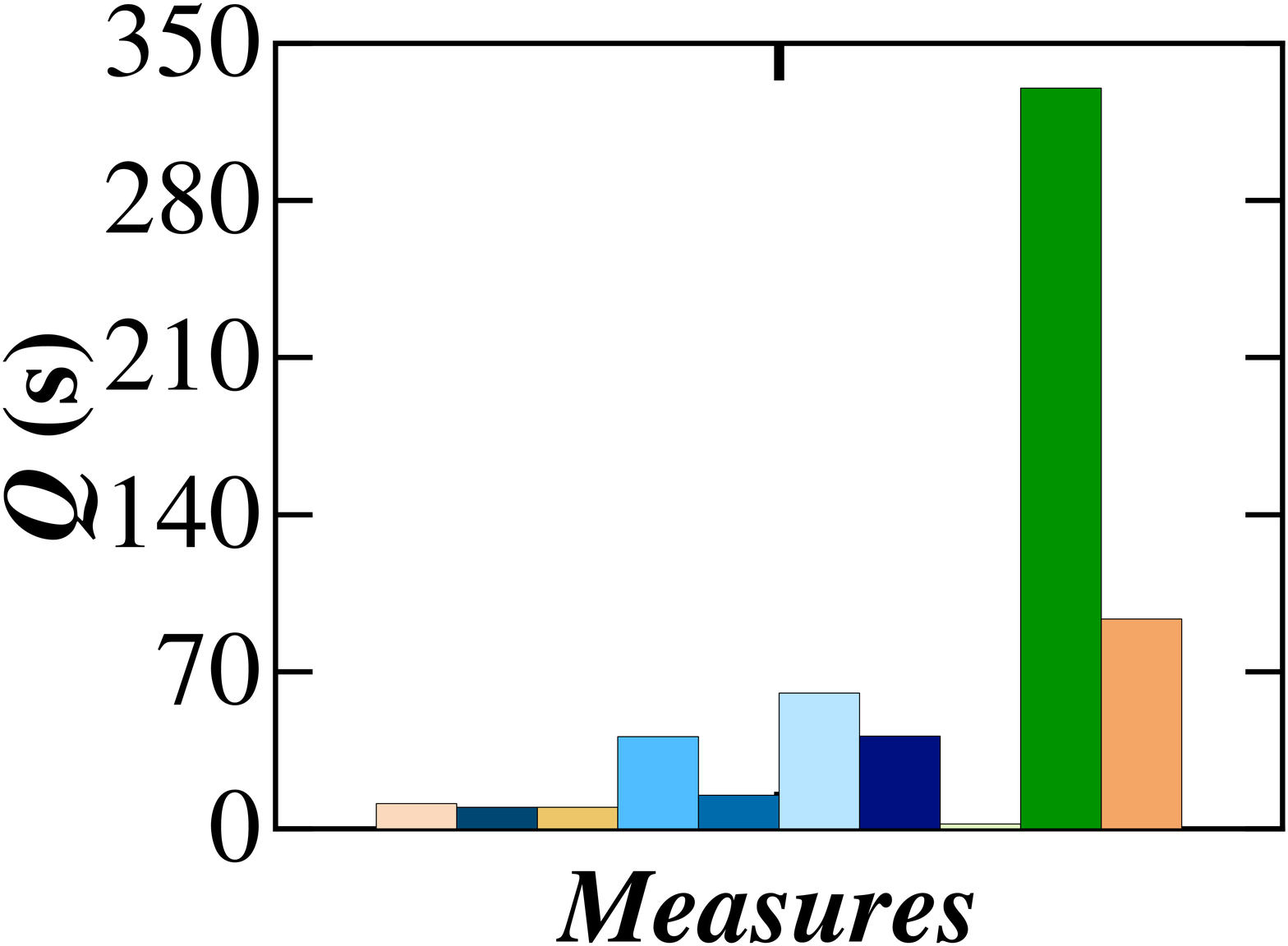}}
	\subfigure[$O_r$=100\% (AIS)]{
		\includegraphics[width=0.16\textwidth]{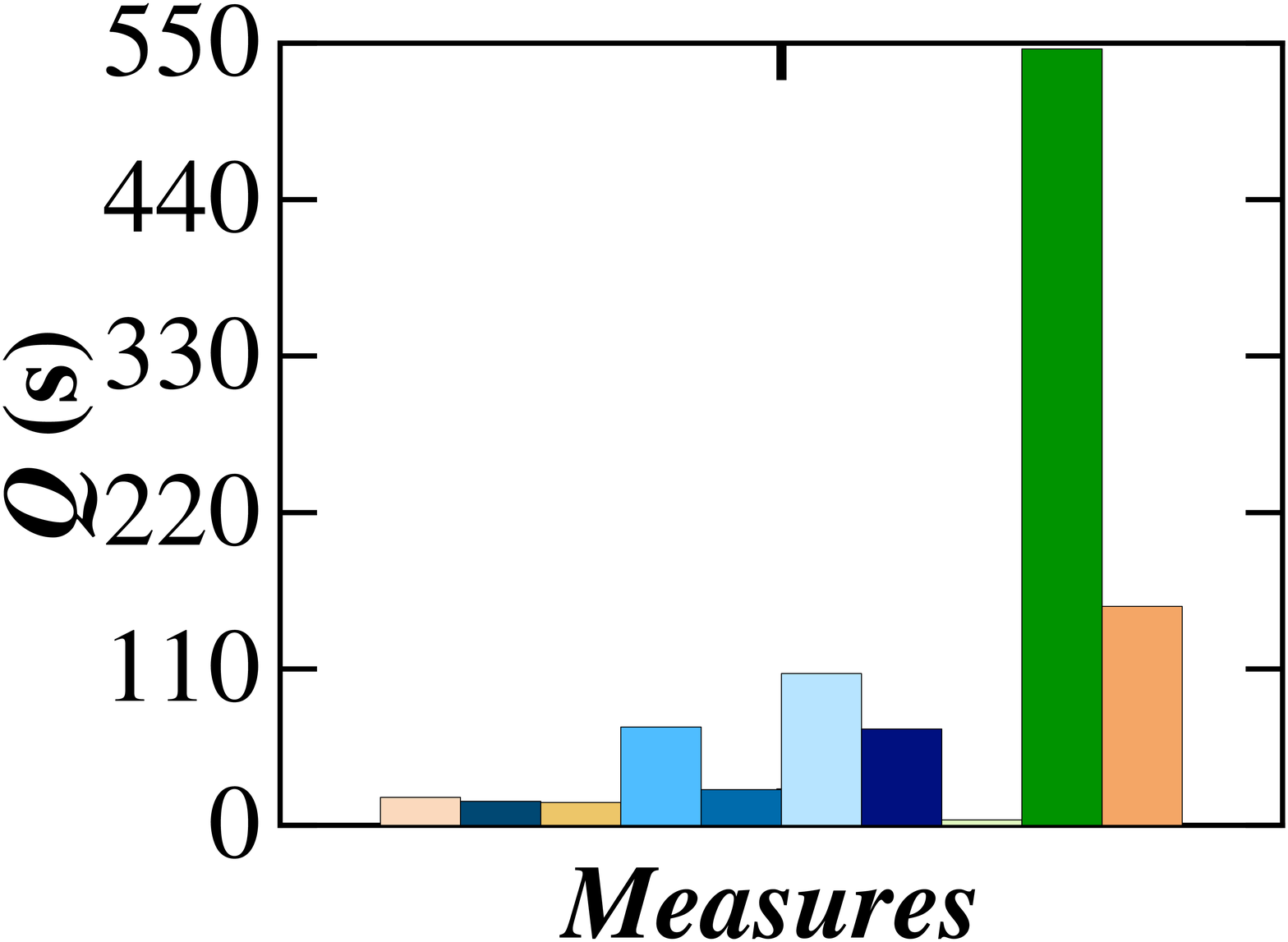}}
		\subfigure[$O_r$=20\% (Geolife)]{
		\includegraphics[width=0.16\textwidth]{Figs/Exp/Scalability/AF1.eps}}
	\subfigure[$O_r$=60\% (Geolife)]{
		\includegraphics[width=0.16\textwidth]{Figs/Exp/Scalability/AF2.eps}}
	\subfigure[$O_r$=100\% (Geolife)]{
		\includegraphics[width=0.16\textwidth]{Figs/Exp/Scalability/AF3.eps}}\\
	
	\vspace{-2mm}
	\hspace{-4mm}
	\subfigure[$O_r$=20\% (T-Drive)]{
		\includegraphics[width=0.16\textwidth]{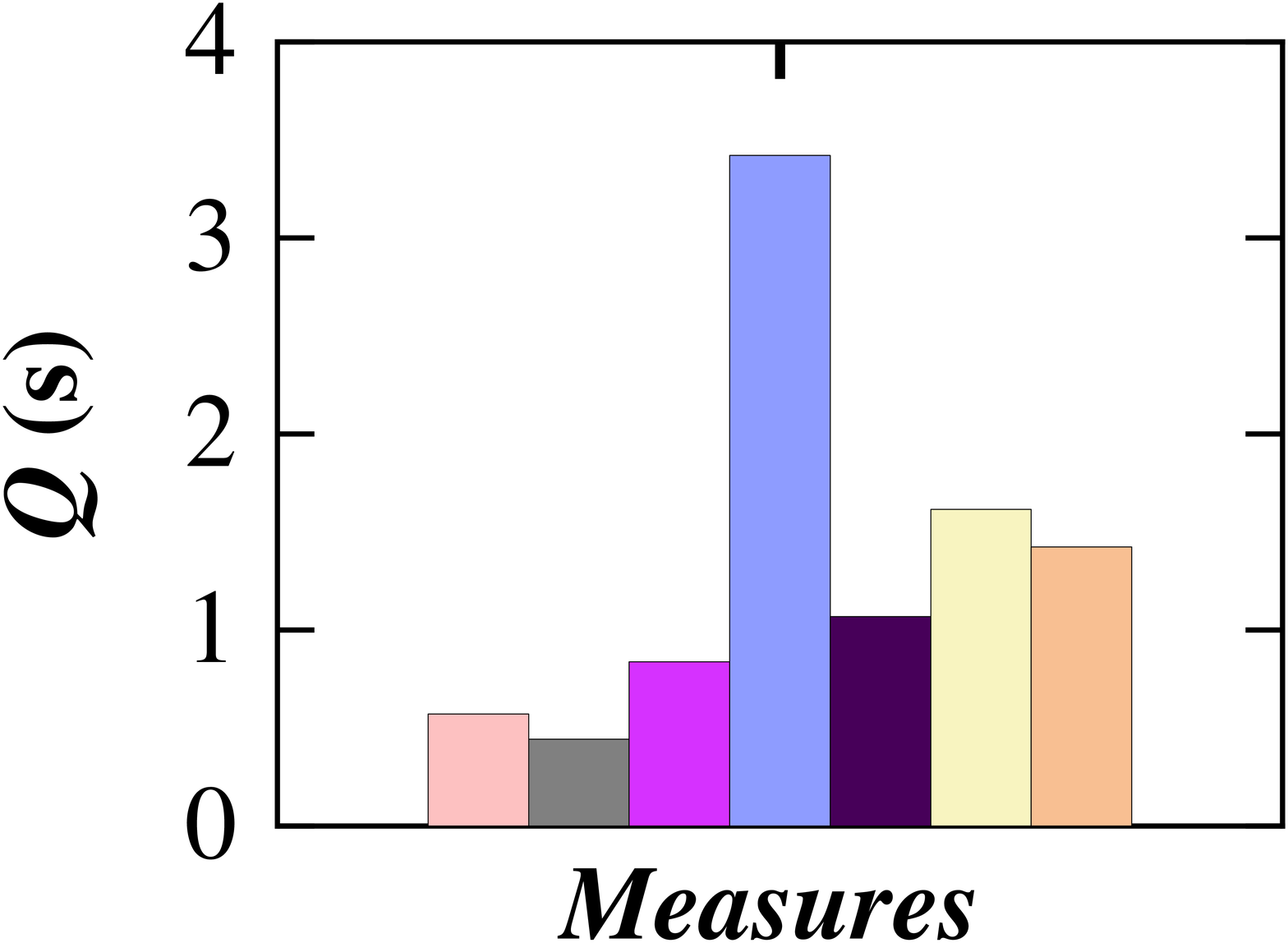}}
	\subfigure[$O_r$=60\% (T-Drive)]{
		\includegraphics[width=0.16\textwidth]{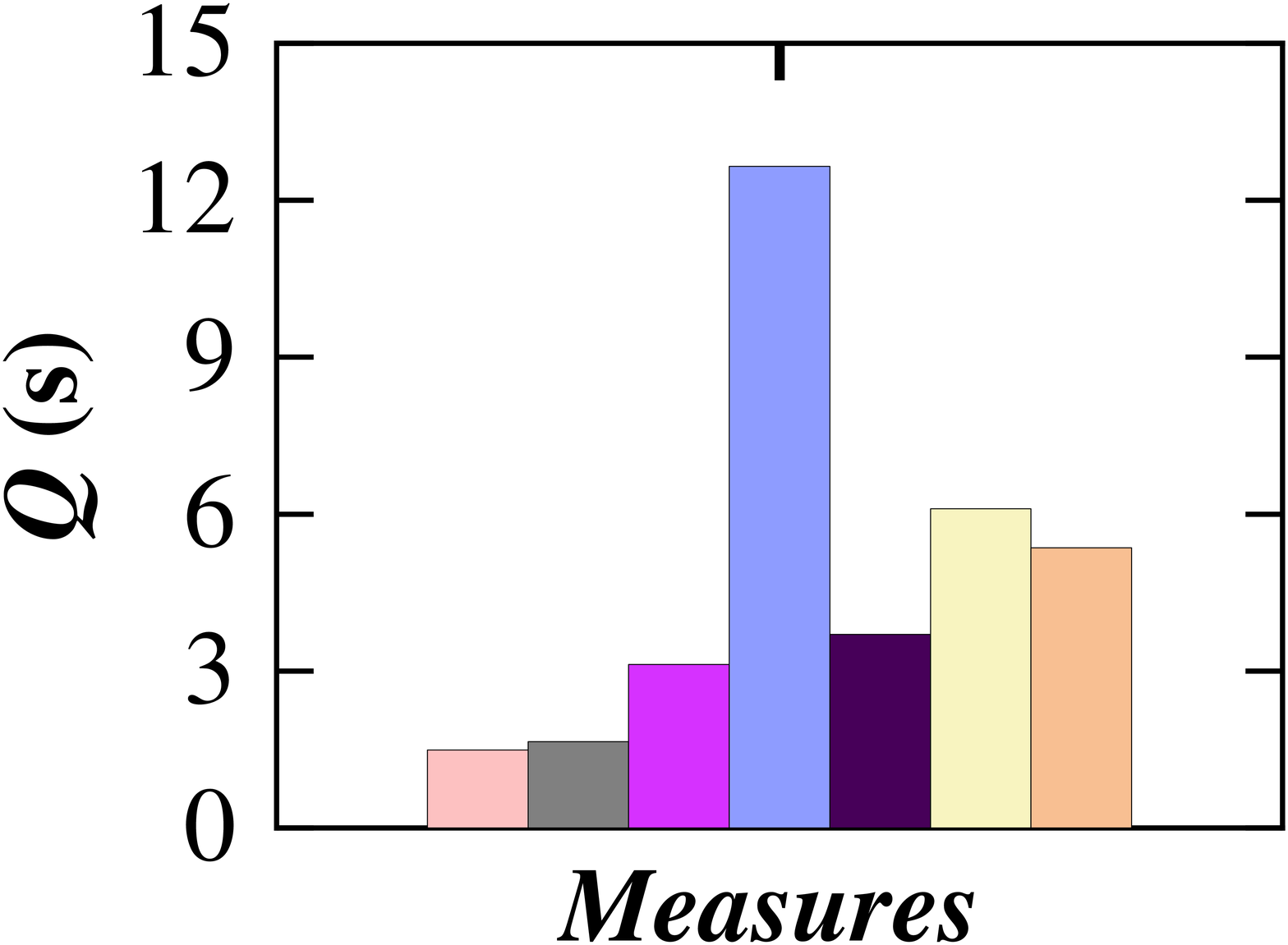}}
	\subfigure[$O_r$=100\% (T-Drive)]{
		\includegraphics[width=0.16\textwidth]{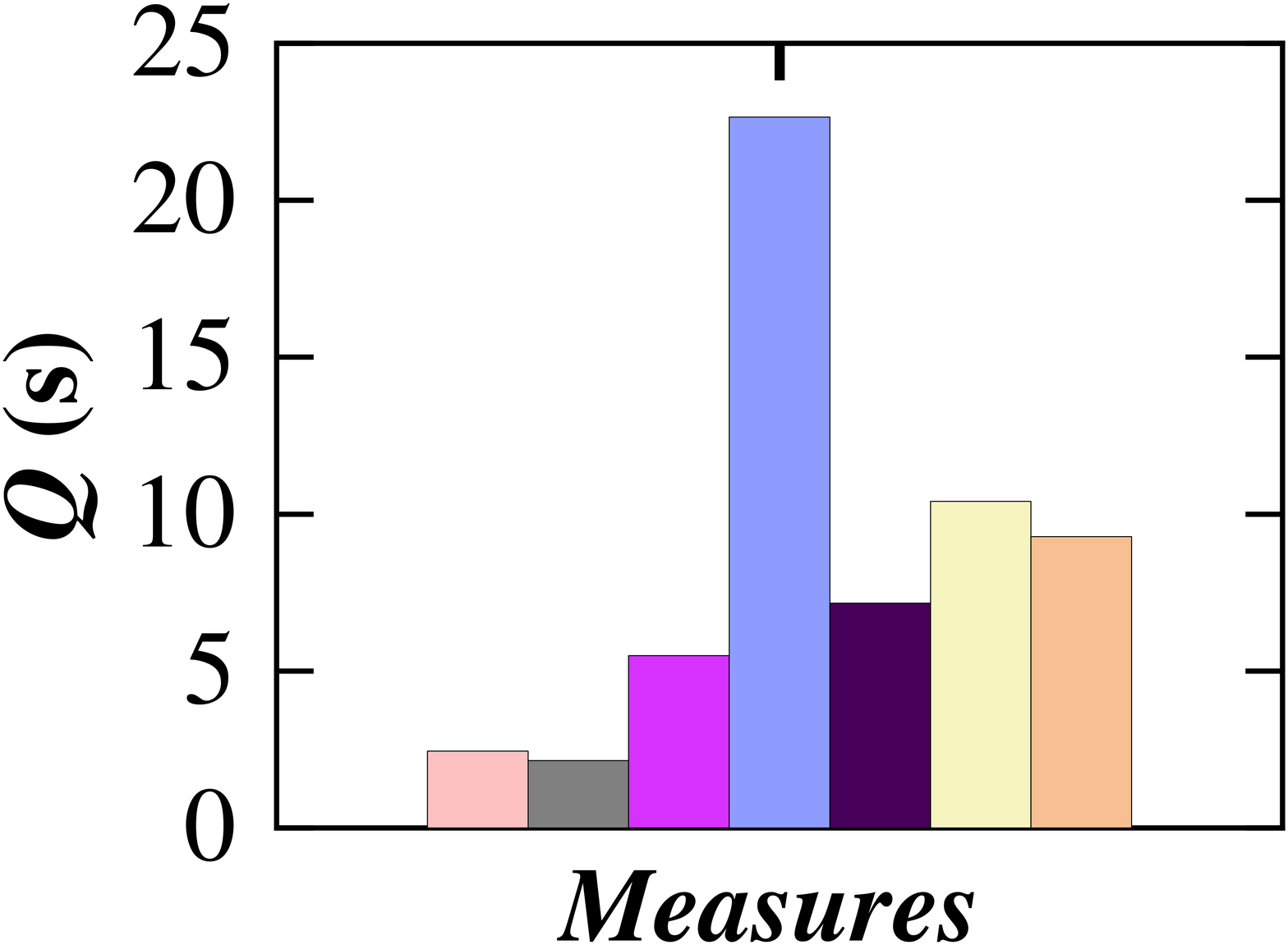}}
	\subfigure[$O_r$=20\% (Porto)]{
		\includegraphics[width=0.16\textwidth]{Figs/Exp/Scalability/AR1.eps}}
	\subfigure[$O_r$=60\% (Porto)]{
		\includegraphics[width=0.16\textwidth]{Figs/Exp/Scalability/AR2.eps}}
	\subfigure[$O_r$=100\% (Porto)]{
		\includegraphics[width=0.16\textwidth]{Figs/Exp/Scalability/AR3.eps}}\\
    \up
	\caption{Scalability Evaluation of Non-leaning based Measures vs. Data Cardinality}
	\label{fig:scalabilityA}
	\vspace{-2mm}
\end{figure*}

\subsubsection{Learning-based measures}Given a query trajectory $QT$, we record the training time (denoted as $T_{tra}$) and query time (denoted as $Q$) of using free space oriented and road network constrained measures for querying on four datasets in Table~\ref{tab:efficiency}.

\begin{table*}[]
\caption{Scalability Evaluation of Learning-based Measures on Training Time and Query Time vs. Trajectory Length (AIS \& T-Drive)}
\vspace{-3mm}
\hspace{1.5mm}
\small
\begin{tabular}{|c|cccc|cccc|cccc|}
\hline
                                                                                     \makebox[0.1\textwidth][c]{\textbf{$L(\%)$}}                     & \multicolumn{4}{c|}{{ \makebox[0.25\textwidth][c]{$L$=20}}}                                       & \multicolumn{4}{c|}{{ \makebox[0.25\textwidth][c]{$L$=60}}}                                        & \multicolumn{4}{c|}{{ \makebox[0.25\textwidth][c]{$L$=100}}}                                        \\ \hline
{\color[HTML]{000000} }                                                                                    & \multicolumn{2}{c|}{NEUTRAJ}                               & \multicolumn{2}{c|}{Traj2SimVec}      & \multicolumn{2}{c|}{NEUTRAJ}                               & \multicolumn{2}{c|}{Traj2SimVec}      & \multicolumn{2}{c|}{NEUTRAJ}                               & \multicolumn{2}{c|}{Traj2SimVec}      \\ \cline{2-13}
\multirow{-2}{*}{{ \begin{tabular}[c]{@{}c@{}}Measures in\\ Free Space\end{tabular}}}   & \multicolumn{1}{c|}{\makebox[0.03\textwidth][c]{$T_{tra}$(s)}} & \multicolumn{1}{c|}{\makebox[0.02\textwidth][c]{$Q$(ms)}} & \multicolumn{1}{c|}{\makebox[0.03\textwidth][c]{$T_{tra}$(s)}} & \makebox[0.02\textwidth][c]{$Q$(ms)} & \multicolumn{1}{c|}{\makebox[0.03\textwidth][c]{$T_{tra}$(s)}} & \multicolumn{1}{c|}{\makebox[0.02\textwidth][c]{$Q$(ms)}} & \multicolumn{1}{c|}{\makebox[0.03\textwidth][c]{$T_{tra}$(s)}} & \makebox[0.02\textwidth][c]{$Q$(ms)} & \multicolumn{1}{c|}{\makebox[0.03\textwidth][c]{$T_{tra}$(s)}} & \multicolumn{1}{c|}{\makebox[0.02\textwidth][c]{$Q$(ms)}} & \multicolumn{1}{c|}{\makebox[0.03\textwidth][c]{$T_{tra}$(s)}} & \makebox[0.02\textwidth][c]{$Q$(ms)} \\ \hline
DTW                                                                                                        & \multicolumn{1}{c|}{101.12}       & \multicolumn{1}{c|}{42.19}     & \multicolumn{1}{c|}{45.36}        & 4.85     & \multicolumn{1}{c|}{114.06}       & \multicolumn{1}{c|}{43.72}     & \multicolumn{1}{c|}{91.92}        & 6.82     & \multicolumn{1}{c|}{125.01}       & \multicolumn{1}{c|}{46.80}     & \multicolumn{1}{c|}{135.19}       & 6.59     \\ \hline
LCSS                                                                                                       & \multicolumn{1}{c|}{105.10}        & \multicolumn{1}{c|}{40.36}     & \multicolumn{1}{c|}{44.68}        & 4.40     & \multicolumn{1}{c|}{108.21}       & \multicolumn{1}{c|}{45.33}      & \multicolumn{1}{c|}{96.79}        &  5.90   & \multicolumn{1}{c|}{107.87}       & \multicolumn{1}{c|}{42.59}     & \multicolumn{1}{c|}{143.94}       & 5.97     \\ \hline
EDR                                                                                                        & \multicolumn{1}{c|}{100.77}       & \multicolumn{1}{c|}{41.67}      & \multicolumn{1}{c|}{44.92}        & 4.37     & \multicolumn{1}{c|}{111.34}       & \multicolumn{1}{c|}{44.40}     & \multicolumn{1}{c|}{92.74}        & 5.73     & \multicolumn{1}{c|}{126.02}       & \multicolumn{1}{c|}{45.89}     & \multicolumn{1}{c|}{138.36}       & 6.13     \\ \hline
ERP                                                                                                        & \multicolumn{1}{c|}{103.86}       & \multicolumn{1}{c|}{41.56}      & \multicolumn{1}{c|}{42.34}        & 4.82     & \multicolumn{1}{c|}{113.05}       & \multicolumn{1}{c|}{42.80}     & \multicolumn{1}{c|}{90.34}        & 5.82     & \multicolumn{1}{c|}{111.30}       & \multicolumn{1}{c|}{47.14}     & \multicolumn{1}{c|}{141.94}       & 5.99     \\ \hline
Frechet                                                                                                    & \multicolumn{1}{c|}{103.86}       & \multicolumn{1}{c|}{41.33}      & \multicolumn{1}{c|}{44.89}        & 4.80     & \multicolumn{1}{c|}{114.11}       & \multicolumn{1}{c|}{41.48}     & \multicolumn{1}{c|}{96.56}        & 6.83      & \multicolumn{1}{c|}{104.01}       & \multicolumn{1}{c|}{42.80}     & \multicolumn{1}{c|}{136.37}       & 7.46     \\ \hline
Hausdorff                                                                                                  & \multicolumn{1}{c|}{103.87}       & \multicolumn{1}{c|}{41.67}      & \multicolumn{1}{c|}{44.78}        & 4.87     & \multicolumn{1}{c|}{113.76}       & \multicolumn{1}{c|}{42.96}     & \multicolumn{1}{c|}{93.47}        & 7.28     & \multicolumn{1}{c|}{113.38}       & \multicolumn{1}{c|}{46.89}     & \multicolumn{1}{c|}{135.45}       & 6.61     \\ \hline

{\color[HTML]{000000} }                                                                                    & \multicolumn{2}{c|}{GTS}                                           & \multicolumn{2}{c|}{ST2Vec}                   & \multicolumn{2}{c|}{GTS}                                           & \multicolumn{2}{c|}{ST2Vec}                   & \multicolumn{2}{c|}{GTS}                                           & \multicolumn{2}{c|}{ST2Vec}                   \\ \cline{2-13}
\multirow{-2}{*}{ \begin{tabular}[c]{@{}c@{}}Measures in\\ Road Network\end{tabular}} & \multicolumn{1}{c|}{\makebox[0.03\textwidth][c]{$T_{tra}$(s)}} & \multicolumn{1}{c|}{\makebox[0.02\textwidth][c]{$Q$(ms)}} & \multicolumn{1}{c|}{\makebox[0.03\textwidth][c]{$T_{tra}$(s)}} & \makebox[0.02\textwidth][c]{$Q$(ms)} & \multicolumn{1}{c|}{\makebox[0.03\textwidth][c]{$T_{tra}$(s)}} & \multicolumn{1}{c|}{\makebox[0.02\textwidth][c]{$Q$(ms)}} & \multicolumn{1}{c|}{\makebox[0.03\textwidth][c]{$T_{tra}$(s)}} & \makebox[0.02\textwidth][c]{$Q$(ms)} & \multicolumn{1}{c|}{\makebox[0.03\textwidth][c]{$T_{tra}$(s)}} & \multicolumn{1}{c|}{\makebox[0.02\textwidth][c]{$Q$(ms)}} & \multicolumn{1}{c|}{\makebox[0.03\textwidth][c]{$T_{tra}$(s)}} & \makebox[0.02\textwidth][c]{$Q$(ms)} \\ \hline
NetLCSS                                                                                                    & \multicolumn{1}{c|}{3.08}         & \multicolumn{1}{c|}{0.79}      & \multicolumn{1}{c|}{27.13}        & 0.81      & \multicolumn{1}{c|}{5.84}     & \multicolumn{1}{c|}{2.21}   & \multicolumn{1}{c|}{84.69}          & \multicolumn{1}{c|}{1.93}             & \multicolumn{1}{c|}{8.62}         & \multicolumn{1}{c|}{2.67}      & \multicolumn{1}{c|}{171.84}       & 3.61     \\ \hline
NetDTW                                                                                                     & \multicolumn{1}{c|}{2.98}         & \multicolumn{1}{c|}{0.77}      & \multicolumn{1}{c|}{25.58}        & 0.82      & \multicolumn{1}{c|}{5.35}    & \multicolumn{1}{c|}{1.36}     & \multicolumn{1}{c|}{87.90}      & \multicolumn{1}{c|}{1.97}            & \multicolumn{1}{c|}{8.66}         & \multicolumn{1}{c|}{2.66}      & \multicolumn{1}{c|}{166.04}       & 2.56      \\ \hline
TP                                                                                                         & \multicolumn{1}{c|}{3.06}         & \multicolumn{1}{c|}{0.65}      & \multicolumn{1}{c|}{27.21}        & 0.92      & \multicolumn{1}{c|}{5.29}    & \multicolumn{1}{c|}{1.71}     & \multicolumn{1}{c|}{84.04}     & \multicolumn{1}{c|}{1.97}           & \multicolumn{1}{c|}{8.64}         & \multicolumn{1}{c|}{2.63}      & \multicolumn{1}{c|}{165.75}       & 3.14     \\ \hline
NetERP                                                                                                     & \multicolumn{1}{c|}{3.06}         & \multicolumn{1}{c|}{0.81}      & \multicolumn{1}{c|}{26.17}        & 0.94      & \multicolumn{1}{c|}{5.45}    & \multicolumn{1}{c|}{1.69}     & \multicolumn{1}{c|}{82.56}     & \multicolumn{1}{c|}{1.66}          & \multicolumn{1}{c|}{8.83}         & \multicolumn{1}{c|}{2.82}     & \multicolumn{1}{c|}{212.68}       & 12.32     \\ \hline

\end{tabular}
\label{tab:scalabilityL}
\vspace{-4mm}
\end{table*}

\begin{table*}[]
\caption{Scalability Evaluation of Learning-based Measures on Training Time and Query Time vs. Data Cardinality (AIS \& T-Drive)}
\vspace{-3mm}
\hspace{1.5mm}
\small
\begin{tabular}{|c|cccc|cccc|cccc|}
\hline
                                                                                                     \makebox[0.1\textwidth][c]{$O_r(\%)$}      & \multicolumn{4}{c|}{{ \makebox[0.25\textwidth][c]{$O_r$=20}}}                                        & \multicolumn{4}{c|}{{ \makebox[0.25\textwidth][c]{$O_r$=60}}}                                        & \multicolumn{4}{c|}{{ \makebox[0.25\textwidth][c]{$O_r$=100}}}                                        \\ \hline
{\color[HTML]{000000} }                                                                                    & \multicolumn{2}{c|}{NEUTRAJ}                               & \multicolumn{2}{c|}{Traj2SimVec}      & \multicolumn{2}{c|}{NEUTRAJ}                               & \multicolumn{2}{c|}{Traj2SimVec}      & \multicolumn{2}{c|}{NEUTRAJ}                               & \multicolumn{2}{c|}{Traj2SimVec}      \\ \cline{2-13}
\multirow{-2}{*}{{ \begin{tabular}[c]{@{}c@{}}Measures in\\ Free Space\end{tabular}}}   & \multicolumn{1}{c|}{\makebox[0.03\textwidth][c]{$T_{tra}$(s)}} & \multicolumn{1}{c|}{\makebox[0.02\textwidth][c]{$Q$(ms)}} & \multicolumn{1}{c|}{\makebox[0.03\textwidth][c]{$T_{tra}$(s)}} & \makebox[0.02\textwidth][c]{$Q$(ms)} & \multicolumn{1}{c|}{\makebox[0.03\textwidth][c]{$T_{tra}$(s)}} & \multicolumn{1}{c|}{\makebox[0.02\textwidth][c]{$Q$(ms)}} & \multicolumn{1}{c|}{\makebox[0.03\textwidth][c]{$T_{tra}$(s)}} & \makebox[0.02\textwidth][c]{$Q$(ms)} & \multicolumn{1}{c|}{\makebox[0.03\textwidth][c]{$T_{tra}$(s)}} & \multicolumn{1}{c|}{\makebox[0.02\textwidth][c]{$Q$(ms)}} & \multicolumn{1}{c|}{\makebox[0.03\textwidth][c]{$T_{tra}$(s)}} & \makebox[0.02\textwidth][c]{$Q$(ms)} \\ \hline
DTW                                                                                                        & \multicolumn{1}{c|}{100.92}   & \multicolumn{1}{c|}{18.31} & \multicolumn{1}{c|}{45.26}    & 1.30  & \multicolumn{1}{c|}{122.88}   & \multicolumn{1}{c|}{29.11} & \multicolumn{1}{c|}{106.01}   & 3.32 & \multicolumn{1}{c|}{125.01}   & \multicolumn{1}{c|}{46.80} & \multicolumn{1}{c|}{135.19}   & 6.59 \\ \hline
LCSS                                                                                                       & \multicolumn{1}{c|}{109.90}   & \multicolumn{1}{c|}{18.26} & \multicolumn{1}{c|}{29.64}    & 1.28  & \multicolumn{1}{c|}{125.88}   & \multicolumn{1}{c|}{29.26} & \multicolumn{1}{c|}{107.02}   & 3.39  & \multicolumn{1}{c|}{107.87}   & \multicolumn{1}{c|}{42.59} & \multicolumn{1}{c|}{143.94}   & 5.97 \\ \hline
EDR                                                                                                        & \multicolumn{1}{c|}{100.79}   & \multicolumn{1}{c|}{18.08} & \multicolumn{1}{c|}{44.52}    & 1.33  & \multicolumn{1}{c|}{123.67}   & \multicolumn{1}{c|}{29.69} & \multicolumn{1}{c|}{106.56}   & 3.21 & \multicolumn{1}{c|}{126.02}   & \multicolumn{1}{c|}{25.89} & \multicolumn{1}{c|}{138.36}   & 6.13 \\ \hline
ERP                                                                                                        & \multicolumn{1}{c|}{101.95}   & \multicolumn{1}{c|}{18.48} & \multicolumn{1}{c|}{45.35}    & 1.33  & \multicolumn{1}{c|}{123.46}   & \multicolumn{1}{c|}{29.56} & \multicolumn{1}{c|}{106.58}   & 3.46 & \multicolumn{1}{c|}{111.30}   & \multicolumn{1}{c|}{47.14} & \multicolumn{1}{c|}{141.94}   & 5.99 \\ \hline
Frechet                                                                                                    & \multicolumn{1}{c|}{102.10}   & \multicolumn{1}{c|}{18.32} & \multicolumn{1}{c|}{43.69}    & 1.32  & \multicolumn{1}{c|}{126.28}   & \multicolumn{1}{c|}{29.26} & \multicolumn{1}{c|}{107.24}   & 4.55 & \multicolumn{1}{c|}{104.01}   & \multicolumn{1}{c|}{42.80} & \multicolumn{1}{c|}{136.37}   & 7.46 \\ \hline
Hausdorff                                                                                                  & \multicolumn{1}{c|}{101.10}   & \multicolumn{1}{c|}{18.39} & \multicolumn{1}{c|}{44.58}    & 1.30  & \multicolumn{1}{c|}{122.48}   & \multicolumn{1}{c|}{29.20} & \multicolumn{1}{c|}{106.65}   & 3.34 & \multicolumn{1}{c|}{113.38}   & \multicolumn{1}{c|}{46.89} & \multicolumn{1}{c|}{135.45}   & 6.61 \\ \hline
{\color[HTML]{000000} }                                                                                    & \multicolumn{2}{c|}{GTS}                                   & \multicolumn{2}{c|}{ST2Vec}           & \multicolumn{2}{c|}{GTS}                                   & \multicolumn{2}{c|}{ST2Vec}           & \multicolumn{2}{c|}{GTS}                                   & \multicolumn{2}{c|}{ST2Vec}           \\ \cline{2-13}
\multirow{-2}{*}{{ \begin{tabular}[c]{@{}c@{}}Measures in\\ Road Network\end{tabular}}} & \multicolumn{1}{c|}{\makebox[0.03\textwidth][c]{$T_{tra}$(s)}} & \multicolumn{1}{c|}{\makebox[0.02\textwidth][c]{$Q$(ms)}} & \multicolumn{1}{c|}{\makebox[0.03\textwidth][c]{$T_{tra}$(s)}} & \makebox[0.02\textwidth][c]{$Q$(ms)} & \multicolumn{1}{c|}{\makebox[0.03\textwidth][c]{$T_{tra}$(s)}} & \multicolumn{1}{c|}{\makebox[0.02\textwidth][c]{$Q$(ms)}} & \multicolumn{1}{c|}{\makebox[0.03\textwidth][c]{$T_{tra}T$(s)}} & \makebox[0.02\textwidth][c]{$Q$(ms)} & \multicolumn{1}{c|}{\makebox[0.03\textwidth][c]{$T_{tra}$(s)}} & \multicolumn{1}{c|}{\makebox[0.02\textwidth][c]{$Q$(ms)}} & \multicolumn{1}{c|}{\makebox[0.03\textwidth][c]{$T_{tra}$(s)}} & \makebox[0.02\textwidth][c]{$Q$(ms)} \\ \hline
NetLCSS                                                                                                    & \multicolumn{1}{c|}{1.46}     & \multicolumn{1}{c|}{0.41}  & \multicolumn{1}{c|}{39.70}    & 0.64  & \multicolumn{1}{c|}{4.71}     & \multicolumn{1}{c|}{1.46}  & \multicolumn{1}{c|}{84.95}    & 1.83  & \multicolumn{1}{c|}{8.62}     & \multicolumn{1}{c|}{2.67}  & \multicolumn{1}{c|}{171.84}   & 3.61 \\ \hline
NetDTW                                                                                                     & \multicolumn{1}{c|}{1.49}     & \multicolumn{1}{c|}{0.50}  & \multicolumn{1}{c|}{28.98}    & 0.52  & \multicolumn{1}{c|}{4.72}     & \multicolumn{1}{c|}{1.28}  & \multicolumn{1}{c|}{84.58}    & 1.52  & \multicolumn{1}{c|}{8.66}     & \multicolumn{1}{c|}{2.66}  & \multicolumn{1}{c|}{166.04}   & 2.56  \\ \hline
TP                                                                                                         & \multicolumn{1}{c|}{1.49}     & \multicolumn{1}{c|}{0.51}  & \multicolumn{1}{c|}{27.07}    & 0.50  & \multicolumn{1}{c|}{4.66}     & \multicolumn{1}{c|}{1.28}  & \multicolumn{1}{c|}{86.04}    & 2.04  & \multicolumn{1}{c|}{8.64}     & \multicolumn{1}{c|}{2.63}  & \multicolumn{1}{c|}{165.75}   & 3.14 \\ \hline
NetERP                                                                                                     & \multicolumn{1}{c|}{1.49}     & \multicolumn{1}{c|}{0.52}  & \multicolumn{1}{c|}{30.40}    & 0.59  & \multicolumn{1}{c|}{5.33}     & \multicolumn{1}{c|}{1.24}  & \multicolumn{1}{c|}{90.59}    & 1.88  & \multicolumn{1}{c|}{8.83}     & \multicolumn{1}{c|}{2.82} & \multicolumn{1}{c|}{212.68}   & 12.32 \\ \hline
\end{tabular}
\label{tab:scalabilityA}
\vspace{-4mm}
\end{table*}

First, the average query time of the learning based methods is 1-2 orders of magnitude lower than that of the non-learning based measures (cf. Figures~\ref{fig:efficiency}(a) and \ref{fig:efficiency}(b)). Next, in road network, ST2Vec consumes more training time than GTS, while their query time is similar. This is because ST2Vec considers temporal information, which needs to be computed during model training. Finally, ST2Vec spends more training time on NetERP than other measures. In particular, NetERP is designed specifically for spatial data. When NetERP is adapted to process spatio-temporal data, it requires additional time to compute the sum of temporal distances between road vertices and reference points. In contrast, other measures are designed for spatio-temporal data, and does not leverage the reference points when calculating similarities.

\subsection{Scalability Evaluation}
\label{subsec:scalable}
We verify the scalability of each measure by varying the lengths and cardinality of trajectories. We report the index building time (denoted as $T_{idx}$) of distributed measures, training time (denoted as $T_{tra}$) of learning-based measures, and Top-$50$ query time (denoted as $Q$) of all measures. Note that, DISON (a distributed version of LCRS), which is the distributed version of LCRS and the single road network oriented measure designed for distributed environment, only supports threshold-based similarity queries. Specifically, given a query trajectory $QT$, DISON needs to specify a threshold $\tau$, and returns a trajectory set, such that the similarity between each two trajectories in the set and $QT$ are larger than $\tau$. Thus, DISON does not support Top-$k$ similarity queries. In this case, we set $\tau\textit{=}$0.8, where the trajectories returned by DISON are 50, i.e., it is the same as the number of trajectories returned by other distributed measures. In addition, we implement the local and global indexes assocaited with the distributed measures (i.e., an R-Tree index of DFT, an R-Tree and a Trie-like indexes of DITA, an RP-Tree of REPOSE, and a two-layer index with hashmap of DISON) for a fair and objective comparison.

\subsubsection{Non-learning based measures}

\textbf{Standalone methods.} Figures~\ref{fig:scalabilityL} and~\ref{fig:scalabilityA} illustrate the query time of standalone measures when varying trajectory lengths and data cardinality. First, the query time of most standalone measures (except for LIP, LCRS, and LORS) increases significantly as trajectory length or data cardinality grows, which has poor scalablity. Next, as depicted in Figure~\ref{fig:scalabilityA}, the query time of LIP increases slightly with the growth of data cardinality. This is because the time complexity of LIP is $\textit{O((m+n)log(m+n))}$, which is lower than that of other standalone measures. LCRS's and LORS's query time remain stable when data cardinality rises from 60\% to 100\%. The reason is that both LCRS and LORS compute the similarity between a pair of trajectories by counting the number of their overlapping road segments, instead of calculating pairwise distances between their points or road vertices. This enables LCRS and LORS to process large datasets.



\begin{figure}[tb]
	\centering
	\vspace{-1mm}
	\hspace{-4mm}
	\subfigure[Query (AIS)]{
		\includegraphics[width=0.16\textwidth]{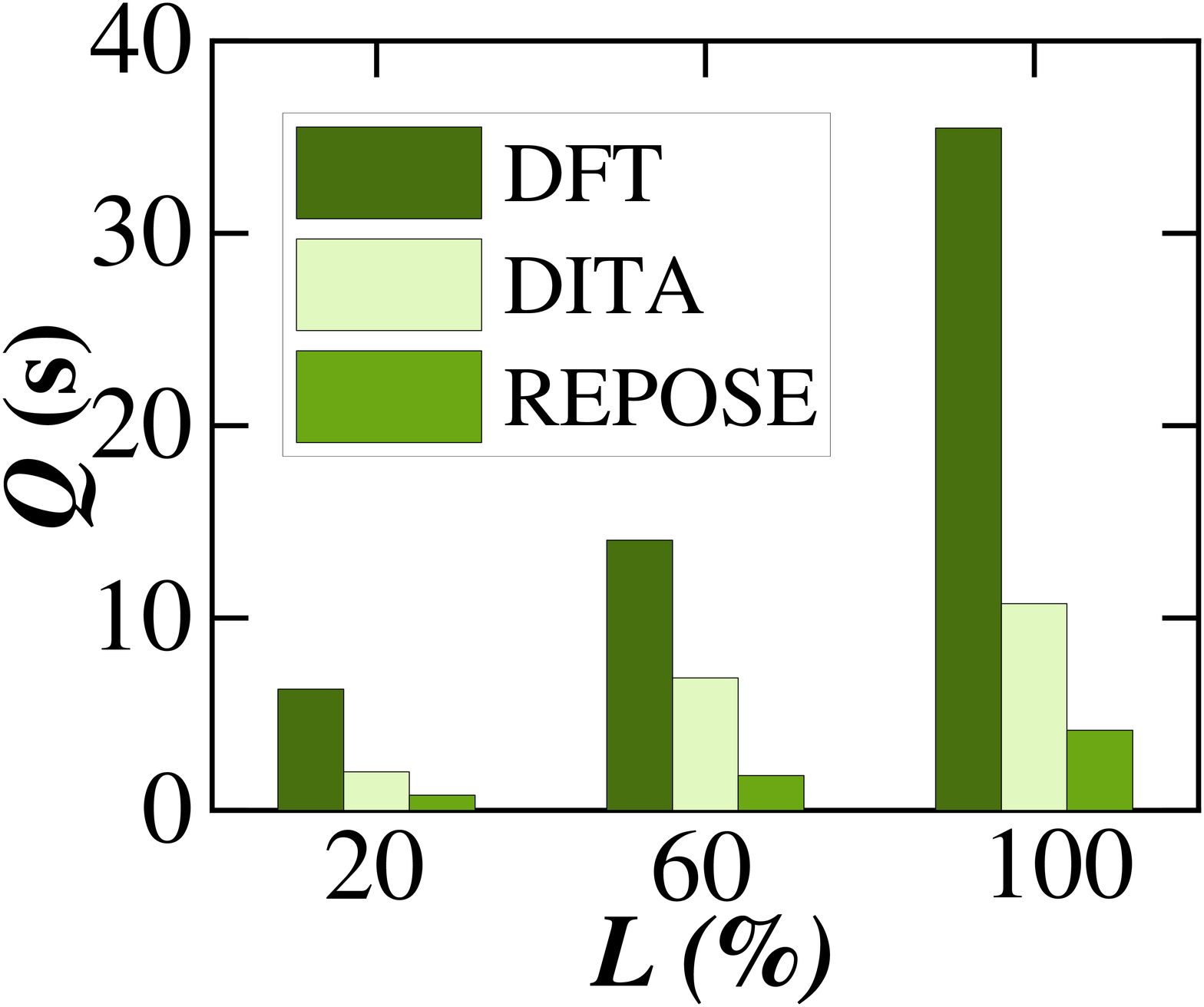}}
	\subfigure[Indexing (AIS)]{
		\includegraphics[width=0.16\textwidth]{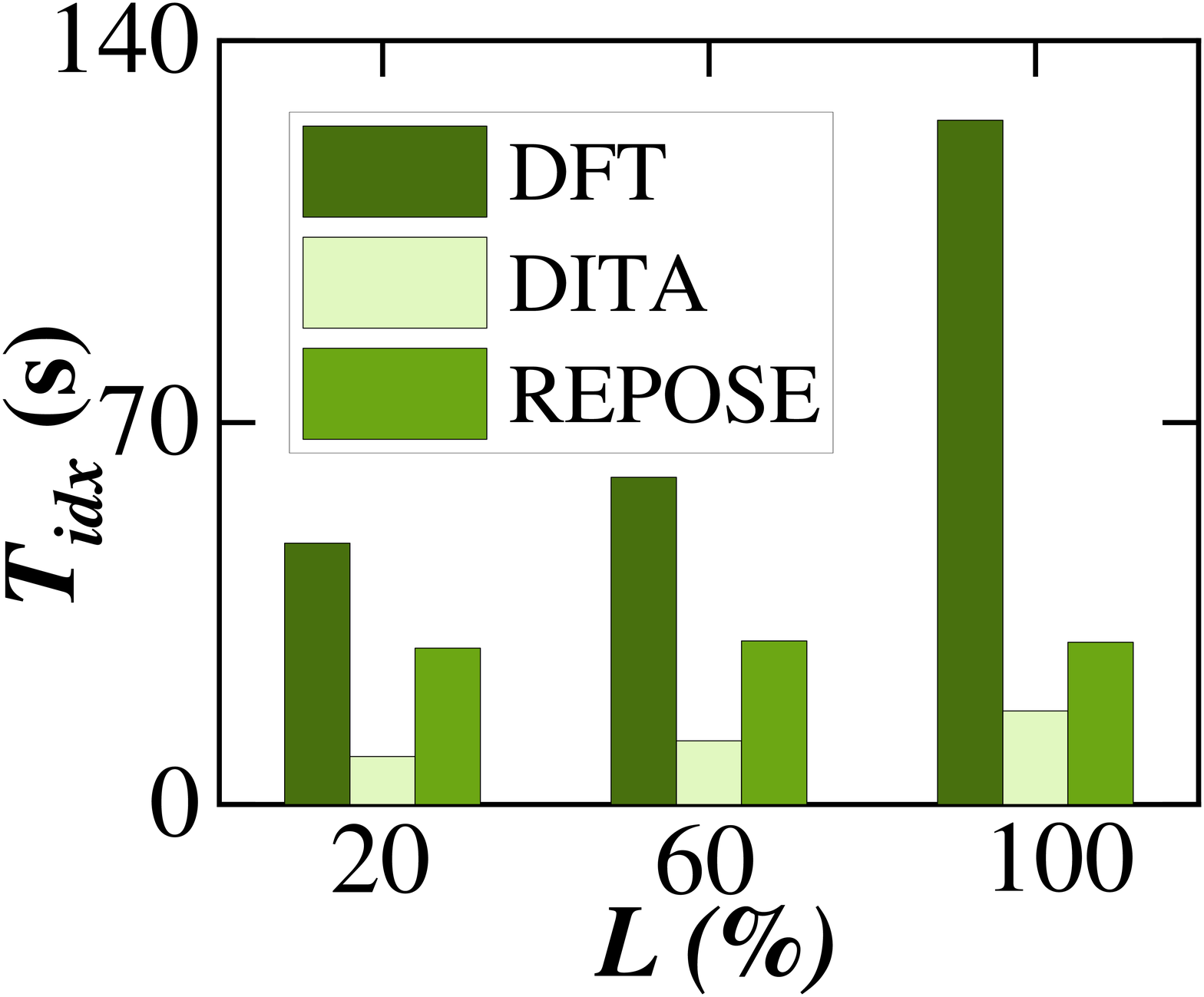}}
	\subfigure[DISON (T-Drive)]{
		\includegraphics[width=0.16\textwidth]{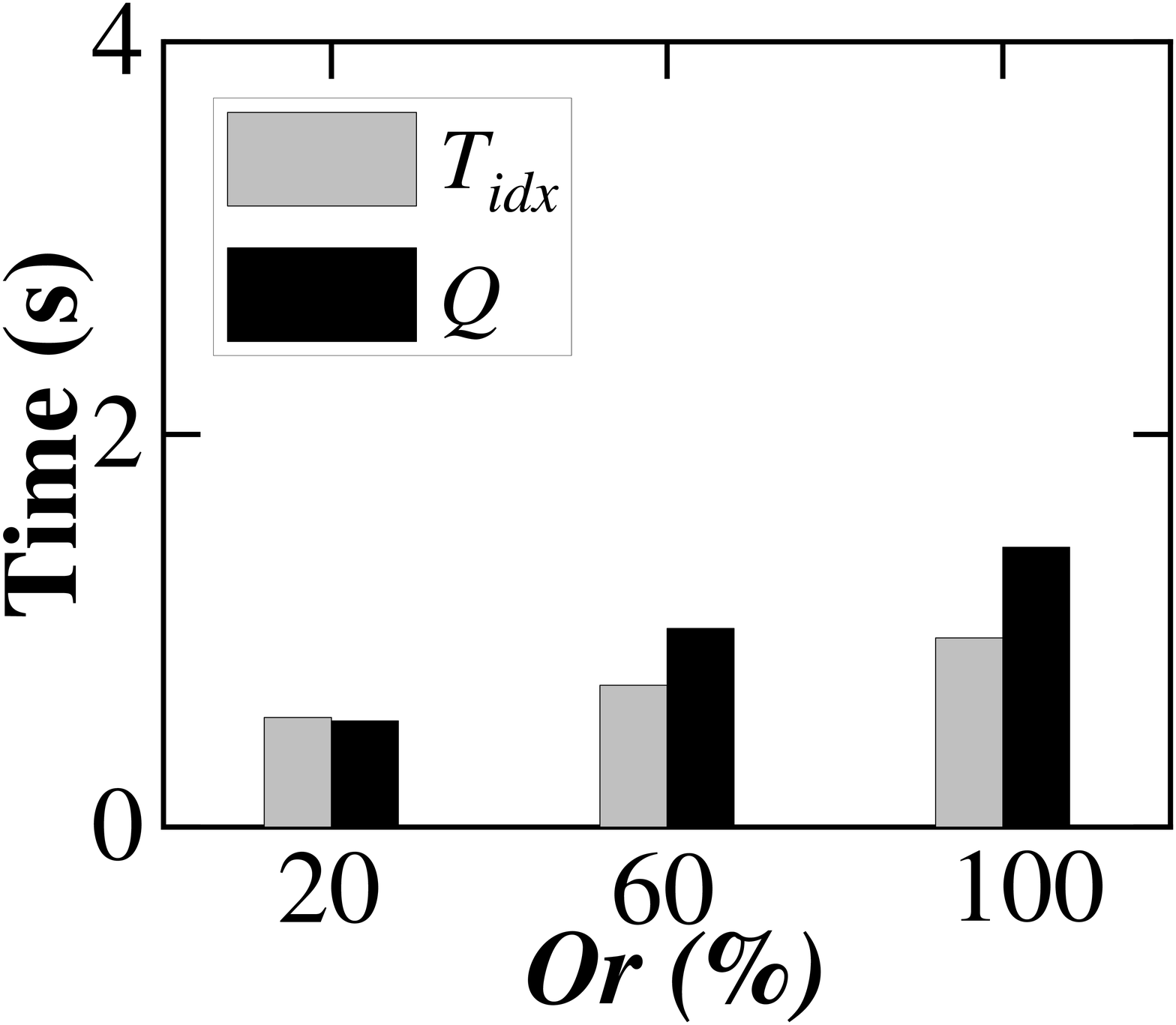}}\\
	
	\vspace{-2mm}
	\hspace{-4mm}
	\subfigure[Query (Geolife)]{
		\includegraphics[width=0.16\textwidth]{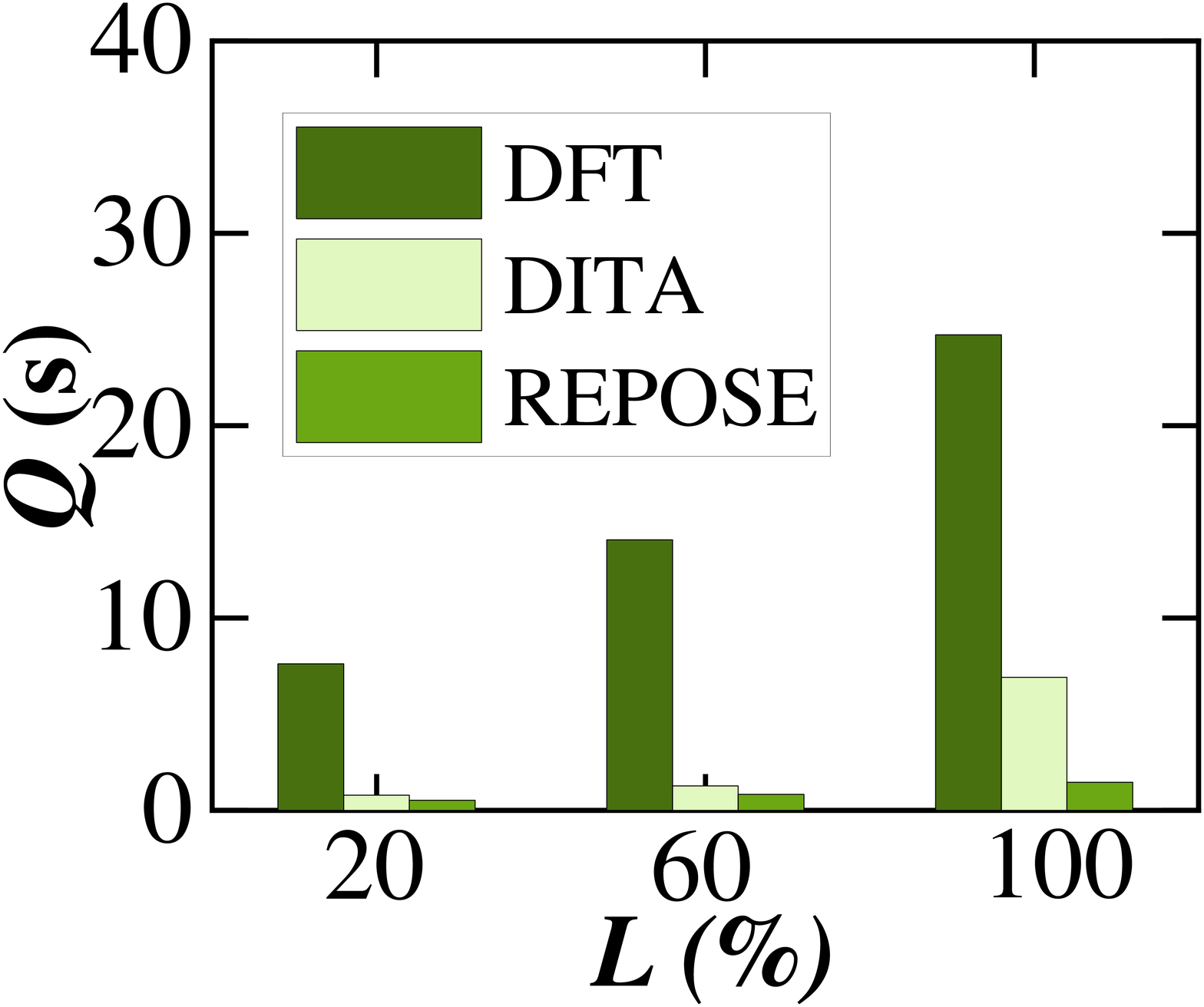}}
	\subfigure[Indexing (Geolife)]{
		\includegraphics[width=0.16\textwidth]{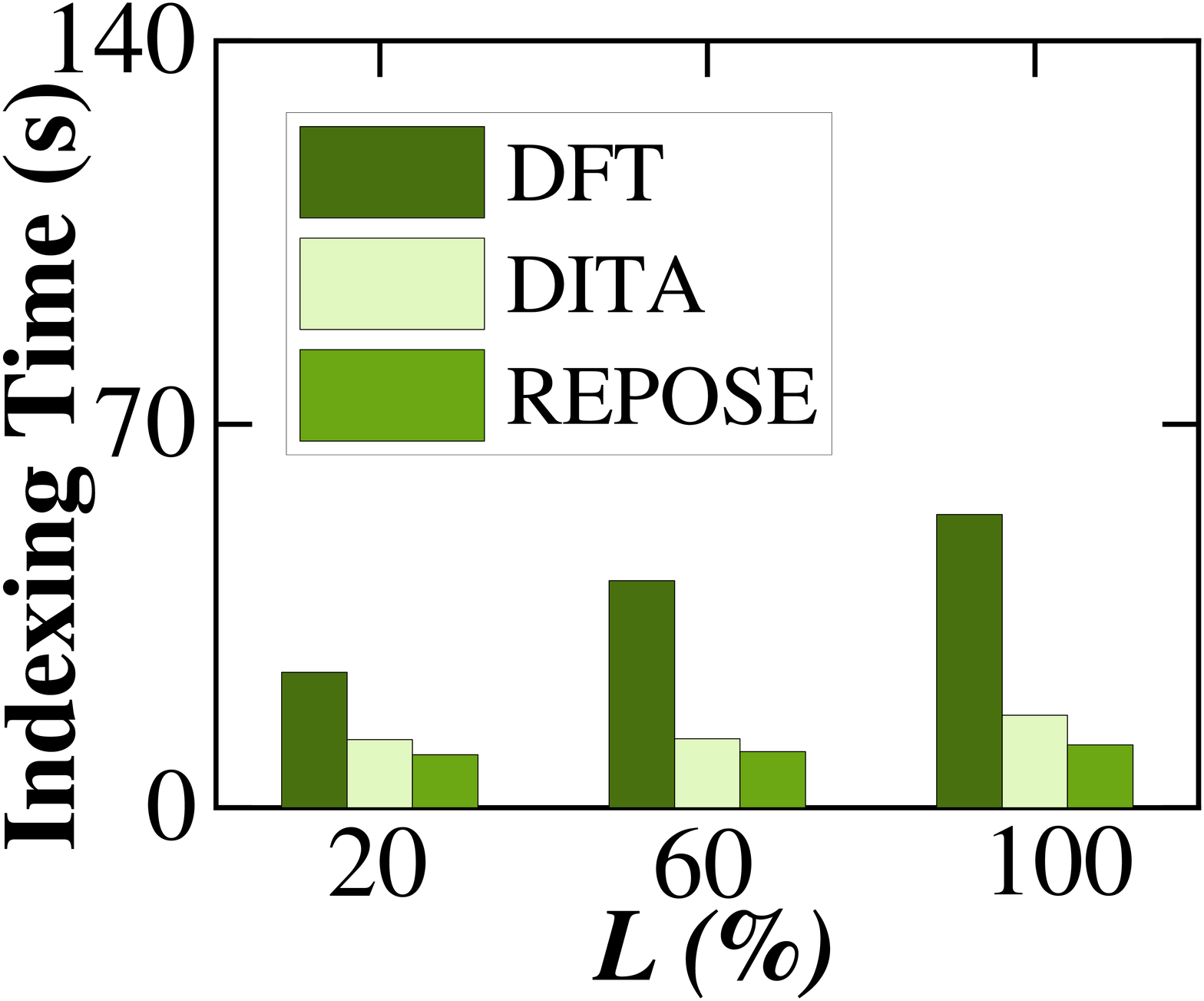}}
	\subfigure[DISON (Porto)]{
		\includegraphics[width=0.16\textwidth]{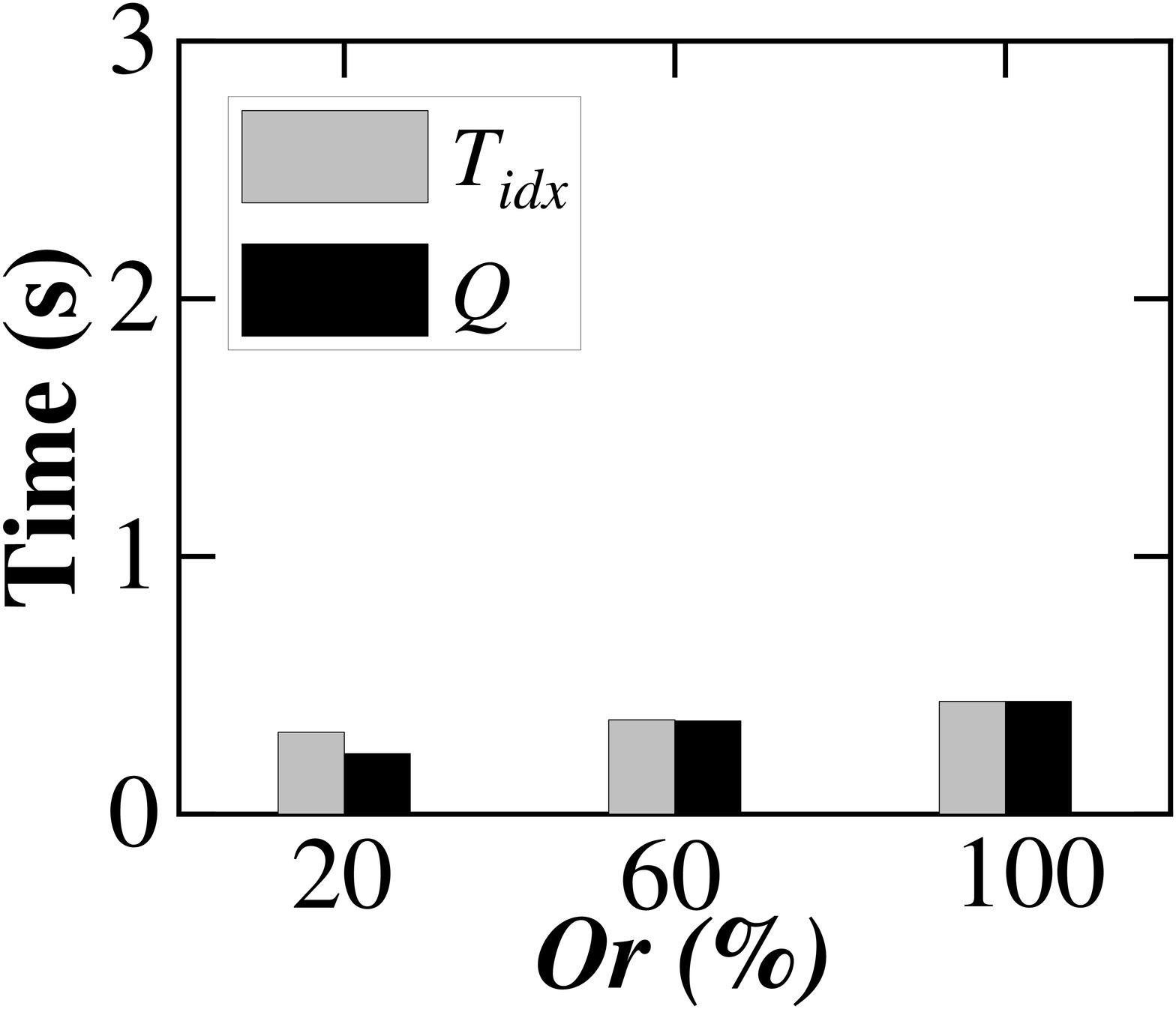}}\\
    \up
	\caption{Scalability of Distributed Measures vs. Trajectory Length} 
	\label{fig:scalabilityD}
	\vspace{-4mm}
\end{figure}

\begin{figure}[tb]
	\centering
	\vspace{-1mm}
	\hspace{-4mm}
	\subfigure[Query (AIS)]{
		\includegraphics[width=0.16\textwidth]{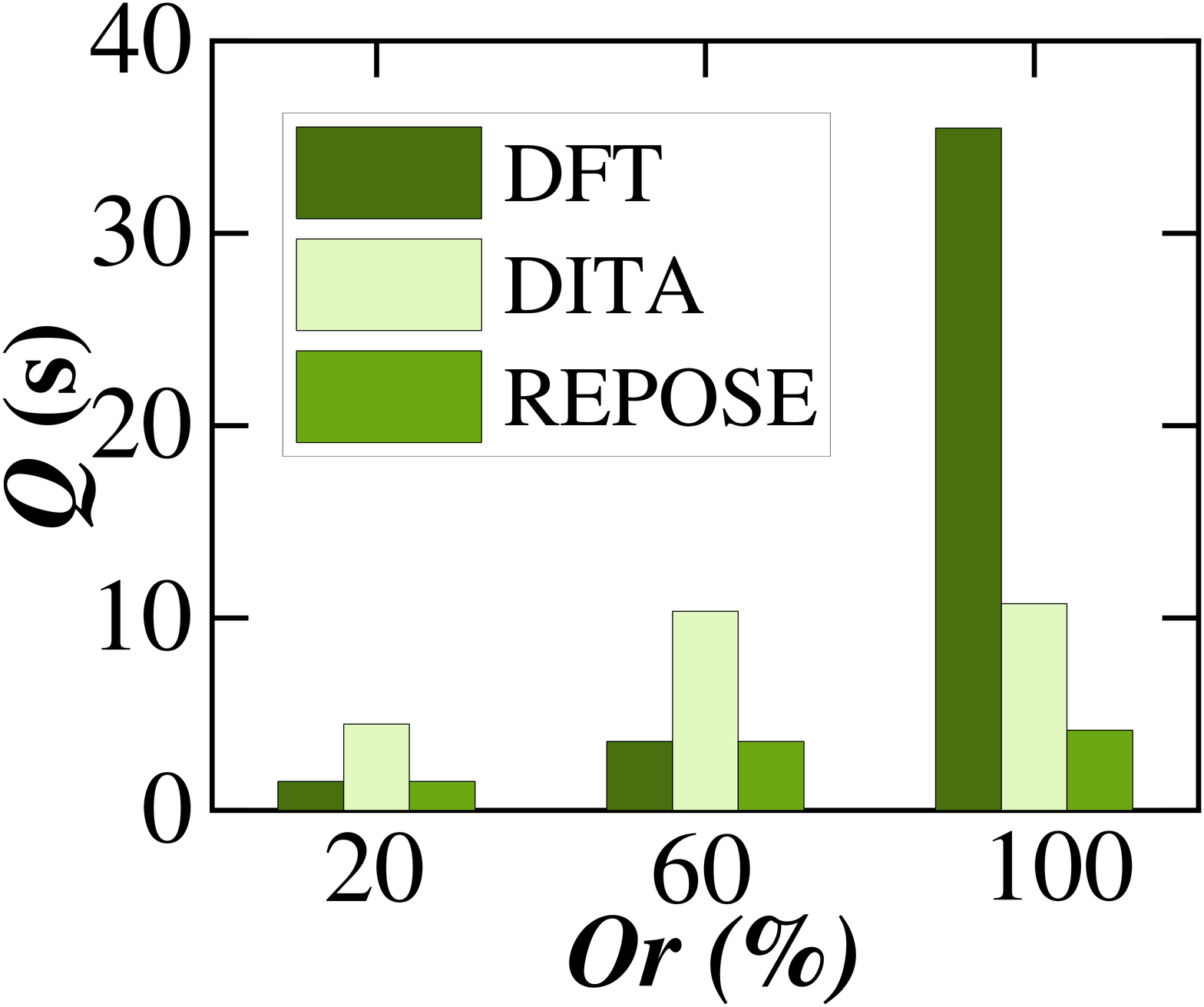}}
	\subfigure[Indexing (AIS)]{
		\includegraphics[width=0.16\textwidth]{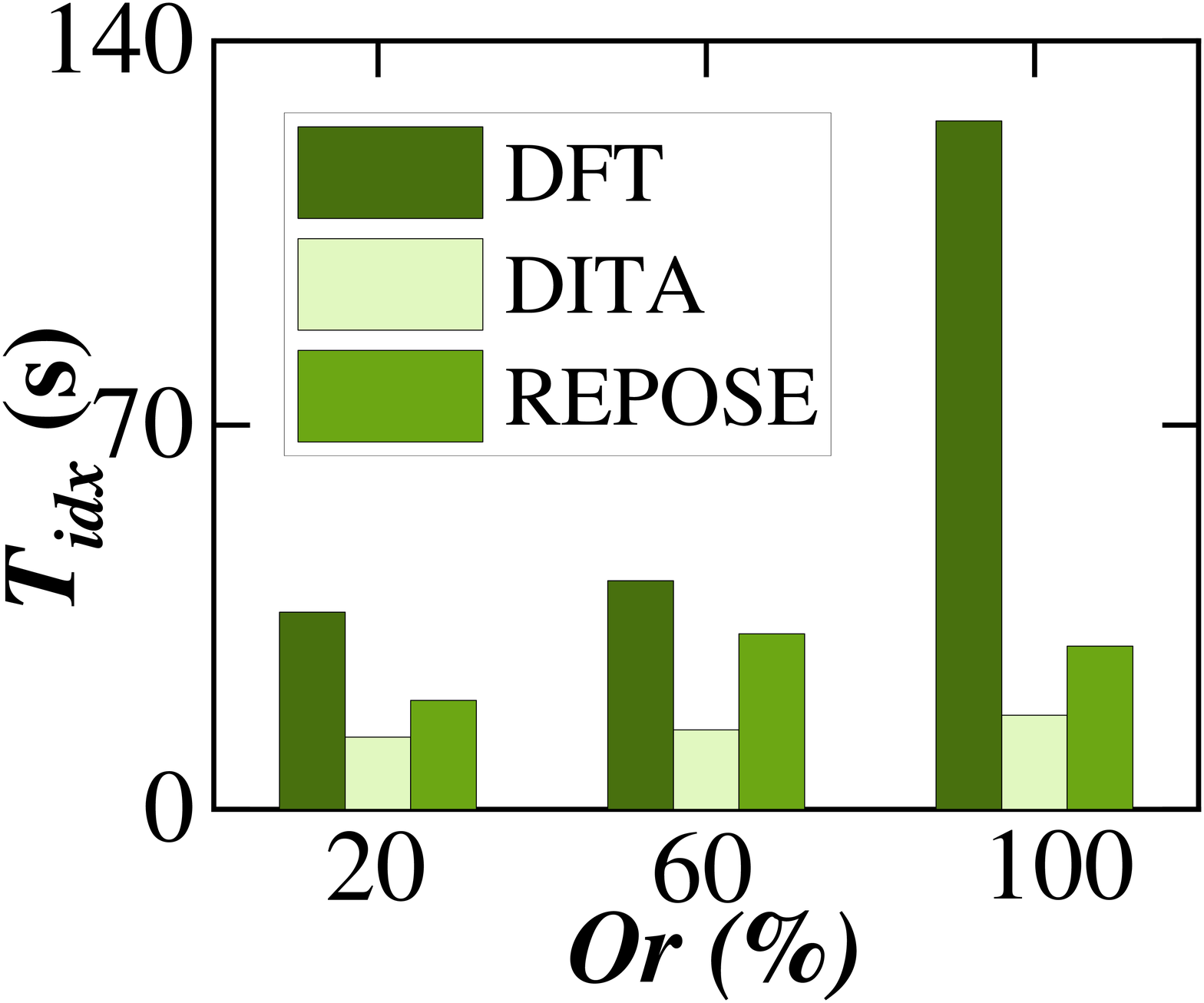}}
	\subfigure[DISON (T-Drive)]{
		\includegraphics[width=0.16\textwidth]{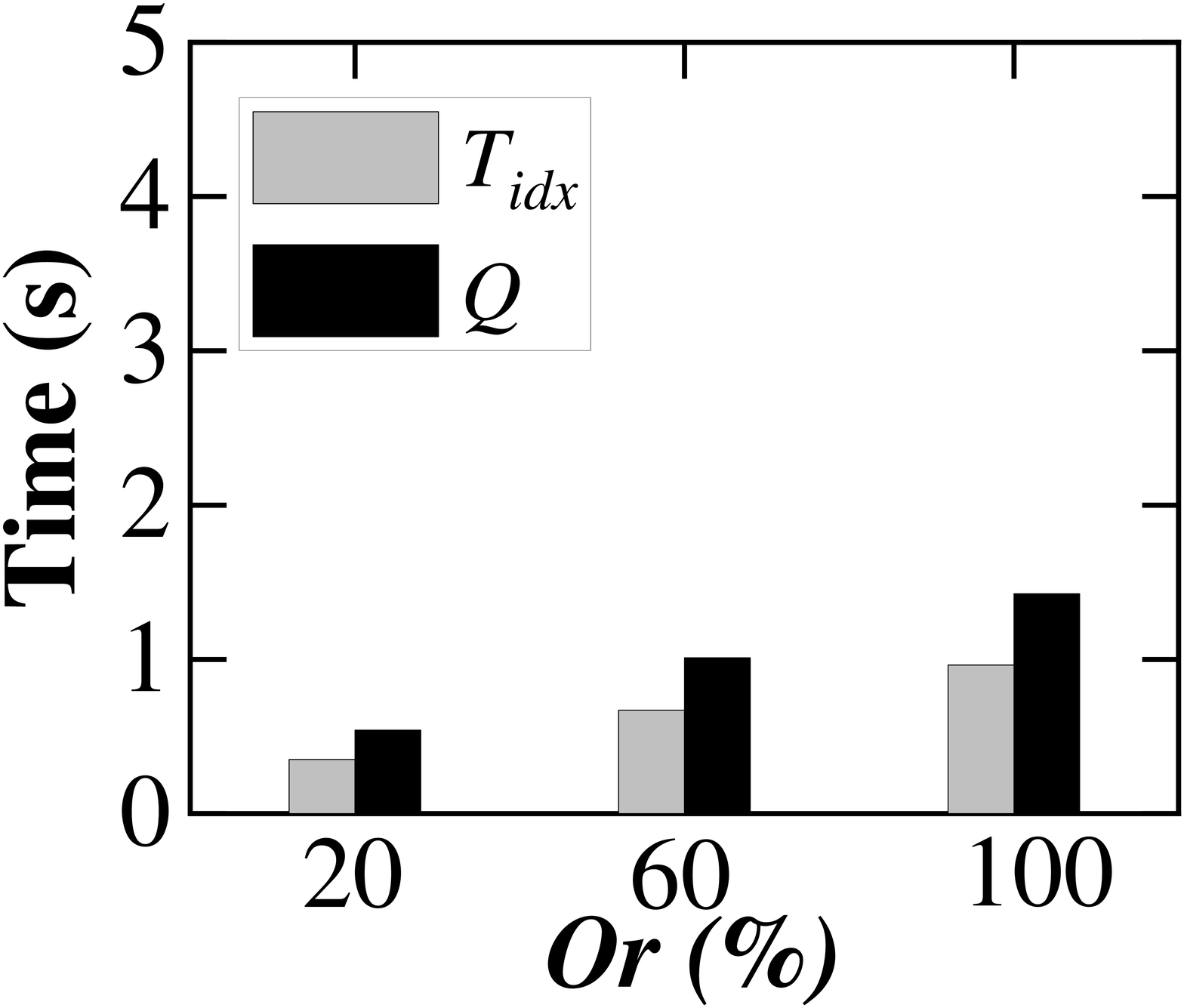}}\\
	
	\vspace{-2mm}
	\hspace{-4mm}
	\subfigure[Query (Geolife)]{
		\includegraphics[width=0.16\textwidth]{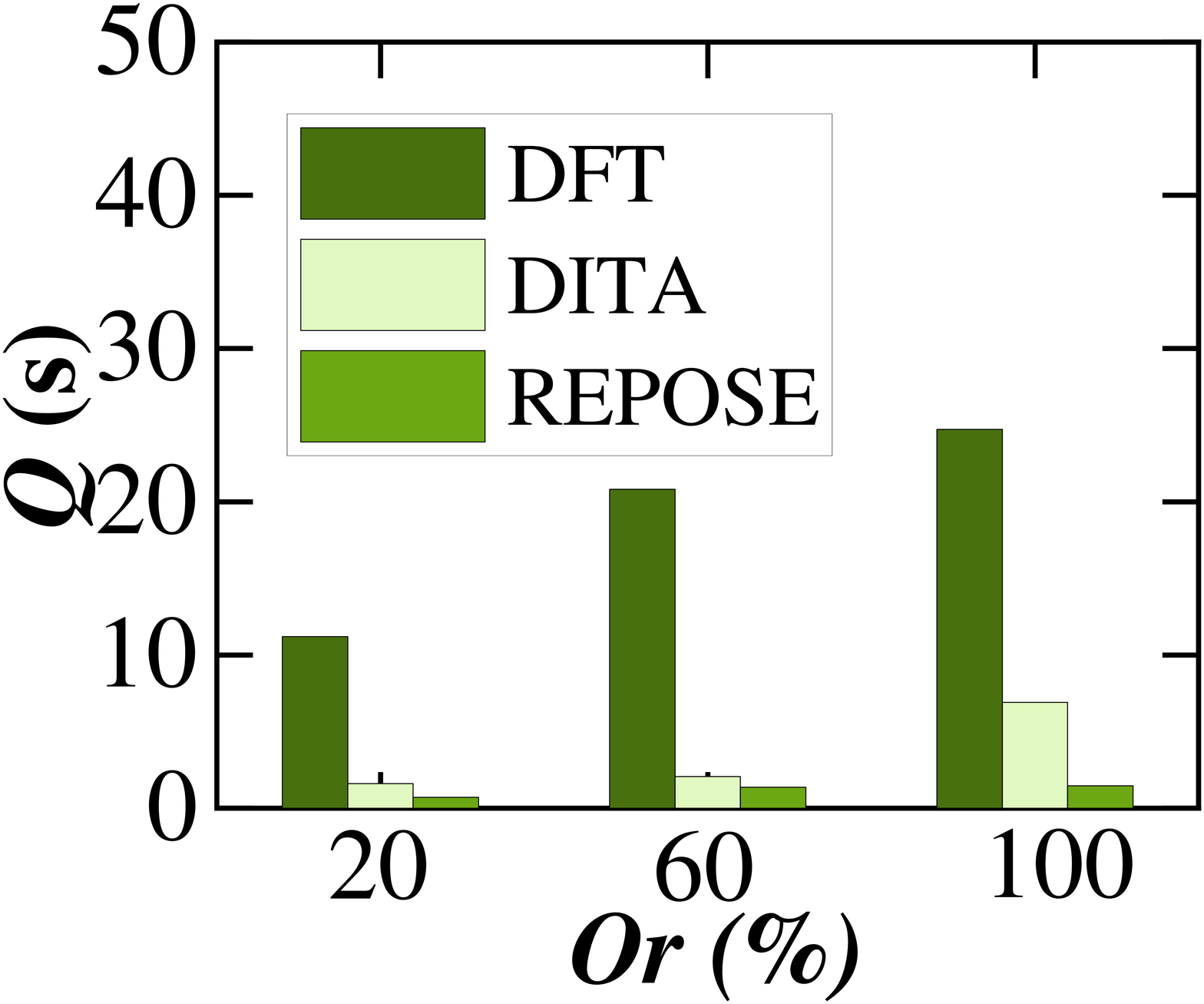}}
	\subfigure[Indexing (Geolife)]{
		\includegraphics[width=0.16\textwidth]{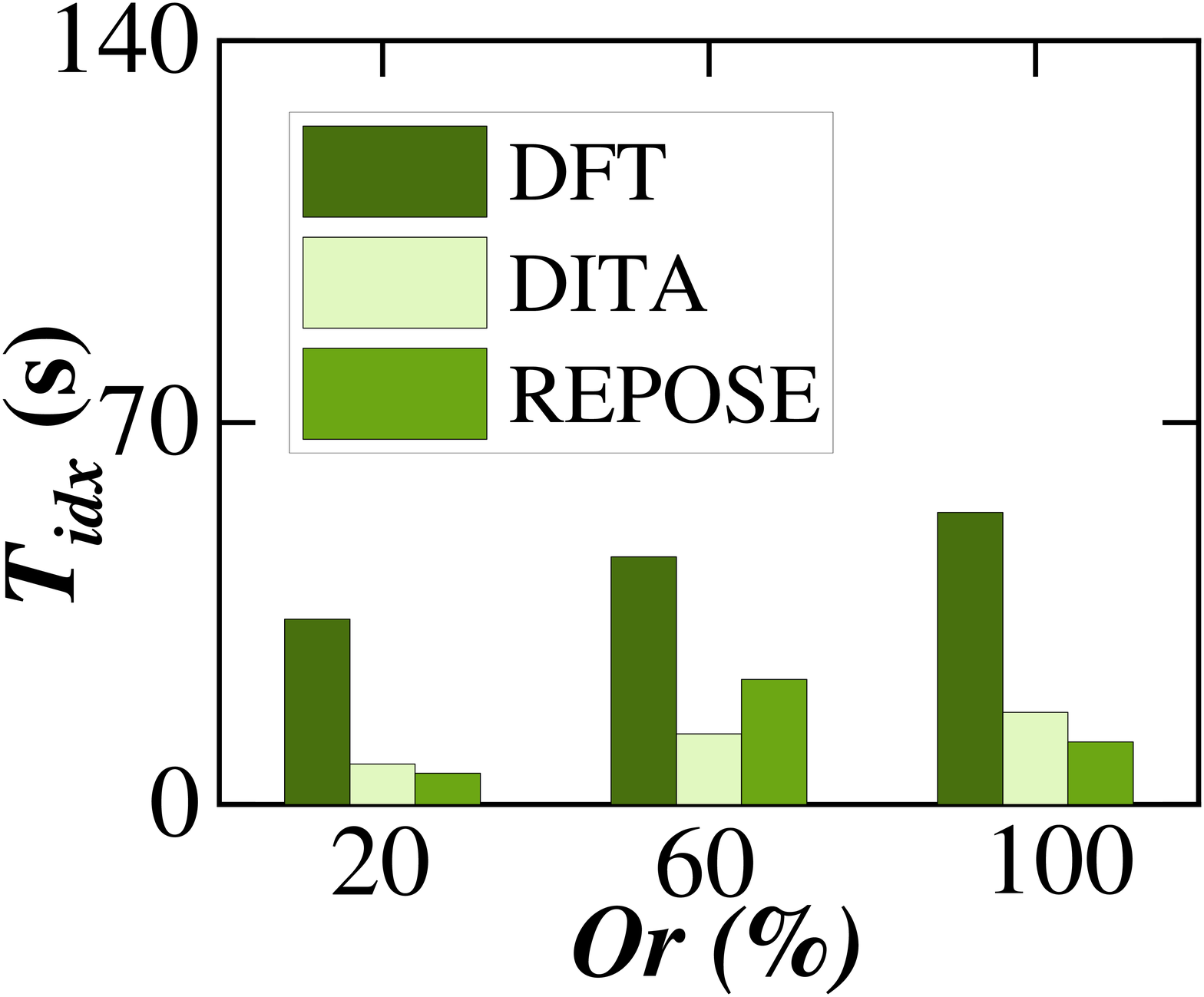}}
	\subfigure[DISON (Porto)]{
		\includegraphics[width=0.16\textwidth]{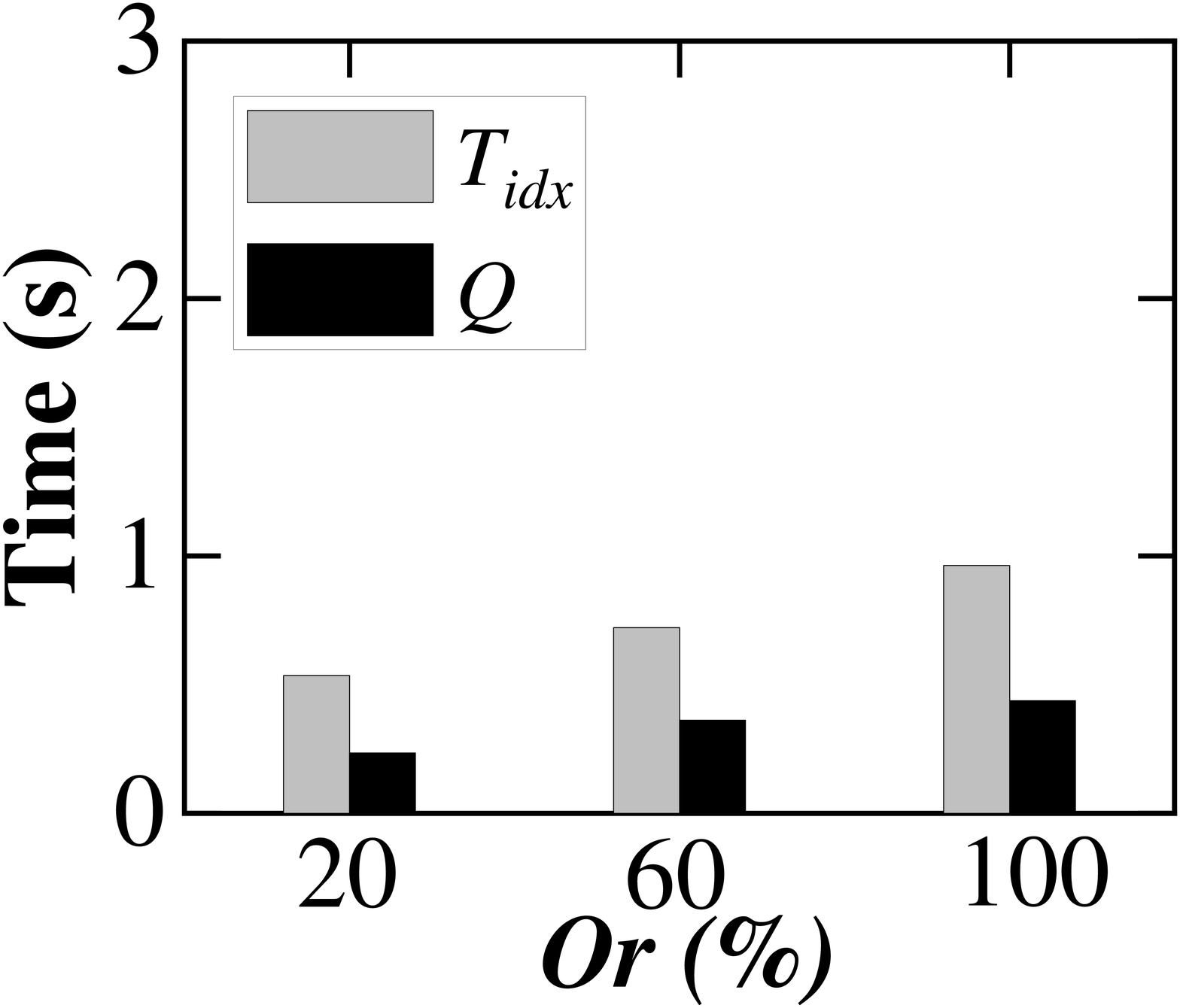}}\\
    \up
	\caption{Scalability of Distributed Measures vs. Data Cardinality} 
	\label{fig:scalabilityDA}
	\vspace{-4mm}
\end{figure}

\noindent\textbf{Distributed methods.}  Figures~\ref{fig:scalabilityD} and~\ref{fig:scalabilityDA} show the index building time and query time of distributed measures when varying trajectory lengths and data cardinality. First, compared with Figures~\ref{fig:scalabilityL} and~\ref{fig:scalabilityA}, distributed measures achieve much higher query efficiency than standalone measures, when varying both trajectory lengths and data cardinality on both datasets.

Second, the query time and index building time of DITA, REPOSE, and DISON vary gently with both trajectory lengths and cardinality on both datasets; while those of DFT increase rapidly with the growth of trajectory lenghs and cardinality (cf. Figures~\ref{fig:scalabilityL} and \ref{fig:scalabilityA}). Specifically, DFT (i) collects the trajectories' ids in each node of the index; and (ii) calculates the distances between each data partition and query trajectory $QT$ in order to select a threshold for pruning. Clearly, the time costs of these operations increase when the trajectory length and data cardinality grow.

Next, Figures~\ref{fig:scalabilityD}(a) and (d) show that REPOSE and DITA always perform better than DFT in terms of query time, when varying trajectory length and data cardinality on both AIS and Geolife. DFT uses R-tree for indexing while others use trie-like index. 
Trie-like index is a multi-level structure where each level stores a set of trajectory points, with the points that have earlier timestamps located at higher levels of the index, i.e., the root of the index represents the starting point of a trajectory, and the leaf nodes represent the end point. The similarity computation of both DITA and DFT requires to calculate the distance between each pair of GPS points in a chronological order. Thus, DITA's trie-like index allows for efficient computation of similarity between GPS points while the data is being indexed. In contrast, the R-tree does not organize the points based on their chronological order. Thus, DFT is less efficient than DITA.




Finally, as depicted in Figures~\ref{fig:scalabilityD}(c), \ref{fig:scalabilityD}(f), \ref{fig:scalabilityDA}(c), and \ref{fig:scalabilityDA}(f), the query time $Q$ of DISON is always less than 2 seconds on T-Drive, and is always less than 0.2 seconds on Porto, and slightly increases with the growth of trajectory length and data cardinality. On the contrary, the query time of LCRS is more than 3 seconds on T-Drive and 0.4 seconds on Porto, and drastically increases with trajectory length and data cardinality (cf. Figures~\ref{fig:efficiency}(c) and \ref{fig:efficiency}(d)).
This means that the distributed techniques are enable to enhance the scalability of similarity computation in  road networks.

\subsubsection{Learning-based methods}
We evaluate the scalability of learning-based methods only on AIS and T-Drive, due to the space limitation and their larger sizes than Geolife's and Porto's.
Tables~\ref{tab:scalabilityL} and~\ref{tab:scalabilityA} list the training time and query time of learning-based measures when varying trajectory lengths and data cardinality. First, the query time of all measures slightly increase with the growth of trajectory length and data cardinality; while the training time increases significantly. This is because when trajectory lengths or data cardinality rises, learning-based methods need more time to capture the information of sampling points for model training; then they just apply the well-trained model to obtain the Top-$50$ results in a constant time. 
Finally, the query time of learning-based measures are much lower than that of non-learning based measures. This indicates that learning-based measures are able to accelerate the similarity computation.


\subsection{Study of Metric Similarity Measures}
Triangle inequality and indexing can be exploited to filter unqualified results and accelerate similarity computation, respectively, for metric similarity measures (cf. Section~\ref{sec:traditional}), and thus can affect the performance of them significantly.
In this subsection, we evaluate the impacts of triangle inequality-based pruning and different indexing structures on the performance of three metric measures: ERP, Frechet, and Hausdorff. Specifically, given a query trajectory $QT$, we apply the three measures to perform top-50 queries, and study their index building time (denoted as $T_{idx}$), query time (denoted as $Q$), and Prune rate (denoted as $\textit{Prune Rate}$) when the three indexing structures are used, respectively. Here, $\textit{PruneRate}=\frac{|\textit{TrajNum}_{pruned}|}{|\textit{TrajNum}_{all}|}$ is designed to verify the pruning performance, where $\textit{TrajNum}_{pruned}$ is the number of filtered trajectories and $\textit{TrajNum}_{all}$ is the total number of trajectories. Clearly, the higher the $\textit{PruneRate}$, the higher the pruning performance. 
Note that, three representative pivot-based indexes LEASA~\cite{LEASA}, MVP-Tree (MVPT)~\cite{MVPT}, and PM-Tree (PMT)~\cite{PMT} are selected, while HF method~\cite{HF} is used to select pivot trajectories, where the pivot number is set to 5 following the previous survey~\cite{IndexSurvey}.



\begin{figure}[tb]
	\centering
	\vspace{-1mm}
	\hspace{-4mm}
	\subfigure[Indexing (AIS)]{
		\includegraphics[width=0.16\textwidth]{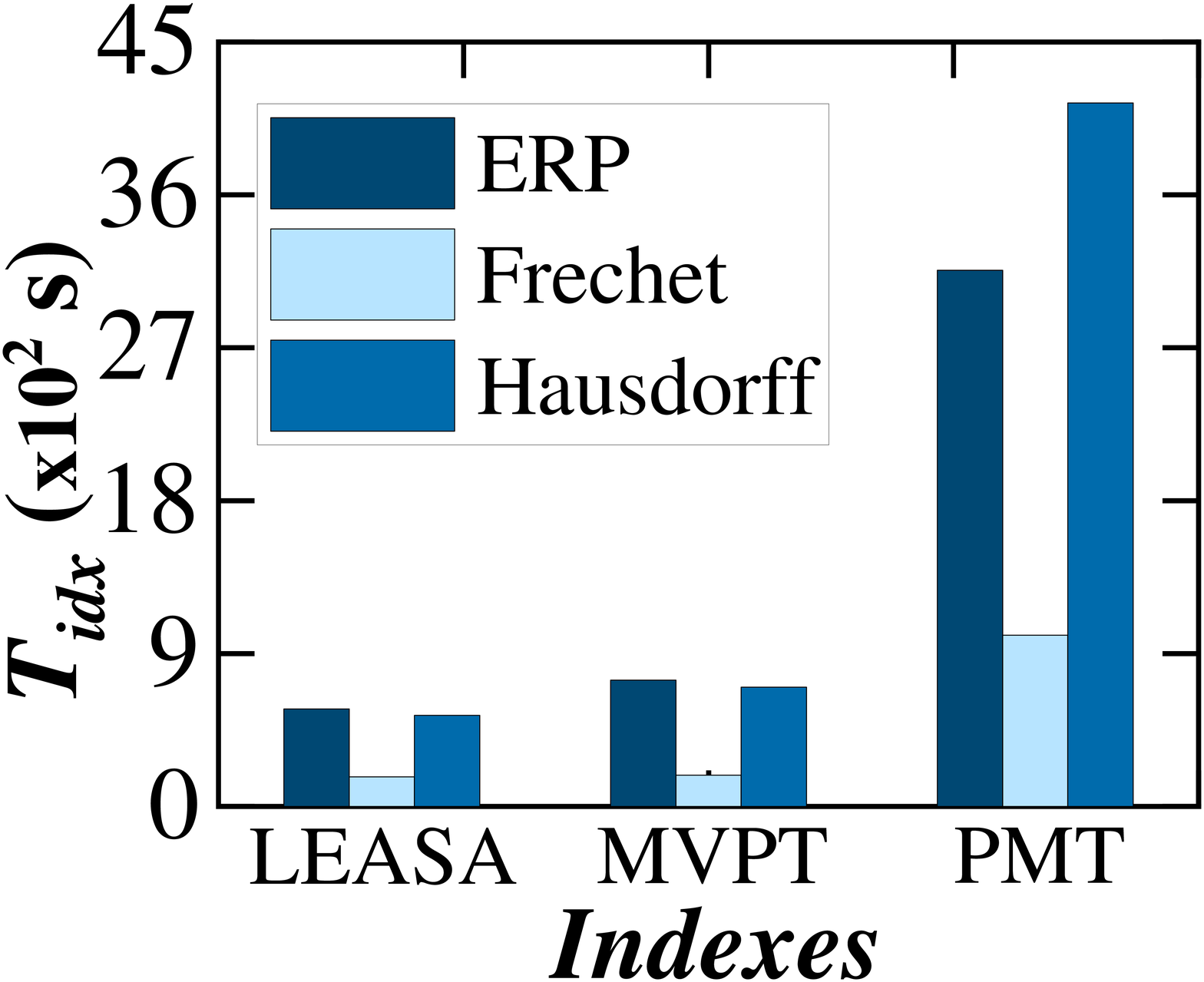}}
	\subfigure[Query (AIS)]{
		\includegraphics[width=0.16\textwidth]{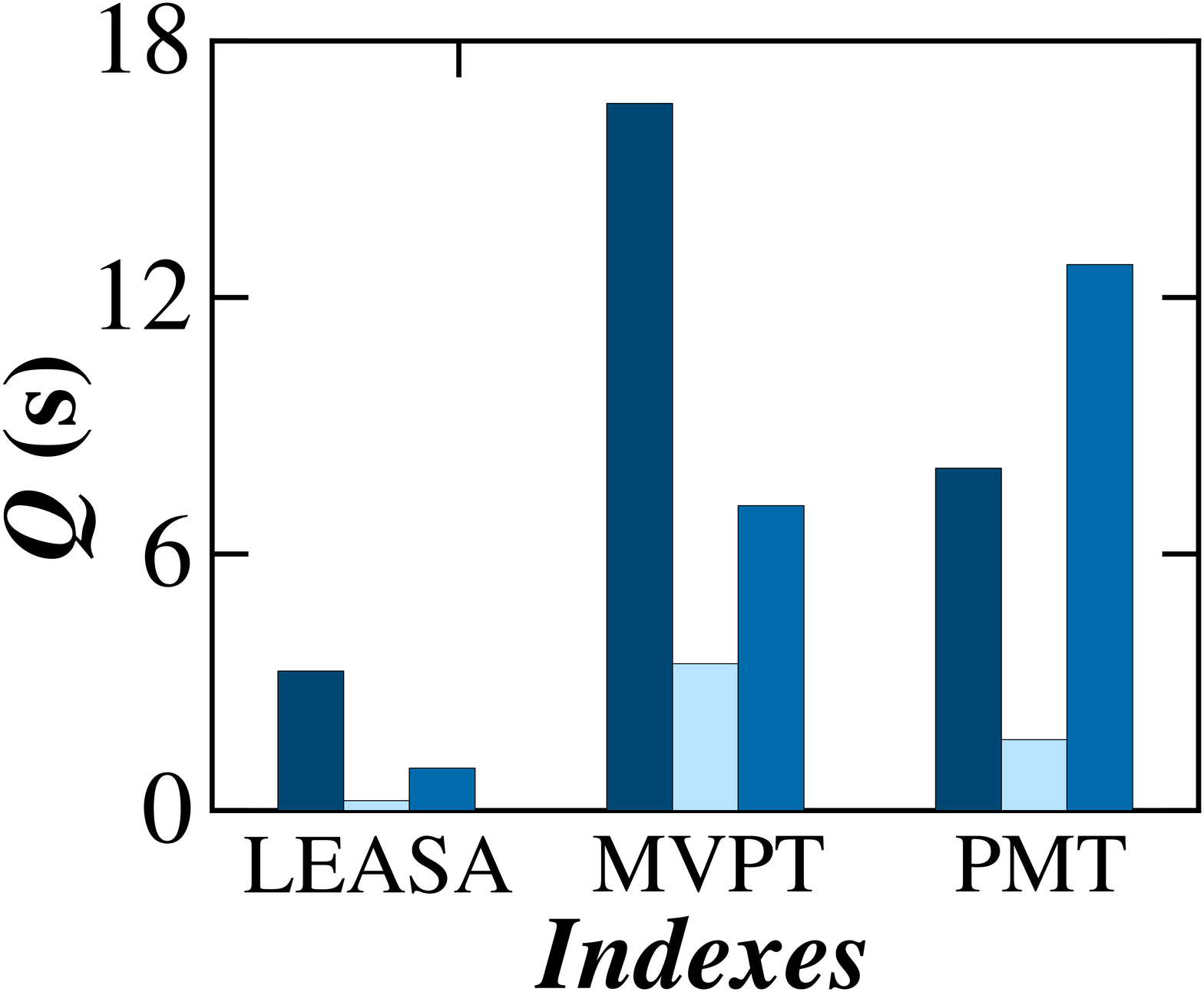}}
	\subfigure[Prune Rate (AIS)]{
		\includegraphics[width=0.16\textwidth]{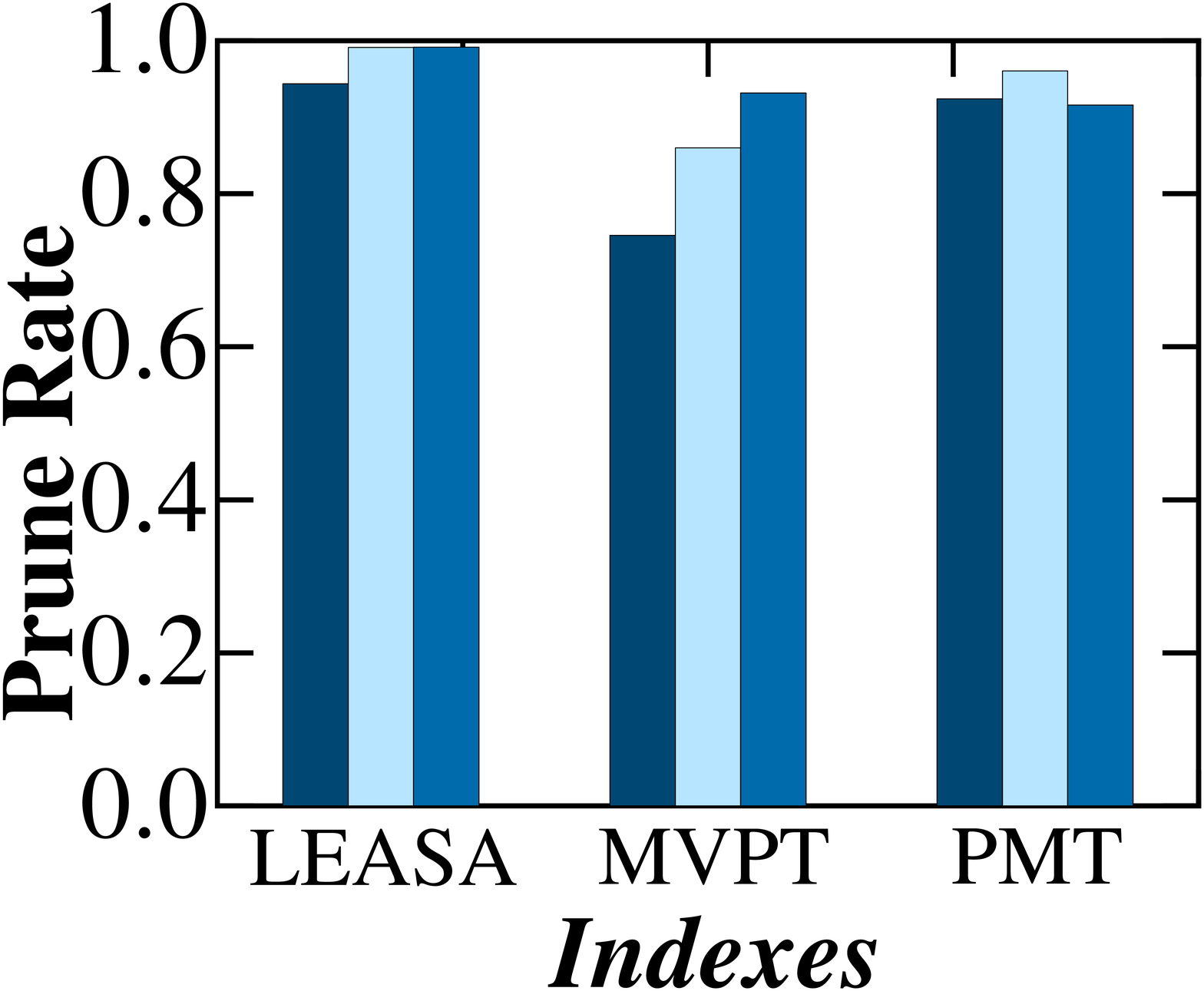}}\\
	
	\vspace{-2mm}
	\hspace{-4mm}
	\subfigure[Indexing (Geolife)]{
		\includegraphics[width=0.16\textwidth]{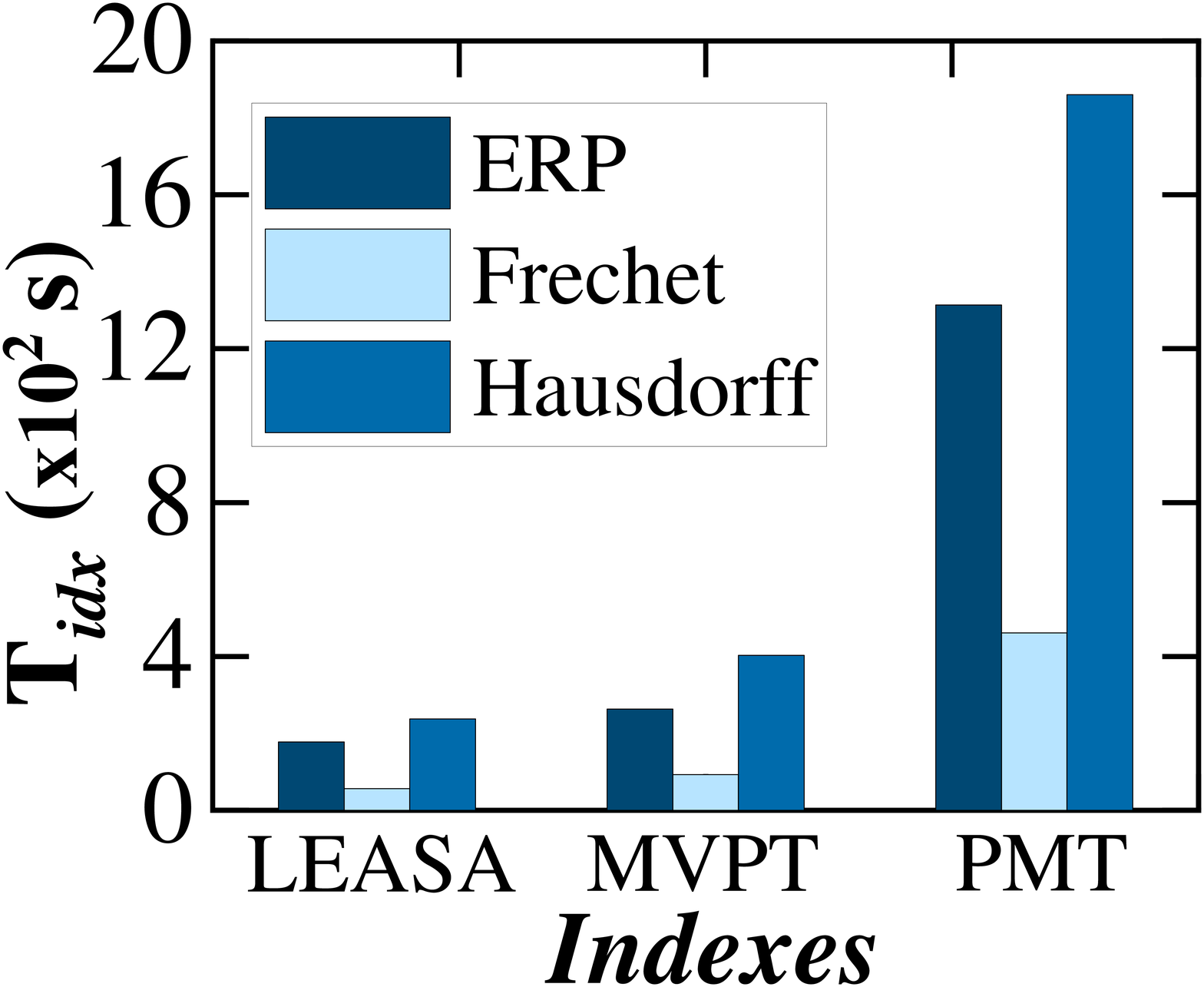}}
	\subfigure[Query (Geolife)]{
		\includegraphics[width=0.16\textwidth]{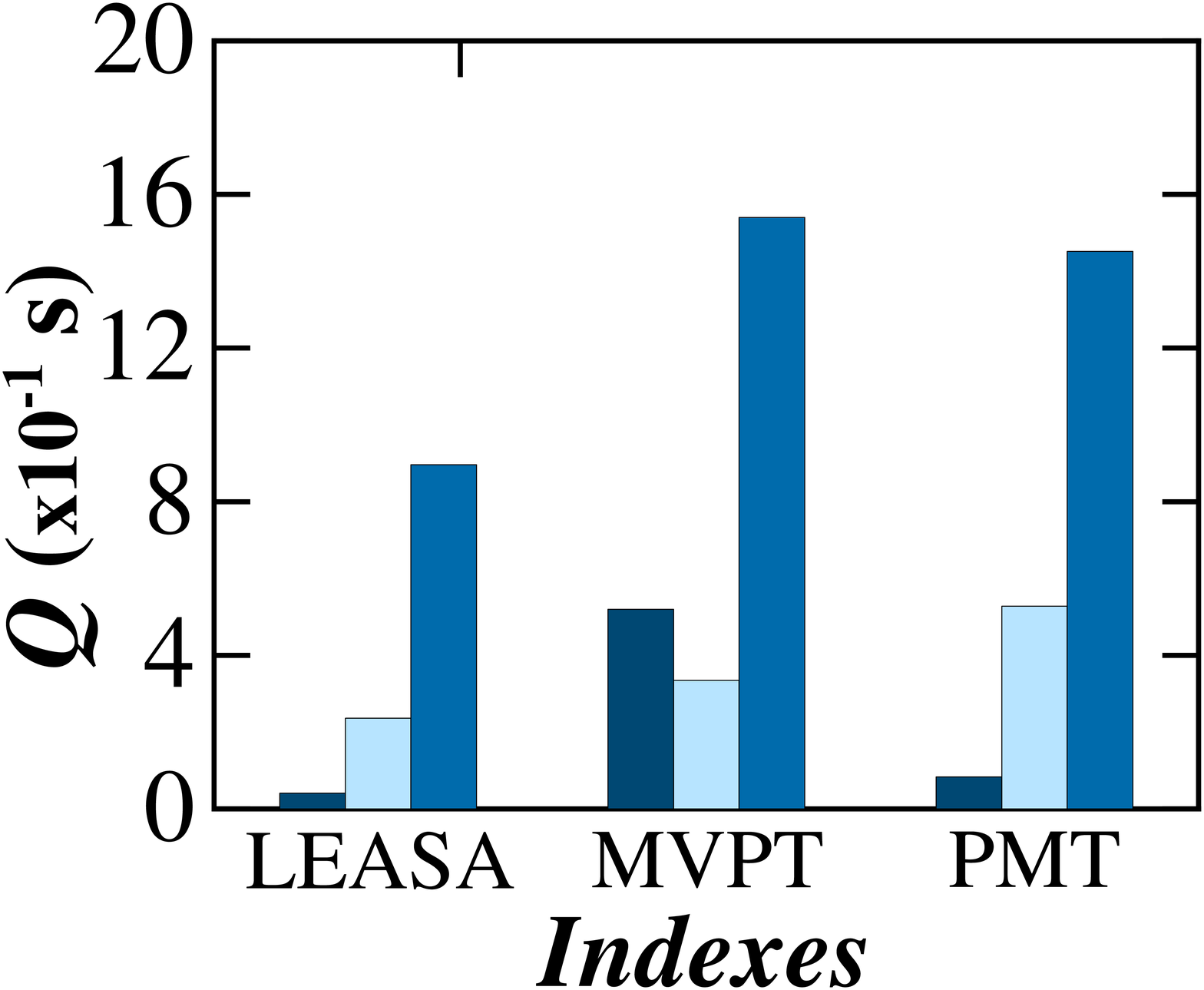}}
	\subfigure[Prune Rate (Geolife)]{
		\includegraphics[width=0.16\textwidth]{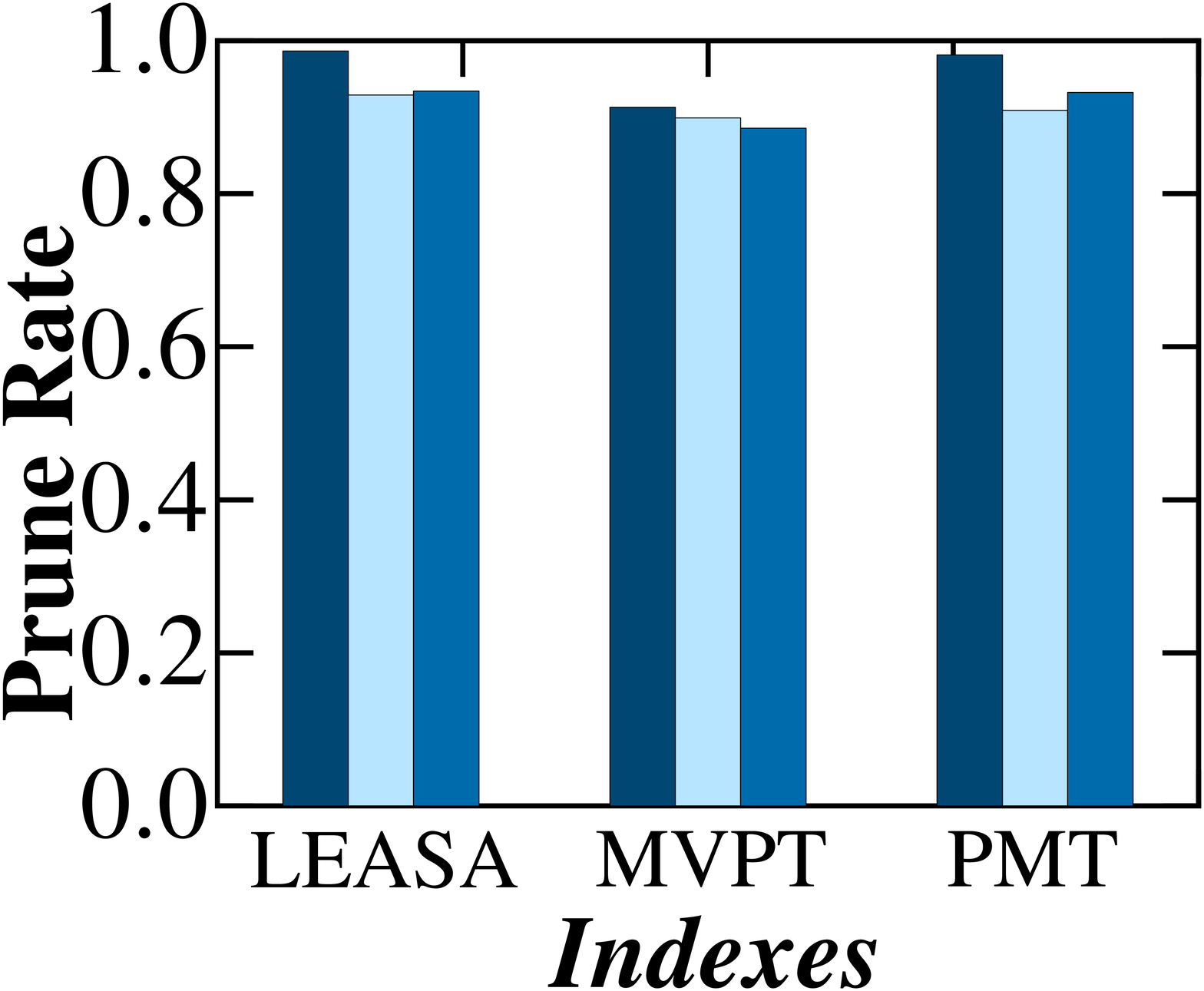}}\\
    \up
	\caption{Study of Metric Measures}
	\label{fig:Index}
	\vspace{-4mm}
\end{figure}





Figure~\ref{fig:Index} shows the results.
First, the indexing helps to reduce the query time of all metric measures, i.e., the lower the query time, the higher the $\textit{PruneRate}$ (cf. Figures~\ref{fig:efficiency}(a)--(b)).
Second, the rank of the search time among three measures changes when applying indexing structures. For example as shown in Figure~\ref{fig:efficiency}(a), among Frechet, ERP, and Hausdorff, Frechet achieves the highest efficiency and Hausdorff performs worse than ERP on AIS when there is no indexing structure applied; however, Hausdorff achieves higher efficiency than ERP on AIS when using LEASA and MVPT, as illustrated in Figure~\ref{fig:Index}(b).
Finally, the index building time and pruning performance of different measures with the same indexing structure are different. As shown in Figure~\ref{fig:Index}(a) and \ref{fig:Index}(c), Frechet spends less time on building index, and has more stable and higher $\textit{PruneRate}$ than ERP and Hausdorff.

Overall, it is necessary to build indexes and explore triangle inequality-based pruning before evaluating the performance of a metric similarity measure, as those auxiliary techniques may enhance the effectiveness of the metric similarity measures greatly.

\subsection{Experimental Takeaways} We perform an independent and comprehensive experimental study on trajectory similarity measures, and analyze their effectiveness, robustness, efficiency, and scalability. The insights are summarized below.

\begin{itemize}\setlength{\itemsep}{-\itemsep}
    \item The effect of most trajectory similarity measures is greatly affected by the distributions and features of datasets. Specifically, Frechet, Hausdorff, LCRS, and LORS always perform well and stable in different datasets; while ERP performs the worst.
    \item Most measures only compute the spatial distance of trajectories to measure the similarity; however, they perform effectively in spatio-temporal scenarios when being directly extended to temporal distance (even though the time cost is high).
    \item The point-based measures perform better in free space; while the segment-based measures are more effective in road network. In addition, some measures (i.e., DTW and LCSS) in free space can be directly extended to the road networks and preserve good performance; while others (i.e., EDR and ERP) are not suitable for being adjusted to the road networks, which affects the robustness.
    \item For metric similarity measures, building indexes and exploring triangle inequality-based pruning may help to enhance their efficiency significantly.
    \item In most cases, learning based measures are more robust in time consumption over different-scale datasets. Standalone-based measures have obviously lower efficiency and scalability but higher effectiveness. Overall, distributed-based measures enable to improve time performance while ensuring high effectiveness of standalone-based measures, which are promising in the study of trajectory similarity measures.
\end{itemize}



\section{Future Directions}
\label{sec:futurework}

Although massive efforts are devoted to spatio-temporal trajectory similarity/distance measures, many challenges still exist to be addressed. In this section, we provide the following potential future research directions.


\textbf{Temporality.} Trajectory data includes both spatial and temporal information. Nonetheless, most of the existing trajectory similarity measures only consider spatial information, while only TP~\cite{TP} considers both spatial and temporal information. In addition, the time complexity of TP is high, which cannot be efficiently applied to downstream tasks that require to capture time-dependent features (e.g., traffic flow forecasting). Hence, it is of interest to design efficient trajectory similarity measures that consider both spatial and temporal information.

\textbf{Timeliness.} With the wide applications of location services and positioning technologies, a large amount of GPS trajectory data is continuously collected as streaming data, making trajectory similarity computation more challenging. However, existing measures are very costly, failing to meet the real-time requirements of downstream tasks (e.g., ridesharing).  Thus, real-time or online trajectory similarity computation is also a potential research direction.

\textbf{Privacy.} Due to the sensitive of private location information, the privacy protection is required when processing the trajectories. None of existing trajectory similarity measures consider the data privacy, and it is inefficient to directly apply existing measures to support high query efficiency/quality while achieving privacy protection. 
Therefore, it is also of interest to achieve effective and efficient similarity computation while preserving the privacy.


\textbf{Effectiveness.} The goal of trajectory similarity computation is to benefit the downstream trajectory analytics. As various  similarity measures exist, how to select proper measure for different analysis tasks is important. For example, most of the existing similarity measures (e.g., DTW, NetERP, etc.) are designed for trajectory retrieval and clustering, which are ineffective to the tasks such as anomaly detection. Consequently, it is a promising way to combine similarity measure selection with downstream tasks to enhance the effect of trajectory analytics.

\section{Conclusions}
\label{sec:conclusion}

In this paper, we review spatio-temporal trajectory distance measures from three-dimensional perspectives. Then, we offer an evaluation benchmark to examine the effectiveness, robustness, efficiency, and scalability of each measure. According to experimental results, we give objective insights for trajectory measure selection in varying scenarios. Based on the issues to be addressed in the community, we believe effective/interpretable AI-driven similarity analysis as well as distributed similarity analysis in road networks or high-dimensional embedding space, are promising directions.


\bibliographystyle{abbrv}
\bibliography{refer}

\end{document}